\documentclass[
11pt, 
english, 
singlespacing, 
nohyperref, 
headsepline, 
]{MastersDoctoralThesis} 

\usepackage[utf8]{inputenc} 

\usepackage[T1]{fontenc} 

\usepackage{palatino} 

\usepackage{comment} 

\usepackage[autostyle=true]{csquotes} 

\usepackage{empheq}

\usepackage[usenames, dvipsnames]{color}

\usepackage{amsmath,amssymb}
\usepackage{bbm}

\usepackage[LGR,T1]{fontenc}
\newcommand{\textgreek}[1]{\begingroup\fontencoding{LGR}\selectfont#1\endgroup}

\usepackage{dirtytalk}

\newcommand{\dd}{ \mathrm{d}} 

\def\sl2c{{ SL(2,\mathbb{C})}} 
\def\slt{{ $SL(2,\mathbb{C})\ $}} 
\def\cp1t{{ $\mathbb{CP}^1\ $. }} 
\newcommand{\nn}{ {\tt n}} 

\newcommand{\threej}[6]{\begin{pmatrix}  #1&#3&#5\\#2&#4&#6\end{pmatrix}}

\newcommand{\tar}{\,\kern-0.14em{\rightarrow}\!\bullet}
\newcommand{\sou}{\,\kern-0.14em{\leftarrow}\!\bullet}

\def\vol{ {\bf{vol}^4}} 

\def\erf{{\rm erf}} 
\def\tr{{\rm Tr}} 
\def\diag{{\rm diag}} 
\newcommand{\ident}{ \mathbbm{1}} 
\def\re{{\rm Re}} 
\def\im{{\rm Im}} 
\def\sign{{\rm sign}} 
\def\arccosh{{\rm Ach}} 
\def\spn{{\rm span}} 

\def\bra#1{\mathinner{\langle{#1}|}}
\def\ket#1{\mathinner{|{#1}\rangle}}
\def\braket#1{\mathinner{\langle{#1}\rangle}}

\def\braY#1{{}_\gamma \!\mathinner{\langle{#1}|}}
\def\ketY#1{\mathinner{|{#1}\rangle}_\gamma }

\def\braketY#1{{}_\gamma \!\mathinner{\langle{#1}\rangle}_\gamma}

\thesistitle{ Transition de g\'eom\'etrie en Gravit\'e Quantique \`a Boucles Covariante }
\supervisor{\Large Carlo \textsc{Rovelli}} 
\examiner{} 
\degree{} 
\author{\Large Marios \textsc{Christodoulou}} 
\addresses{} 

\subject{} 
\university{Aix-Marseille Universit\'e} 
\department{\Large Centre de Physique Th\'eorique} 
\group{\Large L' \'Equipe de Gravit\'e Quantique,} 
\faculty{Centre National de la Recherche Scientifique} 

\usepackage[colorlinks,citecolor=blue,linkcolor=black,urlcolor=blue]{hyperref}
\usepackage{cite}

\begin{document}

\frontmatter 

\pagestyle{plain} 


\begin{titlepage}
\begin{center}

{\scshape\Large \univname\par}\vspace{0.3cm} 
{\scshape\large \'Ecole Doctorale de la Physique et Sciences de la Matière \par}\vspace{0.6cm} 
{\scshape\Large Centre National de la Recherche Scientifique \par}\vspace{0.3cm} %
{\scshape\large Centre de Physique Th\'eorique \par}\vspace{0.3cm} %
{\scshape\large \'Equipe de Gravit\'e Quantique \par}\vspace{1.2cm} %
\textsc{\LARGE Th\`ese de Doctorat}\\[0.5cm] 

\HRule \\[0.4cm] 
{\huge \bfseries \ttitle\par}\vspace{0.4cm} 
\HRule \\[1.3cm] 
 
\begin{minipage}[t]{0.45\textwidth}
\begin{flushleft} \large
\emph{Auteur:}\\
\authorname 
\end{flushleft}
\end{minipage}
\begin{minipage}[t]{0.4\textwidth}
\begin{flushright} \large
\emph{Directeur de recherche:} \\
\supname 
\end{flushright}
\end{minipage}\\[1.5cm]
 
\large \textit{Thèse présentée pour obtenir le grade \\ universitaire de docteur \degreename}\\[0.2cm] 
\textit{en}\\[0.2cm]
\Large Physique Théorique et Mathématique\\[1cm]
{\large \textit{Soutenue le 23 Octobre, 2017, \\ devant le jury compos\'e de }}\\[0.55cm] 
\begin{minipage}[c]{.5\textwidth}
\begin{tabular}{l l}
\hspace*{-1cm} Rapporteurs : &  Aurelien \textsc{Barrau}  \\
& Karim \textsc{Noui} \\ & \\
\hspace*{-1cm} Examinateurs :
& Aurelien \textsc{Barrau} \\
& Eugenio \textsc{Bianchi} \\
 & Hal \textsc{Haggard}  \\
& Karim \textsc{Noui} \\
& Alejandro \textsc{Perez}
\end{tabular}
\end{minipage}%
 \\[1cm]

\vfill
\end{center}
\end{titlepage}

\newpage\null\thispagestyle{empty}\newpage

\vspace*{0.2\textheight}

\noindent\enquote{\itshape I have one good characteristic: I'm a pessimist, so I always imagine the worst --- always. To me, the future is a black hole.
}\bigbreak

\hfill Krzysztof Kieslowski

\newpage\null\thispagestyle{empty}\newpage

\begin{abstractFrench}
\addchaptertocentry{Resum\'e} 

Dans ce manuscrit, nous présentons un mise en place et calcul d'un observable physique dans le cadre de la Gravité Quantique à Boucles covariante, pour un processus physique mettant en jeu la gravité quantique de façon non-perturbatif. Nous considerons la transition d'une région de trou noir à une région de trou blanc, traitée comme une transition de géométrie  assimilable à un effet de tunnel gravitationnel. L'observable physique est le temps caractéristique dans lequel ce processus se déroule.

 L'accent est mis sur le cas pertinent de géometrie Lorentzienne à quatre dimensions. Nous commençons par une dérivation formelle de haut--en--bas, allant de l'action de Hilbert-Einstein au ansatz qui definit les amplitudes de l'approche covariante de la Gravité Quantique à Boucles. Nous prenons ensuite le chemin de bas--en--haut, aboutissant à l'image d'une integrale de chemin du type somme-de-geometries qui émerge à la limite semi-classique, et discutons son lien étroite avec une integrale de chemin basé sur l'action de Regge.
 
 En suite, nous expliquons comment construire des paquets d'ondes décrivant des géométries spatiales quantiques, plongées dans un espace-temps quantique de signature Lorentzienne, à partir d'une hypersurface continue plongée dans un espace-temps Lorentzienne.

Nous montrons que lors de la mise en œuvre de ces outils, nous avons une estimation simple des amplitudes decrivant des transitions de géometrie de façon probabiliste. 

Nous fournissons une formulation indépendante de choix d'hypersurface de l'espace-temps Haggard-Rovelli, qui modélise l'espace--temps classique entourant la région de transition de géométrie. Nous construisons un mise en place basée sur l'espace-temps HR, où une approche d'intégrale de chemin peut être appliquée naturellement. Nous proposons une interprétation des amplitudes de transition et définissons les observables classiques et quantiques pertinents pour ce processus.

Nous procédons à une derivation d'une expression explicite, analytiquement bien--définie et finie, pour une amplitude de transition décrivant ce processus. Nous utilisons ensuite l'approximation semi-classique pour estimer la 
dépendance, vis-à-vis de la masse, du temps characteristique de la processus, pour une classe d'amplitudes et pour un choix arbitraire des conditions de borne. Nous terminons en discutant l'interprétation physique de nos résultats et de futurs directions de recherche.

\end{abstractFrench}


\begin{abstract}
\addchaptertocentry{\abstractname} 

In this manuscript we present a calculation from covariant Loop Quantum Gravity, of a physical observable in a non-perturbative quantum gravitational physical process. The process regards the transition of a trapped region to an anti--trapped region, treated as a quantum geometry transition akin to gravitational tunneling. The physical observable is the characteristic timescale in which the process takes place.

 Focus is given to the physically relevant four--dimensional Lorentzian case. We start with a top--to--bottom formal derivation of the ansatz defining the amplitudes for covariant Loop Quantum Gravity, starting from the Hilbert-Einstein action. We then take the bottom--to--top path, starting from the Engle-Perreira-Rovelli-Livine ansatz, to the sum--over--geometries path integral emerging in the semi-classical limit, and discuss its close relation to the naive path integral over the Regge action. We proceed to the construction of wave--packets describing quantum spacelike three-geometries that include a notion of embedding, starting from a continuous hypersurface embedded in a Lorentzian spacetime. 

We derive a simple estimation for physical transition amplitudes describing geometry transition and show that a probabilistic description for such phenomena emerges, with the probability of the phenomena to take place being in general non-vanishing.

The Haggard-Rovelli (HR) spacetime, modelling the spacetime surrounding the geometry transition region for a black to white hole process, is presented and discussed. We give the HR-metric in a form that emphasizes the role of the bounce time as a spacetime parameter and we give an alternative path for its construction. We define the classical and quantum observables relevant to the process, propose an interpretation for the transition amplitudes and formulate the problem such that a path-integral over quantum geometries procedure can be naturally applied. 

We proceed to derive an explicit, analytically well-defined and finite expression for a transition amplitude describing this process. We then use the semi--classical approximation to estimate the amplitudes describing the process for an arbitrary choice of boundary conditions. We conclude that the process is predicted to be allowed by LQG, with a crossing time that is linear in the mass. The probability for the process to take place is suppressed but non-zero. We close by discussing the physical interpretation of our results and further directions.  

\end{abstract}


\begin{acknowledgements}
\addchaptertocentry{\acknowledgementname} 

I gratefully acknowledge the hospitality and support through my PhD, from the University of Aix-Marseille, the École Doctorale de la Physique et de Sciences de la Matière, the Centre National de la Recherche Scientifique, the Centre de Physique Theorique, the Leventis A.G. Foundation. 

I especially thank Samy Maroun and the Samy Maroun center for Space, Time and the Quantum for invaluable support.

\medskip

I would like to thank the professors that welcomed me to their research groups and institutes in visits undertaken during my PhD, from which I have gained many insightful comments and received constructive criticism: Abhay Ashtekar and Eugenio Bianchi at the Pennsylvania State University, Jorge Pullin, Ivan Agullo and Parampreet Singh at the Louisiana State University, Jonathan Engle, Muxin Han and Chris Miller at the Florida Atlantic University, Daniele Oriti at the Albert Einstein Institute.

I acknowledge contribution in this work and thank for the many stimulating discussions during these years, from: Simone Speziale, Alejandro Perez, Hal Haggard, Eugenio Bianchi, Jonathan Engle, Abhay Ashtekar, Francesca Vidotto,  Aurelien Barrau, Daniele Sudarsky, Robert Oeckl, Muxin Han, Thomas Krajewski and Marco Vojinovi\'c.

I thank Louis Garay and Raul Carballo for their hospitality and discussions during a recent visit in Universidad Complutense Madrid. Their open-mindedness and insights helped clarify the physical interpretation of the results in this manuscript.

\medskip

I want to thank Beatrice Bonga and acknowledge insightful discussions during my visit in LSU, when the ideas leading to the reformulation of the Haggard-Rovelli spacetime as presented in this manuscript were first formed.

I thank my colleagues Tommaso de Lorenzo, Boris Bolliet, Thibaut Josset and Ilya Vilensky, for all the fun and the discussions we had in our parallel journey through our theses, and for being supportive friends. 

I especially thank Fabio d'Ambrosio for bringing invaluable momentum to this project. Crucial results presented in this manuscript reflect our close and continued collaboration.

\medskip

Lina, thank you, for everything. 

\medskip

I want to thank my family, my brothers, Vasilis and Dimitris, and my parents, Koula and Takis, for all the support they have given me through these years and without which I would have been unable to complete this project. 

To my supervisor, Carlo Rovelli, I express my gratitude for making this journey possible, and for being my teacher, collaborator and friend.  

\end{acknowledgements}


\tableofcontents 

\listoffigures 

\listoftables 


\dedicatory{\textgreek{Sthn L'ina.} } 

\newpage\null\thispagestyle{empty}\newpage

\chapter*{Resum\'e de th\`ese}
\begin{figure}
\centering
\includegraphics[scale=0.45]{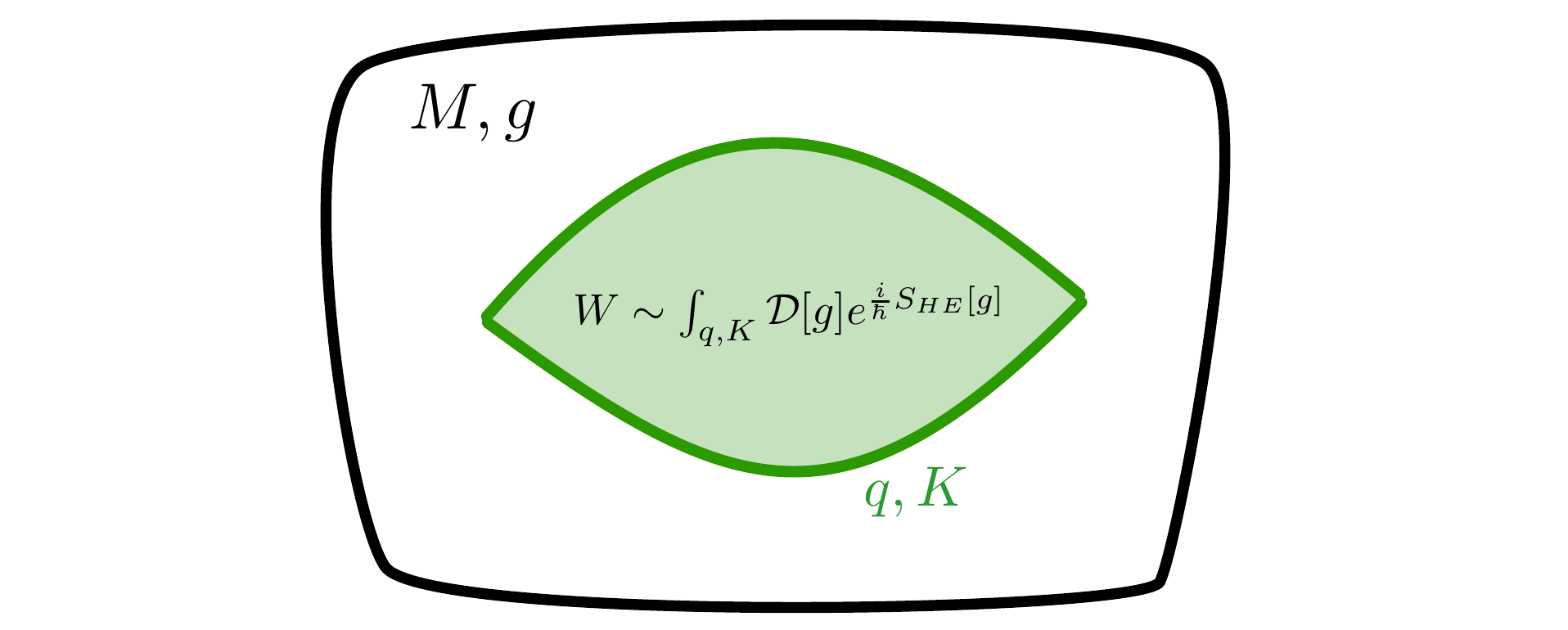} 
\caption[]{Transition de géometrie schematiquement. L'integrale de chemin sur geometries du type Wheeler-Misner-Hawking et \emph{emergeant} dans l'approche covariante de la Gravité Quantique à Boucles. La théorie est defini en l'absence d'un notion d'espace-tempd ou de fond fixe (``background-free'').} 
\label{resume:geomTransition}
\end{figure}

\addchaptertocentry{Resum\'e de th\`ese}
Dans ce manuscrit, nous présentons une première application du formalisme de la Gravité quantique de boucles covariante, à un processus gravitationnel quantique non-perturbative, traitée comme une transition de géométrie, voir Figure \ref{resume:geomTransition}. Nous étudierons la transition d'une région de trou noir (trapped region) à une région de trou blanc (anti--trapped region) et définirons et estimerons un observable physique: l'échelle de temps caractéristique, vue depuis l'infini, dans lequel ce processus se déroule.

\medskip

Nous commençons dans Chapitre \ref{ch:topBottom}, par une dérivation formelle de haut--en--bas, à partir de la Relativité Générale, de l'ansatz définissant la fonction de partition de la Gravité Quantique à Boucles covariante.  Nous nous focalisons sur le cas Lorentzienne à quatre dimensions physiquement pertinent, et nous expliquons les motivations qui conduisent au modèle EPRL Lorentzienne, à partir de l'action de Hilbert-Einstein. 

Dans la Section \ref{sec:dictionary}, nous présentons un dictionnaire, Table \ref{tab:aimEPRL}, qui souligne l'objectif du Chapitre \ref{ch:topBottom}. Ainsi, nous esquissons un chemin  à partir de la phrase  
\begin{eqnarray} \label{resume:WMHsumovergeom}
W \sim \int \mathcal{D} [g] \; e^{\frac{i}{\hbar} S_{HE}[g]  }
\end{eqnarray}
dont $S_{HE}$ est l'action de Hilbert-Einstein 
\begin{equation} \label{resume:HEaction}
S_{HE}[g] = \frac{1}{16 \pi G} \int \dd x^4 \sqrt{\det g(x)} R(g(x)) 
\end{equation}
et $g(x)$ est la métrique de l'espace-temps, à l'ansatz qui définit le modèle EPRL, et qui est le point liminaire du Chapitre \ref{ch:AsympAnalysis},

\begin{eqnarray} 
W_\mathcal{C} = \sum_{j_f} \mu(j_f) \; \int_{\sl2c}  \! \mu(g_{ve}) \; \prod_f A_f(g_{ve},j_f) \label{resume:EPRLampWMH} \\
A_f(g_{ve},j_f) = \int_{SU(2)} \! \mu(h_{vf}) \; \delta{(h_f)} \tr^{j_f} \left[\prod_{v \in f} Y_\gamma^\dagger g_{e'v}g_{ve} Y_\gamma  h_{vf} \right] \label{resume:faceAmpWMH}.
\end{eqnarray}
L'amplitude $W_\mathcal{C}$ est comprise ici comme une fonction de partition, qui se distingue des amplitudes de transition dont on introduira dans le Chapitre \ref{ch:gravTunneling} et qui comprendront une dépendance aux variables de borne.

Dans la Section \ref{sec:gameOfActions}, nous commençons par l'action de Hilbert-Einstein et nous introduisons le formalisme des repères mobiles. Nous soulignons la disparition des coordonnées dans ce formalisme. Nous dérivons la forme Palatini et BF de l'action Hilbert Einstein et passons à l'action de Palatini, où la connexion est considérée comme une variable indépendante. Nous présentons l'action de Holst et la formulation de la Relativité Générale comme une théorie topologique BF contrainte. Nous discutons les contraintes de simplicité qui réduisent l'action de la théorie BF à la Relativité Générale. Ce voyage est résumée dans le Tableau \ref{tab:actionsTable}. L'action de Holst est le point de départ pour le programme de quantization de spinfoam: elle fournit une manière de définir l' integrale de chemin en comparaison avec les théories topologiques BF, et constitue un pont formel vers la théorie canonique et l'espace de Hilbert cinématique. Ce dernière relation fera l'objet de la Section \ref{sec:relationToCanonical}.

Dans la Section \ref{sec:spinStateSumFormEprl}
 nous commencons par introduisant l'holonomie, le transporteur parallele, qui representera la version discrete de la courbure de la connection. Par la suite, nous décrivons comment l'integrale de chemin de la théorie BF peut être definit par une régularization effectué sur une structure topologique combinatoire, le ``2-complex''. Nous expliquons la stratégie et les principales motivations de l'ansatz definissant le modèle EPRL: postuler que l' intégrale de chemin pour la gravité peut être réalisée en modifiant la quantization tentative d'une théorie topologique, en imposant les contraintes de simplicité au niveau quantique. Nous verrons comment en-suite deduire la forme ``spin state-sum'' des amplitudes EPRL qui definit le modèle.

\medskip

 Nous prenons ensuite dans Chapitre \ref{ch:AsympAnalysis} le chemin du bas vers le haut. Le point de depart sera l'EPRL ansatz et nous verrons comment la somme--sur--géométries émerge dans la limite semi-classique, et discutons sa relation étroite avec l'intégrale de chemin naïf basé sur l'action  de Regge. Une image approximative de la façon dont l'intégrale du chemin du type Wheeler-Misner-Hawking émerge de modèles de ``spinfoams'' est donnée dans la section \ref{sec:emergenceOfGR}, pour servir de guide pour la présentation du matériel des sections suivantes. Nous introduisons la quantité minimale d'outils nécessaires afin d' amener les amplitudes EPRL à la forme appropriée pour l'analyse asymptotique dans la Section \ref{sec:prelim}.

En suite, nous présentons notre dérivation de la représentation Krajewski-Han. Nous fournissons une dérivation indépendante de la forme des amplitudes EPRL adaptées à l'analyse asymptotique dans la Section \ref{sec:ampsInAsAnform}, dans la représentation Krajewski-Han. Ceci est une manière peu connue et plus directe d'arriver à ce point, qui ne consiste pas à insérer des résolutions de l'identité sur des états cohérents $ SU (2) $. Le``face-amplitude'' prends la forme
\begin{equation} 
A_f(j_f,\{g_f\}) = \int \! \mu(g_{ve}) \int \! \mu(z) \; e^{j_f F_f\left(\{g_f\},\{z_f\}\right)}
\end{equation} 
Par la suite, nous examinons la définition des angles Lorentziennes et l'action correspondante de Regge dans la section \ref{sec:lorReggeAction}.

Dans la Section \ref{sec:overviewAsAnEPRL} nous examinons les résultats de l'analyse asymptotique du modèle EPRL, qui sera utilisé dans le Chapitre \ref{ch:gravTunneling} pour montrer comment estimer les amplitudes décrivant des transitions de géometrie. Ces outils seront finalement mis en œuvre dans la Section \ref{sec:lifetimeEstim}, dans le cadre d'une transition de trou noir à trou blanc. Dans ce chapitre nous aurons vu que la fonction de partition du modèle EPRL se comport comme l'integrale de chemin naïf de Regge, avec la difference que les variables de base seront les aires au lieu des longeurs. En sommaire, 

\begin{eqnarray}
W_{WMH} &\sim& \int\! \mathcal{D} [g] \ \ \ \ e^{\frac{i}{\hbar} S_{HE}[g]  } \nonumber \\ \nonumber \\
W_{NR} &\sim& \int\! \mu(\ell_s) \ \ \ \ e^{\frac{i}{\hbar} S^\mathcal{C^\star}_{R}[\ell_s]  } \nonumber \\ \nonumber \\
W^{EPRL}_\mathcal{C}  &\sim & \int \mu(a_f) \ \ e^{\frac{i}{\hbar} S^\mathcal{C^\star}_R[a_f]} \nonumber
\end{eqnarray}

\bigskip

Dans le Chapitre \ref{ch:gravTunneling} nous entrons dans la partie principale de ce manuscrit. D'abord, nous abordons les états cohérents de réseau de spins, qui serviront comme des états de borne décrivant les paquets d'ondes d'une géométrie trois-dimensionnelle plongée dans un espace-temps. Nous allons combiner ces états avec une amplitude de ``spinfoam'' pour définir une amplitude de transition physique. Nous utiliserons les résultats de l'analyse asymptotique pour des spins fixes présentés dans le chapitre précédent et expliquerons comment le sommes de spins, qui represente l'intégrale de chemin au limite semi--classique, peut être effectué.

Les états cohérents de réseau de spins sont introduits dans la Section \ref{sec:wavepacketOfGeometry}. Nous donnons d'abord un aperçu de l'espace cinématique de Hilbert de GQB au niveau d'un graphe fixe. Nous présentons ensuite des outils utiles pour la compréhension géométrique des différentes constructions qui suivreront, ainsi que pour la réalisation de calculs. Par la suite, nous expliquons comment une superposition de états de reseau de spins (``spin-networks'') résulte à un paquet d'ondes décrivant une géométrie trois-dimensionnelle dans la Section \ref{sec:intrinsicCoherentStates}. Ces états décrivent la géometrie intrinsèque d'une triangulation de l'espace.

Nous présentons par la suite, 
dans la Section \ref{sec:boundaryState}, comment une deuxième superposition $\Psi_{\Gamma}$ de ces états intrinsèques, peut être interpretée comme un état representant une géometrie discrète trois-dimensionelle semi--classique, intégrée dans un espace-temps de signature Lorentzien.  Ces états servireront des états de borne pour une amplitude de transition physique
\begin{equation}
\Psi^{t}_\Gamma(h_\ell;\eta_\ell,\zeta_\ell,k_{\ell\nn}) = \sum_{\{j_\ell\}} \prod_\ell d_{j_\ell} e^{-(j_\ell-\omega_\ell\!(\eta_\ell,t)\,)^2 t + i \zeta_\ell\!\, j_\ell} \;  \psi (h_\ell,j_\ell; k_{\ell\nn}) 
\end{equation}
\begin{figure}
\centering
\includegraphics[scale=0.45]{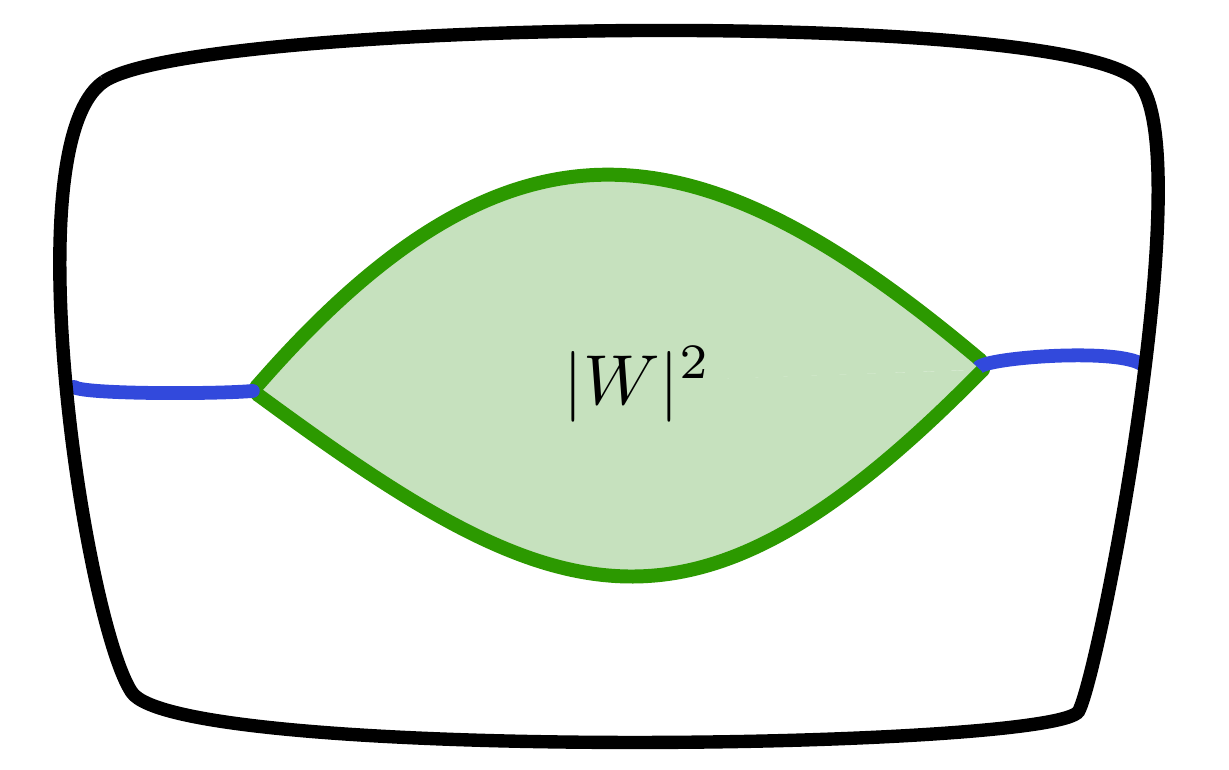}
\caption[] {Les amplitudes de GQB covariante comme probabilités pour le des transition de géometrie. Résoudre un problème de valeur initiale pour les équations d'Einstein, avec les données de Cauchy étant la géométrie intrinsèque et extrinsèque de l'hypersurface formée par les hypersurfaces bleues et la surface limite supérieure, et évoluant vers la direction dans laquelle le temps de foliation augmente, resultera à la moitié supérieure de la espace-temps. Ceci est relié à la moitié inférieure, qui est l'évolution dans le passé, avec les données de Cauchy étant la géométrie intrinsèque et extrinsèque de l'hypersurface formée par les hypersurfaces bleues et la surface limite inférieure. 
}
\label{resume:conectingGRsolutions}
\end{figure}
Par la suite, dans la Section \ref{sec:gravTunneling}, nous introduisons des amplitudes des spinfoam definies sur un ``2-complex'' avec borne. Combiné avec un état de borne, ceci definie un amplitude de transition physique.

Le spin-sum est en suite effectué. Nous montrons qu'en général, les amplitudes décrivant de transitions entre des géometries macroscopiques, admettent une estimation simple. Ils decroissent exponentiellement dans la limite semi-classique $\hbar\rightarrow 0$. Le facteur de supression exponentielle est le désadaptement entre la géométrie extrinsèque discrète $\zeta$ de l'état de borne, et la géométrie extrinsèque $\phi(\omega;\vec{k})$ de la configuration geometrique qui domine l'integrale de chemin. Ce derniere est calculé en fonction de donnés d'aire $\omega$ et diretionnelle trois-dimensionelle $k_{\ell\nn}$ de la géometrie discrète spatiale. Nous avons l'estimation suivante pour les amplitudes de transition physiques; un produit de termes du type 
\begin{equation} 
W(\omega,\zeta,\vec{k},t) \sim \prod_f e^{- ( \zeta -\phi(\omega_f;\vec{k} )\,)^2 /{4t}} e^{ i \omega ( \zeta_f -\phi_f(\omega;\vec{k})\,) } \nonumber
\end{equation}
Ainsi, nous deduirons que la Gravité Quantique à Boucles designe une proabilité finit à des transition des géometries, voir Figure \ref{resume:conectingGRsolutions}.

\medskip

Dans Chapitre \ref{ch:exteriorSpacetime} nous présenterons l'espace-temps extérieur, que nous nommerons l'espace-temps de Haggard-Rovelli (HR). Ceci fournit un modèle de  l'espace-temps entourant la région de la transition. Nous fournissons une formulation simple et complète de cet espace temps et discutons ses proprietes.

\begin{figure}
\centering
   \includegraphics[scale=0.6]{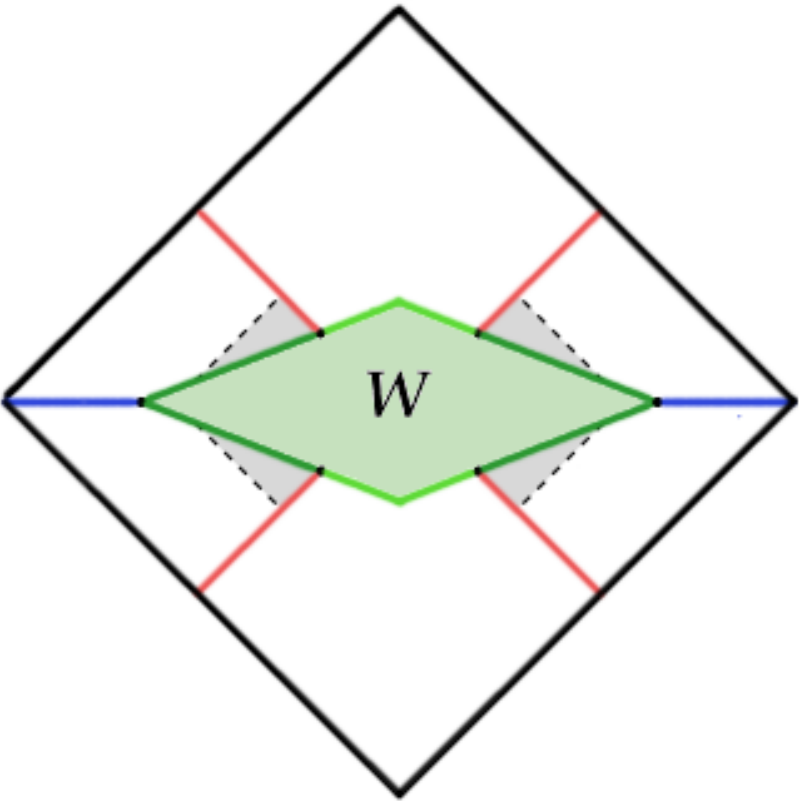}
\caption[]{
L'espace-temps de Haggard-Rovelli comme un exemplaire prototype d'une configuration bien définie pour étudier une transition de géométrie.} 
\label{resume:fireworksComparisonPic}
\end{figure}

\begin{figure}[h]
\centering
   \includegraphics[scale=0.8]{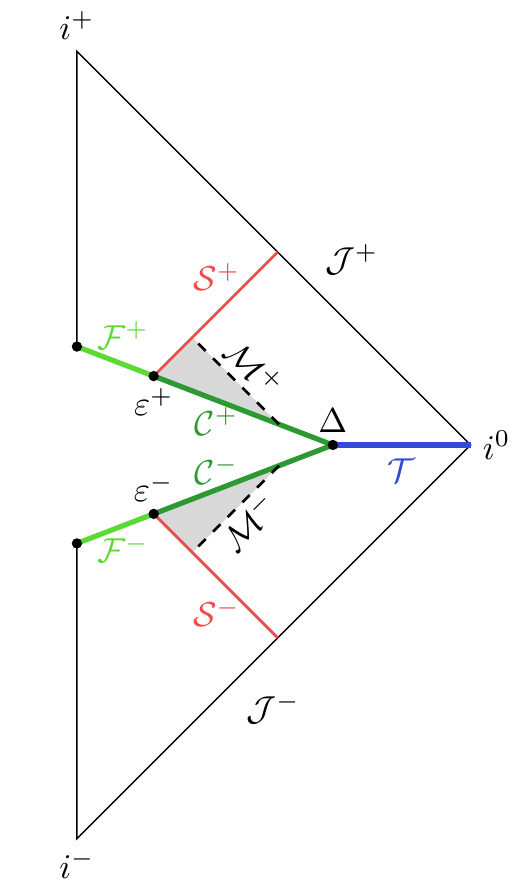} \caption[]{L'espace--temps de Haggard-Rovelli. } 
\label{resume:fireworksPic}
\end{figure}

Nous discutons d'abord dans la Section  \ref{sec:blackholesAge} le contexte relatif à l'idée que les trous noirs ne sont pas des objets éternels et au manque de consensus sur la façon dont un trou noir termine sa vie. Nous examinons brièvement les idées récentes et des résultats sur le rôle du volume intérieur dans les trous noirs. L'espace-temps de Haggard-Rovelli, représenté ici dans la figure \ref{resume:fireworksPic}, est ensuite construit dans la section \ref{sec:RH}.

Le phénomène physique modélisé par un espace-temps HR est \emph {la transition, par des effets gravitationnels quantiques qui ne sont pas négligeables que dans une région compacte de l'espace-temps, d'une région de trou noir formée par effondrement, dans une région de trou blanc, de laquelle la matière est en suite diffusée}. Une région compacte est excisée de l'espace-temps, en introduisant un borne interieure de genre espace. En dehors de cette région qui isole des effets non--classiques, l'image spatio-temporelle habituelle s'applique et la métrique résout les équations de champs d'Einstein.

Le temps de rebond $ T $ est discuté dans la section \ref{sec:BounceTime}, où nous expliquons sa signification comme une échelle de temps qui caractérise la géométrie de l'espace-temps HR. Nous montrons que le temps de rebond peut être compris comme un concept équivalent au temps d'évaporation Hawking cet espace-temps.

Dans la Section section \ref{sec:Lifetime}
nous proposons une interprétation probabiliste des amplitudes de transition attribuées par le modèle EPRL. La relation de notre construction de l'espace--temps HR avec la construction originale, est discutée dans la section \ref{sec:crossedFingers}. La construction que nous donnons évite l'utilisation d'hypersurfaces spécifiques et contribue à révéler les idées de la construction originale.

\medskip

Nous procedons dans le Chapitre \ref{ch:calculationOfAnObservable} à appliquer tout ce qui preced pour estimer l'amplitude de transition decrivant ce processus. Nous commençons par une une discussion de la procédure de discrétisation. Nous prouvons une relation intrigante entre la courbure extrinsèque $K$ d'une surface et la rapidité $\zeta$, reliant le vecteur normal à l'hypersurface à son transport parallèle le long d'une courbe $ \Upsilon $ 
\begin{equation}
\zeta = \int_{\Upsilon} \; K
\end{equation}  
Ce résultat sert de justification de la procédure de discrétisation présentée ici et est intéressant à part entière. Par la suite, nous proposons une procédure de discrétisation afin de construire un état de borne semi--classique à partir d'une hypersurface continue intégrée dans un espace-temps lorentzien.

\begin{figure}[h]
\centering
\begin{tabular}{rcl}
  \includegraphics[scale=0.12]{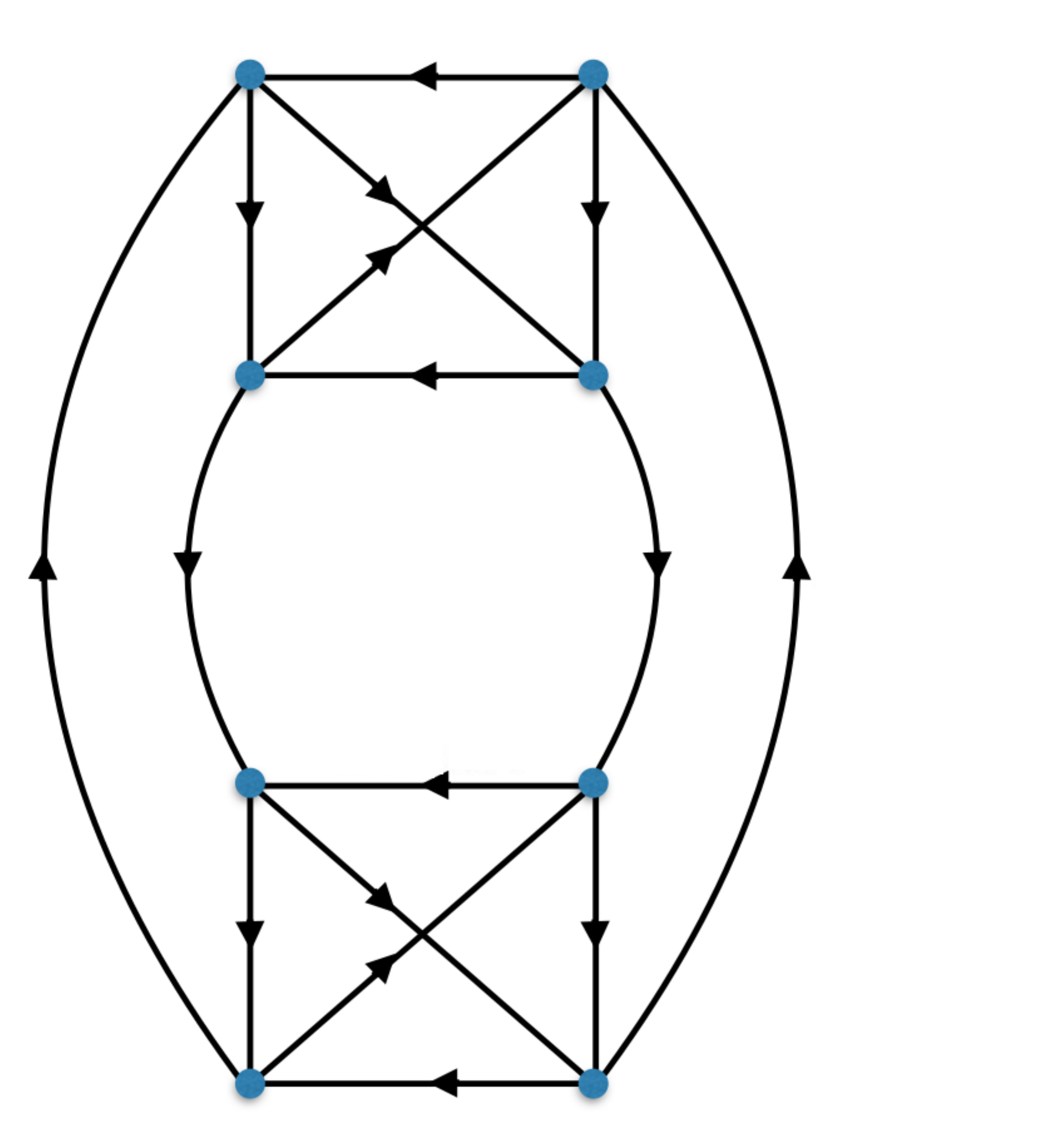} &   \includegraphics[scale=0.15]{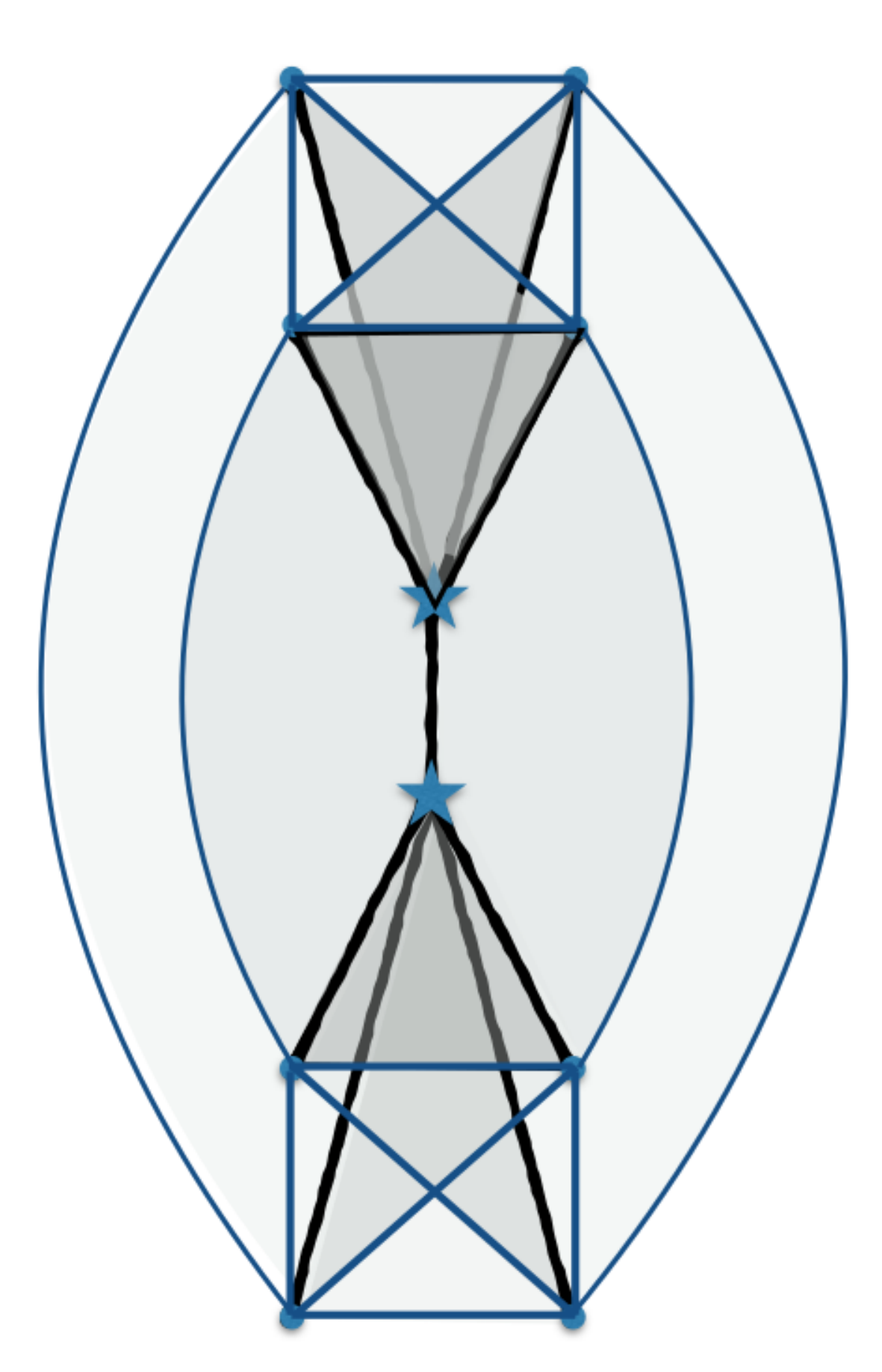} & \includegraphics[scale=0.090]{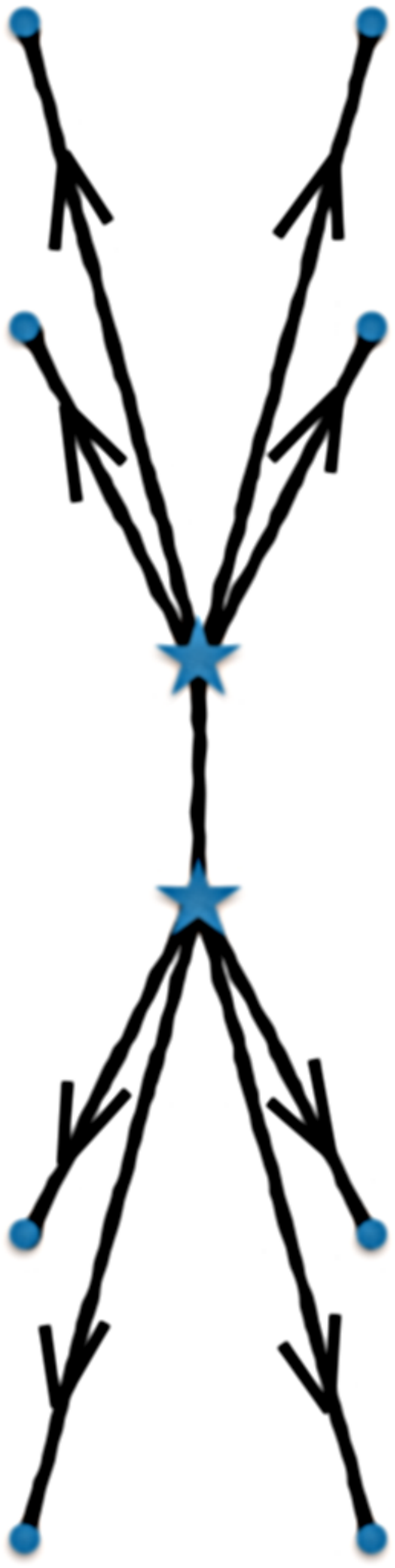} \\
\end{tabular}
\caption[]{``Spinfoam 2-complex'' (milieu) dont l'amplitude $W_\mathcal{C}$ soit défini, avec son graph de borne $\Gamma=\partial \mathcal{C}$ (gauche) et son ``spinfoam skeleton'' (droite).}
\label{resume:spinfoamPic} 
\end{figure}

Dans la Section \ref{sec:explicitAmp} nous donnons une forme explicite de l'amplitude de transition définissant la durée de vie pour un choix de surface de borne explicite:
\begin{eqnarray}
\hspace*{-1cm} W(m,T) \hspace*{-0.2cm}&=&\hspace*{-0.4cm}\sum_{\{ j_a,j^\pm_{ab} \}}\hspace*{-0.2cm} w(z_0,z_\pm,j_a,j^\pm_{ab})
\times \hspace*{-0.4cm} \sum_{\{J_a^\pm,{K}_a^\pm ,l_a ,l_{ab}^\pm\}} \left( \bigotimes _{a,\pm}N^{J_a^\pm}_{\{j_a^\pm\}}(\nu_{\ell \in a^\pm}) \; f^{J_a^\pm,K_a^\pm}_{\{j_a^\pm\}\{l_a^\pm\}}
\right) \nonumber \\
&{}&\left(  \bigotimes_{a,\pm} i^{K_a^\pm,\{l_a^\pm\}}\right)_\Gamma.
\end{eqnarray}
Le ``2-complex'' dont l'amplitude de transition $ W_\mathcal{C}(m, T) $ soit définie, est représentée dans la Figure \ref{resume:spinfoamPic}. L'amplitude de transition physique est fini,
définie explicitement en termes de fonctions analytiques connues. Par la suite, nous donnons une estimation simple de ce type d'amplitude en utilisant les resultats de chapitres preécedants.

On arrive au résultat
\begin{equation} 
\tau(m) = m \frac{ 
\int \dd x \; x \,  F(x) \prod_\ell e^{-(\gamma \zeta_\ell(x) - \gamma \phi_\ell(x) + \Pi_\ell)^2/4 t} 
}
{
\int \dd x \; F(x) \prod_\ell e^{-(\gamma \zeta_\ell(x) - \gamma \phi_\ell(x)^2 + \Pi_\ell)^2/4 t} 
}
\end{equation}
Ensuite, à partir de cet stimation, nous deduirons que le temps characteristique $\tau(m)$ de la processus est de l'ordre
\begin{equation}
\tau(m) = \frac{m}{\gamma}
\end{equation}
avec de corrections sous-dominants qui contients $\hbar$. Nous concluons que la transitions aura lieu, sans que les effets quantiques qui sont sous-dominants ne l'empeche. La probabilité de ce phénomène est supprimé de façon exponentielle. Par la suite, nous discutons ces résultats et futurs directions de récherce.

\newpage\null\thispagestyle{empty}\newpage


\mainmatter 

\pagestyle{thesis} 


\chapter[Introduction]{Introduction} \label{ch:Intro}

\begin{figure} 
\centering
\includegraphics[scale=0.6]{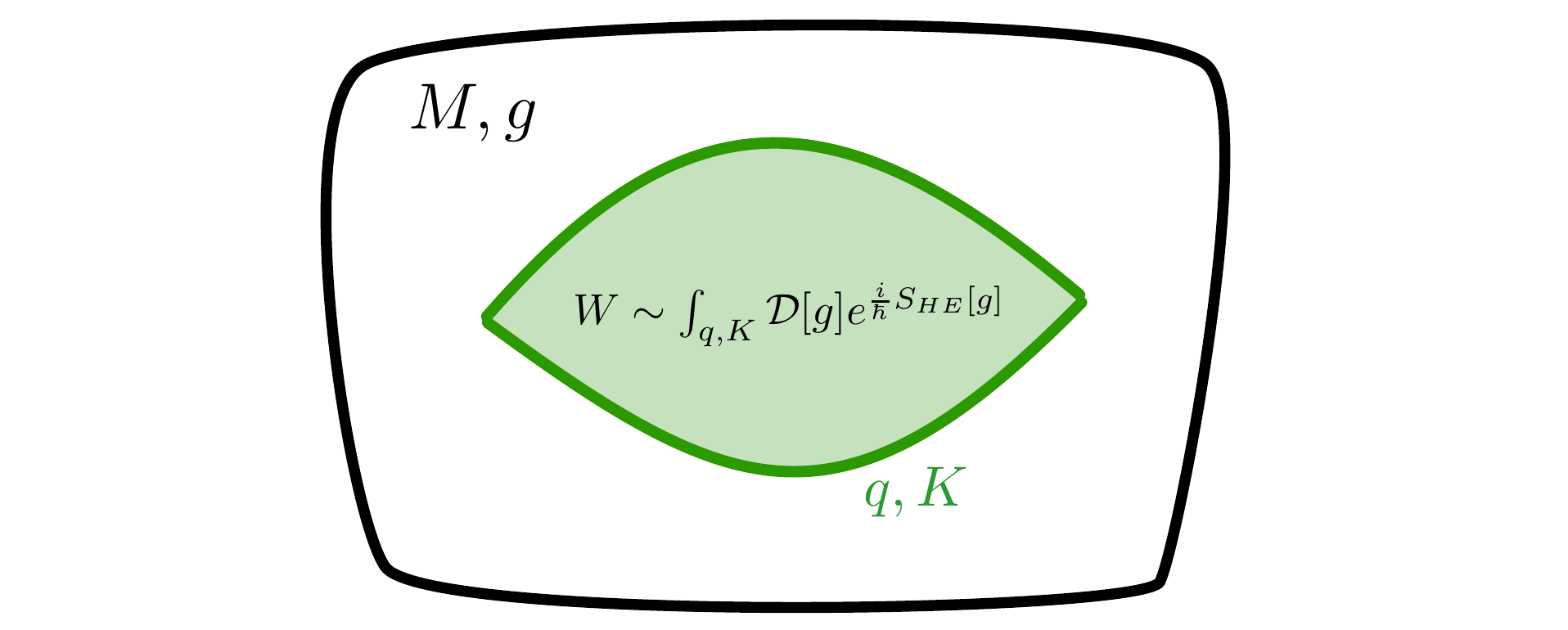} 
\caption[Geometry transition viewed as a path-integral over geometries.]{Geometry transition viewed as a path-integral over geometries. The boundary surface (dark green) separates the parts of the system treated as classical and quantum. The exterior spacetime is classical with a metric $g$ solving Einstein's field equations. A path-integral $W$ is performed in the interior with the metric fixed to the intrinsic metric $q$ and extrinsic curvature $K$ of the boundary surface. We emphasize that the interpretation of the amplitude as a path-integral over geometries is \emph{emergent} in covariant LQG, in the semiclassical limit of large quantum numbers. The theory is defined in the absence of any notion of classical metric, or indeed, spacetime. }
\label{fig:lens}
\end{figure}

In this manuscript, we present an  application of the formalism of covariant Loop Quantum Gravity in a non-perturbative quantum gravitational process understood as geometry transition.

We will show that gravitational tunnelling emerges naturally in covariant LQG. We will use these results to study the transition of a black to a white hole geometry and define and estimate a physical observable: the characteristic physical time--scale in which the process takes place. 

\bigskip

Quantum Gravity refers to an open problem: the quest for a physical theory that describes the phenomenon of gravity and incorporates the notions of time and space, in a way compatible with the 20th century and contemporary understanding, that all natural phenomena obey the principles of Quantum Mechanics. 

We expect a theory of Quantum Gravity to be compatible with previously established physical theories, and to be able to provide definite predictions for phenomena that do not admit a description by its predecessors. 

The theory and the mathematical objects defining it must be accompanied by an interpretation, a correspondence to measurements, so that it may be used unambiguously. 

\bigskip

 Research in the past century has overwhelmingly focused on laying the foundations for Quantum Gravity. This was partly due to the lack of experimental guidance and partly due to the formidable nature of the technical and conceptual issues that had to be overcome in order to achieve meaningful progress.

Loop Quantum Gravity refers to the body of results, and the corresponding community of researchers, from the research program for the canonical quantization of General Relativity, based on the parametrization of the phase space of General Relativity by the Ashtekar \cite{ashtekar_new_1986} and subsequently the Ashtekar-Barbero \cite{barbero_g._real_1995} variables. 

General Relativity formulated in terms of these variables resembles, at the kinematical level, an $SU(2)$ Yang-Mils theory with an $SU(2)$ gauge field as configuration variable, and a densitzied triad field playing the role of the conjugate momentum, the ``electric field''. 

The Hamiltonian constraint, encoding the dynamics, was much simplified in this reformulation, circumventing technical difficulties of Wheeler's geometrodynamics \cite{wheeler_nature_1957}. This led to the loop representation of quantum general relativity \cite{rovelli_loop_1990,rovelli_knot_1988}, introduced by C. Rovelli and L. Smolin, and subsequently to Loop Quantum Gravity.

The research program of Loop Quantum Gravity attempts and has set as its ultimate goal to complete Dirac's program for the canonical quantization of General Relativity. The recipe, or algorithm, to be defined and completed, was laid out in Dirac's influential lectures on constrained Hamiltonian systems \cite{p.a.m._dirac_lectures_1964}. 

Dirac believed it is of utmost importance to found an approach to Quantum Gravity on the Hamiltonian or canonical formulation, which, if completed successfully, guarantees that we have at least a first approximation of the correct theory. In his words, on approaches that are not based on a canonical quantization program
\begin{quote}
\say{I feel that there will always be something missing from them which we can only get by working from a Hamiltonian, or maybe some generalization of the concept of a Hamiltonian. So, I take the point of view that the Hamiltonian is really very important for quantum theory.}
\end{quote}

In this manuscript, we will not be dealing directly with the Hamiltonian for General Relativity \cite{thiemann_phoenix_2006,thiemann_anomaly_1996,thiemann_loop_2007}. We will present and use the covariant approach to Loop Quantum Gravity. 

\bigskip

The covariant approach to Loop Quantum Gravity, or, spinfoam quantization program, is interlinked, inspired and based on the results of its parent theory, Loop Quantum Gravity. The Hamiltonian will in this sense be implicit. 

The program of covariant LQG can be placed historically as an attempt to realize the Feynman Quantization of General Relativity within the context of LQG. This avenue for a quantum treatment of gravity was beautifully outlined in Misner's eponymous article \cite{misner_feynman_1957}. 

Misner emphasized the merits of working directly with an action principle that has the relevant symmetries. A covariant formulation provides a way to circumvent, at least temporarily, the difficulties inherent in a rigorous Hamiltonian treatment.

The spirit of Wheeler's geometrodynamics \cite{wheeler_nature_1957}, and of Hawking's subsequent attempts \cite{hawking_path_1980} to define the path--integral for gravity as a Wheeler--Misner--Hawking (WMH) sum--over--geometries is sketched in Figure \ref{fig:lens}

\bigskip

The wider LQG community shares a common set of convictions that translate into basic principles on which the theory is built. 

At the heart of the approach, is the insistence on background independence. This is also the central departure from a Yang-Mills field theory, since it implies that the dynamics are coded in a Hamiltonian constraint, a vanishing Hamiltonian, with the physical states being those in its kernel. 

Backround independence can be stated as demanding the following: no physical entity can be modelled in a theory of quantum gravity by a mathematical object that is fixed à priori, does not interact with other physical entities and is in this sense non-dynamical. The quantum theory must be background--free, as is the classical theory, General Relativity. 

This is the ontology of physical reality according to LQG; that is to say, of the tangible physical entities that may leave a mark in experimental apparatuses. Rovelli's way of putting this world--view is to say: ``covariant quantum fields on top of covariant quantum fields'' \cite{rovelli_quantum_2004}. 

The ``covariant'' emphasizes another principle taken as fundamental in LQG: Lorentz invariance is accepted as a fundamental symmetry of nature, not to be broken even at the Planck scale. The description of the covariant quantum gravitational field is four--dimensional.  

LQG follows in the footsteps of Einstein's conceptual revolution that geometry is key for gravity, in the sense that can be summarized as

\begin{quote}
Gravity is Geometry; Quantum Gravity is Quantum Geometry.
\end{quote}

In turn, as emphasized by Wheeler \cite{wheeler_nature_1957}, we do not expect quantum geometry to resemble a smooth classical geometry. The latter should be an appropriate semi-classical limit of the former. The right intuition for the deep quantum gravitational regime may then be one of quantum fluctuations of the quantum field to which the notions of time and space are ultimately attributed, a ``foamy spacetime''.

\bigskip

Spinfoams are a fusion of ideas from topological quantum field theories and covariant lattice quantization, the quantization of geometrical shapes \cite{haggard_pentahedral_2013,
barbieri_quantum_1998,baez_quantum_1999,bianchi_polyhedra_2011} and the canonical quantization program of LQG.  

The transition amplitudes of covariant LQG provide a definition for the regularized path integral over histories of the quantum spatial geometries predicted by LQG to be the states of the quantum gravitational field. The Lorentzian spinfoam model consists of an ansatz for the definition of the regularized partition function, with the regularization effectuated by a skeletonization on a 2-complex $\mathcal{C}$, a sort of topological 2-dimensional graph. 
 
The 2-complex $\mathcal{C}$ serves as a combinatorial account--keeping device, arranging a sense of adjacency for a finite subset, a truncation, of the degrees of freedom of the quantum gravitational field. The amplitudes $W_\mathcal{C}$ of covariant LQG, are defined by a state-sum spin model. They are partition functions over spin configurations, colouring the faces of a 2-complex and its boundary graph. These quantum numbers label irreducible unitary representations of the Lorentz group, and recoupling invariants intertwining between them. They are interpreted as the degrees of freedom of the quantum gravitational field. 

State-sum spin models have been studied in a variety of simplified settings, topological, Euclidean, in three and four (or more dimensions), with compact and q-deformed groups and in the context of tensor models. These toy--models served as a testing--ground and a stepping stone until a state-sum spin model relevant for gravity became available. This pre-history of the spinfoam program can be summarized as a progression through the Ponzano-Regge-Tuarev-Viro-Ooguri-Boulatov models \cite{ooguri_partition_1992,rovelli_basis_1993,turaev_state_1992,barrett_ponzano-regge_2009,ponzano_semiclassical_1968,boulatov_model_1992}. 

Efforts over the past two decades within the context of LQG culminated in what has become known as the EPRL model \cite{engle_flipped_2008, engle_spin-foam_2013-2,baratin_group_2012,
dupuis_holomorphic_2012,
freidel_new_2008-1,barrett_lorentzian_2000}. Several variations of this model have been proposed and a version treating the physically relevant Lorentzian case is currently available. A naming attributing credit to the main developments leading to its present form could have been the Barret--Crane--Freidel--Krasnov--Speziale--Livine--Perreira--Engle--Rovelli--\ldots \ model.

Although many fundamental issues remain open, the spinfoam quantization program has seen significant advances over the past decade. 
 
The EPRL model has had some important successes, possessing a well-studied semi--classical limit closely related to General Relativity
\cite{han_einstein_2017,engle_lorentzian_2016,
han_path_2013,han_semiclassical_2013,
han_asymptotics_2013,engle_spin-foam_2013-2,han_asymptotics_2012,perini_einstein-regge_2012,magliaro_curvature_2011,magliaro_emergence_2011,
barrett_lorentzian_2010,barrett_quantum_2010,barrett_asymptotic_2009,
bianchi_semiclassical_2009},
and reproducing the two-point function of quantum Regge calculus \cite{bianchi_lorentzian_2012, shirazi_hessian_2016,bianchi_lqg_2009,
alesci_complete_2008,alesci_complete_2007,
bianchi_graviton_2006} at the level of the vertex--amplitude.

\bigskip

The Lorentzian EPRL model can be understood as a tentative attempt at a spinfoam quantization of the Holst action for General relativity. The Holst action provides a way to define the path-integral by means of a modification of the well--understood spinfoam quantization for topological theories. The central observation is that the Holst action, and related actions, is equivalent to a topological BF-action when imposing certain constraints, called the simplicity constraints.

The Holst action also provides the main bridge with the canonical theory and the Hamiltonian. The real Ashtekar-Barbero variables arise naturally in its 3+1 split, the canonical quantization of which results in the kinematical Hilbert space of LQG. 

In turn, the path--integral is conditioned on the requirement that states live in the kinematical Hilbert space of LQG.

Covariant LQG can be related to the body of results of Loop Quantum Gravity by understanding the spinfoam amplitudes as projectors on the physical states of LQG, annihilated by the Hamiltonian.

\subsubsection*{Towards contact with measurements}

 Loop Quantum Gravity, is maturing to the point where the extraction of experimental predictions of which the phenomenology can be understood and be testable in the foreseeable future is no longer seen as a remote possibility \cite{ashtekar_quantum_2016,agullo_detailed_2015,
 ashtekar_loop_2015-1,barrau_bouncing_2017,
ashtekar_loop_2014,barrau_fast_2014,
barrau_planck_2014-1,barrau_phenomenology_2016-1}

Efforts to calculate observable quantities necessarily employ approximations and simplified models inspired from the full edifice of LQG. A significant part of the community is currently participating in the development of such approaches.

These efforts are bringing LQG to a new phase of development. They teach us about the theory itself and lead us to view persistent issues from different angles. The results, detailed or qualitative, can turn out to be in contradiction with reality or yield inconsistencies, pointing to problems in theory and/or in interpretation. 
They can be used to compare different models and approaches within LQG and help to understand their equivalence or lack thereof. 
Of course, the hope is that one of these calculations leads to a prediction that turns out to be correct and measurable in practice.

  The most developed and detailed results in this direction come from the canonical formulation of the theory and the program of Loop Quantum Cosmology \cite{ashtekar_loop_2011,
  banerjee_introduction_2012,bojowald_absence_2001}. Having produced strong results on the resolution of the cosmological singularity, LQC is now starting to produce predictions for its pre-inflationary dynamics that could be put to the test in the coming years. 
  
  We appear to still be far from a falsifiable prediction, but the prospect of \emph{any} experimentally relevant prediction from LQG, within a well--defined set of assumptions, must have been hardly imaginable twenty years ago.

\bigskip

  Covariant approaches to quantum gravity share a basic intuitive picture: that of a quantum gravitational process that is spatiotemporally confined, enclosed in a lens-shaped boundary composed of two spacelike surfaces, see Figure \ref{fig:lens}. The remaining spacetime is taken to be adequately approximated by Einstein's theory. 

The covariant approach to LQG was partly developed with this intuitive picture in mind. It is designed to provide the physical transition amplitudes between semi-classical quantum states describing spatial geometries. The tools developed for the study of the semi--classical limit of the EPRL model are tailor--made to describe non-perturbative quantum gravitational phenomena as a quantum geometry transition.

What has been missing until recently was a concrete setting for a physical problem in the spirit of Figure \ref{fig:lens}. 

Cosmological singularities, are different than black hole singularities in one simple aspect: the former concern all space and the latter are confined in space. Thus, the picture for a geometry transition resolving the cosmological singularity will not be that of Figure \ref{fig:lens}, but one of a quantum transition between two disjoint classical spacetimes. This is the basic reason it is currently unclear how to define a relevant observable in the nascent field of spinfoam cosmology \cite{henderson_local_2011,bianchi_towards_2010,
bianchi_cosmological_2011,vilensky_spinfoam_2016,
bahr_towards_2017}

During the past few years, an idea that has raised much interest is the possibility of non-perturbative quantum-gravity effects that can cause the mass in the interior of a black hole to bounce out of the (would--be) horizon, in a time scale shorter than the Hawking evaporation time \cite{haggard_black_2014,rovelli_planck_2014,
de_lorenzo_improved_2016,christodoulou_realistic_2016,barrau_fast_2014,
barrau_bouncing_2017-1,barrau_planck_2014-1}, resulting in what would look from far away as a highly intense one--off explosion. 
 
\bigskip

\begin{figure}
\centering
   \includegraphics[scale=0.8]{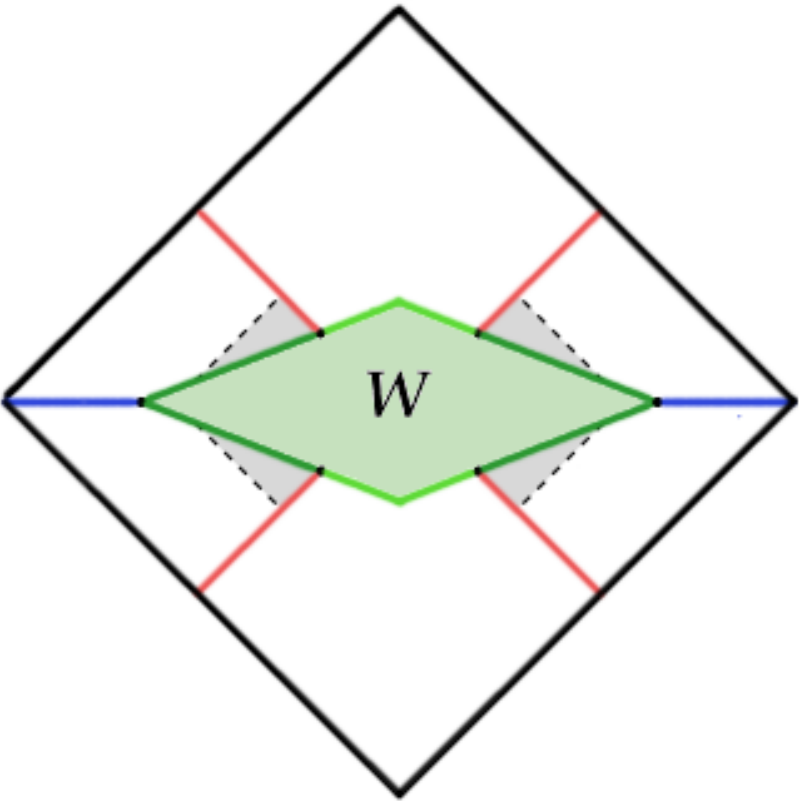}
\caption[The Haggard-Rovelli spacetime as a prototype well--defined setup for geometry transition.]{The Haggard-Rovelli spacetime as a simple, well--defined setup for geometry transition. Depicted is a cross--section of the rotated Penrose diagram, for easier comparison with Figure \ref{fig:lens}. For the Penrose diagram see Figure \ref{fig:ansatz}. The red lines are collapsing and anti--collapsing null shels. The shaded regions are trapped and anti--trapped regions. The definition and properties of the spacetime are developed in Chapter \ref{ch:exteriorSpacetime}. }
\label{fig:fireworksAsWMHsum}
\end{figure}

 The introduction of the spacetime pictured in Figure \ref{fig:fireworksAsWMHsum} by H. Haggard and C. Rovelli proposes a mechanism for this to happen, and a prototype minimalistic setup in the spirit of a WMH sum--over--geometries as in Figure \ref{fig:lens}. 
 
 Depicted, is a simple model of the following phenomenon: the formation of a black hole (trapped region) by collapsed matter, that transitions to a white hole (anti-- trapped region) and results in an explosion. Similar ideas have been independently and concurrently been put forward by L. Garay, C. Barcelo and R. Carballo \cite{barcelo_exponential_2016,barcelo_black_2016,
barcelo_lifetime_2015,barcelo_mutiny_2014}, with the important difference that an interpolating (Euclidean) family of metrics is used to carry out the path--integral. 

 The spacetime features a spacelike compact interior, chosen to a large degree arbitrarily, serving as a  separation of the parts of the system treated as quantum and as classical.

A theory for quantum gravity should be able to predict the characteristic time scale, as measured from far away, at which the quantum fluctuations of the metric make a geometry transition from a black to a white hole geometry likely. 

We will argue in this manuscript that we are presented with an \`a priori well-defined setup for studying geometry transition in a Feynman quantization of General Relativity, in the spirit of a WMH sum-over-geometries \cite{misner_feynman_1957,hawking_path_1980,
wheeler_nature_1957}. 

We are looking to estimate the scaling of a characteristic time of the process as measured from infinity with the mass of the collapsed object. That is, we are looking to get a rough estimate from the quantum theory of the correlation between two prominent physical scales involved in this phenomenon. 

\bigskip

Singularity resolution in black holes has been studied in the canonical framework over the past decade, in the context of a canonical quantization of the interior of a Schwartzschild black hole and the quantization of null shells \cite{husain_quantum_2005,modesto_disappearance_2004,
modesto_kantowski-sachs_2006, modesto_semiclassical_2010,modesto_loop_2006,ashtekar_quantum_2006,modesto_black_2008,
modesto_gravitational_2008,campiglia_loop_2008,
gambini_quantum_2014,corichi_loop_2016-1,
gambini_loop_2013,gambini_introduction_2015,
perez_black_2017-1}. With much remaining to be understood in the quantization procedure,  current investigations suggest singularity resolution through a bounce, although the physical picture for the phenomenon from the exterior spacetime is unclear. 

The two Paradigms, the canonical and covariant, serve as complementary: in covariant LQG, an effective description of the interior process is strictly speaking irrelevant for the extraction of the desired observable. In the canonical approach, a proper quantization of the interior geometry is desired and hoped that it can lead to an effective description. Understanding of the latter would provide crucial insights into the proper treatment of the phenomenon in the covariant approach, and vice--versa.
  
 It has been suggested that this hypothesis can account for the GeV excess observed from the galactic center by the Fermi satellite, and that the phenomenon will exhibit a characteristic redshift-dependence, a signal, that would distinguish it from other astrophysical phenomena that could account for this anomaly \cite{barrau_bouncing_2017}. It has also been been argued to be a plausible source of fast radio bursts, short and highly energetic low frequency signals of a yet to be understood astrophysical origin  \cite{barrau_fast_2014,barrau_phenomenology_2016,barrau_planck_2014,vidotto_quantum-gravity_2016}.

\bigskip

In this work, we complete a first attempt at calculating the lifetime of the HR-spacetime in the context of the EPRL model of covariant Loop Quantum Gravity. We arrive at an explicit estimate for a transition amplitude describing this process.

 The physical interpretation of these recent results are currently the subject of intense discussions. This manuscript reflects a snapshot of the evolution of our understanding of this phenomenon at the time of writing.  

 Whether the simple setup of the Haggard-Rovelli spacetime is sufficient to capture the desired physics, and whether covariant LQG at its current stage of development has the necessary tools to give a robust prediction remains to be seen. An optimist would hope that this simple model turns out to be the square potential barrier tunneling or the harmonic oscillator for black hole geometry transition. 

The set-up for the calculation presented here was first given in \cite{ christodoulou_planck_2016}. Unpublished results in this thesis will appear in \cite{mariosFabio} and \cite{mariosFabioCarlo}, in collaboration with Fabio d'Ambrosio.

 \subsection*{Outline of thesis}

The results and method presented in this manuscript are restricted to a certain set of simplifications and assumptions. Some will be justified with varying degrees of rigour as approximations. 

However, we stress that the main motivation of the simplifications taken is that with current results and techniques available, this was the ``best we could do'' to arrive at an explicit estimation; at least, the best the author could do. The reasons are laid out as we proceed. Many technical developments are currently under way and we may see much activity in refinements, extensions and alternative strategies of attacking this problem in the coming years.
  
   In summary, the setting will be the following:	 we will present results relevant to spinfoam amplitudes defined on a fixed 2-complex $\mathcal{C}$ without interior faces and dual to a simplicial complex. The boundary of the 2-complex coincides with the fixed graph on which the boundary state is defined, i.e.\! $\partial \mathcal{C}=\Gamma$. 

We do not take into account the graph changing nature of the Hamiltonian constraint, and do not consider a sum or refinement over spinfoams. 

We do not address the relevance of the continuum limit debate and the degree to which the truncated set of degrees of freedom should be understood as fundamental or collective, coarse grained degrees of freedom \cite{dittrich_discrete_2012,dittrich_coarse_2016,
oriti_group_2014}.

\bigskip

This thesis is organized in three main parts. The first part consists of Chapters \ref{ch:topBottom} and \ref{ch:AsympAnalysis}.

In Chapter \ref{ch:topBottom} we present a top--down path from General Relativity to the ansatz defining the amplitudes of covariant LQG. 

We focus on the physically relevant four--dimensional Lorentzian case and present intuition and motivations leading to the definition of the EPRL model, starting from the Hilbert-Einstein action. 
  
For lack of space we do not start by introducing   the canonical quantization of General Relativity in terms of Ashtekar-Barbero variables. We also do not present Euclidean 3D and 4D spinfoam models. We close the chapter by outlining how the spinfoam program is placed in Loop Quantum Gravity in Section \ref{sec:relationToCanonical}. 

In Chapter \ref{ch:AsympAnalysis} we take the bottom--up path, starting from the definition of the EPRL amplitudes and discuss the sense in which  General Relativity and the WMH sum-over--geometries emerges in the semi-classical limit. 

 We provide an independent derivation of the Krajewski-Han representation \cite{han_path_2013} in Section \ref{sec:ampsInAsAnform}, which brings the amplitudes in the form suitable for asymptotic analysis without using the coherent state representation. We proceed to review the fixed--spins asymptotics of the Lorentzian EPRL model. 
\bigskip

The second part consists of Chapter \ref{ch:gravTunneling} where a boundary is introduced. We explain how to build a boundary state corresponding to a wavepacket of an embedded quantum 3-geometry. We proceed to combine these states with a spinfoam amplitude to define a transition amplitude.

We show how the sum over spins (spin--sum) can be performed in Section \ref{sec:gravTunneling} and see that a gravitational tunnelling picture emerges naturally when considering physical transition amplitudes describing geometry transition in the spirit of Figure \ref{fig:lens}.

The third part consists of Chapters \ref{ch:exteriorSpacetime} and \ref{ch:calculationOfAnObservable}. In Chapter \ref{ch:exteriorSpacetime} we introduce the exterior spacetime. We formulate the Haggard--Rovelli spacetime starting from the Penrose diagram and give the exterior metric in a simpler form that emphasizes the role of the bounce time as a spacetime parameter. We avoid the use of specific hypersurfaces to implement the junction conditions. The relation to the original construction, based in part on a one to many (one to two) mapping from the Kruskal manifold, is explained in Section \ref{sec:crossedFingers}. 

We introduce the relevant classical and quantum observables, the mass $m$ and the bounce time $T$, and the lifetime $\tau(m)$ respectively. We propose an interpretation for the transition amplitudes and formulate the problem such that a path-integral over quantum geometries procedure can be naturally applied.

In Chapter \ref{ch:calculationOfAnObservable} we start by discussing how to construct a transition amplitude by discretizing a given choice of boundary. We proceed to make an explicit boundary and 2-complex choice and define explicitly a physical transition amplitude for this process. 

We will then use the results of Chapter \ref{ch:gravTunneling} to estimate the lifetime from this amplitude. We also provide a calculation valid for an arbitrary choice of boundary and a class of spinfoam amplitudes. We close the chapter by discussing the interpretation of our results and open issues.

\chapter[From General Relativity to the amplitudes of covariant Loop Quantum Gravity]{ From General Relativity \\ to the amplitudes of covariant Loop Quantum Gravity } \label{ch:topBottom}

The amplitudes of the EPRL model of covariant LQG represent the state-of-the-art understanding and current best guess of this research program, for the definition of the path integral of Quantum Gravity.

The main results that support the EPRL model as a bona fide candidate for a Feynman Quantization of General Relativity, come from studying its semiclassical limit, where a Wheeler-Misner-Hawking sum-over-geometries picture emerges. That is, the definition of the model is physically justified à posteriori, starting from a mathematically well-defined and well-motivated ansatz. This will be the subject of Chapter \ref{ch:AsympAnalysis}.

In this chapter, we present a bottom--up path to the EPRL amplitudes of covariant LQG starting from the Einstein-Hilbert action for General Relativity. Spinfoam models in the context of the covariant approach to LQG are treated in the introductory and review articles \cite{perez_spin_2012,rovelli_zakopane_2011,perez_introduction_2004,engle_spin_2014-1,
dona_introductory_2010,baez_introduction_1999}. Two main  references for this chapter are the recent text by C. Rovelli and F. Vidotto \cite{rovelli_quantum_2004} and the review by A. Perez \cite{perez_spin_2012}. For an introduction to LQG, see also \cite{gambini_first_2011
,ashtekar_introduction_2013} and for a more complete treatment, the textbooks \cite{rovelli_covariant_2014,thiemann_modern_2007}.

\bigskip

We do not follow the historical development leading to the EPRL and related models. We start directly from the physically relevant 4D Lorentzian case and follow the steps leading to the ansatz defining the EPRL model. We focus on the covariant actions pertinent to covariant LQG. For lack of space, we mention relevant results from the canonical analysis of these actions in passing.

The logic of how covariant LQG is placed into the Loop Quantum Gravity Program is outlined in Section \ref{sec:relationToCanonical}. We do not introduce a boundary until Chapter \ref{ch:gravTunneling}. To avoid confusion, we note that the amplitudes $W_\mathcal{C}$ in this chapter and the next chapter are to be understood as partition functions. Their meaning as physical transition amplitudes, when combined with a semi--classical boundary state, is treated in Chapter \ref{ch:gravTunneling}.
  
\section{A dictionary}  
\label{sec:dictionary}
 The aim of this chapter is to sketch a path from the phrase
\begin{eqnarray} \label{eq:WMHsumovergeom}
W \sim \int \mathcal{D} [g] \; e^{\frac{i}{\hbar} S_{EH}[g]  }
\end{eqnarray}
where $S_{EH}$ is the Einstein-Hilbert action
\begin{equation} \label{eq:HEaction}
S_{EH}[g] = \frac{1}{16 \pi G} \int \dd x^4 \sqrt{\det g(x)} R(g(x)) 
\end{equation}
and $g(x)$ is the spacetime metric, to the ansatz defining the EPRL model, and which is the starting point of the next chapter:

\begin{eqnarray} 
W_\mathcal{C} = \sum_{j_f} \mu(j_f) \; \int_{\sl2c}  \! \mu(g_{ve}) \; \prod_f A_f(g_{ve},j_f) \label{eq:EPRLampWMH} \\
A_f(g_{ve},j_f) = \int_{SU(2)} \! \mu(h_{vf}) \; \delta{(h_f)} \tr^{j_f} \left[\prod_{v \in f} Y_\gamma^\dagger g_{e'v}g_{ve} Y_\gamma  h_{vf} \right] \label{eq:faceAmpWMH}
\end{eqnarray}

Prima facie, the defining expressions \eqref{eq:EPRLampWMH} and \eqref{eq:faceAmpWMH} for the amplitudes $W_\mathcal{C}$ of covariant LQG bear no resemblance to General Relativity. That is, other than the use of the letter $g$ for both the metric and the \slt group elements $g_{ve}$. 

By the end of this chapter, we will have seen a path motivating equations  \eqref{eq:EPRLampWMH} and \eqref{eq:faceAmpWMH} from \eqref{eq:WMHsumovergeom} and \eqref{eq:HEaction}. To facilitate the discussion, we give a dictionary in Table \ref{tab:aimEPRL} of the correspondence we are aiming to establish.

\begin{table}[h]

\begin{eqnarray}
W &\sim & W_\mathcal{C} \nonumber \\ \nonumber \\
\mathcal{D} [g] &\sim & \sum_{j_f} \mu(j_f) \nonumber \\ \nonumber \\
e^{\frac{i}{\hbar} S_{EH}[g]  } &\sim & \;\int_{\sl2c} \!\! \! \mu(g_{ve}) \;   \prod_f \; A_f(g_{ve},j_f) \nonumber 
\end{eqnarray}
\caption[Comparison of a Misner-Hawking-Wheeler sum-over-geometries and the EPRL model state-sum ansatz]{Comparison of a Misner-Hawking-Wheeler sum-over-geometries and the EPRL model spin state-sum ansatz.} 
\label{tab:aimEPRL}
\end{table} 

 In Chapter \ref{ch:AsympAnalysis}, we will proceed from this point to see how in the semi-classical limit of large spins $j_f$, the \slt integrations can be performed using a stationary phase approximation. This will result in the behaviour 
\begin{eqnarray}
\int_{\sl2c} \!\! \! \mu(g_{ve}) \;   \prod_f \; A_f(g_{ve},j_f)  \sim \prod_f \;  e^{i j_f \phi_f(j_f)} 
\end{eqnarray}
The spins $j_f$ will be interpreted as quanta of areas $a_f$ , the functions $\phi_f(j_f)$ will be the deficit angles and the product over faces results in the Regge action, introduced in Section \ref{sec:lorReggeAction}, 
\begin{equation} \label{eq:reggeInExponent}
e^{\frac{i}{\hbar} \sum_f a_f \, \phi_f(a_f)} = e^{\frac{i}{\hbar} \, S_R(a_f)}
\end{equation}
The Regge action $S_R(l)$ describes discrete General Relativity, with the degrees of freedom being edge lengths $l$ of a triangulation, a piecewise-flat manifold. We caution that in \eqref{eq:reggeInExponent} the Regge action has instead the areas $a_f$ as variables. This is closely related to the fact that the spins $j_f$ will not be the only quantum numbers capturing degrees of freedom of the quantum gravitational field. The asymptotic form \ref{eq:reggeInExponent}, with some further details and features, will be essential in estimating a physical observable in Chapter \ref{ch:calculationOfAnObservable}. 

Throughout this manuscriptint, based on these results, we  will present and use the EPRL model, putting it to the test in a physical application, as a proposal for the regularized path-integral for Quantum Gravity, in the spirit of a MHW sum-over-geometries.

\section{A game of actions}
\label{sec:gameOfActions}
In this section we start from the Einstein-Hilbert action and introduce the tetrad and differential-forms formalism. We will derive the Palatini and BF form of the Einstein-Hilbert action and then pass to the tetradic Palatini action, the Holst action and the formulation of General Relativity as a constrained topological BF-theory. We discuss the simplicity constraints, reducing BF-theory to General Relativity.

We introduce the holonomy as a parallel transporter and discuss its meaning as a discretization of the curvature of the connection. We proceed to explain the quantization strategy leading to the EPRL model as a modification of the well-understood path--integral quantization of BF-theory for compact groups, regularized by a skeletonization on a topological lattice.  We then discuss the strategy for the imposition of the simplicity constraints at the quantum level.

We outline how covariant LQG is placed within the body of results of Loop Quantum Gravity in Section \ref{sec:relationToCanonical}. A hint for this is already present in equations \eqref{eq:EPRLampWMH} and \eqref{eq:faceAmpWMH}, where we see that both the Lorentz group and the rotation group appear through their double covers, $\sl2c$ and $SU(2)$. The kinematical state-space is that resulting from the canonical quantization of General Relativity when written in 3+1 form in terms of the Ashtekar-Barbero variables, in which it resembles an $SU(2)$ Yang-Mills theory.

\subsection{The dissappearance of coordinates: the tetrad and Einstein-Hilbert action in Cartan's formalism} \label{sec:HEtoCar}
The first step in our journey is to rewrite the Einstein-Hilbert action $S_{EH}[g]$ for General Relativity using Cartan's formalism. Thus, for the moment, the connection determining the covariant derivative is fixed to be the Levi-Civita connection.\footnote{The Levi-Civita connection is unique. It is compatible with the metric, meaning that the covariant derivative $\nabla$ it defines, satisfies $\nabla g =0$. An equivalent necessary and sufficient condition is that $\nabla$ be torsion--free: the difference between the directional derivatives of two vector fields $V$ and $U$, taken with the Levi-Civita covariant derivative, is their Lie bracket, $\nabla_V U-\nabla_U V = \mathcal{L}_V U = -\mathcal{L}_U V\equiv [V,U] $. This is the content of the fundamental theorem of Riemannian geometry.} We will ``free'' the connection in Section \ref{sec:HEtoPal}. 

 The departure from the formulation of differential geometry that Einstein used to define his theory in this context, is to introduce non-coordinate vector bases for the tangent space. This is what we call the tetrad.  

The tetrad is also known as the moving frame, the English translation of what Cartan called it, r\'ep\'ere mobile, and as vierbein, its German name. The idea is very simple, the consequences are profound. Let us start from the pre-Cartan story.

\medskip

At its most elementary level, General Relativity  is the theory that describes the motion of test particles in a curved spacetime. Different possibilities of curved spacetimes, metrics that solve the field equations, result in different behaviours for the test-particle trajectories, accounting for the phenomenon we call gravity. 

This elementary image carries through intact from Newtonian to Einsteinian Gravity: the history of a particle is modelled as a curve in a four-dimensional spacetime. But how do we model its instantaneous velocity? For that, we need a vector at each point on the curve, pointing in the instantaneous direction of movement and with a magnitude giving the speed.

A differentiable manifold $M$ is a natural arena for realising such a model of nature, as it comes equipped with a tangent space $TM(p)$ at each point $p \in M$. 

 This is the usual mathematical construction: a vector $v(p)$ at a point $p$ corresponds to an equivalence class of functions from the reals to the manifold, that is, curves, that traverse $p$ at the same speed and in the same direction.     
 
 To make this explicit, and to have objects we can do calculations with, we introduce coordinate systems covering the manifold. The basis of tangent vectors for $TM(p)$ is provided by considering four such equivalence classes, associated to the coordinates at $p$. 
 
 We usually denote these four basis vector fields $\partial_\alpha$, which denote four vector fields in the tangent space $TM \equiv \cup_{p \in M} TM(p)$. The notation $\partial_\alpha$ is meant to imply that this is the vector field pointing in the direction where, locally, only the coordinate $\alpha$ changes (increases), with $\alpha$ being any one of the four coordinates.
 
 Vector fields decompose in this basis as $V=V^\alpha \partial_\alpha$ or are written simply as their components $V^\alpha$, leaving the coordinate--based vector field basis implicit. 

This is fine for describing velocities for instance. There is however a second reason we use differentiable manifolds as a fiducial canvas for spacetime. Locally, a Lorentzian manifold is Minkowski. 

In Minkowski, a convenient choice of basis is to use Cartesian coordinates and take the orthonormal vector field basis associated to these coordinates. That is, to take as a basis the set of normalized vectors $\hat{x}_\alpha$ pointing in the positive $x_\alpha$-axis direction. We know that there is a tangent space structure on a differentiable manifold. Why not do the same in General Relativity?

There is no reason not to. The caveat is, that this orthonormal basis of vector fields \emph{cannot} correspond to \emph{any} global coordinate system for a curved spacetime. If it did, then the spacetime would be flat, since it would admit a global Cartesian or Minkowski $t-x-y-z$ coordinate system. We can always choose such coordinate systems locally. 

However, there is nothing wrong with choosing arbitrarily four orthonormal vectors at each point $p$ of $M$.

In summary, in General Relativity, by tetrad we mean an arbitrary choice of a set of four (smooth) vector fields, unrelated to \emph{any} coordinate system, that are furthermore taken to be orthonormal with respect to the Minkowski metric. This construction captures directly Einstein's intuition that General Relativity reduces to Special Relativity locally. 

 Any vector field can now be written in this basis as $V=V^I e_I$. Capital latin letters, called Minkowski or internal indices, are used to index the tetrad basis. This is used to indicate that no coordinate--based bases, usually indexed by lower-case case greek letters, are involved. 
 
When one sticks to the tetrad, coordinates entirely disappear from the game. The arbitrariness in choosing a coordinate system is replaced by the arbitrariness in choosing the tetrad field. Changing from one tetrad field to another, amounts to ``rotating`` one orthonormal frame into another at each point, which is achieved by a Lorentz transformation, the symmetry group sending an orthonormal frame to another in Minkowski. 

To be precise, the tangent space in which these vectors live is not $TM$. There is a linear map between the tangent space $TM$, which is constructed for instance as above using equivalence classes of functions, and the vector bundle which has Minkowski space as a fiber at each point $p$. For a given choice of tetrad $e_I$, this map is the matrix\footnote{
 The inverse matrix $e^I_\alpha$, relating the dual one-form bases $e_I$ and $\dd x_\alpha$, is defined by 
\begin{eqnarray}
e^\alpha_I e^J_\alpha =\delta_I^J \nonumber \\
e^\beta_I e^I_\alpha =\delta_\alpha^\beta
\end{eqnarray}
see next Section.} $e^\alpha_I$ , which can be used to translate between a coordinate basis $\partial_\alpha$ of vector fields and the orthonormal basis $e_I$:
\begin{equation}
e_I = e^\alpha_I \partial_\alpha
\end{equation}
Let us consider $e_{\alpha I} \equiv g_{\alpha\beta}e^\beta_I$, with which we may rewrite the metric as 
\begin{equation} \label{eq:metricInTetradComps}
g = g_{\alpha \beta}(x) \dd x^\alpha \dd x^\beta  = e_{I \alpha}(x) e_{J\beta}(x) \, \eta^{IJ} \dd x^\alpha \dd x^\beta
\end{equation}

Often, $e^\alpha_I(x)$ is also called the tetrad field. While mixed--index objects such as $e^\alpha_I(x)$ are of their own importance and central objects in the canonical analysis, until we relate the topological BF-theory to General Relativity, we do not need to consider coordinates. It will be conceptually more illuminating to skip the intermediate step of introducing explicitly a coordinate--based basis of vector fields and one-forms, and write directly the metric $g$ as 
\begin{equation} 
g(p) = e_I(p) e_J(p) \eta^{IJ}
\end{equation}
\medskip
No coordinates are involved in this expression, which is purely geometric.   

A second important thing to notice: the tetrad field $e_J$ completely replaces the metric $g$. This is surprising. The metric encodes all the geometry. How can an arbitrary assignment of four orthonormal vectors $e_J(p)$ at each point $p$ encode the geometry? The answer is, it doesn't. The geometry is now purely encoded through the metric in two objects: the dual basis $e^J$ of one-forms, introduced in the next section, and the parallel transport. The dual basis $e^J$ encodes (infinitesimal) geometrical magnitudes such as lengths, areas and volumes, while the parallel transport, determined by the connection, codes the curvature. In particular, parallel transporting the tetrad will be very convenient. This will be the subject of Section \ref{sec:holonAsParTransport}. 

	Finally, the arbitrariness in choosing the tetrad field is manifested by the possibility to rotate it \emph{at each point}  (so long as it remains smooth), by a Minkowski rotation. That is, there is a local Lorentz gauge invariance, $e_I(p) \rightarrow e_L(p) \, \Lambda^L_I(p)$,
\begin{equation} 
g(p) = \Lambda^K_I(p) \, e_K(p) \, \Lambda^L_J(p) \, e_L(p) \eta^{IJ} = e_I(p) e_J(p) \eta^{IJ}
\end{equation}
by the definition of Lorentz transformations as preserving the Minkowski metric, $\Lambda^K_I \Lambda^L_J \eta_{KL}=\eta_{IJ}$.

\medskip

Let us move to General Relativity.

The dynamics of General Relativity are encoded in the Einstein-Hilbert action, 
\begin{equation}
S_{EH}[g] =  \int \dd x^4 \, \sqrt{-\det g(x)} R[g(x)] 
\end{equation}
The Einstein--Hilbert action in terms of the tetrad, by direct substitution of \eqref{eq:metricInTetradComps}, reads
\begin{equation} \label{eq:HEtetrad}
S_{EH}[e] =  \int \dd x^4 \, \vert\det e(p)\vert \; R[e(p)] 
\end{equation}
and variation with respect to the tetrad leads again to Einstein's field equations. 

There is something odd in \eqref{eq:HEtetrad}. The coordinates reappear as an integration measure, but the integrand of the action depends only on the tetrad field, which knows nothing of coordinates. We get rid of this tension in the following section.

As a final note, keep in mind that we had to take the absolute value of $\det e(x)$ to make the two actions match. We will henceforth ignore the absolute value, which does not affect the field equations and return to this point in Section \ref{sec:emergenceOfSimpGeoms} where we will see that it appears as the ``cosine feature'' in the asymptotic analysis of the EPRL model. In a work by Rovelli, Riello and the author, it was shown that when considering the coupling with fermion fields, the actions with and without the absolute value have in principle measurable observable differences \cite{christodoulou_how_2012}. 

\subsection{The tetradic Einstein-Hilbert action}
\label{sec:tetradicHE}
 The measure $\dd x^4$ in \eqref{eq:HEtetrad}, denotes an anti--symmetrized product between the coordinate-based basis of one-forms $\dd x^\alpha$. That is,
\begin{eqnarray}
\dd x^4 & \equiv & \dd x^0 \wedge \dd x^1 \wedge \dd x^2 \wedge \dd x^3 
\end{eqnarray}

In coordinate language, $\dd x^4$ times $\sqrt{-\det g(x)}$ is the integration measure that is invariant under change of coordinates: $\sqrt{-\det g(x')}$ is $\sqrt{-\det g(x)}$ times the Jacobian between $x$ and $x'$.

More geometrically, it is the 4-volume form $\vol$ of the manifold, and the Einstein--Hilbert action \ref{eq:HEtetrad} reads
\begin{eqnarray}
S_{EH}[e] &=&  \int \vol(e) \; R(e) 
\end{eqnarray}
The idea now is to write $\vol(e)$ in terms of the co-tetrad and see what are the objects that arise. Naturally, we define the dual basis $e^I$ to the tetrad, providing an orthonormal basis of 1-form fields and which we call the co-tetrad. It is defined via the metric by 
\begin{equation}
g(e_I, \cdot ) = \eta_{JI} e^J 
\end{equation}
or in mixed-index notation
\begin{equation}
g_{\alpha \beta} e^\beta_J \dd x^\alpha= \eta_{JI} e_\alpha^J  \dd x^\alpha
\end{equation}
The 4-volume form in terms of the co-tetrad reads\footnote{To show this compute
\begin{eqnarray}
e^1\wedge \ldots \wedge e^n &=& (e^1)_{k_1}\ldots (e^n)_{k_n} \; \dd x^{k_1}\wedge \ldots \wedge \dd x^{k_n} \nonumber \\
&=&(e^1)_{k_1}\ldots (e^n)_{k_n} \; \dd x^{1}\wedge \ldots \wedge \dd x^{n}\;  \epsilon^{k_1 \ldots k_n} \nonumber \\
&=& (\det e)  \; \dd x^{1}\wedge \ldots \wedge \dd x^{n} \nonumber
\end{eqnarray}
 }
\begin{equation}
\vol = e^0 \wedge e^1 \wedge e^2 \wedge e^3 = \det(e) \, \dd x^4
\end{equation}
and the Einstein-Hilbert action becomes
\begin{equation} \label{eq:HEfirstTetradStep}
S_{EH}[e] = \int e^1 \wedge e^0 \wedge e^2 \wedge e^3  \; R(e) 
\end{equation}
We write the Riemann tensor with Minkowski indices as
\begin{equation} \label{eq:RicciEpsilonContraction}
R_{\ \ \ KL}^{IJ}(e) \equiv R_{\ \ \ \mu\nu}^{ \rho\sigma} (e) \, e^\mu_K \,  e^\nu_L \, e_\rho^I \,  e_\sigma^J
\end{equation}
Using the identity $2 (\delta^K_I \delta^L_J - \delta^L_I \delta^K_J ) = \epsilon_{IJCD} \epsilon^{CDKL}$ and the antisymmetry of the Riemann tensor in its first or second pair of indices, we write the Ricci scalar as
\begin{eqnarray} \label{eq:RiemannMink}
R(e)=R_{\ \ \ IJ}^{IJ}(e) &=& \delta^K_I \delta^L_J R_{\ \ \ KL}^{IJ} \nonumber \\ 
 &=& \frac12 \left(\delta^K_I \delta^L_J - \delta^L_I \delta^K_J \right)R_{\ \ \ KL}^{IJ} \nonumber \\
&=& \frac{1}{4} \epsilon_{IJCD} \epsilon^{CDKL} R_{\ \ \ KL}^{IJ} \nonumber \\
\end{eqnarray}
where we introduced the Levi-Civita symbol $\epsilon^{CDKL}$ (antisymmetric in all its indices). Next, we replace \eqref{eq:RiemannMink} in \eqref{eq:HEfirstTetradStep}, and notice that we may write 
\begin{equation} \label{eq:4volForm}
\epsilon^{CDKL} \, \vol = \epsilon^{CDKL} e^0 \wedge e^1 \wedge e^2 \wedge e^3 =  e^C \wedge e^D \wedge e^K \wedge e^L 
\end{equation}
Thus, we get
\begin{equation}
S_{EH}[e] =  \frac{1}{4} \int  \epsilon_{IJKL} e^I \wedge e^J \wedge e^A \wedge e^B \; R_{\ \ \ AB}^{KL}(e)
\end{equation}

Now, we \emph{define} the curvature or field strength,
\begin{equation}
F^{IJ}(e) \equiv R_{\ \ \ KL}^{IJ} (e)  \, e^K \wedge e^L 
\end{equation}
which brings the Einstein-Hilbert action in its tetradic form  
\begin{equation} \label{eq:tempHEtetrad}
S_{EH}[e] = \frac{1}{4} \int \epsilon_{ABIJ} \; e^A \wedge e^B \wedge F^{IJ}(e) 
\end{equation}
In the next section, we will see the close relation of this form of the Einstein-Hilbert action with the Palatini action.

The curvature, $F^{IJ}$ is usually called the curvature two-form and refers to the fact that it is a Lie-algebra valued two-form, with the algebra being that of the symmetry group, the Lorentz group. In this sense, $IJ$ are understood as component indices, a pair of anti-symmetric Minkowski indices, that can be naturally contracted with a basis of the Lie algebra of the Lorentz group. This is a necessary abstraction when working for instance with topological theories, where no notion of geometry is present, and for the quantum theory. 

Until we introduce the topological BF-theory in Section \ref{sec:simplicityConstraintsBFtheory}, we adopt a different point of view, and think of the curvature $F^{IJ}$ as a matrix of 2-forms. In the end of this section, we will see that thought of this way, the geometrical meaning is transparent and allows us to make a parallel to the Regge action. We take a moment to emphasize this alternative way to understand the curvature two--form.

In the previous section, we introduced the tetrad $e_I$ as a set of four orthonormal vector fields. The $I$ index in $e_I$ is not a component index, it means $e_I(p)=(\vec{e_0}(p),\vec{e_1}(p),\vec{e_2}(p),\vec{e_3}(p))$, a set of four orthonormal vectors at each point $p$. Similarly for the set of one-forms $e^I(p)$. Thus, when we write $e^I \wedge e^J $, we may understand this as a set of orthonormal two-form fields. 

In particular, there are six of them because they are arranged in an antisymmetric four by four matrix, which is the dimension of the space of two-forms. Geometrically, they correspond to the number of planes defined by arranging four linearly independent vectors in all possible pairs. 

Thus, we think of the curvature $F^{IJ}(e) \equiv R_{\ \ \ KL}^{IJ} (e)  \, e^K \wedge e^L$ as a matrix, with entries for fixed $IJ$ the two form $R_{\ \ \ KL}^{IJ} (e)  \, e^K \wedge e^L$ constructed by adding the basis two forms $e^K \wedge e^L$ with coefficients the componentes $R_{\ \ \ KL}^{IJ}(e)$ of the Riemann tensor.

Let us return to the Hilbert-Einsten action. Since all indices are contracted in \eqref{eq:tempHEtetrad}, we may write more compactly
\begin{equation} 
S_{EH}[e] = \frac{1}{4}\int \tr \; e \wedge e \wedge F(e)  
\end{equation}
This is a notation that will be useful throughout this manuscript: whenever we write the trace $\tr$, we mean that the objects inside are contracted in all their indices in an appropriate way depending on the context. Here, $\tr$ means contraction with the Levi-Civita symbol. In general, $\tr$ will be a multi-linear form that is invariant under the pertinent symmetry group, here, the Lorentz group.

We introduce a second notation that will be useful. Using the Hodge dual\footnote{The definition of the Hodge dual in general is with the Levi-Civita tensor, which includes a square root of the metric, not the Levi-Civita symbol. However, we are using tetrads and the indices are Minkowski. Thus the absolute value of the determinant of the metric is unit. $\epsilon_{ABCD}$ here is understood as a tensor with respect to the Minkowski metric, indices may be freely raised and lowered with $\eta_{IJ}$. 
 } or $\star$  operation, we write 
\begin{equation}
(\star F)_{AB} \equiv \frac{1}{2} \epsilon_{ABIJ } F^{IJ}
\end{equation}
and the Einstein-Hilbert action is written as
\begin{equation}  \label{eq:PalatiniformHE}
S_{EH}[e] =  \frac{1}{2} \int e \wedge e \wedge \star F(e)  
\end{equation}
Now, let us give a name to the basis of two-forms $e^I \wedge e^J$ and call it $B^{IJ}(e)$.
\begin{equation}
B(e) \equiv e \wedge e
\end{equation}
This is the BF-form of the Einstein-Hilbert action
\begin{equation} \label{eq:BFformHE}
S_{EH}[e] = \frac12 \int B(e) \wedge \star F(e) 
\end{equation}
Let us now go back to the original form of the Hilbert Einstein action. The curvature $F(e)$ reads
\begin{equation}
F(e)^{IJ} =  R_{\ \ \ KL}^{IJ} (e)  B^{KL} 
\end{equation}
thus
\begin{eqnarray} 
S_{EH}[e] &=&  \frac{1}{4}\int \epsilon_{IJAB} R_{\ \ \ KL}^{IJ} B^{KL} \wedge B^{AB}
\end{eqnarray}
we use the same trick as \ref{eq:4volForm} and write
\begin{equation} \label{eq:preSimplicityConstraintHE}
B^{KL} \wedge B^{AB} = \epsilon^{KLAB} vol^4
\end{equation}
The two $\epsilon$ contract the Riemann tensor as in \eqref{eq:RicciEpsilonContraction}, and we are back to
\begin{eqnarray}
S_{EH}[e] &=&  \int vol^4 \; R(e) 
\end{eqnarray}
We will find again equation  \eqref{eq:preSimplicityConstraintHE} in Section \ref{sec:simplicityConstraintsBFtheory}, where we will see that when treating $B^{AB}$ as an abstract Lie-algebra valued two-form, the simplicity constraint, reducing the topological BF-action to GR, ensures that
\begin{equation} 
\epsilon_{KLAB} B^{KL} \wedge B^{AB} = \epsilon_{KLAB} \epsilon^{KLAB} vol^4 = 4!\; vol^4
\end{equation}

\begin{center}
\rule{7cm}{0.05cm}
\end{center}
To close this section, we discuss the geometrical meaning of the tetradic and BF forms of the Einstein-Hilbert action, in equations \eqref{eq:PalatiniformHE} and \eqref{eq:BFformHE}. 

 Locally and for fixed $A$ and $B$, $e^A \wedge e^B$ corresponds to a plane  (in the Minkowski fiber at $p$). By construction, $e^A(p) \wedge e^B(p)$ is a 2-form in the plane defined by the span of $e_A(p)$ and $e_B(p)$ at $p$.
 
 Consider dimensions in $\hbar$ (geometrical units, $G=c=1$). The term $e^A \wedge e^B$ goes as $\hbar$ and when $e_A(p)$ and $e_B(p)$ are taken spacelike (we may gauge fix the tetrad for it to be so) it corresponds to an area. 

Let us fix for definitenes $e_0$ to be timelike and $e_1$,$e_2$,$e_3$ to be spacelike and take this plane to be given by $e_1(p)$ and $e_2(p)$. Then, $e^1(p) \wedge e^2(p)$ is the invariant area form on this (local) spacelike plane, that is, the infinitesimal proper area. 

Now, the remaining term, $\star F^{IJ}$, must be related to the curvature since it includes the Ricci scalar. What are its dimensions? It is dimensionless, the Riemann tensor has inverse area dimensions, $\hbar^{-1}$. What are then the geometrical objects that are dimensionless: angles. Compare this with the Regge action in Table \ref{tab:actionsTable}, that is of the form ``face area times deficit angle around that face'', with the curvature encoded in the deficit angle. 

As a final remark, $\star F^{IJ}$ is the Ricci scalar \emph{times the Hodge dual} of the plane corresponding to $e^I \wedge e^J $. It thus determines its co-plane (in a Minkowski fiber). This is also the construction in the discretization of General Relativity given by the Regge action, explained in Section \ref{sec:lorReggeAction}. The deficit angle encodes the curvature and is defined in the plane normal to a (spacelike) triangular face.

\subsection{The Palatini action} \label{sec:HEtoPal}
We saw above that when writing the Einstein-Hilbert action as\footnote{We are henceforth neglecting the $\frac{1}{2}$ factor.}
\begin{equation}
S_{EH}[e] =  \int e \wedge e \wedge \star F(e)  
\end{equation}
the area element $e \wedge e$ and the curvature $\star F(e)$ are in a sense ``decoupled'', with $\star F(e)$  living in the co-plane of $e \wedge e$. This can be taken as a hint that there is a second object, in addition to the tetrad $e$, that may be treated as an independent variable and that is directly related to the curvature. We will see in Section \ref{sec:holonAsParTransport}, that when working with tetrads, the curvature is coded in the parallel transport. The parallel transport along a curve connecting two points is a mapping between the tangent spaces, determined by the connection. In particular, we will need to define a covariant derivative that works for mixed-index objects. This is the spin-connection, defined below, which we call $\omega$. Let us put the dependence on $\omega$ explicitly in the curvature
\begin{equation}
S_{EH}[e] =  \int e \wedge e \wedge \star F(\omega[e])  
\end{equation}
where $\omega[e]$ is the Levi-Civita connection.
 
 The Palatini action is defined as
\begin{equation} \label{eq:Palatini}
S_{P}[e,\omega] =  \int e \wedge e \wedge \star F(\omega) 
\end{equation}
where $\omega$ is an independent variable, that is, we have left the covariant derivative arbitrary.

Variation of $S_{P}[e,\omega]$ with respect to  $\omega$, fixes $\omega[e]$ to be the Levi-Civita connection, thus $S_{P}$ reduces to $S_{EH}$ when evaluated on the equation of motion for $\omega$:
\begin{equation}
S_{P}[e,\omega[e]] = S_{EH}[e]
\end{equation}

We may then proceed and take the variation of $S_{P}[e,\omega[e]]$ with the tetrad, \emph{after}\footnote{Note that this remarkable property of the relation between the Palatini and Einstein actions is not in general a valid procedure when deriving equations of motion from an action. } having replaced the solution of the variation with respect to $\omega$. This will yield Einstein's field equations since it is the same as varying $S_{EH}[e]$ with the tetrad.

  Writing General Relativity this way is called a first order formalism, because the variables $\omega$ and $e$ appear in the action up to their first derivative, see equations \eqref{eq:riemannChristoffel} and  \eqref{eq:connectionChristoffel}.
  
   We recall the form of the Riemann tensor written in terms of the Christoffel symbols
\begin{equation} \label{eq:riemannChristoffel}
\Omega^\rho_{\ \ \sigma \mu \nu}(\Gamma) = \partial_\mu \Gamma^\rho_{\nu \sigma} - \partial_\nu \Gamma^\rho_{\mu \sigma} +\Gamma^\rho_{\mu \lambda} \Gamma^\lambda_{\nu \sigma} -\Gamma^\rho_{\nu \lambda}
\Gamma^\lambda_{\mu \sigma}
\end{equation}
where notice that we have written $\Omega$ for the curvature tensor (of the connection) instead of $R$ for the Riemann tensor. When the connection is the Levi-Civita connection, the Christoffel symbols are given with respect to first order derivatives of the metric (or tetrad, by replacing $g_{\alpha \beta}=e^I_\alpha e^J_\beta \eta_{IJ}$)
\begin{equation} \label{eq:ChristoffelSymbols}
\Gamma^\sigma_{\mu \nu}(g) = \frac12 g^{\sigma \rho}\left(\partial_\nu g_{\mu \rho} + \partial_\mu g_{\rho \nu} - \partial_\rho g_{\mu \nu} \right)
\end{equation}
and the curvature tensor $\Omega^\rho_{\ \ \sigma \mu \nu}$ becomes the Riemann tensor
\begin{equation}
\Omega^\rho_{\ \ \sigma \mu \nu}(\Gamma(g))=R^\rho_{\ \ \sigma \mu \nu}(g)
\end{equation}
That is, the familiar definition for the curvature in terms of the connection carries through when leaving the covariant derivative $\nabla$ arbitrary, except that the metric compatibility and the torsionless condition are not satisfied ($\nabla g \neq 0$). When they are, the connection is Levi-civita. 

To act on objects with also Minkowski indices, we need a covariant derivative that can also act on them. This is provided by the spin-connection, which we called $\omega$, related to $\Gamma$ via the tetrad by 
\begin{equation} \label{eq:connectionChristoffel}
\omega^{IJ}_\mu = e^I_\nu \Gamma^\nu_{\sigma \mu}e^{\sigma J} - e^{\nu I} \partial_\mu e_\nu^J
\end{equation}
where $\omega^{IJ}_\mu$ should be understood as components here. We define the curvature 1-form by
\begin{equation}
\omega^{IJ}= \omega^{IJ}_K e^K = \omega^{IJ}_\mu \dd x^\mu
\end{equation}
As with the field strength $F^{IJ}$, we may understand $\omega^{IJ}$ either as a matrix of one forms, with entries for fixed $IJ$ the forms defined by the sum $\omega^{IJ}_K e^K = \omega^{IJ}_0 e^0 + \omega^{IJ}_1 e^1 + \cdots$, or as a Lie algebra valued one-form. The connection $\omega$ is not a tensorial object, as with Christoffel symbols. 

There are two tensorial invariants of the connection. The curvature $F(\omega)$, defined by Cartan's second structure equation,
\begin{equation}
F^{IJ}(\omega) = \dd \omega^{IJ} + \omega^I \wedge \omega^J
\end{equation} 
where $\dd$ here is the exterior derivative, and the torsion
\begin{equation}
T^{I}(\omega) = \dd e^{I} + \omega^I_J \wedge e^J
\end{equation}
defined by Cartan's first structure equation. 

 An alternative expression for the field strength $F^{IJ}(\omega)$, that allows for easier comparison with the actions presented in this chapter, summarized in Table \ref{tab:actionsTable}, is with respect to the curvature tensor written with Minkowski indices
\begin{equation}
F^{IJ}(\omega) =\Omega_{\ \ \ KL}^{IJ} (\omega)  \, e^K \wedge e^L
\end{equation}
where
\begin{equation}
\Omega_{\ \ \ KL}^{IJ}(\omega) = \Omega_{\ \ \ \mu\nu}^{ \rho\sigma} (\Gamma) \, e^\mu_K \,  e^\nu_L \, e_\rho^I \,  e_\sigma^J
\end{equation}
and $\Gamma$ is related to $\omega$ through \eqref{eq:connectionChristoffel}.

As a last note, we could have started from the following form of the Palatini action
\begin{equation} 
S_{P}[\omega,e] = \int \dd x^4 \sqrt{\det e(x)} \Omega(\omega) 
\end{equation}
resembling the usual form of the Einstein-Hilbert action as in equation \ref{eq:HEaction}, and with identical manipulations as in the previous section, arrive to the BF--form of the Palatini action
\begin{equation} \label{eq:BFformPalatini}
S_{P}[e,\omega] = \int B(e) \wedge \star F(\omega) 
\end{equation}

\subsection{Holst action and the Immirzi parameter} \label{sec:PaltoHolst}
 From equation \ref{eq:BFformPalatini}, we see that there is a natural term that we may consider to add to the action: to contract $ B^{AB} = \left( e^A \wedge e^B \right)$ directly with $F(\omega)_{AB}$. $F(\omega)_{AB}$ is the field strength with both indices lowered with the Minkowski metric, instead of its Hodge dual $(\star F(\omega))_{AB}$.

The action of the Hodge dual on objects with two lower indices, such as $(\star F)_{AB}$ is given by 
\begin{equation}
\star (\star F)^{KL} = \frac12 \epsilon^{ABKL}(\star \Omega)_{AB}
\end{equation}
 The well known property that $\star \star = 1$ follows from 
\begin{eqnarray}
\star \star \Omega^{KL} &=& \frac12 \epsilon^{KLAB}\star \Omega_{AB} \nonumber \\
&=& \frac{1}{4} \epsilon^{KLAB}\epsilon_{ABIJ} \Omega^{IJ} \nonumber \\
&=&  \Omega^{KL}
\end{eqnarray}

We can rewrite things as
\begin{eqnarray}
B(e)^{IJ} \wedge F_{IJ} &=& B(e)^{IJ} \wedge \star \star F_{IJ}\nonumber \\
 &=&\frac12 \epsilon^{MNKL} B(e)^{IJ}\eta_{IM} \, \eta_{IN}  \wedge  \star F_{KL} \; 
\nonumber \\
&\equiv& \star  B(e)  \wedge \star F
\end{eqnarray}
The Holst action then amounts to adding to the Palatini action, the action
\begin{equation}
S_{HI}[e,\omega]= \int  \star B(e)  \wedge \; \star \Omega(\omega) 
\end{equation}
where $HI$ stands for Holst-invariant. The (inverse) ``coupling constant'' for this term, $\gamma$, is called the Immirzi parameter
\begin{equation} \label{eq:HolstAction}
S_H[e,\omega] = S_P[e,\omega] + \frac{1}{\gamma}S_{HI}[e,\omega]
\end{equation}
The Immirzi parameter is a fundamental parameter of Loop Quantum Gravity \cite{rovelli_immirzi_1998}, proportional to to the quantum of area \cite{immirzi_quantum_1997}.\footnote{It is understood to determine the coupling constant of a four-fermion interaction and lead to observable effects \cite{perez_physical_2006}.} We will see the Immirzi parameter appearing in the result of the calculation of the lifetime in Chapter \ref{ch:calculationOfAnObservable}. 

The BF form of the Holst action reads
\begin{eqnarray} \label{eq:HolstActionBFform}
S_H[e,\omega] = \int \left( \; B(e) + \frac{1}{\gamma} \star B(e) \; \right) \wedge \star F
\end{eqnarray}

The equations of motion for the Holst action describe General Relativity. This happens as follows. When $\omega[e]$ is the Levi-Civita (torsionless spin connection), the Bianchi identity ($\dd_\omega F(\omega)=0$) is satisfied and corresponds to the vanishing of the Holst term  
\begin{equation}
 \star B(e)  \wedge \; \star F(\omega[e]) =0
\end{equation}
Thus, the variation of the Holst action with respect to $\omega$ is again solved for $\omega[e]$: we added to the Palatini action a term that vanishes on-shell and therefore the equations of motion are not altered. 

  The spinfoam approach is based on this action, which is a natural generalization of the Einstein-Hilbert action. There are two main reasons for this: first, it allows us to modify the quantization of a topological BF-theory so as to achieve a quantization of general relativity. We see this in Section \ref{sec:BFSpinfoamQuantization}. The second is that the 3+1 split of this action in the time gauge provides the bridge to the canonical theory, outlined in Section \ref{sec:relationToCanonical}. For a review of the canonical analysis of the Holst action in the time gauge and an investigation of the role of the Immirzi parameter see \cite{geiller_note_2013}.

\subsection{Simplicity constraints: General relativity as constrained BF-theory} \label{sec:simplicityConstraintsBFtheory}
Classical BF-theory deals with a further abstraction, that follows naturally from the actions we studied above. 

We will see that in our context it essentially amounts to figuring out a way to remove the dependence on $e$ from $B(e)$, and treat $B$ as an independent variable. This step kills essentially all non-trivial features of General Relativity. This is the usefulness of BF-theory: it is a much simpler theory, straightforwardly related to the physically pertinent theory of General Relativity through the simplicity constraints. 

As in the previous sections, we restrict study to the four dimensional Lorentzian case.

 Up to this point, $B(e)$ was a short-cut notation for $e \wedge e$. We saw that $B(e)^{IJ}=e^I \wedge e^J $ admits a straightforward interpretation as a matrix of orthonormal two-forms, providing the infinitesimal area elements for the six planes defined by the possible pairs of the tetrad basis vectors.
 
  This is a purely geometrical interpretation of $B(e)^{IJ}$. We also mentioned that an object with $n$ antisymmetric Minkowski indices, is naturally interpreted as an $n$ form in the Lie algebra of the Lorentz group.  Equivalently, we may insist on an interpretation of $B^{IJ}$ as a matrix of bi-vector fields, that is, treat $B^{IJ}$ as objects in the exterior algebra of the Minkowski vector bundle over the manifold. 
 
 The idea is to treat $B^{IJ}$ as an independent variable on its own right and not assume a decomposition of the form $B = e \wedge e$. The general form of a bi--vector is $B = a \wedge b + c \wedge d$, and cannot always be decomposed as $B = a \wedge b$. Demanding that there exists an $e$ such that $B^{IJ} = e^I \wedge e^J$ implies that the bi-vector be \emph{simple}, i.e.\! of the form $B = a \wedge b$. 

When deriving the BF form of the Einstein-Hilbert action, we noted in equation \eqref{eq:preSimplicityConstraintHE} that when $B(e) =e \wedge e$ we have
\begin{equation} \label{eq:BBfourvol}
B^{KL}(e) \wedge B^{AB}(e) = \epsilon^{KLAB} vol^4(e)
\end{equation}
We will get rid of the assumption that $B^{KL}(e)$ depends on $e$, by bringing back coordinates and mixed-index objects, in order to express the above relation only in terms of a $B$ field. 

We start from the right-hand side of equation \eqref{eq:BBfourvol}. The four--volume reads
\begin{equation}
vol^4 = \det(e) \, \dd x^1 \wedge \dd x^2 \wedge \dd x^3 \wedge \dd x^4
\end{equation}
and the determinant of the co-tetrad $e^\beta_L$ is
\begin{equation}
\det(e) =  e^\alpha_C e^\beta_D e^\gamma_I e^\delta_J \epsilon_{\alpha \beta \gamma \delta} \epsilon^{CDIJ} \frac{1}{4!}
\end{equation} 
which can be written in terms of the B-field, 
\begin{equation}
\det(e) =  B^{\alpha\beta}_{CD}(e) B^{\gamma \delta}_{IJ}(e) \epsilon_{\alpha \beta \gamma \delta} \epsilon^{CDIJ} \frac{1}{4!}
\end{equation} 
The left hand side of \eqref{eq:BBfourvol} reads
\begin{align} 
B^{KL}(e) & \wedge B^{AB}(e) = B^{KL}_{\mu \nu}(e)  B^{AB}(e)_{\rho \sigma} \dd x^\mu \wedge \dd x^\nu \wedge \dd x^\rho \wedge \dd x^\sigma \nonumber \\&=
B^{KL}_{\mu \nu}(e) B^{AB}(e)_{\rho \sigma} \epsilon^{\mu \nu \rho \sigma} \dd x^0 \wedge \dd x^1 \wedge \dd x^2 \wedge \dd x^3
\end{align}
Equating the two sides, we have
\begin{equation}
 B^{KL}_{\mu \nu}(e) B^{AB}(e)_{\rho \sigma} \epsilon^{\mu \nu \rho \sigma} =  \frac{1}{4!} B^{\alpha\beta}_{CD}(e) B^{\gamma \delta}_{IJ}(e) \epsilon_{\alpha \beta \gamma \delta} \epsilon^{CDIJ} \epsilon^{KLAB} 
\end{equation}
Now, contracting both sides with $\epsilon_{\mu \nu \rho \sigma}$ and $\epsilon_{KLAB}$, we get
\begin{equation} \label{eq:simplicityBofE}
\epsilon_{KLAB} B^{KL}_{\mu \nu}(e)  B^{AB}(e)_{\rho \sigma} = \frac{1}{4!}  B^{\alpha\beta}_{CD}(e) B^{\gamma \delta}_{IJ}(e) \epsilon_{\alpha \beta \gamma \delta} \epsilon^{CDIJ}  \epsilon_{\mu \nu \rho \sigma}
\end{equation}

\medskip

Here we have shown that equation \eqref{eq:simplicityBofE} is always satisfied when $B=e \wedge e$. Now, we may \emph{remove} the dependence on $e$, and write the same equation for an arbitrary bi-vector. We define the difference of the two sides of equation \eqref{eq:simplicityBofE}, to be equal to a set of constraints $C_S[B]$
\begin{align} \label{eq:simpconstraints}
C_S[B]^{\mu \nu \rho \sigma} \equiv \epsilon_{KLAB} B^{KL}_{\mu \nu} \wedge B^{AB}_{\rho \sigma} - \frac{1}{4!}  B^{\alpha\beta}_{CD} B^{\gamma \delta}_{IJ} \epsilon_{\alpha \beta \gamma \delta} \epsilon^{CDIJ}  \epsilon_{\mu \nu \rho \sigma}
\end{align}
 This is called the simplicity constraint(s). We emphasize that $e$ does not appear in this expression. It turns out \cite{livine_consistently_2008, engle_spin_2014-1}, that $C_S[B]=0$ is a necessary and sufficient condition (almost \footnote{There is another solution, $B= \star e \wedge e$, which we do not discuss here. }), ensuring that the bi-vector $B$ comes from a tetrad and is of the form $B^{IJ}=e^I \wedge e^J$. Similar considerations hold when considering the Holst action, by taking the $B$ bi-vector field to have the split
\begin{equation}
B= \tilde{B} + \frac{1}{\gamma} \star \tilde{B}
\end{equation}
in terms of a bi-vector $\tilde{B}$\footnote{We are neglecting the existence of different Plebanski sectors, see for instance \cite{engle_spin_2014-1,engle_proposed_2013,
engle_spin-foam_2013-1}}. Remarkably, if we introduce Lagrange multipliers $\lambda_{\mu\nu\rho\sigma}$ (scalar densities of weight one), imposing $C_S[B]=0$, General Relativity may be written in the form \cite{freidel_bf_1999,baez_spin_1998}
\begin{equation}
S[B,\lambda,\omega] = \int B \wedge \star F(\omega) + \lambda \; C[B]
\end{equation}
The action for classical BF-theory reads
\begin{equation}  \label{eq:BFaction}
S_{BF}[B,\omega] = \int B \wedge \star F(\omega)
\end{equation}
Thus, General Relativity may be written as a constrained BF-theory. As we saw, the simplicity constraint enforces the geometricity of the bi-vector $B^{IJ}$, essentially that it be, for fixed $IJ$, an infinitesimal proper area element.

The beauty of the BF formalism is that the notion of the connection $\omega$, which geometrically we may understand as the parallel transport of vectors in a Riemannian context, is abstracted to a connection in the (principal) bundle of the Lorentz group over the manifold and defines a notion of covariant derivative with respect to the natural derivatives in the group, the right or left actions. Thus, we need not define the connection $\omega$ through the tetrad, as in \eqref{eq:connectionChristoffel}.

The interest in doing all this is that classical $BF$ theory turns out to have no local degrees of freedom at all, it has only global or topological degrees of freedom. All solutions $B$ and $\omega$ to the equations of motions are locally pure gauge: they are all related by symmetries of the action, and the gauge invariant space of solutions, for a given manifold, reduces to a single point, for instance $B=0$ and $\omega=0$. No local ``physical'' information remains.

 BF-theory then resembles more a quantum mechanical system than a quantum field theory, and as such can be quantized with relative ease. Indeed, BF-theory admits an exact spinfoam quantization \cite{ooguri_partition_1992} for the case of compact groups, such as $SU(2)$ and $SO(4)$, pertinent for the 3D and 4D Euclidean gravity toy models. For the Lorentzian case, since the group is non-compact, divergences appear as infinite volume factors. 

\section{Spin-state sum form of EPRL amplitudes}
\label{sec:spinStateSumFormEprl}

In this section we start by introducing the holonomy as the parallel transporter and as the exponential of the connection. We then explain the strategy and main motivations for the EPRL ansatz, as modifying the naive skeletonized path-integral for a discrete topological theory with \slt as a gauge group. We arrive at the EPRL ansatz, defining the model.

\subsection{The holonomy}
\subsubsection{The holonomy as the parallel transporter} \label{sec:holonAsParTransport}
In the first half of this section, we introduce holonomies as parallel transporters $P \equiv P(\omega[e])$, when the connection is the Levi-Civita. The geometrical meaning of the tetrad is straightforward. Holonomies, on the other hand, can appear at first sight as arcane mathematical objects, often defined as solutions to the differential equation solved by the path ordered exponential. This definition is introduced in the second half of this section.

 The holonomy $P_\Upsilon$ corresponds to a simple geometric picture: the Lorentz transformation rotating two tetrad frames at two different points of the manifold to one another, when parallel transporting one of them along a path $\Upsilon$. The transformation property of the holonomy under the Lorentz group will then follow. 

\begin{figure} 
\centering
\includegraphics[scale=0.6]{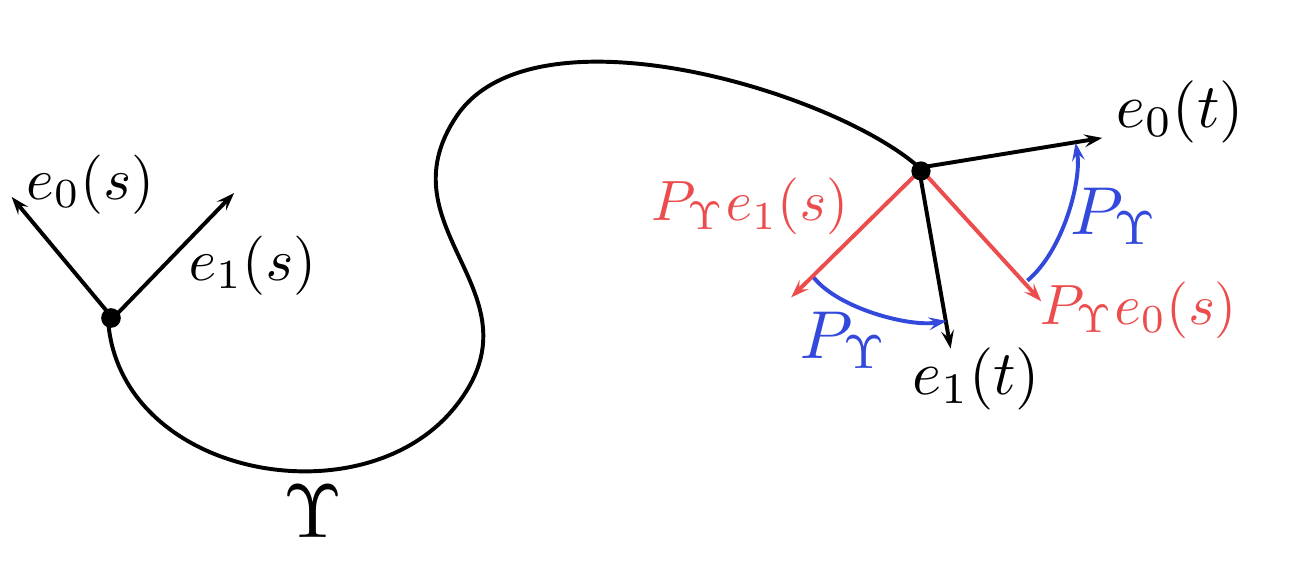} 
\caption[Holonomy as the parallel transporter.]{Holonomy as the parallel transporter. Note the special role of the holonomy $P_\Upsilon$ when acting on the tetrads. $P_\Upsilon \equiv P_\Upsilon(t\!\rightarrow\! s)$ gives the parallel transport for \emph{any} vector from a point $s$ to a point $t$ along a curve $\Upsilon$. Pictured, in red, are the parallel transports $P_\Upsilon e(s)$ at the point $t$, of the tetrad basis vectors at the point $s$, $e_I (s)$. In blue, the role of the holonomy as the Lorentz transformation rotating (and boosting) $P_\Upsilon e(s)$ to the tetrad vector basis $e_I (t)$.}
\label{fig:holonomy}
\end{figure}

We refer to Figure \ref{fig:holonomy}. Consider two arbitrary points on the manifold, connected by a spacetime curve $\Upsilon$. Anticipating notation in following chapters, we call the two points $s$ and $t$, for source and target. 

 The holonomy defines a vector field along $\Upsilon$, for any given vector $V_s$ at $s$, the vector field generated by parallel transporting $V_s$ along $\Upsilon$. Given a vector $V_s$ at $s$, we can parallel transport it along $\Upsilon$ to any other point $t$ on $\Upsilon$, using the covariant derivative. This will define a vector on $t$, which we denote $P_\Upsilon V_s \equiv P_\Upsilon(s\! \rightarrow\!t) V_s$. 
 
 $P_\Upsilon(s\! \rightarrow\!t)$ is the parallel transporter, or, the holonomy of the connection. It is a map from the tangent space at $s$ to the tangent space at $t$
\begin{equation}
P_\Upsilon(s\! \rightarrow\! t) : TM(s) \rightarrow TM(t)
\end{equation}
Thus, in component notation, it will be a matrix. We hereafter suppress the dependence of $P$ on $s$ and $t$, which can be taken to be any two arbitrary points connected by some curve.

Writing both vectors in the corresponding tetrad basis of vectors at $s$,  $V_s=V_s^I e_I(s)$ and $V_s=V_t^I e_I(t)$, we have 

\begin{equation} \label{eq:parTransporter}
V_t^I = P_{\Upsilon \; J}^I \; V_s^J
\end{equation}

 For definiteness, let us choose say $e_0(s)$ at $s$ and $e_0(t)$ at $t$. Tetrad bases $e_I(s)$ are themselves a set of four vectors, that are furthermore normalized. The parallel transport of $e_0(s)$ at $t$ will again be a normalized vector (when the connection is the Levi-Civita). Then, $P_\Upsilon e_0(s)$ and $e_0(t)$ are two normalized vectors of the same Minkowski space. Thus, by definition since it preserves the Minkowski norm, there exists a Lorentz transformation $\Lambda$ which turns one into the other:

\begin{equation} \label{eq:LorentzHolonomy}
\Lambda(t) \, e_0(t) = P_\Upsilon e_0(s)
\end{equation}

The components $e_0(s)$ on the tetrad basis $e_I(s)$ at $s$ may be written as $\delta^I_0$. Similarly, the components of $e_0(t)$ in the basis $e_I(t)$ at $t$ read $\delta^I_0$. Thus,
\begin{equation} 
\Lambda^K_{\ \ I}\, \delta^I_0  = P^K_{\; \Upsilon I} \,\delta^I_0
\end{equation}
Now let us ``free'' the $0$ index and allow it to run over the entire set of four basis vectors,
\begin{equation} 
\Lambda^K_{\ \ I} \, \delta^I_J  = P^K_{\Upsilon\, I}\, \delta^I_J
\end{equation}

Thus, we have shown that, seen as elements of the Lorentz group, 
\begin{equation} 
\Lambda  = P_\Upsilon
\end{equation}
Thus, the holonomy of the Levi-Civita connection is a Lorentz transformation. 

The parallel transports of \emph{all} vectors from $s$ to $t$, are given by $P_\Upsilon$. 
We have seen that, furthermore, $P_\Upsilon$ is \emph{the} Lorentz transformation $\Lambda$ that relates the tetrad frames at two points, when one frame is parallel transported onto the other via some curve
\begin{equation} \label{eq:HolonomyTetrad}
\Lambda(t) \, e(t) = P_\Upsilon e(s)
\end{equation}
This is illustrated in Figure \ref{fig:holonomy}.

 The transformation property of the holonomy, when rotating the tetrad at $t$ and $s$, follows from equation \eqref{eq:parTransporter}.  Rotating arbitrarily the vector basis $e(t)$ and $e(s)$ by some $\tilde{\Lambda}(t)$ and $\tilde{\Lambda}(s)$, gives
\begin{equation}
V_t^I \tilde{\Lambda}(t)^J_I = \tilde{P}_{\Upsilon \; I}^J \; \tilde{\Lambda}(s)^I_K V_s^K 
\end{equation}
Thus, in matrix notation
\begin{equation}
V_t \, =\tilde{\Lambda}^{-1}(t)  \tilde{P}_\Upsilon \tilde{\Lambda}(t)\, \,  V_s 
\end{equation}
from which we conclude by comparison with \eqref{eq:parTransporter} that
\begin{equation}
P_\Upsilon =\tilde{\Lambda}^{-1}(t)  \tilde{P}_\Upsilon \tilde{\Lambda}(s)
\end{equation}
giving the transformation of the holonomy under a  change of a tetrad gauge 
\begin{equation} \label{eq:holonomyGaugeTransformation}
P_\Upsilon \rightarrow \tilde{\Lambda}(t)  \; P_\Upsilon \; \tilde{\Lambda}^{-1}(s)
\end{equation}

\subsubsection{The holonomy as the exponential of the connection}
As we saw above, the holonomy is the parallel transporter. That is, $P_\Upsilon V_s$ is the \emph{solution} of the differential equation giving the parallel transport
\begin{equation}
\frac{D}{\dd \lambda } P_\Upsilon V_s=0
\end{equation}
where $\frac{D}{\dd \lambda }$ is the directional covariant derivative along $\Upsilon$ and $\lambda$ an affine parameter along $\Upsilon$.

Let us write this explicitly. We call $\lambda_s$ the value of the affine parameter at the initial point $s$ and $\lambda$ at $t$. The curve $\Upsilon$ is represented parametrically by $x^\alpha(\lambda)$. The parallel transported vector field generated by $P_\Upsilon$,
\begin{equation} \label{eq:tempParTr}
V^\mu(\lambda)=P^\mu_\rho(\lambda,\lambda_s) V^\rho (\lambda_s)
\end{equation}
solves
\begin{equation} \label{eq:tempParTr2}
\frac{\dd}{\dd \lambda} V^\mu(\lambda) = - A^\mu_\rho(\lambda) V^\rho(\lambda) 
\end{equation}
where we defined 
\begin{equation}
A^\mu_\rho(\lambda) =\Gamma^\mu_{\rho \sigma} \dot{x}^\sigma
\end{equation}
We supress for clarity the dependence of $P$ on the initial value of the affine parameter $\lambda_s$.

Inserting \eqref{eq:tempParTr} to \eqref{eq:tempParTr2}, and since $P$ satisfies this equation for any $V_s$, we see that 
\begin{equation}
\frac{\dd}{\dd \lambda} P^\mu_\rho(\lambda) = - A^\mu_\nu(\lambda) \; P^\nu_\rho(\lambda)
\end{equation}
Suppressing the indices and writing
\begin{equation}
\frac{\dd}{\dd \lambda} P(\lambda) = - A(\lambda) \; P(\lambda)
\end{equation}
we almost have the differential equation which has the exponential of $A$ as a solution, but  $A$ now depends on $\lambda$. 

The formal solution to this differential equation, which we get by iteration, is called the path-ordered exponential, familiar from quantum field theory. It is denoted as 
\begin{equation}
P(\lambda) = \mathcal{P} e^{\int^\lambda\!\dd \lambda'\, A(\lambda')  }  
\end{equation}
In Minkowski indices, using equation \eqref{eq:connectionChristoffel} and after some algebra, we can write a similar expression in terms of the spin-connection, with 
\begin{equation}
A^I_J (\lambda) = \dot{x}^\sigma \omega^{I}_{\sigma J}
\end{equation}
More abstractly, when we understand the spin connection as a Lie algebra valued form, we may write
\begin{equation}
A = A^I \mathcal{J}_I \equiv \epsilon^{IJL} A_{JL} \mathcal{J}_I = \dot{x}_\sigma \omega_{I \sigma J} \epsilon^{IJL} \mathcal{J}_I
\end{equation}
with $\mathcal{J}_I$ a basis of the Lie algebra,
and
\begin{equation} \label{eq:holExpConnection}
P = \mathcal{P} e^{\int A  }  
\end{equation}

\subsection{Spinfoam quantization of BF-theory}
\label{sec:BFSpinfoamQuantization}

The action for classical BF theory, equation \ref{sec:simplicityConstraintsBFtheory}, reads (equation \eqref{eq:BFaction})
\begin{equation} 
S_{BF}[B,\omega] = \int B \wedge F(\omega)
\end{equation}
We discussed in the end of Section \ref{sec:simplicityConstraintsBFtheory}, that thanks to its simplicity and the lack of local degrees of freedom, BF-theory admits an exact spinfoam quantization (when the group is compact). For an introduction to this topic see Section 6 of the review by Perez \cite{perez_spin_2012} and the lectures \cite{baez_introduction_1999} by Baez.

The quantum theory is defined by regularizing (skeletonizing) the path integral for BF-theory on a lattice. The partition function for the continuous theory is formally written as
\begin{eqnarray} \label{eq:formalBFpathintegral}
W^{BF} &=& \int \mathcal{D}[B] \mathcal{D}[\omega] e^{i B \wedge \star F(\omega)} \nonumber \\
&=& \int \mathcal{D}[\omega] \;  \delta\left( F[\omega] \right)
\end{eqnarray}
where we performed a formal integration (like $\delta(x) = \frac{1}{2\pi} \int \dd p \; e^{ipx}$) over the B-field. Notice that this implies that the partition function is the volume of the space of flat connections of $\omega$.

 To give a meaning to the formal expression \ref{eq:formalBFpathintegral}, the manifold $M$ is replaced by an arbitrary cellular decomposition $\mathcal{C}^\star$. We restrict to the case where $\mathcal{C}^\star$ is a (\emph{topological}) triangulation, a  topological simplicial manifold. The notation $\mathcal{C}^\star$ anticipates that the spinfoam amplitudes will be built on the two-complex $\mathcal{C}$ dual to $\mathcal{C}^\star$. The relation between a 2-complex $\mathcal{C}$ and  its dual simplicial triangulation $\mathcal{C}^\star$ in three and four dimensions is explained in Table \ref{tab:triangulation}.
 
  Henceforth, starred quantities correspond to the duals to the elements of the 2-complex and its boundary, which may be geometrical or topological depending on the context. That is, $v^\star$ is the dual 4-simplex to the vertex $v$, $e^\star$ is the dual tetrahedron to the edge $e$ and $f^\star$ is the triangle dual to the face $f$. On the boundary, $\nn^\star$ is the tetrahedron dual to the node $\nn$ and $\ell^\star$ is the triangle dual to the link $\ell$.
 
\begin{table}
\makebox[\textwidth]{
\includegraphics[scale=0.8]{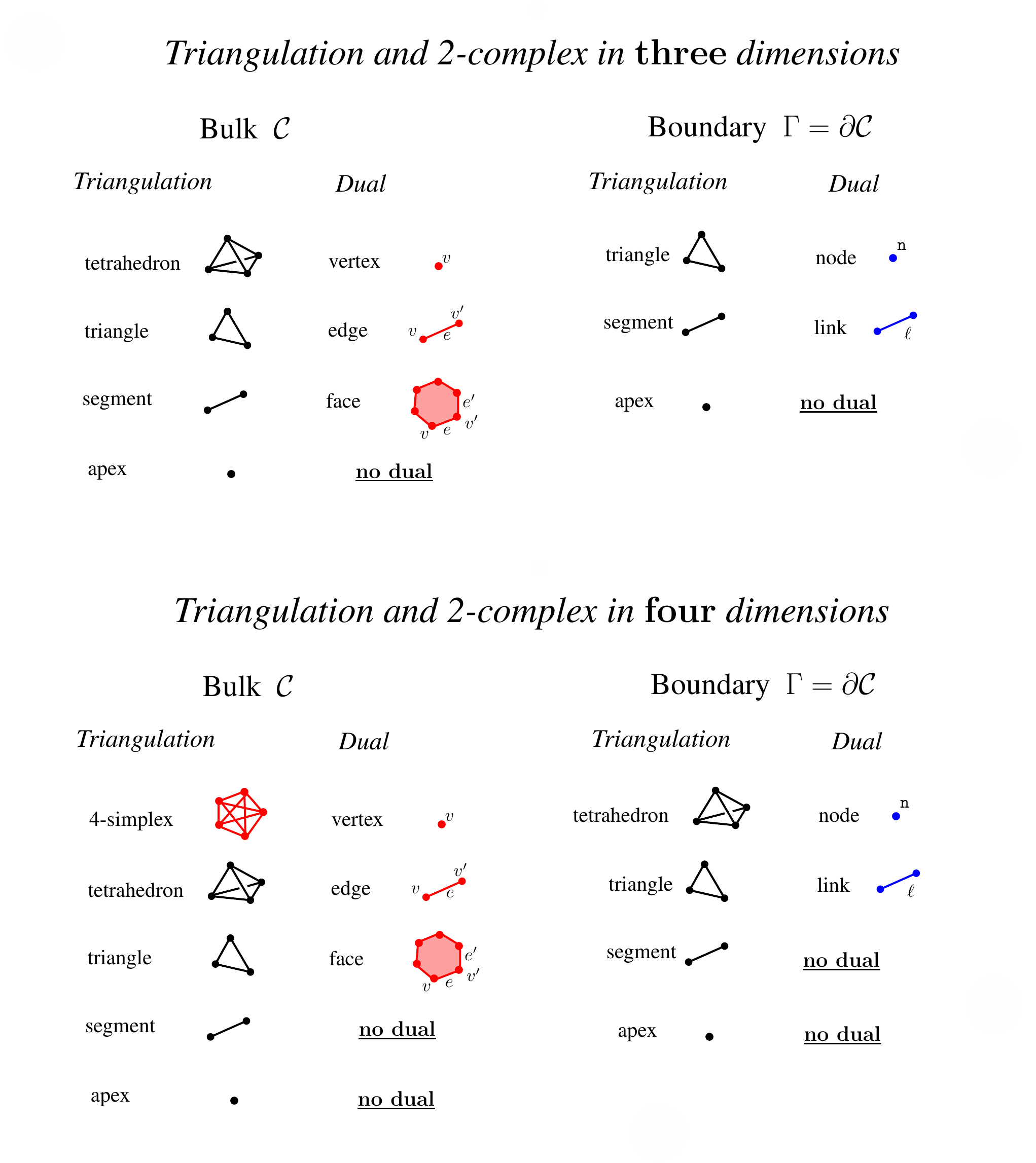}
}
\caption[Triangulation and Dual 2-complex in three and four dimensions.]{Triangulation and Dual 2-complex in three and four dimensions. The 2-complex $\mathcal{C}$ has up to two dimensional objects, the faces $f$, regardless of the dimension of the triangulation. The boundary of $\mathcal{C}$ is a graph $\Gamma$.  }
\label{tab:triangulation}
\end{table}

The variables $B$ and $\omega$ are discretized as follows. We discussed in Section \ref{sec:gameOfActions} that the $B$-field, thought of here as a Lie algebra valued two-form, ``wants'' to be integrated on a two-dimensional submanifold.

 We introduce the variable $B_f$, a Lie algebra element associated to each dual face $f$, which formally corresponds to a smearing of $B$, a discretization, on the triangle $f^\star$ of the triangulation $C^\star$  
\begin{equation}
B_f \sim \int_{f^\star} B
\end{equation}
 
\begin{figure} 
\centering
\includegraphics[scale=0.6]{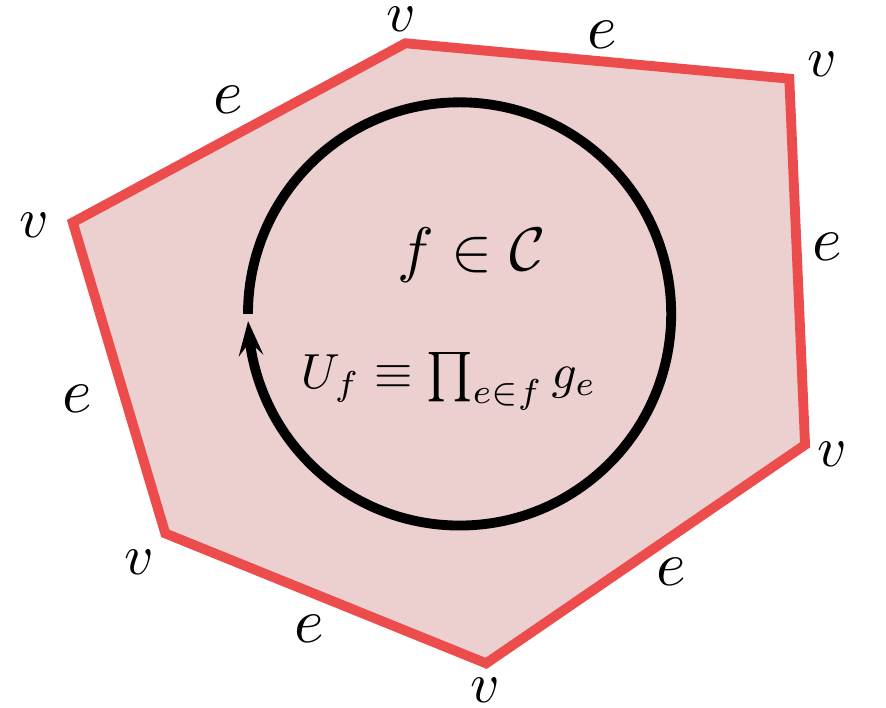} 
\caption[The face holonomy.]{The face holonomy along a dual face $f$ of equation \eqref{eq:faceHolonomy}. Notice the abuse of notation: each edge should be labelled by a different $e$, that is, $e,e',e'',\ldots$, and similarly for the vertices, $v,v',v''\ldots$.}
\label{fig:faceHolonomy}
\end{figure}

The connection $\omega$ is discretized by a group element, formally corresponding to the holonomy $g$ of the connection $\omega$, between the two topological four-simplices $v^\star$ sharing a tetrahedron $e^\star$. Thus, the holonomy is associated to the dual edge $e$ 
\begin{equation}
g_e \sim \mathcal{P}e^{\int_e \mathcal{\omega}},
\end{equation}
see also Figure \ref{fig:faceHolonomy}.

The skeletonized partition function for BF-theory, in analogy with the continuous version \eqref{eq:formalBFpathintegral}, is defined as 

\begin{eqnarray} \label{eq:BFtheoryDiscrPathIntegral}
W^{BF}_\mathcal{C} &=& \int \mu(g_e) \; \delta(U_f)
\end{eqnarray}
where we defined the face holonomy $U_f$  
\begin{equation} \label{eq:faceHolonomy}
U_f \equiv \prod_{e \in f}  g_e
\end{equation}
The integration measure $\mu(g_e)\equiv \prod_e \dd g_e $ is a product of Haar measures $\dd g_e$ over \slt, see Appendix \ref{appsec:sl2cHaar} for details. This construction is sketched in Figure \ref{fig:faceHolonomy}.

\medskip

Let us pause to make a comment. Recalling the discussion at the end of Section \ref{sec:tetradicHE}, in the tetradic and the BF-form of the Hilbert Einstein action, as well as in the tetradic Palatini action, the curvature $F[\omega]$ lives in the co-plane of the $B(e)$-field.

We caution that here we are in a pre-geometric setting and there is no notion of co-plane to a topological triangle $f^\star$ yet. The idea is that when we impose the geometricity of $B_f$ through the simplicity constraints, the curvature of the connection will be encoded in the face holonomies, that can then be interpreted as living in the co-plane associated to $f^\star$. That is, as the holonomy along a closed loop around $f^\star$, in the co-plane of $f^\star$.

We will see in Section \ref{sec:overviewAsAnEPRL} that this is exactly what happens at the critical points of the EPRL partition function that correspond to simplicial geometries. We pointed out that this construction is very similar to the construction behind the Regge action, which we introduce in Section \ref{sec:lorReggeAction}. In turn, this geometrical interpretation of the semiclassical regime of the EPRL model will allow us to perform a path--integral in the spirit of Figure \ref{fig:fireworksAsWMHsum} in Chapter \ref{ch:gravTunneling}.

\medskip

\begin{figure} 
\centering
\includegraphics[scale=0.6]{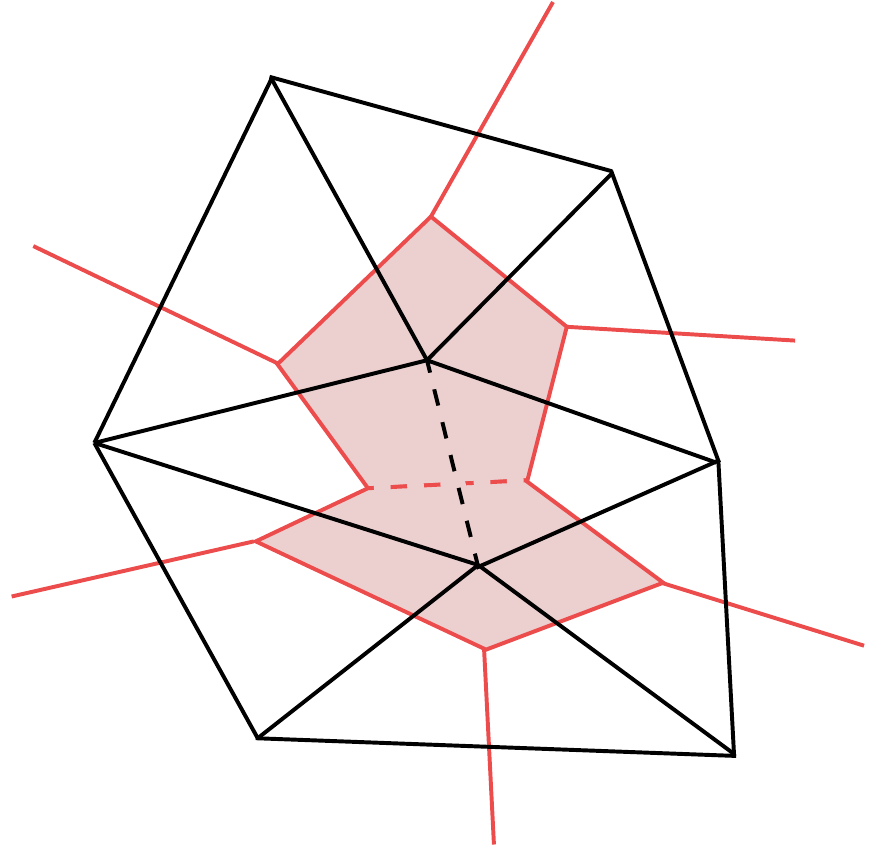} 
\caption[A triangulation in two dimensions.]{A triangulation in two dimensions. Each edge of the dual graph, shown in red, is common to two faces. As an example, the segment in dotted black is dual to the edge in dotted red, which is common to the two faces in pale red. This generalizes in $d$ dimensions: an edge is common to exactly $d$ faces. }
\label{fig:2Dtriang}
\end{figure}

Let us continue. In a simplicial triangulation in $d$ dimensions, each dual edge is connected to exactly $d$ faces.  In two dimensions, an edge is dual to a segment and faces are dual to triangles. A segment is common to exactly two triangles, so the dual edge will be common to the two faces dual to these two triangles. This is shown in Figure \ref{fig:2Dtriang}.

For four dimensions, we proceed analogously. The edge $e$ is dual to a tetrahedron, which is bounded by four triangles. Faces are dual to triangles and thus there are four faces that share the edge $e$.
Thus, for 4D BF-theory with the group being \slt \! , the double cover of the Lorentz group, each \slt group element $g_e$ in  
\eqref{eq:BFtheoryDiscrPathIntegral} appears four times, in four different delta functions, that is, face holonomies. 

The integrations over the group elements $g_e$ can then be performed by expanding the delta functions of the holonomies in unitary representations, using the Peter-Weyl theorem. 

The central idea here is that performing the integrations over the group elements, is like performing the integration over the configuration variable. We will end up with the ``conjugate representation'', where the integration (summation) is now over spin variables (\emph{discrete} variables) playing the role of the conjugate momentum. For \slt \!, because it is a non-compact group, there will also be some labels that are continuous.

These quantum numbers will label irreducible representations of the Lorentz group and intertwiners, equivariant maps under the action of the group, among representations. The principal series of unitary representations of \slt is introduced in Section \ref{sec:elementsOfSL2CrepTheory}.

However, doing all this would only give us a tentative quantization of BF-theory with \slt as a gauge group. Our aim is to impose the simplicity constraints at the quantum level. For this, we need to manipulate the BF-partition function \eqref{eq:BFtheoryDiscrPathIntegral} in a form that brings in also the group $SU(2)$, on which the canonical theory is based. This is done in the following section.

\begin{center}
\rule{0.5 \textwidth}{0.4pt}
\end{center}

For lack of space and to avoid technical details not necessary for what follows, we do not present recoupling techniques relevant to \slt here. They are presented in Appendix \ref{app:recoupTheory}. 

To get a feeling of how the integration over the group elements gives rise to recoupling invariants see Appendix \ref{appsec:integratingDeltas}, where we work out a simple example and introduce the four valent intertwiners of $SU(2)$, intertwining between the four representations of the four faces living at an edge $e$. In turn, the invariant integration measure, the Haar measure over \slt, can be decomposed in two copies of an $SU(2)$ Haar measure and an integration over a real parameter that is understood as a boost. This allows us to write explicitly the recoupling invariants in terms of analytic functions. In Appendix \ref{appsec:sl2cHaar} we introduce the \slt Haar measure and its decomposition and in Appendix \ref{appsec:CGsl2c} we show the precise relation of the Clebsch-Gordan coefficents of \slt and the recoupling invariants with which we can explicitly express EPRL, as was done in \cite{christodoulou_planck_2016} for the transition amplitude summarized in Section \ref{sec:explicitAmp}. In Appendix \ref{appsec:sl2cHaar}, we also show the explicit sense in which the simplicity constraints are imposed at the quantum level in this decomposition. For further details see the work
\cite{speziale_boosting_2017-1} by Speziale where these techniques were recently introduced.

\subsection{The EPRL ansatz}
\label{sec:eprlAnsatz}

In this section we introduce the spin state-sum form of the EPRL amplitudes, completing the correspondence sketched in Table \ref{tab:aimEPRL}. We will arrive at the form of the EPRL amplitudes as in equations \eqref{eq:EPRLampWMH} and \eqref{eq:faceAmpWMH} that are the starting point of the next chapter.

We start from equation \eqref{eq:BFtheoryDiscrPathIntegral}, defining the regularized path integral for BF-theory, repeated here for convenience
\begin{eqnarray} 
W^{BF}_\mathcal{C} &=& \int \mu(g_e) \; \delta(U_f)
\end{eqnarray}
We remind that the integration measure $\mu(g_e)$ is a product of \slt Haar measures, defined in Appendix \ref{appsec:sl2cHaar},
and we are integrating over $E$ copies of \slt, where $E$ is the number of edges in $\mathcal{C}$.

We would now like to  arrive at 
a spin state-sum form of the amplitudes, with an overall sum over spin configurations, as in \eqref{eq:EPRLampWMH}.

The prototype inspiration for spin state--sum models is the three-dimensional Euclidean Ponzano--Regge model, whose partition function reads \cite{barrett_ponzano-regge_2009,ponzano_semiclassical_1968}

\begin{equation} \label{eq:ponanoReggePartitionFunction}
W^{PR}_\mathcal{C} = \sum_{\{j_f\}}\, \mu(j_f) \;\prod_v \begin{Bmatrix}  j_1&j_2&j_3\\j_4&j_5&j_6\end{Bmatrix}
\end{equation}
where the symbol in curly brackets is the 6j-symbol, an invariant object in the tensor product of six representations of $SU(2)$. The summation measure $\mu(j_f)$, is a product of $d_j=2j+1$ factors and sign factors, see \cite{barrett_ponzano-regge_2009,ponzano_semiclassical_1968}. We will not need its explicit form here. Since we are in three dimensions, the vertices $v$ here are dual to tetrahedra and the faces$f$ are dual to segments. 

The 6j-symbol is then understood as a quantum tetrahedron, in the sense that in an appropriate semiclassical limit of large spins, we may understand them as the edge--lengths of a classical tetrahedron. 

A classical Euclidean tetrahedron is completely determined by its six edge lengths (up to inversion). In the limit of large spins, the 6j--symbol becomes the exponential of the Regge action $S_R(v;l_{f^\star})$ for the four simplex $v$, and we have the behaviour \cite{barrett_ponzano-regge_2009,ponzano_semiclassical_1968}
\begin{equation}
W^{PR}_\mathcal{C} \sim \int \! \dd{l_{f^\star}} \; \mu(l_{f^\star}) \;\prod_v e^{S_R(v;l_{f^\star})} = \int \! \dd{l_{f^\star}} \; \mu(l_{f^\star}) e^{S^\mathcal{C}_R(l_{f^\star})}
\end{equation}
where we have turned the sums into integrations and identified the segment lengths $l_{f^\star}$ with the spins $j_f$ on the face $f$ dual to the segment $f^\star$. The sum of the Regge action for each vertex gives the Regge action for the 2-complex $S^\mathcal{C}_R(l_{f^\star})$.

This is the sense in which we will recover the path integral over the Hilbert--Einstein action in Chapter \ref{ch:AsympAnalysis}. We will be using similar manipulations in Chapter \ref{ch:gravTunneling} to see how a gravitational tunneling picture emerges when considering a semiclassical boundary state; which serves as a boundary condition for the analogous naive path integral in four dimensions. We caution that the quantum numbers in the EPRL model are not only the spins $j_f$, there are also the intertwiner degrees of freedom and the continuous boost labels, see Appendix \ref{appsec:CGsl2c}. 

We are aiming to arrive at a similar expression to equation \ref{eq:ponanoReggePartitionFunction}, of the form\footnote{The summation measure $\mu(j_f)$ is not of much importance in what follows. The precise choice can be important for the finiteness of the amplitudes, see \cite{mikovic_finiteness_2013}.} 
\begin{equation} \label{eq:statesum}
W_\mathcal{C} = \sum_{j_f} \mu(j_f) \prod_v A_v(j_f)
\end{equation}

The idea here is two--fold: the first, is to introduce variables that are local to a vertex $v$, in this way we can write the amplitude as a product of terms local to the vertices, the vertex amplitudes. The second, is to introduce fiducial boundaries, isolating each vertex from the others, where the $SU(2)$ group elements will live. 

This is to allow contact with the boundary space of LQG, which is based on $SU(2)$, and to provide a way to impose the simplicity constraint. 

Some steps in the manipulations that follow work only for the case of compact groups. We point them out as we go. These manipulations lead to well-defined and finite expressions when compact groups are involved. The EPRL ansatz can be understood as carrying out formally these manipulations, ignoring infinite volume factors, and is justified à posteriori by the fact that the amplitudes are finite \cite{baez_integrability_2001,crane_finiteness_2001,
 perez_spin_2001,crane_perturbative_2001, mikovic_finiteness_2013,kaminski_all_2010}. 
 
 \bigskip

The first step is to double the number of group variables, by splitting $g_e$ in two, called $g_{ev}$ and $g_{ev'}$, where $v$ and $v'$ are the vertices joined by the edge $e$. 

For a compact group, we may do this as follows. We write $g_e =g_{ve} g_{ev'}$, where $g_{ve}=g_{ev}^{-1}$. Using the invariance of the Haar measure, we change the integration from $\dd g_{e}$ to say $\dd g_{ve}$. Notice that this is only an integration over half the variables, we left out all the $g_{ev'}$. We now have
\begin{equation}
W_C = \int  \mu(g_{ve}) \prod_f \delta\!\left( \prod_{v\in f} g_{ev}g_{ve'} \right) 
\end{equation}
where the integration is over $E$ copies of the group. Now, in a compact group $G$, we may write
\begin{equation}
1 = \frac{\prod_e \int \dd g_{ev'}}{\mathcal{V}_G^E}
\end{equation}
where $\mathcal{V}_G$ is the volume of $G$, and absorb the volume factors in a normalization constant $\mathcal{N}_C$
\begin{equation}
W_C = \mathcal{N}_C \int  \mu(g_{ve},g_{ev'}) \prod_f \delta\!\left( \prod_{v\in f} g_{ev}g_{ve'} \right) 
\end{equation}
where the integration is over $2E$ copies of $G$. 

Let us focus again on the Lorentzian case and ignore the infinite volume factors. We write simply $\mu(g_{ev})$ to imply an integration over all variables, that is, over $2E$ copies of \slt
\begin{equation}
\mu(g_{ev}) = \prod_{ev} \dd g_{ev}
\end{equation}
 Some of these integrations, one per vertex, turn out to be superfluous, leading to infinite volume factors, and have to be dropped. This can be seen by working out an example for a single vertex. Henceforth, by $\mu(g_{ve})$ we mean that one of the integrations over $g_{ve}$ per vertex has been dropped.
 
  We have
\begin{equation}
W_C = \mathcal{N}_C \int_\sl2c  \mu(g_{ve}) \prod_f \delta\!\left( \prod_{v\in f} g_{ev}g_{ve'} \right) 
\end{equation}

The next step is to introduce a fiducial boundary around each vertex. This is done as follows. We introduce another set of group variables, that are now $SU(2)$, equal to the number of pairs of vertices and faces $V \times F$, which we denote $h_{vf}$.

We impose that they be equal to the product of the two \slt elements that are attached to the same vertex $v$ in the face $f$, by inserting a delta $\delta(g_{ev}g_{ve'}h_{vf})$. Thus, the holonomy around the face will now read 
\begin{equation} \label{eq:faceHolonomySU2}
h_f \equiv  \prod_{vf} h_{vf}
\end{equation}
Putting things together, we have 
\begin{equation}
W_\mathcal{C}= \int_{SU(2)} \mu(h_{vf})
 \prod_f \! \delta(h_f) \; \prod_{vf} \int \mu(g_{ev}) \delta(g_{ev}g_{ve'}h_{vf}) 
\end{equation}

We have almost arrived at equation \eqref{eq:faceAmpWMH} and \eqref{eq:EPRLampWMH}. The idea is now to expand the delta function over \slt in unitary representations. However, remember that we have not yet imposed the simplicity constraints at the quantum level.

The EPRL ansatz is based on the following conjecture: the imposition of the simplicity constraints at the quantum level is equivalent to expanding the delta function in only the so--called \emph{simple} unitary representations of the principal series of \slt. 

 The motivations for this strategy come from the canonical analysis of the Holst action in the time--gauge and are discussed in Sections \ref{sec:elementsOfSL2CrepTheory} and  \ref{sec:weakImpositionSimpConstr}. See also Appendix \ref{appsec:CGsl2c}. 

We thus arrive at equations \eqref{eq:faceAmpWMH} and \eqref{eq:EPRLampWMH}
\begin{eqnarray}  \label{eq:theEprlAnsatz}
W_\mathcal{C} = \sum_{\{j_f\}} \mu(j_f) \; \int_{\sl2c}  \! \mu(g_{ve}) \; \prod_f A_f(g_{ve},j_f)  \\
A_f(g_{ve},j_f) = \int_{SU(2)} \! \mu(h_{vf}) \; \delta{(h_f)} \tr^{j_f} \left[\prod_{v \in f} Y_\gamma^\dagger g_{e'v}g_{ve} Y_\gamma  h_{vf} \right] 
\end{eqnarray}
This definition of the EPRL model is the starting point of the following chapter. The trace notation $\tr^{j_f}$ means that the expansion is only over simple irreducible representations of the principal series of \slt, and will be given a precise meaning in Section \ref{sec:weakImpositionSimpConstr}. 

To finish this section, we give the standard EPRL model definition  \cite{engle_flipped_2008, engle_spin-foam_2013-2,baratin_group_2012,dupuis_holomorphic_2012,
freidel_new_2008-1,barrett_lorentzian_2000,rovelli_new_2011,rovelli_simple_2011}
in its product over vertex amplitudes form, which is a rearrangement of the equations above:

\begin{align} \label{eq:theEprlAnsatzVertexForm}
& W_\mathcal{C} = \int_{SU(2)} \!  \mu(h_{vf}) \;\prod_f  \delta{(h_f)} \;  \prod_v A_v(h_{vf},g_{ve},j_f)  \\
& A_v(h_{vf},g_{ve},j_f) = \sum_{\{j_f\}}  \int_{\sl2c}  \! \!\!\: \mu(g_{ve})\,  \prod_f d_{j_f}   \tr^{j_f} \left[\prod_{v \in f} Y_\gamma^\dagger g_{e'v}g_{ve} Y_\gamma  h_{vf} \right]  \nonumber \\
\end{align}
from which we also read the summation measure
\begin{equation}
\mu(j_f) = \prod_f d_{j_f}
\end{equation}

\section{Relation to the canonical theory}
\label{sec:relationToCanonical}
In this Section we outline how covariant LQG is placed within the body of results of Loop Quantum Gravity. The main insight behind the spinfoam quantization program, is that the physical inner product, at least formally, may be written as an inner product in the kinematical space, which puts the physical inner product at a similar footing as a propagator.  

The manipulations that follow are formal. For a more detailed discussion see Thiemann's contribution in \cite{giulini_quantum_2003} and the influential work by Reisenberger and Rovelli  \cite{reisenberger_spin_2000,rovelli_quantum_1997,reisenberger_sum_1997,reisenberger_space-time_2001,rovelli_spacetime_2000}.

The Einstein-Hilbert action, in canonical form as a fully constrained system and in terms of Ashtekar variables reads
\begin{equation} \label{eq:canonicalGRAshtVars}
S[E;A,N,N^a,\Lambda^i] = \int \dd t \, \dd x^3 \int_\Sigma   E^a_j\dot{A}^j_a - \left( \Lambda^i G_i + N^a D_a + N H \right)
\end{equation}
The conjugate variables here are the densitized triads $E^a_j$ and the connection $\dot{A}^j_a$, see also Section \ref{sec:kinematicsLQG}.

Implied above is that $\dot{A}^j_a(E^a_j,A^j_a)$, similarly for $G_i$, $D_a$ and $H$ and that $E^a_j=E^a_j(t,x)$,  $A^j_a(x,t)$, similarly for $\Lambda^i$,$N^a$ and $N$ . 

Compare with the Hamiltonian form of the action for a 1D system 
\begin{equation}
S[p;q] = \int \dd t \;  p\,\dot{q} - \mathcal{H}(q,p)
\end{equation}
where it is implied, as in \eqref{eq:canonicalGRAshtVars}, that $\dot{q} = \dot{q}(q,p)$, $p=p(t)$,  $q=q(t)$. The semicolon in these actions separates the momenta from the configuration variables. 

The conjugate momenta for the variables $\Lambda^i$, $N^a$ and $N$ do not appear in the action for General Relativity, they vanish identically. The Euler-Lagrange equations for the configuration variables $\Lambda^i$, $N^a$ and $N$ simply impose the condition that $G_i$, $D_a$ and $H$ vanish. These are the constraints. The Hamiltonian is a sum of constraints multiplied by Lagrange multipliers. This is the general form of a fully constrained system. That is, on shell (when the constraints are satisfied) the Hamiltonian vanishes.    

The picture is that of a phase space spanned by all configuration variables and their conjugates. The conjugate of the Lagrange multipliers vanishes, which trivially reduces the phase space. Then, the constraints further reduce the phase space dimension on the constraint surface. In addition, we need to gauge--out the orbits generated by these constraints. The orbits of the constraints are the physical states.

The inability to explicitly describe the physical space resulting from this last step is what has become known as the problem of observables in general relativity \cite{bergmann_observables_1961, torre_gravitational_1993,
dittrich_chaos_2015,
dittrich_perturbative_2007,dittrich_partial_2006-1,
rovelli_partial_2002,dittrich_partial_2006}, an issue brought up as fundamental by Bergmann in 1961 \cite{bergmann_observables_1961}. 

This is the underlying reason that although a precise and well--defined proposal for the Hamiltonian constraint of LQG exists through Thiemann's master constraint program \cite{thiemann_phoenix_2006}, the existence of the full space of solutions is shown by invoking the axiom of choice \cite{thiemann_loop_2007}. As we discussed in the Introduction, ``sidestepping'' these profound technical difficulties of the complete Hamiltonian formulation, are one of the main motivations for introducing the covariant approach to LQG.

 Holst showed \cite{holst_barberos_1996} that the action \eqref{eq:canonicalGRAshtVars} is the 3+1 split of the action \eqref{eq:HolstAction}, which then took his name.\footnote{Second class  constraints  must be satisfied, in order for \eqref{eq:HolstAction} to reduce to \eqref{eq:canonicalGRAshtVars}.  }

This is the main link between the canonical and the covariant formulations of Loop Quantum Gravity. We now identify the reason for the naming ``covariant'' approacht to LQG: the Holst action is manifestly covariant. To be precise, all objects appearing in the Holst action transform covariantly (or contra-variantly) and the action is invariant under a simultaneous action of the Lorentz group on all indices. This is in contrast with the canonical form \eqref{eq:canonicalGRAshtVars}, in which we have objects with only space indices and manifest Lorentz invariance is lost.

The idea behind spinfoams can then be sketched as follows. The kinematical space of General Relativity, written in terms of Ashtekar's variables, is known explicitely and is well understood. It is a separable Hilbert space with a well-defined inner product, given by the Ashtekar-Lewandowski measure \cite{ashtekar_differential_1995}. The states in this Hilbert space solve the operator form of the Gauss and diffeomorphism constraints. 

Let us call the space of states that \emph{also} solve the Hamiltonian constraint, the physical Hilbert space $\mathcal{H}^{phys}$. Then, we are interested in general to calculate inner products, amplitudes, between states in $\mathcal{H}^{phys}$. In principle, the modulus squared of these contains the physically observable information of quantum gravity. In this manuscript, we attempt to calculate such an amplitude. 

Consider a state $\ket{\psi_{kin}}$ in $\mathcal{H}_{kin}$. Then, in general, $\ket{\psi_{kin}}$ is not in $\mathcal{H}_{phys}$, which is a subspace of $\mathcal{H}_{kin}$. However, generically, $\ket{\psi_{kin}}$ will also not be orthogonal to $\mathcal{H}_{phys}$. Thus, a component of $\ket{\psi_{kin}}$, is in $\mathcal{H}_{phys}$, that is, the projection of $\ket{\psi_{kin}}$ under the Hamiltonian constraint $\hat{H}$ is a state in $\mathcal{H}_{phys}$.

A kinematical state $\ket{\psi_{kin}}$ can be mapped to a (generalized) physical state by  	
\begin{equation}
\ket{\psi_{kin}} \rightarrow \delta{\hat{H}} \ket{\psi_{kin}} \in \mathcal{H}_{phys}^*
\end{equation}
where $\mathcal{H}_{phys}^*$ is the rigged physical Hilbert space and this is achieved via the so-called rigging map. Loosely speaking and to give an idea, the rigged Hilbert space construction is meant to generalise the standard notion of Hilbert space in order to allow for non-square integrable states that are physically interesting and the distributions that arise when fourier transforming them. For instance, the plane waves $e^{ipx}$ of quantum mechanics are eigenstates of the position and momentum operators but are not square integrable. The fourier transform in position space of $e^{ipx}$, thought of as an eigenstate $\ket{p}$ of the momentum operator, is $\delta(x)= \int \dd e^{ipx}$.  See
\cite{rovelli_quantum_2004,thiemann_modern_2007} for a detailed explanation.

The idea here is that if $\delta(\hat{H})$ is projecting $\ket{\psi_{kin}}$ to the physical Hilbert space, then the kinematical inner product of another state $\ket{\psi'_{kin}}$ with $\ket{\psi'_{kin}}$, is in effect the inner product of the component of $\ket{\psi'_{kin}}$ that is in $H_{phys}$ with the component of $\ket{\psi_{kin}}$ that is in $H_{phys}$. That is, writing 
\begin{equation}
\bra{\psi'_{kin}} = \bra{\psi^K_{kin}} + \bra{\psi^P_{kin}}
\end{equation}
with $\bra{\psi^P_{kin}} \in \mathcal{H}^{phys}$ and $\bra{\psi^K_{kin}} \perp \mathcal{H}^{phys}$, and similarly for $\ket{\psi_{kin}}$, then 
\begin{equation} 
\braket{\psi'_{kin} \vert \delta(\hat{H}) \vert\psi_{kin} } = \braket{\psi'^P_{kin} \vert \psi^P_{kin}}
\end{equation}
This is a central idea to the spinfoam program: if one gives a meaning to $\delta(\hat{H})$, it is formally possible to compute physical transition amplitudes, using the kinematical inner product. 

The next step is to write $\braket{\psi'_{kin} \vert \delta(\hat{H}) \vert\psi_{kin} }$ as a path integral over the lapse $N$
\begin{equation}
\delta{\hat{H}} = \int \mathcal{D N} e^{i \hat{H}[N]}
\end{equation}
with the integration taken over two time slices and the boundary conditions being that the lapse is constant on these slices. We have
\begin{equation}
\braket{\psi'^P_{kin} \vert \psi^P_{kin}}  =\int \! \mathcal{D N} \braket{\psi'_{kin} \vert e^{i \hat{H}[N]}
 \vert\psi_{kin} }
\end{equation}
which now looks like a propagator formula between times $t_i$ and $t_f$, with a multi-fingered time evolution. That is, we integrate over all intermediate choices of time slicings,  choices of the lapse function. We may formally turn this into an integral over the entire classical phase space and write 

\begin{equation}
\int \mathcal{D [\cdots]} \braket{\psi'_{kin} \vert e^{i S_{EH}[A,E,N,\vec{N},\Lambda]}
 \vert\psi_{kin} }
\end{equation}
where $\mathcal{D [\cdots]}$ is a path-integral measure over $A,E,N,\vec{N},\Lambda$.

But, we saw earlier in this section that $S_{EH}$ written in Ashtekar variables is the 3+1 split of the Holst action. Thus, we may write

\begin{equation}
\int \mathcal{D}[\omega]\, \mathcal{D}[e]\; \braket{\psi'_{kin} \vert e^{i S_{H}[\omega,e]}
 \vert\psi_{kin} }
\end{equation}
From this point, one can write the Holst action as a constrained BF-theory, as we did in Section \ref{sec:simplicityConstraintsBFtheory}, and continue with the BF-theory spinfoam quantization of Section \ref{sec:BFSpinfoamQuantization}.

\section{Summary} \label{sec:summaryJourney}
In this introductory chapter we presented main ideas and motivations behind the EPRL ansatz. The steps we took can be understood as a formal derivation of the ansatz. In the next chapter, we will take the partition function \ref{eq:theEprlAnsatz} as a starting point and study its semiclassical limit.

The continuous and discrete covariant actions at the basis of the covariant approach to LQG are summarized in Table \ref{tab:actionsTable}, where we included the dimensions of each term for easier comparison. We have included the Wilson--Lattice Yang-Mills action, to be compared with the discretized BF-theory action. Note that the canonical form of General Relativity, equation \ref{eq:canonicalGRAshtVars}, resembles a Yang-Mills theory with gauge group $SU(2)$. 

We have emphasized the role of the plane $e \wedge e$, which has dimensions of an area, and the co-plane where the curvature term $\star F[\omega]$ lives, which is dimensionless and can be discretized by a deficit angle. In the next chapter, we will see this geometrical construction emerging in the semiclassical limit of the EPRL model, giving rise to the Regge action.

\begin{table}
\begin{equation}
\def\arraystretch{2}
\begin{array}{@{}llll@{}}
\toprule
\\
{\scriptstyle [\hbar]}\, S_{EH}[g]\,= & {\scriptstyle [\hbar^2]} \, \int \dd x^4  & {\scriptstyle [\hbar^0]} \, \sqrt{-\det g(x)} \delta^\alpha_\gamma \; \delta^\beta_\delta \ \ \ & R^{\gamma \delta}_{\ \ \ \alpha \beta}(g(x)) \,  {\scriptstyle [\hbar^{-1}]}\\  \\ 

{\scriptstyle [\hbar]}\,S_{EH}[e]\,= & {\scriptstyle [\hbar^2]} \, \int \dd x^4 & {\scriptstyle [\hbar^0]} \, \vert \det e(x) \vert \; e^\alpha_{\ \ I} e^\beta_{\ \ J} & R^{I J}_{\ \ \ \alpha \beta}(e(x)) \,  {\scriptstyle [\hbar^{-1}]} \\  \\ 

{\scriptstyle [\hbar]}\,S_{P}[e,\omega]\,= & {\scriptstyle [\hbar^2]} \, \int \dd x^4  & {\scriptstyle [\hbar^0]} \, \det e(x) \;  e^\alpha_{\ \ I} e^\beta_{\ \ J} & \Omega^{I J}_{\ \ \ \alpha \beta}(\omega(x))\,  {\scriptstyle [\hbar^{-1}]} \\ \\

{\scriptstyle [\hbar]}\,S_{P}[e,\omega]\,= & {\scriptstyle [\hbar^0]}\,\int    & {\scriptstyle [\hbar]} \, e \wedge e & \wedge \star F(\omega)\,  {\scriptstyle [\hbar^0]}  \\ \\

{\scriptstyle [\hbar]}\,S_{H}[e,\omega]\,= & {\scriptstyle [\hbar^0]}\, \int  & {\scriptstyle [\hbar]} \, e \wedge e \left( 1 + \frac{\star}{\gamma} \right) & \wedge \star F(\omega)\,  {\scriptstyle [\hbar^0]} \\  \\

{\scriptstyle [\hbar]}\,S_{H}[e,\omega]\,= & {\scriptstyle [\hbar^0]} \, \int  & \tr[\; {\scriptstyle [\hbar]} \, e \wedge e \left( 1 + \frac{\star}{\gamma} \right) & \wedge F(\omega)\,  {\scriptstyle [\hbar^0]} \; ] \\  \\

\midrule
\\
{\scriptstyle [\hbar]}\,S_{R}[l]\,= & {\scriptstyle [\hbar^0]} \, \sum_f & {\scriptstyle [\hbar]} \, A_f(l) & \phi_f(l)\,  {\scriptstyle [\hbar^0]} \\ \\
\midrule
\\
{\scriptstyle [\hbar]}\,S_{BF}[B,\omega]\,= & {\scriptstyle [\hbar^0]} \, \int & \!\! \tr[\ \ \;{\scriptstyle [\hbar]} \, B & \wedge F\,  {\scriptstyle [\hbar^0]} \;] \\ \\ 

{\scriptstyle [\hbar]}\,S_{WL}[L,U]\,= & {\scriptstyle [\hbar^0]} \, \sum_f & \!\! \tr[\ \ \; {\scriptstyle [\hbar]} \, L_f & U_f \,  {\scriptstyle [\hbar^0]}\;] \\ \\
\bottomrule
\end{array}
\nonumber
\end{equation}
{\small
\caption[Comparison of actions and formalisms leading to the definition of EPRL amplitudes.]{Comparative table of actions and formalisms leading to the definition of EPRL amplitudes. Dimensions in geometrical units ($G=c=1$) are indicated in powers of $\hbar$ next to each term for easier comparison. 
} 
\label{tab:actionsTable}
}
\end{table}
\chapter{Emergence of the sum over geometries} \label{ch:AsympAnalysis}
In the previous chapter we have seen how the spin state-sum form of the EPRL amplitude is related to General Relativity. In this chapter we take the top--down path, starting from the EPRL ansatz and see how the sum--over--geometries path--integral picture that emerges in the semi-classical limit. 

 We will work in the Han--Krajewski representation \cite{han_path_2013}. This is a little-known and more direct way of getting to the form of the EPRL amplitudes suitable for asymptotic analysis, that does not involve coherent states. We provide an independent and more concise derivation of this representation. 
 
 We discuss the geometrical interpretation of the critical point equations, summarize the asymptotic analysis of the EPRL model and discuss the sense in which the Wheeler--Misner--Hawking sum--over--geometries is recovered. Unpublished results in this and the following chapter will appear in a work \cite{mariosFabio} by the author in collaboration with Fabio d'Ambrosio.  

\section{The rough picture}
\label{sec:emergenceOfGR}
 In this section we give the rough picture of how the path--integral over the Regge action emerges from the EPRL ansatz, to be used as a guide for the presentation of the material in the following sections. 

In Section \ref{sec:spinStateSumFormEprl}, we saw that the amplitudes\footnote{We have not yet introduced a boundary. This is done in the following chapter. $W_\mathcal{C}$ here is understood as a partition function.} of covariant LQG can be defined by the EPRL ansatz, written in the product--over--face--amplitudes form of equation \eqref{eq:theEprlAnsatz}, repeated here for convenience
\begin{equation} 
W_\mathcal{C} = \sum_{j_f} \mu(j_f) \; \int_{\sl2c}  \! \mu(g_{ve}) \; \prod_f A_f(g_{ve},j_f) 
\end{equation}
with the face amplitude given by
\begin{equation}
A_f(g_{ve},j_f) = \int_{SU(2)} \! \mu(h_{vf}) \; \delta{(h_f)} \tr^{j_f} \left[\prod_{v \in f} Y_\gamma^\dagger g_{e'v}g_{ve} Y_\gamma  h_{vf} \right] 
\end{equation}
In the following section we use spinor technology to bring the face amplitude to the form
\begin{equation}
A_f(g_{ve},j_f) = \int \! \mu(z_{vf})\, \prod_f e^{j_f F_f(\{g_f\},\{z_f\})}
\end{equation} 
with the integration being over spinor variables $z_{vf}$. The notation $\{g_f\},\{z_f\}$ implies dependence on the $g_{ve}$ and $z_{vf}$ with $v \in f$. For conciseness, we will also write  $F_f(g,z)$ instead of $F_f(\{g_f\},\{z_f\})$.

 The main point is that the EPRL amplitudes take the form
\begin{equation}
W_\mathcal{C} = \sum_{\{j_f\}} \mu(j_f) I(j_f)
\end{equation} 
with $I(j_f)$ given by 
\begin{equation} \label{eq:partialAmplitude}
I(j_f) = \int \! \mu(g) \mu(z) \, e^{\sum_f j_f F_f(g,z)}
\end{equation}
where the product over faces is written as a sum in the exponent. The function $I(j_f)$ is the so-called ``partial amplitude'' or fixed-spins amplitude. Keep in mind that despite the name, $I(j_f)$ does not have a direct interpretation as an amplitude of any sort. Thus, the results presented here regard the asymptotic analysis at \emph{fixed spins}. To get a physically meaningfull amplitude, we must include a boundary and perform the spin--sum. This is the subject of Chapter \ref{ch:gravTunneling}.

The partial amplitude $I(j_f)$ is then the function of the spins that turns out to behave as the exponential of the Regge action in the semi-classical limit. The exponent is \emph{linear} in the spins, with the function $F_f(g,z)$ not depending on the spins. 

 What roughly happens after the asymptotic analysis is carried through, is that at a critical point admitting a geometrical interpretation as a simplicial geometry, $F_f$ becomes the deficit angle $\phi_f$, which is now a function of the spins evaluated at the critical point. 
 
  The spins are identified with physical areas $a_f$ by $a_f/\hbar \equiv j_f$

\begin{equation}
F_f(g,z) \; \stackrel{\text{crit. point}}{\sim} \; \phi_f(a_f)
\end{equation}
Then, the argument of the exponential in  \eqref{eq:partialAmplitude} becomes
\begin{equation}
\sum_f j_f F_f(g,z) \sim \frac{1}{\hbar} \sum_f a_f \; \phi_f(a_f)
\end{equation}
where we recognize the Regge action, with the important difference that the variables here are the areas $a_f$, not the segment lengths.

\section{Embedding $SU(2)$ in \slt and a taste of spinors}
\label{sec:prelim}
In this section we introduce the minimal amount of tools needed to bring the EPRL amplitudes to a form suitable for asymptotic analysis and explain the strategy for the imposition of the simplicity constraints at the quantum level. 

We first introduce some basics for the principal series of unitary irreducible representations of \slt, in order to give meaning and explicitly define the expression 
\begin{equation} \label{eq:toDefine}
\tr^{j_f} \left[\prod_{v \in f} Y_\gamma^\dagger g_{e'v}g_{ve} Y_\gamma  h_{vf} \right] 
\end{equation}
appearing in the face amplitude of equation \eqref{eq:theEprlAnsatz}. For a complete treatment of the representation theory of \slt see the textbook \cite{ruhl_lorentz_1970}. See also \cite{conrady_unitary_2011} for a summary of basic properties and the relation to $SU(2)$ and $SU(1,1)$.  

We proceed to briefly discuss the postulate for the imposition of the simplicity constraints at the quantum level, see also Appendix \ref{appsec:CGsl2c}. We end this section by introducing the map $H$ from normalized spinors to $SU(2)$, a simple tool central to deriving the Han-Krajewski path-integral representation for the EPRL amplitudes in the following section.

\subsection{Elements of \slt principal series representations}
\label{sec:elementsOfSL2CrepTheory}
The group \slt is non-compact. Intuitively, this is evident from our understanding of the Lorentz group $O(3,1)$, and the fact that \slt is its universal cover. Lorentz transformations are composed of ordinary three-dimensional rotations and \emph{boosts}, which are rotations in a Minkowski plane. Hyperbolic angles are also called Lorentzian angles, boost angles or rapidities. Similarly to the familiar Euclidean angles, they are defined as inverse \emph{hyperbolic} cosines of inner products between two normalised \emph{timelike} vectors, see section \ref{sec:lorReggeAction} where we define the Lorentzian Regge action. 

Thus, boost angles take arbitrarily large values and it is to be expected that (the unitary irreps of) $SO^+_\uparrow(3,1)$ cannot be parametrized by parameters taking values in compact intervals.

Compare this statement with the Euler angle parametrization of Wigner's D-matrices $D^j(\alpha,\beta,\gamma)$, discussed in Chapter \ref{ch:gravTunneling}. The three Euclidean angles $\alpha$, $\beta$ and $\gamma$, represent ordinary $SO(2)$ two-dimensional rotations by that amount, about an axis, and take values in $(0,\pi)$ and $(0,2\pi)$. Attempting to do a similar geometric parametrization of a Lorentz transformation, which is in general a combination of rotations and boosts, would certainly require the use of hyperbolic angles, that take values in $(1,\infty)$. 

In representation theory, this translates to the fact that the unitary representation spaces of \slt are infinite dimensional. Unitary representations play a central role in the harmonic analysis of functions over groups, see Appendix a discussion of the \ref{app:recoupTheory} for \slt case. 

One possible analogy is to think of unitary representations as playing the role of ``plane waves'' $e^{i p x}$ in Fourier analysis, the latter being Harmonic analysis over the reals. When the group is compact, it can be parametrized by parameters taking values in compact intervals, with a certain periodicity. This is the analogue of doing Fourier analysis over periodic functions, which results in discrete summation over modes. In the $SU(2)$ case, the discrete summation is over the spins and magnetic indices. When we have a non-compact group, it is analogous to having to also do Fourier analysis over non-periodic functions, and thus we expect that there will be continuous integrations involved. 

\medskip

The principal series of unitary irreducible representations of \slt is defined through its action to the space $V$ of functions $f(g)$ where $g \in \sl2c$. These representation spaces are labelled, as anticipated above, by a continuous parameter $p \in \mathbb{R}^+$ and a discrete spin label $k \in \mathbb{N}/2$. 

The decomposition of $V$ into unitary irreducible representations of the principal series is given by 
\begin{equation} \label{eq:fullRepSpaceSL2C}
V=\int_{\mathcal{R}^+} \dd p \sum_k V^{(p,k)}
\end{equation}
where the spaces $V^{(p,k)}$ are realised as spaces of homogeneous complex functions of spinors, see \cite{conrady_unitary_2011}. Furthermore, and very conveniently, each representation space $V^{(p,k)}$ decomposes into irreducible representations of $SU(2)$
\begin{equation} \label{eq:repSpaceSL2C}
V^{(p,k)} = \bigoplus_{j=k}^\infty \mathcal{H}^j
\end{equation}
with the equality to be understood in the sense of an isomorphism between vector spaces. The spaces $V^{(p,k)}$ and $V^{(p,-k)}$ are isomorphic and henceforth we consider only positive values for $k$, see \cite{barrett_asymptotic_2009}.

The space $\mathcal{H}^j$ is the usual spin-j representation space of $SU(2)$, on which  unitary irreducible representations act. These are finite dimensional representations
\begin{equation}
d_j \equiv \mathrm{dim} \mathcal{H}^j = 2j+1 
\end{equation}
 An orthogonal basis of $\mathcal{H}^j$ is the usual $\ket{j,m}$ basis of angular momentum $j$ and magnetic moment $m$. The $SU(2)$ unitary irreducible representations can be explicitly realized through Wigner's D-matrices, defined as the matrix elements
\begin{equation}
D^j_{m m'}(h) \equiv \braket{j,m\vert h \vert j,m'} 
\end{equation}
with $h$ in $SU(2)$. This explicit expression for $D^j_{m m'}(h)$ is well known, see equation \eqref{eq:analyticExtensionWignerD} for an explicit expression of the analytical continuation of Wigner's D-matrices to $h \in GL(2,\mathbb{C})$.

The matrix elements $D^j_{m m'}(h)$ provide a basis of square integrable functions on $SU(2)$
\begin{equation} \label{eq:su2HarmonicDecomp}
f(h) = \sum_{jmm'} f^j_{mm'} \braket{j,m\vert h \vert j,m'} 
\end{equation}
for coefficients $f^j_{mm'}$.

Similarly, $V^{(p,k)}$ has an orthogonal basis, which we denote as 
\begin{equation} \label{eq:basisStatesSL2C}
\ket{p,k,j,m} \nonumber  
\end{equation}
That is, a square integrable function $f(g)$ of \slt, seen as an element of the vector space $V$, decomposes as 
\begin{equation} \label{eq:sl2cHarmonicDecomp}
f(g) = \int_{\mathbb{R}^+} \! \dd p \, \sum_{k = -\infty}^{\infty}\; \sum_{jmj'm'} f^{p,k}_{jmj'm'} \, \braket{p,k,j,m\vert g \vert p,k,j',m'}
\end{equation}
where $k,j,m,j',n$ are spins, that is, half-integers. 

Notice the analogy with the $SU(2)$ case \eqref{eq:su2HarmonicDecomp}. The representation for \slt is labelled by the \emph{pair} $(p,k)$ and a basis vector has a \emph{pair} of ``magnetic indices'' $(j,m)$, that are summed over independently in \eqref{eq:sl2cHarmonicDecomp}.

 The range of summations over the index pairs $j,m$ and $j',m'$ is given explicitly by 
\begin{equation} \label{eq:sl2cSumRange}
\sum_{jmj'm'} \equiv \sum_{j \leq k } \; \sum_{m = - j}^{m = + j} \ \sum_{j' \leq k }  \; \sum_{m' = - j'}^{m' = + j'}
\end{equation}

Similarly to Wigner's D--matrices, the matrix elements for the representation matrices acting on $V^{p,k}$ are defined by 
\begin{equation} \label{eq:innerProductSL2C}
D^{p,k}_{j m j' m'}(g) \equiv \braket{p,k,j,m \vert g \vert p,k,j',m'} 
\end{equation}

Multiplication in $V^{p,k}$ is given by 
\begin{equation} \label{eq:multiplicationSL2C}
D^{p,k}_{j m j' m'}(g_1 g_2) = \sum_{ln} D^{p,k}_{j m l n}(g_1) D^{p,k}_{l n j' m'}(g_2) 
\end{equation}
with the summation ranges as in \eqref{eq:sl2cSumRange}. The matrix elements $D^{p,k}_{j m j' m'}(g)$ are explicitly known, see Appendix \ref{app:recoupTheory}. 

 The delta function over \slt is given by 
\begin{equation} \label{eq:deltaSL2C}
\delta(g) =  \int_{\mathbb{R}^+}\! \dd p \,\sum_k\, (p^2+k^2) \sum_{j,m} D^{p,k}_{jmjm}(g)
\end{equation}
Compare this with the Peter-Weyl expansion of the $SU(2)$ delta function
\begin{equation} \label{eq:deltaSL2C}
\delta(h) =  \sum_j\, d_j D^j_{mm}(h)
\end{equation}
The realisation of the spaces $V^{p,k}$ as spaces of homogeneous functions of spinors will allows us to explicitly express the inner products in $V^{p,k}$ in terms of spinor variables in Section \ref{sec:ampsInAsAnform}. That is, in terms of the matrix elements $D^{p,k}_{jmjm'}(g)$, see equation \eqref{eq:innerProductSL2C}. 

\subsection{Imposition of simplicity constraints at the quantum level and the $Y_\gamma$ map}
\label{sec:weakImpositionSimpConstr}
We arrive at the technical core of the EPRL model: the \emph{conjecture} that the imposition of the simplicity constraints at the quantum level is equivalent to restricting the representation space $V$ to $V^{(p=\gamma j, k=j)}$. We call the representations acting on $V^{(\gamma j, j)}$, \emph{simple}. To simplify notation, we will henceforth write $V^{\gamma j, j}\equiv V^{(\gamma j, j)}$.

That is, for emphasis, we \emph{assume} that the state space for the quantum gravitational field is built from the representations in 
\begin{equation}
V^{\gamma j, j }
\end{equation}
and not that of equation \eqref{eq:fullRepSpaceSL2C}, which we repeat here
\begin{equation}
V=\int_{\mathcal{R}^+} \dd p \sum_k V^{(p,k)}
\end{equation}
Let us remind the discussion in Section \ref{sec:BFSpinfoamQuantization}. The quantum variables associated to the discretized spin-connection $\omega$ are the group elements $g_e$. When we perform harmonic analysis and ``Fourier expand'' in the variable playing the role of the conjugate to the quantum holonomy, the spins and intertwiners, we only consider the quantum numbers labelling the simple representations.\footnote{A remark: the integration over the continuous parameter $p$ \emph{does} come into play, because the quantum equivalents of the discrete holonomies $g_e$ from vertex to vertex, were split into two group elements $g_e=g_{ve} g_{ev'}$. The integration over $p$ in \eqref{eq:multiplicationSL2C} has to be performed. The splitting $g_e=g_{ve} g_{ev'}$ is necessary to get the explicit state-sum form of the amplitude over all representation and intertwiner labels, which is the subject of Section \ref{sec:explicitAmp}. The physical meaning of this integration, and the degree to which it is necessary, is not well understood and is the subject of current investigations. A simplified model has been recently proposed restricting to simple representations also over these interior summations \cite{speziale_boosting_2017-1}. }

The motivation for this restriction is intimately related to the fact that we are attempting to define a regularized path integral for gravity \emph{starting from the Holst action}. We discuss the motivation at the end of this section. 

\subsubsection{$Y_\gamma$ map}
We introduce a shorthand notation for the states $\ket{\gamma j,j,j,m} \in V^{\gamma j, j}$
\begin{eqnarray}
\ketY{j,m} &\equiv& \ket{\gamma j,j,j,m} \nonumber  \\
\braY{j,m} &\equiv &  \bra{\gamma j,j,j,m} \nonumber
\end{eqnarray}
 The spaces $V^{\gamma j, j}$ and $H^j$ are isomorphic, see equation \eqref{eq:repSpaceSL2C}. The notation we use is to reflect this natural one-to-one correspondance, of the basis $\ketY{j,m}$ of $V^{\gamma j,j}$, with the basis $\ket{j,m}$ of $\mathcal{H}^j$. We may then code this in a map between the representation spaces $V^{\gamma j, j}$ to $H^j$.

The map that sends one basis to another is called the $Y_\gamma$ map
\begin{equation} \label{eq:defYmap}
Y_\gamma \; : \; \ket{j,m} \rightarrow \ketY{j,m} 
\end{equation}
That is, $Y_\gamma$ is a map that \emph{embeds} the spin-$j$ representation space $\mathcal{H}^j$ of $SU(2)$ into the representation space $V^{\gamma j,j}$ of \slt
\begin{equation}
Y_\gamma \; : \; \mathcal{H}^j \rightarrow V^{\gamma j,j} \subset V
\end{equation}
In turn, $V^{\gamma j,j}$ is an $SU(2)$ representation subspace in the full representation space $V$ of \slt. We emphasize this point because it provides the intuition for the meaning of the restriction to simple representations. 

We have thus defined completely the EPRL model. 
The expression  \eqref{eq:toDefine} appearing in the face amplitude of equation \eqref{eq:theEprlAnsatz} and  \eqref{eq:faceAmplitude}
is defined as
\begin{equation}
\tr^{j_f} \left[\prod_{v \in f} Y_\gamma^\dagger g_{e'v}g_{ve} Y_\gamma  h_{vf} \right] \equiv \sum_{\{m\}} \prod_{v\in f} \; \braketY{ j_f \, m_{e'} \vert g_{ve'}^{-1} \, g_{ve} \vert j_f \, \tilde{m}_{e}} \braket{j_f \, \tilde{m}_{e} \vert h_{vf} \vert j_f \, m_{e} }
\end{equation}
and can be written explicitly in terms of a decomposition of the matrix elements  \eqref{eq:multiplicationSL2C}, see Appendix \ref{app:recoupTheory}. Each $SU(2)$ group element $h_{vf}$ appears two times, once in the face amplitude and once in the holonomy $\delta_{h_f}$. They can be integrated out using the orthogonality relation for Wigner's D--matrices and the properties of the Haar measure,  see Appendices \ref{app:recoupTheory} and \ref{app:derivationTreeAmplitudes}. After eliminating $h_{vf}$, the face amplitude reads

\begin{equation} 
A_f(j_f,\{g_f\}) = \sum_{\{m\}} \prod_{v\in f} \; \braketY{ j_f \, m_{e'} \vert g_{ve'}^{-1} \, g_{ve} \vert j_f \, m_{e}}
\end{equation}

\subsubsection*{Weak imposition of the linear simplicity constraints}
The quadratic simplicity constraints of equation \eqref{eq:simpconstraints}, reduce in the spinfoam setting to the \emph{linear simplicity constraints} \cite{freidel_new_2008-1,barrett_relativistic_1998,perez_3+1_2001,engle_flipped_2008,
alexandrov_simplicity_2008,han_commuting_2013,livine_consistently_2008,engle_lqg_2008} understood as a discretized version of the former,  imposed on each face $f$ of the two-complex $\mathcal{C}$. This was first shown in \cite{freidel_new_2008-1} by Freidel and Krasnov.

The linear simplicity constraints read 
\begin{equation}
L^i_f - \frac{1}{\gamma} K^i_f \approx 0
\end{equation}
where $L^i_f$ are the rotation generators \emph{of an arbitrary rotation subgroup of \slt} and $K^i_f $ the corresponding boost generators. The approximate equality implies that this condition must be satisfied in an appropriate semiclassical limit.

The linear simplicity constraints arise naturally when we consider the 3+1 split of the Holst action in the time gauge.  The pull-back of the conjugate momentum of the connection on a hypersurface $\Sigma$, can be decomposed in its electric $K^i$ and magnetic $L^i$ parts, which behave as 3D vector fields in a constant time hypersurface, and are related as
\begin{equation}
K^i=\gamma L^i
\end{equation}
see \cite{rovelli_covariant_2014} for a quick derivation. 

We understand then the choice of a rotation subgroup as corresponding to an internal gauge group of the phase space parametrized in terms of the Ashtekar-Barbero variables, in the \emph{time gauge}. We will see this construction appearing more geometrically in the semi-classical limit. 

Different ways to impose the linear simplicity constraints at the quantum level have been proposed. Slight modifications are possible that could be relevant in the deep quantum regime, but are equivalent in the semiclassical limit with which we will be concerned.

The most used criterion for imposing the simplicity constraints comes from studying the spectrum of Thiemmann's master constraint. The Casimirs of the Lorentz group, which can be witten in a simple form in  terms of $K^i$ and $L^i$  are \emph{diagonal} in the basis $\ket{p,k,j,m}$. See Appendix \ref{appsec:CGsl2c} for more details.

All states in $H^{\gamma j,j}$ satisfy the simplicity constraints in the semiclassical limit, that is, in the sense of an expectation value and taking the limit of large quantum numbers\footnote{The limit is $j \rightarrow \infty$ and $\hbar \rightarrow 0$ with $\hbar j$ kept constant.}

\begin{equation}
\braket{L^i_f - \frac{1}{\gamma} K^i_f }_{\Psi \in  V^{\gamma j,j} } \stackrel{\text{sem.lim.}}{\sim} 0
\end{equation}

\subsection{Spinors and the $H$ map}
\label{sec:spinorForKrajHan}
In this section we give some spinor basics and the definition and properties of the map $H$ from normalized spinors to $SU(2)$. We will only need the final equation of this section, equation \ref{eq:HzHwMatrixProduct}, to express the EPRL amplitudes in the Han-Krajewski representation, done in the following section.
 Spinors are further dicussed and used in Section \ref{sec:wavepacketOfGeometry} where we introduce $SU(2)$ coherent states and in Appendix \ref{app:cohStateRepEPRL} where we introduce the coherent state representation. See also Appendix \ref{app:spinorsAndRiemannSphere} for the relation of spinors and the Riemann sphere. 
\subsubsection{Spinors}
Spinors are elements of $\mathbb{C}^2$. We denote one spinor as $\ket{z}$ and for explicit calculations we write 
\begin{equation}
\ket{z} = \begin{pmatrix} z_0 \\ z_1 \end{pmatrix}
\end{equation}
with $z_0$ and $z_1$ complex numbers. The bra is defined as 
\begin{equation}
\bra{z} = \begin{pmatrix} \bar{z}_0 & \bar{z}_1 \end{pmatrix}
\end{equation}
and the inner product follows as
\begin{equation}
\braket{w\vert z} = \bar{w_0}z_0 + \bar{w_1} z_1
\end{equation}

\subsubsection{Parity map}
We introduce the parity\footnote{See Appendix \ref{app:spinorsAndRiemannSphere} for the reason it is called parity.} map in spinor space
\begin{equation}
J \ket{z} \equiv \ket{Jz}= \begin{pmatrix}
-\bar{z}_1 \\ \bar{z}_0
\end{pmatrix}
\end{equation}
Direct computation shows that $J^2\ket{z}=-\ket{z}$,
\begin{equation}
J^2=-\ident_{2 \times 2}
\end{equation}
In other words, it is useful to keep in mind that 
\begin{equation}
J^{-1}=-J
\end{equation}
The identity matrix can be written as
\begin{equation}
\ident_{2 \times 2} = \ket{z}\bra{z} + \ket{Jz}\bra{Jz}
\end{equation}
The following identity is useful
\begin{equation} \label{eq:usefulParityIdentity}
J g = (g^{-1})^\dagger J
\end{equation}
which follows from writing explicitly $g$ as a matrix and acting on a spinor $\ket{z}$ with both sides of the equation. 
We note the relations
\begin{equation}
\braket{Jw \vert Jz }=\braket{w \vert z }
\end{equation}
and thus
\begin{equation}
\braket{Jz \vert Jz }=\braket{z \vert z }
\end{equation}

\subsubsection{$SU(2)$ map $H$}
The matrix $H(z)$ is defined as
\begin{equation} \label{eq:HzMatrix}
H(z) = \begin{pmatrix}
  z_0 & -\bar{z}_1 \\
  z_1 & \bar{z}_0 
 \end{pmatrix}
\end{equation}  
and defines a map from normalized spinors to $SU(2)$. This is easy to see by direct matrix multiplication
\begin{eqnarray} 
H(z) H^\dagger (z) &=& \begin{pmatrix}
  z_0 \bar{z}_0 + \bar{z}_1 z_1 & z_0 \bar{z}_1 - z_0 \bar{z_1} \\
   z_1 \bar{z}_0 - z_1 \bar{z}_0 &  \bar{z}_1 z_1 + z_0 \bar{z}_0 
 \end{pmatrix} \nonumber \\ \\
&=& \begin{pmatrix}
  \vert z_0 \vert^2  + \vert z_1 \vert^2  & 0\\
   0 &  \vert z_0 \vert^2  + \vert z_1 \vert^2 
 \end{pmatrix} \nonumber \\ \\ &=& \braket{z \vert z} \ident_{2 \times 2}
\end{eqnarray}  
Using the shorthand notation 
\begin{equation} 
H(z) = \bigg( \ket{z} \ \ket{Jz}   \bigg)
\end{equation}
the product $H(z)^\dagger H(w)$ is given by 
\begin{eqnarray} \label{eq:HzHwMatrixProduct}
H(z)^\dagger H(w) &=& 
\begin{pmatrix}
  \bra{z} \\
  \bra{Jz} 
 \end{pmatrix} \bigg( \ket{w} \ \ket{Jw}   \bigg) \nonumber \\
\\ &=&
\begin{pmatrix}
  \braket{z\vert w} & \braket{z\vert Jw}\\
  \braket{J z\vert w} & \braket{J z\vert Jw}
 \end{pmatrix} 
\end{eqnarray}
With these tools, we are ready to bring the EPRL amplitudes into a form suitable for asymptotic analysis, in an arbitrarily large 2-complex, dual to a simplicial complex.

\section{The Han-Krajewski path-integral representation} \label{sec:ampsInAsAnform}
We have seen that EPRL amplitudes can be written as a product over face amplitudes
\begin{equation} \label{eq:ampFaceProduct}
W_\mathcal{C} = \sum_{\{j_f\}}\!  \mu(j_f) \int\! \mu(g_{ve}) \,\prod_f A_f(j_f)
\end{equation} 
with
\begin{equation} \label{eq:faceAmplitude}
A_f(j_f) =  \sum_{\{m\}} \prod_{v\in f} \; \braketY{ j_f \, m_{e'} \vert g_{ve'}^{-1} \, g_{ve} \vert j_f \, m_{e}}
\end{equation}
where the notation $\{g_f\}$ means all $g_ve$ such that the half-edge $ve$ is in the boundary of tha face $f$.

Thus, the main object that needs to be computed is   
\begin{equation} 
\sum_{\{m\}} \prod_{v \in f}\braketY{ j_f \, m_{ve'} \vert g_{ve'}^{-1} \, g_{ve} \vert j_f \, m_{ve}} \nonumber 
\end{equation}

The space $V^{\gamma j, j}$ is explicitly realized as a space of homogeneous functions of a spinor $z$ 
  
\begin{equation} \label{eq:spinorRepSl2C}
\braket{z \vert j m }_\gamma \equiv f^j_m(z) = \sqrt{\frac{d_j}{\pi}} \braket{z \vert z}^{j(i \gamma -1) - 1} D^j_{mj}(H(z)) 
\end{equation}  
with $H(z)$ defined in \eqref{eq:HzMatrix}. The matrix elements of the Wigner-D matrix $D^j_{mj}(A)$ can be defined by analytical extension for any $A \in GL(2,\mathcal{C})$\begin{equation}
A = \begin{pmatrix}
  a & b \\
  c & d 
 \end{pmatrix}
\end{equation} 
We have \cite{barrett_lorentzian_2010}
\begin{equation}
 \label{eq:analyticExtensionWignerD}
D^j_{q k}( A ) = \sqrt{\frac{(j+q)! (j-q)!}{(j+k)! (j-k)!}} \sum_n 
\begin{pmatrix}
  j+k  \\
  n  
 \end{pmatrix}
 \begin{pmatrix}
  j-k  \\
  j+q-n  
 \end{pmatrix}
\; a^n\, b^{j+q-n}\, c^{j+k-n}\, d^{n-q-k} 
\end{equation}
with the sum over $n$ being over values such that the binomial coefficients do not vanish. All indices are half-integers. This (long) formula is key for the result we are aiming to arrive at. In particular, for the EPRl amplitudes we will only need $D^j_{jj}(A)$, in which case, \eqref{eq:analyticExtensionWignerD} simplifies to 
\footnote{Replace $q=j$ and $k=j$ in \eqref{eq:analyticExtensionWignerD}. The square root is unit. Inside the sum over $n$, we have the binomial coefficient $\begin{pmatrix}
  0  \\
  2j-n  
 \end{pmatrix}$ which equals $\delta_{2j,n}$. This is because, by definition, $\begin{pmatrix}
  0  \\
  0  
 \end{pmatrix}$ is  unit and $\begin{pmatrix}
  0  \\
  n  
 \end{pmatrix}$ with $n \neq 0$ vanishes. }
 \begin{equation} \label{eq:WignerDjjElement}
D^j_{jj}(A) = a^{2j}
\end{equation}
 
The action of  \slt on $\braket{z \vert j m }_\gamma = f^j_m(z)$ is given by the transpose matrix action in the fundamental representation of \slt (that is, as a transposed $2 \times 2$ matrix) on the spinor argument
\begin{equation}
g \triangleright f^j_m(z) = f^j_m(g^T z)
\end{equation}
where we have introduced the notation $\triangleright$, which will be used whenever we want to denote the action of a group on objects on which it does not act in a straightforward manner, i.e.\! as a multiplication.
 
Treating $g_{ve'}^{-1} \, g_{ve}$ as a single \slt element acting on the right by the transpose action, the inner product
\begin{equation}
\braketY{ j_f \, m_{ve'} \vert g_{ve'}^{-1} \, g_{ve} \vert j_f \, m_{ve}} 
\end{equation}
reads \cite{barrett_lorentzian_2010}
\begin{equation} \label{eq:innerYone}
\braketY{ j_f \, m_{e'} \vert g_{ve'}^{-1} \, g_{ve} \vert j_f \, m_{e}} = \int_{\mathbb{CP}^1} \! \dd \mu(z)\; \overline{f^{j_f}_{m_{e'}}( \, z_{vf} \, )} \; f^{j_f}_{m_{e}}(\, g_{ve}^T (g_{ve'}^{-1})^T \, z_{vf} \, )
\end{equation}
where $\mu(z)$ is the homogeneous \slt--invariant measure on $\mathbb{C}^2$. That is, the integration is effectively over $\mathbb{CP}^1$. The integration measure $\mu(z)$ is given explicitly by 
\begin{equation}
\mu(z) = \frac{i}{2} (z_0 \dd z_1 - z_1 \dd z_0 )\wedge (\bar{z}_0 \dd \bar{z}_1 - \bar{z}_1 \dd \bar{z}_0)
\end{equation}
The spinor variable $z_{vf}$ only appears in the term \eqref{eq:innerYone}. Since the measure $\mu(z)$ is \slt invariant, we then notice that we may make the following handy change of variables
\begin{equation}
z_{vf} \rightarrow (g_{ve'}^{-1})^T z_{vf}
\end{equation}
giving
\begin{equation}
\braketY{ j_f \, m_{e'} \vert g_{ve'}^{-1} \, g_{ve} \vert j_f \, m_{e}} = \int_{\mathbb{CP}^1} \! \dd \mu(z)\; \overline{f^{j_f}_{m_{e'}}( \, g_{ve'}^T z_{vf} \, )} \; f^{j_f}_{m_{e}}(\,g_{ve}^T  \, z_{vf} \, )
\end{equation}
Thus, all variables $g_{ve}$ appear as $g_{ve}^T$ in the amplitude. We may then change variables again using the fact that $\mu(g_{ve})$ is a left and right invariant Haar measure 
\begin{equation}
g_{ve}^T \rightarrow g^\dagger_{ve}
\end{equation}
This choice, to change to $g^\dagger$ instead of say to $g$, is so that we recover the expressions in \cite{han_path_2013,han_asymptotics_2013}, the results of which we will be using. 

Thus, the variables $g_{ve}$ and $z_{vf}$ only appear in the combinations $g_{ve'}^\dagger z_{vf}$ and $g_{ve}^\dagger  \, z_{vf}$. We introduce the notation 
\begin{eqnarray}
Z_{vef}=g_{ve}^\dagger z_{vf} \nonumber \\
Z_{ve'f}=g_{ve}^\dagger z_{vf}
\end{eqnarray}
The face amplitude now reads
\begin{equation} 
A_f(j_f,\{g_f\}) = \int \mu(g_{ve}) \sum_{\{m\}} \prod_{v\in f} \; \int_{\mathbb{CP}^1} \! \dd \mu(z)\; \overline{f^{j_f}_{m_{e'}}( \, Z_{ve'f}\, )} \; f^{j_f}_{m_{e}}(\, Z_{vef} \, )
\end{equation}

Now, in the face amplitude \eqref{eq:faceAmplitude}, we may pull the summation over the magnetic index $m_e$ for the two half-edges $ev$ and $e'v$, and combine the two instances in which $f^{j_f}_{m_{e}}$ appears. 

Thus, we write the face amplitude as
\begin{equation} 
A_f(j_f) = \int \! \mu(g_{ve}) \int_{\mathbb{CP}^1} \!\dd \mu(z)\; \prod_{e\in f} \sum_{m_e} \; f^{j_f}_{m_{e}}(\, Z_{vef} \, ) \; \overline{f^{j_f}_{m_{e}}( \, Z_{v'ef}\, )}
\end{equation}
We need then only calculate one term
\begin{equation} \label{eq:prelinkToCohStateRep}
\sum_{m_e} \; f^{j_f}_{m_{e}}(\, Z_{vef} \, ) \; \overline{f^{j_f}_{m_{e}}( \, Z_{v'ef}\, )}
\end{equation}

 We may now perform the summation over the magnetic index using equation \eqref{eq:spinorRepSl2C}. Let us show the calculation in a lighter notation and give the complete expression afterwards. We have, for two spinors $Z$ and $W$ 
\begin{eqnarray} \label{eq:linkToCohStateRep}
\sum_m D^j_{mj}(H(W)) \overline{D^j_{mj}(H(Z))} &=& D^j_{jm}(H(Z)^\dagger) D^j_{mj}(H(W))  \nonumber \\ 
&=& D^j_{jj}( H(Z)^\dagger H(W)) \nonumber \\
&=& D^j_{jj}( H(Z)^\dagger H(W)) \nonumber \\
&=& \braket{Z\vert W}^{2j}
\end{eqnarray}
where in the first line we used that $D^j_{mj}(A) = D^j_{jm}(A^T)$ for any $A\in GL(2,\mathbb{C})$, which can be checked directly from \eqref{eq:analyticExtensionWignerD}, and in the last we used equations \eqref{eq:HzHwMatrixProduct} and \eqref{eq:WignerDjjElement}.

Thus, setting $Z \rightarrow Z_{v'ef}$ and $W \rightarrow Z_{vef} $  we have shown that 

\begin{eqnarray}
&{}& \sum_{m_e} \; f^{j_f}_{m_{e}}(\, Z_{vef} \, ) \; \overline{f^{j_f}_{m_{e}}( \, Z_{v'ef}\, )} = \nonumber \\ &{}& \frac{d_j}{\pi} \braket{Z_{vef}\vert Z_{vef} }^{j(i \gamma-1)-1} \braket{Z_{v'ef}\vert Z_{v'ef} }^{j(-i \gamma-1)-1} \braket{Z_{v'ef} \vert Z_{vef}  }^{2j}
\end{eqnarray}
 
The next thing to notice is that appart from a factor $\braket{Z_{vef}\vert Z_{vef} }^{-1} \braket{Z_{v'ef}\vert Z_{ev'f} }^{-1}$ which can be absorbed in a redefinition of the measure $\mu(z)$, see for instance \cite{barrett_lorentzian_2010}. All other terms are raised to the power $j$. This is the important result that allows us to use the stationary phase method. Thus, up to constants that can be absorbed in the overall normalization of the amplitude and terms that can be absorbed in the integration measures ($d_j$ goes into $\mu(j_f)$), we have shown that 

\begin{equation} 
\sum_{\{m\}}\prod_{v \in f}\braketY{ j_f \, m_{e'} \vert g_{ve'}^{-1} \, g_{ve} \vert j_f \, m_{e}} = e^{j_f F_f(\{g_f\},\{z_f\})}
\end{equation}
where 
\begin{equation} \label{eq:Ffunction}
F_f(\{g_f\},\{z_f\}) = \sum_{e \in f} \log \frac{ \braket{Z_{v'ef}\vert Z_{vef} }^2}{\braket{Z_{vef}\vert Z_{vef} }  \braket{Z_{v'ef}\vert Z_{v'ef} }} +i \gamma \log \frac{\braket{Z_{vef}\vert Z_{vef}}}{\braket{Z_{v'ef}\vert Z_{v'ef}} } 
\end{equation}
and thus the face amplitude \eqref{eq:faceAmplitude} takes the form

\begin{equation} \label{eq:faceAmpSpinors}
\boxed{
A_f(j_f) =  \int \! \mu(z_{vf}) \; e^{j_f F_f\left(\{g_f\},\{z_f\}\right)}
}
\end{equation}
where factors of $d_j$ have been absorbed in the integration measure $\mu(j_f)$ and constants are absorbed in the overall normalization $\mathcal{N}_\mathcal{C}$.

\begin{center}
\rule{0.5 \textwidth}{0.7pt}
\end{center}

This representation was derived in \cite{han_path_2013} by Han and Krajewski. It has the novel feature that $SU(2)$ coherent states have not been used to bring the spinfoam amplitudes to this form. 

The derivation we present here is more straightforward and allows for an easy passage to the coherent state representation.

To get to the coherent state representation, it suffices to insert the resolution of the identity in equation \eqref{eq:prelinkToCohStateRep}, in the sense of equations \eqref{eq:linkToCohStateRep}, see Appendix \ref{app:cohStateRepEPRL}.

When coherent states are used, there are $E \times F$ extra spinor variables in the path--integral representation, where $E$ is the number of edges and $F$ the number of faces of $\mathcal{C}$. The resulting critical points equations, are identical to those resulting from the coherent state representation. Thus, the $E \times F$ spinor variables are superfluous. 

The reconstruction of simplicial geometries at geometrical critical points is possible without the use of the extra spinor variables. This in turn allows for a more straightforward interpretation of the critical point equations,  discussed in Section \ref{sec:overviewAsAnEPRL}. 

Another advantage, and the reason we use this representation, is that the detaild works  \cite{han_path_2013} by Krajewski and Han, and the earlier analysis in the coherent state representation \cite{han_asymptotics_2013} by Han and Shang, are currently the only available studies for the semiclassical limit of the Lorentzian partial amplitude for an arbitrary 2-complex. The detailed analysis by the same authors for the Euclidean case is in \cite{han_asymptotics_2012}. 

In turn, the ability to go back and forth easily between the two representations, allows to contract a spinfoam amplitude with a boundary with a semiclassical boundary state, as we do in Chapter \ref{ch:gravTunneling} and use the results from both works.

The detailed procedure will appear shortly in a work \cite{mariosFabio} by the author and Fabio d'Ambrosio.

\section{Lorentzian Regge action }
\label{sec:lorReggeAction}
In this section we review the definition of the  Regge action and discuss the analogy with the tetradic form of the actions for General Relativity, studied in Chapter \ref{ch:topBottom}. 
 We are following \cite{barrett_semiclassical_1994-1} for the definition of the Lorentzian Regge action.

\subsection{The co-plane in Regge calculus and relation to tetradic General Relativity}
\label{sec:coplaneRegge}
The Regge action is defined as 
\begin{equation}
S_R(l) = \sum_f A_f(l) \phi_f(l)
\end{equation}
where $l$ are the lengths of the segments of a simplicial triangulation $\mathcal{C}^*$. The tetrahedra of $\mathcal{C}^*$ are all taken to be spacelike (and thus also all triangles and segments). 

No coordinates are involved in the definition of the Regge action. This was Regge's main motivation, reflected in the title of the work \cite{regge_general_1961} where the Regge action was first introduced
\begin{quote}
\centering
\say{General relativity without coordinates}.
\end{quote} 
Referring to the comparison Table \ref{tab:actionsTable} and the discussion in Section \ref{sec:HEtoCar}, we remind that the same coordinate disappearance happens for the Einstein--Hilbert action when written in its tetradic form. The two terms $e\wedge e$ and $\star F$ have the same dimensions as the areas and the deficit angles here. 

 The similarity between the tetradic form of the Einstein--Hilbert and Palatini actions with the Regge action goes further.  

In Regge calculus\,---the definition and study of discrete General Relativity and Euclidean toy models by the corresponding Regge action---, the construction of the deficit angle in both Euclidean and Lorentzian case is delegated at the level of a \emph{plane}, perpendicular to the triangle $f$.

 Since all four dimensional normals to the tetrahedra of the triangulation that share the triangle $f$, are also normal to $f$, the plane perpendicular to $f$ is spanned by the normal to any one tetrahedron sharing the face $f$ (this normal is timelike in the Lorentzian case) and the normal to the face, which can be taken to point in any of the tetrahedra that share $f$.
 
This the analogue of what we called the co-plane $\star e^I \wedge e^J$, where the curvature two-form lives, for the Einstein--Hilbert, Palatini and Holst (in this case, \emph{on--shell}) actions.

 Effectively, as we do in the following sections, the definition of deficit angles in Regge calculus comes down to reasoning locally in a Euclidean plane, for a Euclidean gravity toy model, or in a Minkowski plane for discrete General Relativity. 
 \bigskip

 Covariant LQG is an attempt to define a path integral for gravity starting from the Holst action, with the regularization effected by discretizing on a simplicial manifold, much like in Regge calculus. Given the analogy between the Holst action and the Regge action, one would hope and expect to recover discrete classical General Relativity from the Quantum Theory in the form of the Regge action, in an appropriate semiclassical limit. 
 
\subsection{Euclidean deficit angles}
\label{sec:defAngles4DEucdlidean}
We start by reviewing the definition of deficit angles $\phi_f(\ell)$ in the four dimensional Euclidean case, and compare with the Lorentzian version in the next section.

 The deficit angle is defined as
\begin{equation} \label{eq:EuclideanDeficitAngle}
\phi_f(l) = 2 \pi - \sum_{v \in f} \Theta_{vf}(\ell) 
\end{equation} 
where $\Theta_{vf}(\ell)$ are the dihedral angles between two tetrahedra that belong to the 4-simplex $v$ and bound the face $f$. 

Let us call these two tetrahedra $e_1$ and $e_2$ and the four dimensional normals (of unit norm) to them $N_{e_1}$ and $N_{e_2}$

The dihedral angle of a 4-simplex, between two of its tetrahedra $e_1$ and $e_2$, at the face $f$ shared by them, is defined as 
\begin{equation}
\Theta_{vf}(\ell) =\arccos{\left( N_{ve_1} \cdot N_{ve_2} \right)}
\end{equation}
This implies that 
\begin{equation}
\Theta_{vf}(\ell) \in [0, \pi]
\end{equation}
because $N_{ve_1} \cdot N_{ve_2} \in [-1,1]$; the angle between $N_{ve_1}$ and $N_{ve_2}$ can be at most $\pi$.

 In particular \emph{the Euclidean dihedral angles are non negative}.

\medskip

This is the role of the $2\pi$ in the definition \eqref{eq:EuclideanDeficitAngle} of the deficit angle. The sum of the Euclidean dihedral angles is always positive. 

When $\sum_{v \in f} \Theta_{vf}(\ell)$ equals $2 \pi$, the deficit angle vanishes and the local curvature vanishes (flat).  When it is less than $2 \pi $, the deficit angle is positive, and the curvature local to the face is positive (stretched). When it is larger than $2 \pi $, the deficit angle is negative and the curvature local to the face is negative (pinched).

\medskip

The Lorentzian deficit angle has the same interpretation, with the difference that the value for the flat case is zero instead of $2 \pi$.
\subsection{Lorentzian Deficit angles: thin-wedges and thick-wedges}
\label{sec:defAngles4DLorentzian}
The deficit angle for the Lorentzian case is defined as 
\begin{equation} \label{eq:LorentzianDeficitAngle}
\phi_f(\ell) = \sum_{v \in f} \Theta_{vf}(\ell) 
\end{equation} 
Notice the missing $2 \pi$ and the sign difference from this formula with respect to the Euclidean deficit angle \eqref{eq:EuclideanDeficitAngle}. 

Also, notice the following: the main idea in Regge calculus, is that the discrete local curvature is encoded in the deficit angles, with a vanishing $\phi_f(\ell)$ corresponding, locally to $f$, to flat spacetime. Inspection of equation \eqref{eq:LorentzianDeficitAngle} then reveals that some dihedral angles around the hinge must be negative.

This point can often be a source of confusion and we take a moment to explain how the dihedral angles $\Theta_{vf}(\ell_i) $ are defined in Lorentzian Regge calculus.

We need an analogous construction to the Euclidean case for the Lorentzian case. The difficulty, is that there is no natural notion corresponding to the phrase  ``sum of dihedral angles around a hinge'' $f$ for Minkowski space. 

The requirement that all tetrahedra are spacelike implies that at least one of them has to be time oriented in the opposite direction. The possibilities for each 4--simplex are four, 4--1 and 3--2 in either time direction. That is, four tetrahedra past--oriented and one future--oriented, and vice--versa, and similarly for the 3--2 case. Timelike normals with opposite time orientation, are not related by hyperbolic rotations, i.e.\! pure boosts, that carry a natural notion of hyperbolic angle, see discussion below. 

One may try to remedy this by taking intermediate angles, between timelike and null or timelike and spacelike vectors, trying to ``go around'' the face $f$. Such constructions are possible, but require one to introduce complex angles and appropriately deal with infinities that can arise.

Thus we are left without an obvious typical value for the ``sum of dihedral angles around a hinge'' corresponding to the flat case, like $2\pi$ for Euclidean geometry.  

The idea then is to take the characteristic value for the flat case to be an angle around the hinge that sums up to zero. This implies that \emph{some dihedral angles must be negative}. There are indeed two kinds of dihedral angles that may arise: those between faces that are shared between tetrahedra that have the same, or, the opposite time orientation. 

To deal with this, we remind the structure of the Lorentz group $O(3,1)$ in terms of the parity $P$ and the time reversal $T$ in Table \ref{tab:LorGroup}.

\begin{table}
\centering
\renewcommand{\arraystretch}{4}
\begin{tabular}{c|c} 
 $ {}^+_\uparrow$ & $\ \ \ P \triangleright \ {}^+_\uparrow \;=\; {}^-_\uparrow$  \\[0.4cm] \hline
  $T \triangleright \ {}^+_\uparrow \;=\; {}^+_\downarrow \ \ \ $ & $\ \ \ PT \triangleright \ {}^+_\uparrow \;=\;{}^-_\downarrow$   \\[0.4cm]

\end{tabular}
\renewcommand{\arraystretch}{1}
\caption[Connected components of the Lorentz group]{ The four connected components of the Lorentz group $O(3,1)$. The component connected to the identity is $SO^+_\uparrow(3,1)$. We get the other three by applying the parity $P$ and time reversal $T$.}
\label{tab:LorGroup}
\end{table}

When the normals have the same time orientation, they are in the same light-cone and are related by a proper orthochronous Lorentz transformation, an element of $SO^+_\uparrow(3,1)$. This case is called a \emph{thick-wedge}. 

In particular, the normals at at a thick--wedge are related by a pure boost. This is because we are restricting to the co-plane, thus, the normals are related by an element of $SO^+_\uparrow(1,1)$, the group of hyberbolic rotations of the Minkowski plane.

When the normals are of opposite time orientation, they are not in the same light-cone and are not related by a pure boost. This case is called a \emph{thin-wedge}.

 Evidently, the time reversal of one of them, will be related by a pure boost to the other. This is the main idea for the definition of the angle at a thin wedge. As we see below, we take both the time reversal \emph{and} parity of one of the normals to define the angle at a thin--wedge.

\medskip

Let us go back for a moment to the Euclidean case and rephrase things in a way that can be carried through by analogy to the Lorentzian case. 
There always exists an element of $SO(4)$ that takes one normal to the other. Since we are restricted to a plane, we may way write it as an $SO(2)$ rotation 
$$
\begin{pmatrix}
\cos \Theta  & -\sin \Theta \\
\sin \Theta  & \cos \Theta 
\end{pmatrix}
$$
and we identify the parameter $\Theta$ as the dihedral angle. 

The Lorentzian angle for a thick wedge is then defined analogously, as the rapidity $\zeta$ of the boost that takes one normal to the other. That is, there exists a hyperbolic rotation in $SO(1,1)^{+ \uparrow}$, in the co-plane, such that
\begin{equation}
N_{e_1} = 
\begin{pmatrix}
\cosh \zeta  & \sinh \zeta \\
\sinh \zeta  & \cosh \zeta 
\end{pmatrix}
 \triangleright N_{e_2}
\end{equation}
where we are understanding $N_{e_1}$ and $N_{e_2}$ as two--dimensional vectors in the co-plane of $f$.

Then, the inner product $N_{e_1} \cdot N_{e_2}$ reads
\begin{equation}
N_{e_1} \cdot N_{e_2} = - \cosh \zeta
\end{equation}
The inner product between two timelike normals in the same light--cone is always negative. In particular, 
\begin{equation}
N_{e_1} \cdot N_{e_2} \in (-1,-\infty)
\end{equation}
with $-1$ being the case when they are parallel and negative infinity reached in the limit where one of them becomes null. Thus, the rapidity $\zeta$ at a thick wedge is always positive  
\begin{equation}
\zeta \in (0,\infty)
\end{equation}

We then \emph{define the dihedral angle on a thick-wedge to be negative} by 
\begin{equation}
\Theta_{thin} \equiv -\zeta  = - \arccosh\left(- N_{e_1} \cdot N_{e_2}\right)
\end{equation}

The choice to reverse the sign appears perverse, since the angles at thick--wedges are the natural hyperbolic angles. This is done to absorb the sign difference between \eqref{eq:EuclideanDeficitAngle} and \eqref{eq:LorentzianDeficitAngle}, so that  the negative and positive local curvatures are associated to negative and positive deficit angles, as in the Euclidean case, see previous section.

We now take the case of a thin wedge. We consider the inner product between one of the normal and the $PT$ transform of the other. The same apply as above, for $N_{e_1}$ and $PT \, N_{e_2}$ or $PT \, N_{e_1}$ and $ N_{e_2}$. Notice that 

\begin{equation}
PT \, N=-\, N
\end{equation}

 We then \emph{define the dihedral angle for a thin wedge to be positive}, by 
\begin{eqnarray}
\Theta_{thin} \equiv \zeta  &=& \arccosh\left(- PT\;N_{e_1} \cdot N_{e_2}\right) \nonumber \\
&=& \arccosh\left(- N_{e_1} \cdot PT\;N_{e_2}\right) \nonumber \\
&=& \arccosh\left(N_{e_1} \cdot N_{e_2}\right)
\end{eqnarray}

\section{Emergence of General Relativity: overview of fixed--spins asymptotics of the EPRL model}
\label{sec:overviewAsAnEPRL}
In this Section we review results from the asymptotic analysis of the EPRL model, that will be used in Chapter \ref{ch:gravTunneling} to show how a gravitational tunneling picture emerges when combined with a boundary state corresponding to a wavepacket of an embedded quantum 3-geometry. These tools are put to work in Section \ref{sec:lifetimeEstim} to calculate the scaling of the lifetime of the Haggard-Rovelli spacetime, which is the subject of the following chapter. 

The asymptotic analysis of the Lorentzian EPRL model was first carried out in \cite{barrett_quantum_2010,
barrett_asymptotic_2009,barrett_lorentzian_2010} by Barett et al, at the level of a single vertex. The results we present here concern an arbitrary 2-complex, dual to a simplicial complex, based on the analysis in \cite{han_path_2013,han_asymptotics_2012,
han_asymptotics_2013}.

\subsection{The critical point equations} \label{sec:critPointEqs}
In this section we discuss the critical points equations resulting from a stationary phase approximation for the partial amplitude, equation \eqref{eq:partialAmplitude}. We skip some details that are not necessary for what follows, see \cite{han_path_2013,han_asymptotics_2012,han_asymptotics_2013} for a complete treatment in this formalism.

 We are interested in the limit of large quantum numbers, the quantum numbers here being the spins $j_f$. We introduce a dimensionless large parameter $\lambda \gg 1$ and define
\begin{equation}
j_f=\lambda \delta_f
\end{equation}
We will call $\delta_f$ the \emph{spin data}. They are taken to be of order unit in $\lambda$.
That is, we take all the spins to scale uniformly. We see in Chapter \ref{ch:gravTunneling} that this regime is sufficient when considering applications to macroscopic geometries, where a physical area scale characterising the system will play the role of $\lambda$. 

We finished Section \eqref{sec:ampsInAsAnform} with equation \ref{eq:faceAmpSpinors}, which we repeat here
\begin{equation} \label{eq:tempFaceAmp}
A_f(j_f) =  \int \! \mu(z_{vf}) \; e^{j_f F_f\left(\{g_f\},\{z_f\}\right)}
\end{equation}
We employ the simplified notation of Section \ref{sec:emergenceOfGR} and define the partial amplitude $I(j_f)$ as the product of the face amplitudes, including also the integrations over \slt
\begin{eqnarray} \label{eq:partialAmplitude}
I(j_f) = I(\lambda,\delta_f) &=& \int \! \mu(g) \mu(z) \, e^{\lambda \sum_f \; \delta_f F_f(g,z)} \nonumber \\
&=& \int \! \mu(g) \mu(z) \, e^{\lambda \Sigma(\delta_f,g,z)}
\end{eqnarray}
where we have defined
\begin{equation}
\Sigma(g,z;\delta_f) \equiv \sum_f \delta_f F_f(g,z)
\end{equation}
and where the semicolon in $\Sigma(g,z;\delta_f)$ reminds us that we are taking the spin data $\delta_f$ to be \emph{fixed}. That is, the spinfoam amplitude of equation \eqref{eq:ampFaceProduct} reads
\begin{equation} \label{eq:ampWithPartAmp}
W_\mathcal{C} = \sum_{\{j_f\}} \mu(j_f) \prod_f A_f(j_f) =\sum_{\{j_f\}} \mu(j_f) \; I(j_f)
\end{equation} 

The (generalised, for complex ``actions'') stationary phase theorem says that when $\lambda$ is large, we may take the following approximation\footnote{For the precise theorems and conditions that need to be satisfied by the integrand see Section 7.7 of the reference textbook by Hormander \cite{hormander_linear_1969}. 
We have not included the determinant of the Hessian, an overall scaling in $\lambda$, and a sign factor, the power of which depends on the combinatorics of the spinfoam. For the full expression see \cite{han_asymptotics_2013}. }
\begin{equation} \label{eqstatPhase}
I_{j_f}(\lambda,\delta_f) \sim \left( \sum_c e^{\lambda \Sigma(g_c(\delta_f),z_c(\delta_f);\delta_f)} \right) \left( 1+ \mathcal{O}(1/\lambda)\right)
\end{equation}
where $(g_c(\delta_f),z_c(\delta_f))$ is the set of critical points for the given $\delta_f$. 
We emphasize that the point of view here is that \emph{both} the spin data $\delta_f$ and the large parameter $\lambda$ are taken to be fixed, we are not considering a limit $\lambda \rightarrow \infty$. The large parameter $\lambda$ will be identified later on with a finite macroscopic area scale. By critical point, we mean that $(g_c(\delta_f),z_c(\delta_f))$ solve the set of equations
\begin{eqnarray} \label{eq:critPointEqs}
 \re\Sigma=0 \nonumber \\ 
 \delta_g\Sigma=0 \nonumber \\
 \delta_z\Sigma=0 \nonumber
\end{eqnarray}
that extremize the exponent.

The group and spinor variables $g_{ve}$ and $z_{vf}$ appear in the face amplitude of equation \eqref{eq:tempFaceAmp} only in the combination  
\begin{equation}
Z_{vef} = g_{ve}^\dagger z_{vf}  
\end{equation}
see equations \eqref{eq:faceAmpSpinors} and \eqref{eq:Ffunction}. Let us define the normalized version of this spinor 
\begin{equation}
\hat{Z}_{vef} \equiv \frac{Z_{vef}}{\vert Z_{vef} \vert}
\end{equation}
with $\vert Z \vert \equiv \sqrt{\braket{Z\vert Z}}$. 

The critical point equations in these variables read (for a derivation, see for instance \cite{han_path_2013} )
\begin{eqnarray} \label{eq:critPointEqs}
\ket{\hat{Z}_{vef}} &=& e^{i \alpha_{vv'}^f } \label{eq:critCommonFrame} \ket{\hat{Z}_{v'ef}} \\
g_{e'e} \; \ket{\hat{Z}_{vef}} &=& \frac{\vert Z_{vef}\vert}{\vert Z_{ve'f}\vert} \ket{\hat{Z}_{ve'f}}  \label{eq:critParTransport} \\ 
\sum_{f \in e} j_f \; \epsilon_{ef} \braket{\hat{Z}_{vef} \vert \vec{\sigma} \vert \hat{Z}_{vef}} &=& 0 \label{eq:critClosure}
\end{eqnarray}
where we have denoted  $g_{e'e} \equiv g_{e'v}g_{ve}$.

These equations are not too complicated. They can appear obscure at first sight, but, after a few comments, their meaning becomes transparent. 
 
We introduced spinors in Section \ref{sec:ampsInAsAnform}, because it is then possible to calculate the \slt inner products explicitly. We now have the critical point equations in terms of spinor variables, from which we can attempt to reconstruct a geometry. In particular, we want to see whether we can recover a simplicial geometry and we expect the reappearance of the plane and co-plane.

Let us start from the third critical point equation, equation \eqref{eq:critClosure}. On the right hand side we have three--dimensional normalized vectors (elements of $S^2$) 
\begin{equation}
\hat{n}_{ef} \equiv \epsilon_{ef}\braket{\hat{Z}_{vef} \vert \vec{\sigma} \vert \hat{Z}_{vef}}
\end{equation}
This is a well known mapping from normalized spinors to $S^2$ ($\mathbb{CP}1$), see Appendix \ref{app:spinorsAndRiemannSphere} for the  explicit Riemann sphere construction. The factor $\epsilon_{ef}$ is a sign ($\pm 1$), that ensures that this vector points in the correct direction (outwards from the tetrahedron $e$). The reason we did not include a vertex subscript for $\hat{n}_{ef}$ will become apparent when we examine the first critical point equation.

The meaning of \ref{eq:critClosure} is follows: it is the \emph{closure condition}, with the spins having the role of areas
\begin{equation} \label{eq:closureCond}
0= \sum_{f \in e} j_f \; \hat{n}_{ef}
\end{equation}
The normals to and areas of the triangles of a  tetrahedron (any flat convex polyhedron), always satisfy this condition. \emph{Conversely, any such set of areas and normals determines uniquely, up to inversion, a tetrahedron}.\footnote{By inversion, we mean the mirroring of the tetrahedron through any of its faces. We have in mind a tetrahedron as a geometrical shape, thus translational and rotational ambiguities in defining the normals are irrelevant here. } This is a point that cannot be stressed enough: one of the critical point equations clearly states that \emph{all} critical points for the path--integral, are collections of tetrahedra, dual to the edges of the two-complex $\mathcal{C}$. We have then already recovered part of the geometry. 

The crucial question will be whether, at least at some of these critical points, the tetrahedra \emph{glue} properly together to form simplicial geometries. 

Let us examine the other two critical point equations. These are equations at the level of a fixed face $f$. To explain the meaning of the remaining critical equations, we need only consider two vertices $v$ and $v'$, the edge that joins them, which we call $e'$, and a half-edge $e$.

The first critical point equation, equation \ref{eq:critCommonFrame} has a phase factor $e^{i \alpha_{vv'}^f} $. This is important to consider when working in the coherent state representation, but we will not need to consider it explicitly here. Thus, we may write   
\begin{equation}
\hat{Z}_{vef} = \hat{Z}_{v'ef} 
\end{equation}
From the third critical point equation, we saw that the spinors $\hat{Z}_{vef}$ define three dimensional normals, that live at the edge $e$, dual, at a critical point, to a geometrical tetrahedron. Then, this equation is telling us that the two vectors constructed from $\hat{Z}_{vef}$ and $\hat{Z}_{v'ef}$ coincide. That is, there is a notion of a \emph{common frame}, shared among the vertices $v$ and $v'$. This common frame is defined on a face $f$ of a tetrahedron $e$ common to $v$ and $v'$

Remember the role of the co-plane in Regge calculus and in the tetradic forms of GR. We expect the reappearance of the plane and co-plane. The co-plane will be spanned by the timelike normal to a tetrahedron and a spacelike normal to the face.    
 
Let us then attempt to define two tetrad directions corresponding to the co-plane of the face $f$. 
The co-plane is spanned by the timelike normal to the tetrahedron $e_{0ef}$, to which the face $f$ belongs, and by a spacelike normal to the face $e_{1ef}$. This corresponds to a gauge fixing, adapted to this tentative common frame of $v$ and $v'$. We may write
\begin{eqnarray}
e_{0ef} \equiv (1,0,0,0) \\ 
e_{1ef} \equiv (0,\hat{n}_f) 
\end{eqnarray}
 
The co-plane, is then described by a bi-vector proportional to $e^0_{ef} \wedge e^1_{ef}$. Indeed, we may recast the critical point equations in terms of bi-vector variables, defined as 
\begin{equation}\label{eq:thinWedgeJustification2}
X_{ef} \equiv 2 \gamma j_f \; e^0_{ef} \wedge e^1_{ef}
\end{equation} 
Notice that the plane dual to $X$ will then be spacelike, that is 
\begin{equation} \label{eq:thinWedgeJustification}
e_{0ef} \cdot \star X_{ef} =0 
\end{equation}
with the inner product taken with the Minkowski metric. This will be the plane of the triangle dual to $f$ (at a critical point).
 
So far, we have tetrahedra dual to edges $e$, and a notion of plane and co-plane for the triangles dual to faces $f$. This settles the candidates for the emergence of the tetrad and co-tetrad at the semi-classical limit. 

So, how about the curvature? It should live in the co-plane, and we expect it to be manifested through the action of the group elements, the quantum equivalent of the discretized spin connection, on $X_e$.

Let us then examine the remaining critical point equation, equation \eqref{eq:critParTransport} 
\begin{equation}
g_{e'e} \; \ket{\hat{Z}_{vef}} = \frac{\vert Z_{vef}\vert}{\vert Z_{ve'f}\vert} \ket{\hat{Z}_{ve'f}}  
\end{equation}
If we ignore the norms of the spinor variables on the right-hand side, this equation looks like a parallel transport, effectuated by $g_{e'e}$, from the middle of the edge $e$ to the middle of the edge $e'$. In other words, it is a parallel transport from a common frame between the vertices sharing $e$ to the common frame between the vertices sharing $e'$. This is good, because the parallel transport encodes the curvature.

Let us see how this works. We bring in Minkowski indices, and pretend that we have indeed a simplicial geometry. Then, the parallel transport by the discrete connection $g$ is given by first parallel transporting to $v$ from the middle of $e$, by $g_{ev}$, and then to the middle of $e'$, by  $g_{ve'}$. Since $g_{ee'}=g_{ev}g_{ve'}$, we would expect to have 
\begin{eqnarray}
 g_{e'e} \triangleright X_{ef} &=& X_{e'f} \nonumber \\ \nonumber \\
 g_{e'e \ K}^{\ \; I} \; g_{e'e \ \; L}^{\ J} X_{ef}^{KL} &=& X_{e'f}^{IJ}  
\end{eqnarray}

Since we are at a pre-geometric level, we give meaning to the above by understanding the bi-vector variables $X_{ef}$ as taking values in the Lie algebra, and the group action to be given by the adjoint action. Using the property of equation \eqref{eq:usefulParityIdentity} for the parity map, it is not hard to see that the third critical point equation equation is precisely this parallel transport condition.

The idea then is to bring all these variables to a vertex $v$, by parallel transporting $X_{ef}$ along the half edge $ev$
\begin{equation}
X_f(v) \equiv g_{ve} \triangleright X_{ef}
\end{equation}
Notice that there is no $e$ subscript in $X_f(v)$. This is because  $g_{ve} \triangleright X_{ef} = g_{e'v} \triangleright X_{e'f}$.

 The critical point equations can be recast as \cite{han_path_2013}
\begin{eqnarray} \label{eq:critPointEqsBivectors}
X_{f}(v)=X_{ef}(v)=X_{e'f}(v) \nonumber \\
\eta_{IJ} N^I_e(v)\star X^{IJ}_f(v)=0 \nonumber \\
\sum_{f \subset t_e} \epsilon_{ef}(v) X_f(v) =0
\end{eqnarray} 
where the normal $N^I_e(v)$ is the parallel transport of $e_{0ef}$, $N_e(v) \equiv g_{ve} \triangleright e_{0ef}$, and
\begin{equation} 
X_f(v) = 2 \gamma j_f \left(Jz_{vf} Jz_{vf}^\dagger  \wedge g_{ve} \hat{Z}_{vef} \hat{Z}^\dagger_{vef} g_{ve}^\dagger \right)
\end{equation}
 To show these relations, it suffices to use repeatedly the properties of the parity map $J$ of Section \ref{sec:spinorForKrajHan}. 

The critical point equations now have a clear meaning. The first is just the expression of the co-plane, defined between two tetrahedra sharing a face $f$, by the normal to the face $f$ and a common timelike normal to the tetrahedra (because it is normal to both $f$ and the normal to $f$ inside either $e$ or $e'$). We stress again that this construction is identical to the Regge action (when the critical point admits an interpretation as a simplicial geometry).
 
The second is saying that all faces $f$ (all planes $\star X_f$) are all normal to the timelike normal of the tetrahedron $e$ to which they belong. And the third is the closure condition, expressed in terms of the co-planes.
 
 Notice that to cast the critical point equations to this form, we set the tentative timelike tetrad vector $e_{0ef}$ to be normal to the face $f$. Thus, in Minkowski components, the normal to the face $f$ reads $(1,0,0,0)$. That is, we rotated the tetrad to be in the \emph{time gauge}, adapted to the face $f$. 
 
\bigskip

 \subsection{Emergence of simplicial geometries} 
 \label{sec:emergenceOfSimpGeoms}
In this section we review how simplicial geometries emerge from the critical point equations \eqref{eq:critPointEqsBivectors}. The complete analysis for the Lorentzian case in an arbitrary simplicial complex is in \cite{han_asymptotics_2013}. 

For a given set of the fixed spin data $\delta_f$, there exist two distinct possibilities. Either there exists a set of critical points $(g_c(\delta_f),z_c(\delta_f))$, or not. In the latter case, the partial amplitude is exponentially suppressed in $\lambda$. 

When a set of critical points $(g_c(\delta_f),z_c(\delta_f))$ does exist, we distinguish again two cases. The first case, is that we have a \emph{geometric} critical point, determining a simplicial geometry.

In turn, three possibilities exist for geometric critical points: the spin data $\delta_f$ either determine uniquely a 4D Lorentzian Regge geometry, or a degenerate 3D Regge geometry (zero four--volume 4D Regge geometry), or a 4D Euclidean Regge geometry. In this manuscript we will be mainly concerned with these three geometric kinds of critical points. 

When the critical point is geometric, we may reconstruct a \emph{frame}, a discrete version of the tetrad, from the critical point equations expressed in terms of bivectors, i.e.\! using equations \eqref{eq:critPointEqsBivectors}.

The frame is related to the bi--vectors by
\begin{equation}
X_f = \epsilon \, \star \left( e_{s_1} \wedge e_{s_2} \right)
\end{equation}
where $e_{s_1}$ and $e_{s_2}$ are simply the dual forms to \emph{segment vectors} which are now interpreted as two of the three segments of the triangle $f^\star$. The sign $\epsilon$ is to be chosen such that the bi--vector points out of the tetrahedron on which we are working. Think of the cross-product in basic 2D geometry and $X_f$ as the oriented area.

The second, non-geometrical, case for a critical point, is that of a vector geometry.\footnote{Barett includes the geometrical 3D degenerate case in the vector geometry case, this is a matter of definition.} A vector geometry is roughly a collection of tetrahedra that need not glue together to form a simplicial complex. The precise characterization and parametrization of vector geometries is a currently undergoing project in the Centre de Physique Th\'eorique in Marseille, led by S.Speziale. 

\bigskip

We will have more to say about these different cases later. Let us now focus on the case where $\delta_f$ are chosen such that the set of critical points $(g_c(\delta_f),z_c(\delta_f))$ corresponds to a 4D Lorentzian Regge geometry.  

First, notice that we claimed that there is a \emph{set} $(g_c(\delta_f),z_c(\delta_f))$ of critical points that \emph{uniquely} determine a Lorentzian Regge geometry. How is this possible?

 It turns out that there are exactly $2^V$ critical points $(g_c(\delta_f),z_c(\delta_f))$,  corresponding to the $2^V$ choices for the frame orientation, the discrete equivalent of the tetrad. 
 
 This was to be expected: remember that at the end of Section \ref{sec:HEtoCar} we neglected the difference between the Einstein--Hilbert action and the Palatini action; the absolute value in the determinant. The \emph{sign} of the determinant encodes whether the tetrad's orientation is left--handed or right--handed. 
 
 We never imposed that absolute value in the spinfoam quantization procedure outlined in this manuscript. It is then not surprising that the geometrical co-frame that emerged in the semiclassical limit has not chosen an orientation for us.
 
Specifically, what happens is the following: the set of $2^V$ critical points $(g_c(\delta_f),z_c(\delta_f))$ constructs (uniquely, see below) a discrete Regge metric  
\begin{equation}
g_{s s'} = e_{Is} e_{Js'} \eta^{IJ} 
\end{equation}
where $e_{Is}$ are Minkowski vectors corresponding to segments $s$ of the \emph{geometric} simplicial triangulation dual to the 2-complex $\mathcal{C}$ at the geometric critical point. 

To define the frame, we need only construct four of these vectors at each vertex $v$ of $\mathcal{C}$, which we can choose for instance to emanate from the same apex of the geometric 4-simplex dual to $v$, and to point away from the apex. In three dimensions, this is equivalent to giving the three segment vectors emanating from an apex of a tetrahedron.

 The same set of vectors is assigned to every point in the interior of the 4-simplex dual to the vertex $v$ by means of the trivial parallel transport in flat space.

So far, this is convention and does not affect physics. But we have not yet arrived at the point where we have a discrete analogue of the tetrad field. We need in addition to chose an \emph{arbitrary labelling}, say from one to four, for the four dual vectors $e^I_s$ (or, of the vectors $e_{Is}$) at each vertex. Equivalently, choose a naming for the four segments $s$. This defines the discrete (distributional) non-coordinate basis dual vector fields $e^I_{s_1}(v)$, $e^I_{s_2}(v)$, $e^I_{s_3}(v)$, $e^I_{s_4}(v)$ on the simplicial complex dual to $\mathcal{C}$. 

The collection of the four co-vector fields $e^I_{s_i}(v)$ is what is called here the \emph{co-frame} and is the precise discrete analogue of the co--tetrad field $e^I$ that we studied in Chapter \ref{ch:topBottom}, for a piecewise flat simplicial manifold. However, here, the frame is not orthonormal, because it carries also the information on the edge lengths and angles, that is, we may reconstruct from it all the geometry.

 The orientation appears in the asymptotic analysis via the \emph{signed 4-volume} $\mathcal{V}^4(v)$ for every 4-simplex dual to a vertex $v$ :
\begin{equation}
\mathcal{V}^4(v) \equiv e^I_{s_1}(v) e^J_{s_2}(v) e^K_{s_3}(v) e^L_{s_4}(v) \epsilon_{IJKL} 
\end{equation}
Had we chosen a different basis of tangent vectors, by exchanging the labelling in any two elements of the co-frame \emph{at a single vertex} $v'$, say $e^I_{s_1}(v') \leftrightarrow e^I_{s_2}(v')$, the 4-volume of the 4-simplex dual to $v'$ would now have the opposite sign : $\mathcal{V}^4(v') \rightarrow -\mathcal{V}^4(v')$. We recognise that this sign simply encodes the left-handedness or right-handedness of the frame and co--frame at each vertex, which is what we call its orientation. 

\bigskip

In summary, any of the critical points $(g_c(\delta_f),z_c(\delta_f))$ for given $\delta_f$ determines the same metric geometry of a simplicial manifold dual to the entire 2-complex $\mathcal{C}$ and each individual critical point uniquely corresponds to the choice of one of the $2^V$ orientation configurations of the basis $e^I_{s_1}(v)$, $e^I_{s_2}(v)$, $e^I_{s_3}(v)$, $e^I_{s_4}(v)$.  The critical points are then distinguished by the sign of $\mathcal{V}^4(v)$ at each vertex and we use the shorthand notation
\begin{equation}
s(v) \equiv \sign{\mathcal{V}^4(v)} 
\end{equation}
The function $s(v)$ takes the values plus or minus one at each vertex and is the discrete analogue of the \emph{sign} of the determinant of the tetrad.  That is, $s : \{v\} \to \{-1,1\}^V$, where $V$ is the number of vertices in the 2-complex $\mathcal{C}$ and $\{v\}$ the set of vertices in the 2-complex $\mathcal{C}$. 

With these comments, we are ready to summarize the results from the fixed-spins asymptotic analysis of the partial amplitude for the EPRL model.

\subsection{Summary of fixed--spins asymptotics}
\label{sec:summaryFixedSpins}
The asymptotic analysis of the EPRL model for the case of geometrical critical points, can be summarized as 
\begin{equation} \label{eq:FPalatini}
F_f(g_c(\delta_f),z_c(\delta_f)) = i \gamma \sum_{v \in f} s(v) \Theta_f(v;\delta_f) \equiv i \gamma \phi_f(v,s(v);\delta_f)  
\end{equation}
where $\Theta_f(v;\delta_f)$ is the dihedral angle at the triangle dual to the face $f$ of the 4-simplex dual to vertex $v$. See Section \ref{sec:lorReggeAction} for its definition in Euclidean and Lorentzian signature. 

When the geometry is 3D (zero four--volume), we may think of it in either signature, and the dihedral angles are either zero or $\pi$.\footnote{ A three dimensional analogy is to think of a tetrahedron squashed to the plane of one of its triangles. The angles between the normals to the triangles are either zero or $\pi$.}

The subscript $c$ has been dropped in the right hand side of the above equation, and will be similarly dropped from now on, since each critical point $(g_c(\delta_f),z_c(\delta_f))$ is in one to one correspondence with an element $s(v)$ in the image $\{-1,1\}^V$ of $s$. That is, the critical points corresponding to the same geometry are labelled by a $\pm$ configuration on each vertex, coding the left-- or right-- handedness of the frame.

We have introduced a function we will call the \emph{Palatini deficit angle} 
\begin{equation} \label{palatiniAngle}
\phi_f(v,s(v);\delta_f) \equiv \sum_{v \subset f} s(v) \Theta_f(v;\delta_f)   
\end{equation}
On the critical point $(g_{c'}(\delta_f),z_{c'}(\delta_f))$ which has $s(v)=1$ at all vertices $v$, we recognise that its Palatini deficit angle is the \emph{Regge deficit angle}, $\phi_f(v,s(v);\delta_f) = \sum_{v \in f} \Theta_f(v;\delta_f) $.

\smallskip

 We have the following sequence of equations for the exponent in the partial amplitude, equation \eqref{eq:partialAmplitude},
\begin{eqnarray}\label{sequ}
\lambda \Sigma(g_{c}(\delta_f),z_{c}(\delta_f);\delta_f) &=&  \lambda \sum_f \delta_f F_f(g_{c}(\delta_f),z_{c}(\delta_f))  \nonumber \\  &=& i  \sum_f (\gamma \lambda \delta_f) \phi_f(v,s(v);\delta_f)  \nonumber \\ &=& \frac{i}{\hbar}  \sum_f A_f \phi_f(v,s(v);\delta_f)      
\end{eqnarray}
where we have introduced the dimensionful area variable $A_f \equiv \gamma j_f \hbar $ (we use geometrical units, i.e.\! $c=G=1$), using the basic result of LQG that the area spectrum of the area operator is given by \cite{rovelli_discreteness_1995}
\begin{equation} \label{eq:areaSpectrum}
A_j= 8 \pi G \, \hbar \gamma \sqrt{j (j+1)} \approx 8 \pi G \, \hbar \gamma j
\end{equation}

On the critical point $(g_{c'}(\delta_f),z_{c'}(\delta_f))$ we recognise the Regge action for the entire 2-complex. That is, since on this critical point $\phi_f(v,s(v);\delta_f)$ is the Regge deficit angle, we have 
\begin{equation}
 \lambda \Sigma(g_{c'}(\delta_f),z_{c'}(\delta_f);\delta_f)=\frac{i}{\hbar} S_R^\mathcal{C^\star}
\end{equation} 	

 It will also be useful to write the exponent in terms of the Regge action $S_R(v;\delta_f)$ for each 4-simplex dual to a vertex $v$, defined as 
\begin{equation}
S_R(v;\delta_f) = \sum_{f \in v} A_f \Theta_f(v;\delta_f)
\end{equation}

In summary, on a geometrical critical point, we have 
\begin{eqnarray} \label{discretePalatini}
\lambda \Sigma(g_{c}(\delta_f),z_{c}(\delta_f);\delta_f) &=& \lambda \sum_f \delta_f F_f(g_{c}(\delta_f),z_{c}(\delta_f)) \nonumber \\ 
&=& i \gamma \lambda \sum_f \delta_f  \sum_{v \in f} s(v) \Theta_f(v;\delta_f) \nonumber \\ 
&=&  i \sum_v s(v)  \sum_{f \ni v}  ( \gamma \lambda \delta_f) \Theta_f(v;\delta_f) \nonumber \\
&=&  \frac{i}{\hbar} \sum_v s(v)  \sum_{f \ni v}   A_f \Theta_f(v;\delta_f) \nonumber \\
&=& \frac{i}{\hbar} \sum_v s(v) S_R(v;\delta_f,\lambda)
\end{eqnarray}
For emphasis, note that we again recognise that for the critical points for which $s(v)=\pm 1$ everywhere, we have recovered the Regge action 
\begin{equation}
\lambda \Sigma(g_{c'}(\delta_f),z_{c'}(\delta_f);\delta_f)= \pm \frac{i}{\hbar}  \sum_v S_R(v;\delta_f)=\pm \frac{i}{\hbar} S_R^\mathcal{C^\star}  
\end{equation}

In all, to the zeroth order in $\lambda$,
\begin{equation} \label{eq:partAmpAsymptotics}
\boxed{
 I(\lambda,\delta_f)  \sim \sum_{\{s(v)\}} e^{\frac{i}{\hbar} \sum_v s(v) S_R(v)}
 }
\end{equation}

\section{Summary: the sum--over--geometries}
\label{sec:summaryChap3}
Let us summarize. 

We have seen that the fixed--spins asymptotics of the partial amplitude in the semi-classical limit of uniformly large spins $j_f$, yield the exponential of the discretized Palatini-Holst action. 

This special limit is generally sufficient for applications to geometry transition between macroscopic geometries, when a physical area scale will be available and will play the role of $\lambda$. See Chapters \ref{ch:gravTunneling} and \ref{ch:calculationOfAnObservable} for further discussion. 

We appear to have a quantum theory corresponding in this semi-classical limit to a classical theory that is described by the Palatini action. 

\emph{Furthermore the co-frame orientation is summed-over in the path-integral}. The sum in equation \eqref{eq:partAmpAsymptotics} has $2^V$ terms, one for each co-frame orientation configuration $s(v)$. The sign function $s(v)$ is the discrete analogue of the sign of the determinant of the tetrad in the Palatini action. We have not at any point imposed an orientation for the tetrad, and it would have been indeed surprising if at the limit of large quantum numbers we somehow ended up with a preferred local frame orientation. 

This intriguing feature is common to state--sum spin models, since the Ponzano-Reggge model. It has been dubbed the ``cosine problem'' \cite{vojinovic_cosine_2014, riello_self-energy_2013,christodoulou_how_2012-1,christodoulou_divergences_2013,immirzi_causal_2016}, although, as we also see in Chapter \ref{ch:gravTunneling} when we consider physical transition amplitudes, there does not appear to be anything problematic about it. In an early work with the participation of the author \cite{christodoulou_divergences_2013}, it has been linked to infrared divergences \cite{riello_self-energy_2013}, arising because the spins are allowed to take arbitrarily large values. These divergences are expected to be regularized from the inclusion of a cosmological constant \cite{haggard_encoding_2015,haggard_four-dimensional_2016}.

We note that the $F(g,z)$ function at a geometrical critical point, which becomes the Palatini deficit angle, depends only on the \emph{spin data} $\delta_f$, it does not depend the overall scale $\lambda$, see equation  \eqref{eq:FPalatini}. This point is useful to keep in mind for the following chapter and is to be expected: angles do not depend on the overall scale, changing $\lambda \rightarrow \lambda'$ is a dilatation of the geometry (a global conformal transformation).

To get a physical transition amplitude, describing geometry transition between macroscopic semi--classical geometries, we need to consider a boundary and an appropriate semi--classical state, and perform the \emph{spin-sum}. This is done in the following chapter. 

When considering a semi--classical state, the critical points giving rise to vector geometries are excluded. There will still be the possibilities of having a geometrical critical point, or no critical point at all.

Finally, the Regge action we have recovered has the \emph{areas} as variables instead of the edge lengths. The study of the dynamics from this form of the Regge action is known as area Regge calculus \cite{barrett_note_1999,neiman_look_2013}. The degree to which it is related to General Relativity is unclear at the moment.\footnote{As mentioned previously, when writing the spinfoam amplitude as a sum over only spin configurations, we are neglecting the intertwiner degrees of freedom. There are then some semi--classical degrees of freedom missing from the emergent sum--over--geometries path--integral. These intuitively correspond to 3D dihedral angles of the tetrahedra, that are encoded in the 3D normals corresponding to the spinors $z_{vf}$ that were integrated out when taking the stationary phase approximation. It would be desirable to recover instead the Regge action in area--angle variables, see \cite{dittrich_area-angle_2008-1,anza_note_2015-1}. }

\medskip

With the above caveats, we have seen in which sense the Wheeler--Misner--Hawking sum--over--geometries emerges from the amplitudes of covariant LQG. Specifically, recall equation \eqref{eq:ampWithPartAmp}
\begin{equation} 
W_\mathcal{C} = \sum_{\{j_f\}} \mu(j_f) \; I(j_f)
\end{equation} 

From equation \eqref{eq:partAmpAsymptotics}, neglecting the cosine feature (setting $s(v)=1$ on all vertices) and the special semi--classical limit, we have, at the level of the partition function, the correspondence of Table \ref{tab:aimEPRL}

\begin{eqnarray}
W_{WMH} &\sim& \int\! \mathcal{D} [g] \ \ \ \ e^{\frac{i}{\hbar} S_{HE}[g]  } \nonumber \\ \nonumber \\
W_{NR} &\sim& \int\! \mu(\ell_s) \ \ \ \ e^{\frac{i}{\hbar} S^\mathcal{C^\star}_{R}[\ell_s]  } \nonumber \\ \nonumber \\
W^{EPRL}_\mathcal{C}  &\sim & \int \mu(a_f) \ \ e^{\frac{i}{\hbar} S^\mathcal{C^\star}_R[a_f]}
\end{eqnarray}
where we have turned the summations over the spins to integrals over areas. We have included for comparison the naive path integral over the Regge action (the abbreviation NR is for Naive Regge). The Regge action is a discretization of the Einstein--Hilbert action, yielding the latter in an appropriate limit of refinements. In turn, the amplitudes $W_\mathcal{C} $ of covariant LQG are understood as a truncation of the degrees of freedom of the quantum gravitational field.

\chapter{Gravitational Tunneling}
\label{ch:gravTunneling}
In this chapter we enter the main part of this manuscript, and explain how gravitational tunneling emerges from covariant Loop Quantum Gravity. We first introduce coherent spin-network states. These are the semiclassical boundary states used in the second part of this chapter and can be thought of as wavepackets for an embedded three-geometry. We proceed to combine the boundary states with a spinfoam amplitude to define a transition amplitude. We will use the results from the asymptotic analysis of the EPRL model presented in the previous chapter and explain how the spin-sum may be performed for a class of spinfoam amplitudes. 

\section{Boundary state : a wavepacket of geometry} 
\label{sec:wavepacketOfGeometry}
In this section we introduce coherent spin-network boundary states, describing wavepackets of a spacelike 3-geometry embedded in a Lorentzian spacetime. We start with an introductory section, discussing the relation of the spin network basis with the gauge-variant semiclassical boundary states that will be used in the rest of this manuscript.

\subsection{Superposing Spin--Networks}
\label{sec:kinematicsLQG}
In this preliminary section we briefly review 
the kinematical Hilbert space of LQG at the level of a fixed graph. For lack of space, we do not give a full presentation of the kinematics of LQG and refer the reader to introductory texts in the literature \cite{rovelli_covariant_2014,
rovelli_zakopane_2011,dona_introductory_2010,
ashtekar_introduction_2013-1}. A good place to start and one that contains material relevant to spinfoams, is the treatment of the three--dimensional case by A.Perez and K.Noui \cite{noui_three-dimensional_2005}. For the definition of the full kinematical Hilbert space see the influential work by A.Ashtekar and J.Lewandowski \cite{ashtekar_differential_1995}.

\bigskip

Loop Quantum Gravity, and spinfoam quantization, are based on the Ashtekar-Barbero variables. We mentioned in Section \ref{sec:relationToCanonical} that they arise naturally in the canonical analysis of the Holst action for General Relativity, written as a fully constrained system. 

The Ashtekar-Barbero variables are the Ashtekar-Barbero connection $A^i_a$ and the densitized triads $E^a_i$, defined as
\begin{eqnarray} \label{eq:AshtekarBarberoVars}
A^i_a & \equiv & \omega^i_a + \gamma K^i_a = \omega_{jak}\epsilon^{jki} + K_{cb} e^i_b q^{ac}   \nonumber \\
E^i_a\nonumber & \equiv & \sqrt{\det{q}} \; e^a_i
\end{eqnarray}
where $\omega$ is a 3D spin connection.

The AB--variables parametrize the phase space of general relativity and are canonical field variables
\begin{eqnarray}
\{E^i_a(x), A^j_b(y) \} &=& \gamma \, \delta^a_b \, \delta^i_j \delta(x,y) \nonumber \\
\{E^i_a(x), E^j_b(y) \} &=& \{A^i_a(x), A^j_b(y) \}=0
\end{eqnarray}

An important point to keep here, is that the ``good'' connection variable for General Relativity, the AB-connection, turns out to be a simple combination of the two objects \emph{fully characterising} the spacetime curvature in a neighbourhood of a hypersurface: the 3D spin connection, from which we may reconstruct the intrinsic curvature, and the extrinsic curvature. 

We will see how semiclassical boundary states in LQG are built only out of $SU(2)$, which is naturally associated to the symmetries of a three-dimensional Euclidean space. The reason we may encode the \emph{spacetime} curvature is the interpretation of the $SU(2)$ group variables as holonomies of the AB--connection. The Levi-Civita 3D spin connection would only know about the intrinsic curvature, because the AB--connection knows of the extrinsic curvature, which codes the embedding of a time slice in spacetime, we will be able to include a notion of ``quantum embedding'' (in a discrete setting) in the boundary state.

The geometric interpretation of the Ashtekar-Barbero variables is along the lines of the discussion in Chapter \ref{ch:topBottom}. The choice of connection corresponds to a choice of a notion of parallel transport. The Ashtekar-Barbero connection $A^i_a$ provides a notion of parallel transport for $SU(2)$ spinors along curves in a hypersurface $\Sigma$, belonging to a foliation of a 3+1 spacetime split, with its holonomy an $SU(2)$ element.

The triad has the same meaning as the tetrad. The tetrad is a four--dimensional Minkowski orthonormal frame
and the triad a three--dimensional Euclidean orthonormal frame.

\subsubsection{Doubly superposing spin networks}

The quantization of General Relativity in terms of Ashtekar-Barbero variables is carried out by first discretizing. The spirit is similar to the skeletonization of BF-theory of Section \ref{sec:BFSpinfoamQuantization}. 

 The discretized triads become a set of variables we call the fluxes, encoding the area and the normal direction to a face, and the connection is discretized as a set of holonomies. The resulting canonical algebra is the holonomy--flux algebra.   
 
 When promoted to operators, the holonomy--flux algebra is realised on the kinematical space at the level of a  fixed graph, $\mathcal{H}_\Gamma$, see below. The continuum kinematical space of LQG is built by taking an appropriate refinement limit (a projective limit, see \cite{ashtekar_differential_1995}). In this manuscript, we will only be concerned with states at the level of a fixed graph $\Gamma$.
 
Since we are working at the level of the boundary and the boundary graph $\Gamma$, we employ the notation for the boundary of Table \ref{tab:triangulation}.

 Links are labelled as $\ell$, nodes as $\nn$, and a half link attached to $\nn$ as $\ell\nn$. The graph $\Gamma$ is oriented and we also employ the notation $\ell_t$ and $\ell_s$ for the half link attached to the target and source of $\ell$ respectively.

An orthonormal basis of $\mathcal{H}_\Gamma$ is given by the spin--network states. These states provide an orthonormal basis and diagonalize simultaneously the two basic operators of the kinematics of LQG: the area and volume operator. 

To build the spin-network states, we introduce an oriented (here, four valent) graph $\Gamma$, and assign an $SU(2)$ element $h_\ell$ at each link. We consider square integrable functions $\phi_\Gamma(h_\ell)$ of the group elements $h_\ell$, elements of $\mathcal{L}(SU(2)^L)$, where $L$ is the number of links in $\Gamma$. This is the group representation. The group elements $h_\ell$ are interpreted as (the quantized version of) the holonomies of the AB-connection and correspond to the configuration variables
\begin{equation} 
h_\ell = \mathcal{P} e^{\int_{\ell} A_{AB}  }  
\end{equation}

To go to the conjugate representation, we expand in unitary irreducible representations of $SU(2)$ using the Peter-Weyl theorem, see also Appendix \ref{app:recoupTheory}.

The wavefunctions become functions of the spins, and we write them as

\begin{eqnarray} \label{eq:spinNetsLikeIntrinsicStates}
\phi_\Gamma(j_\ell) &=& \bigotimes_{\ell }  D^{j_\ell}(h_\ell) \nonumber \\
&=& \sum_{\{m\}}\bigotimes_{\ell }  D^{j_\ell}_{m_{\ell s(\ell)}\,m_{\ell t(\ell)}}(h_\ell)
\nonumber \\
&=& \sum_{\{m\}}\bigotimes_{\ell } \braket{j_\ell,m_{\ell s(\ell)} \vert D^{j_\ell}(h_\ell) \vert j_\ell,m_{\ell t(\ell)}}
\end{eqnarray}
These rewritings are to aid comparison with equation \eqref{eq:intrinsicStatesLikeSpinNets}.

The transformation property of the holonomy under a gauge transformation is equation \eqref{eq:holonomyGaugeTransformation}, where now the group is $SU(2)$
\begin{equation} h_{\ell} \rightarrow h_{\ell_t}
  \; h_{\ell} \; h_{\ell_s}^{-1}
\end{equation}
To get to spin--networks, we gauge average at the nodes. That is, we insert $SU(2)$ group elements $h_{\ell\nn}$ on each half link and we integrate over $SU(2)$. 

The integrations over $SU(2)$ will give rise to four--valent $SU(2)$ intertwiners, invariants recoupling between the four representations on the four half links $\ell\nn$ meeting at a node $\nn$, see also Appendix \ref{app:recoupTheory}. 

Spin networks provide an orthonormal basis of $\mathcal{H}_\Gamma$, which is realised as $\mathcal{L}(SU(2)^L/SU(2)^N)$, where the division by $SU(2)^N$ implies the gauge averaging at the nodes. 

When the graph $\Gamma$ is understood as an abstract combinatorial object\footnote{We are taking the point of view of \cite{rovelli_covariant_2014}. See also \cite{thiemann_modern_2007}. }, $H_\Gamma$ corresponds to the kinematical space of LQG at the level of a fixed graph (the states are in the kernel of the gauss and diffeomorphism constraints of Section \ref{sec:relationToCanonical}).

\bigskip

The boundary states we will present and use correspond to a double superposition of spin--network states. 

Before explaining this point, we recall the discussion in Section \ref{sec:relationToCanonical}: spinfoams are understood as projectors on the physical Hilbert space, thus they impose the Gauss constraint. We do not need to consider a state in the gauge--invariant space $H_\Gamma$, since the gauge--invariance is automatically imposed by the \slt integrations in the spinfoam amplitude.\footnote{This can be checked easily from the invariance of the Haar measure, from the contracted form of the amplitude, equation \eqref{eq:contractedAmplitude}}

Spin--network states defined on a four--valent graph behave like plane waves for the intrinsic geometry of a spacelike tetrahedral triangulation, in the following sense: they are sharply peaked in five of the six classical variables describing the geometry of each tetrahedron and are completely spread in the remaining classical variable. This follows from the fact that spin--networks form a complete orthonormal basis of the Hilbert space at the level of a fixed graph, that is, they  diagonalize a complete set of commuting observables, the area and volume operators. 

They are thus labelled by an area eigenvalue at each link (a spin $j_\ell$), which labels a unitary irreducible $SU(2)$ represendation and corresponds to the classical areas of the corresponding triangle, and one volume eigenvalue at each node (a spin $v_\nn$), which labels a basis element of the intertwiner space and  corresponds to the volume of the tetrahedron.

 Thus, at each node, we have five eigenvalues, four areas corresponding to the four areas of the triangles of a tetrahedron, and the volume eigenvalue corresponding to the volume of the tetrahedron. Attempting to calculate any other observable corresponding to an independent geometrical quantity, for instance a dihedral angle, we find that it is completely spread (infinite uncertainty).

The first step will then be to superpose the states $\phi_\Gamma(j_\ell)$ of equation \eqref{eq:spinNetsLikeIntrinsicStates},  and define states that are peaked, with some quantum uncertainty, on all six classical variables. Imposing gauge-invariance at each node results in the Livine--Speziale states, an overcomplete basis of intrinsic geometry coherent states in $H_\Gamma$.

The second step will be to superpose these states, in order to create states that are peaked also on the extrinsic curvature, encoded in the Ashtekar--Barbero connection. The (gauge--invariant version of the) resulting states will be the semi--classical limit of another overcomplete basis of $H_\Gamma$, Thiemann's heat kernel states.

\section{Intrinsic coherent states}
\label{sec:intrinsicCoherentStates}
In this section we introduce intrinsic coherent states. We start by giving some relevant tools, fixing conventions and explain the interplay between $SO(3)$, $SU(2)$, $S^3$ and $S^2$ when building coherent states. 

\subsection{Preparation: rotations, the three-sphere and the two--sphere}
A useful parametrization for $SU(2)$, is the Euler angle parametrization. This parametrization has a clear geometric interpretation and allows one to translate easily between the rotation group $SO(3)$ and its double cover $SU(2)$. We will be defining semiclassical ``wave--packets of quantum directions'', $SU(2)$ coherent states, that are the building blocks of the full boundary states describing wave--packets of embedded 3-geometries.
These semiclassical directions correspond to points in $S^2$, and we will be understanding $SU(2)$ as the three-sphere $S^3$, which is locally isomorphic to $S^2 \times S^1$. Let us start from $SO(3)$.
\subsubsection{Rotations, $SO(3)$ and Wigner's matrices}

The Euler angle parametrization exploits the fact  that any rotation in three dimensions can be decomposed as an initial and final rotation about an axis by some angles $\alpha$ and $\gamma$, with one intermediate rotation about any other axis, by some angle $\beta$. This gives a simple geometrical meaning to the parametrization. Throughout the manuscript, we use the active interpretation, that is to say, the coordinate system is taken fixed (imagine $x-y-z$ Cartesian axes) and the rotations act on vectors. 

$SO(3)$ unitary representations act on the spin representation spaces $H^j$. We saw in Section \ref{sec:prelim} that the states of the gravitational field in covariant LQG are built from states in the  $V^{\gamma j,j}$ representation space of \slt, which is isomorphic to $H^j$. This is the fact that will allow us to build a semi--classical state peaked on a simplicial geometry, that is then naturally combined with a spinfoam amplitude defined on a 2--complex with a boundary.

 The eigenstates of the angular momentum operator in the $z$ direction $J_z$, provide an orthonormal basis for $H_j$ 
\begin{eqnarray}
 H^j &=& \spn \{ \ket{jm} , m = -j,-j+1, \ldots, j \} \nonumber \\ \nonumber \\
J_z \ket{jm} &=& m \ket{jm} \label{eq:JzEigen} \\ \nonumber \\
J^2 \ket{jm}=J_x^2+J_y^2+J_z^2\ket{jm} &=& j(j+1) \ket{jm} \label{eq:J2Eigen} \\ \nonumber \\
\braket{jm \vert jn} &=& \delta_{mn}  \label{eq:normalizHjBasis}
\end{eqnarray}
 The dimension of $H^j$ can be read off from the first equation above 
\begin{equation}
d_j \equiv \dim H^j = 2j+1  
\end{equation}

The standard Euler angle parametrization considers two rotations along the $z$ axis with an intermediate rotation along the $y$ or $x$ axis. \footnote{We use the $z-y-z$ convention, where the small Wigner-d matrices $d^j_{mn}(\beta)$ are real. However, the manipulations presented here apply also to the $z-x-z$ convention, when $d^j_{mn}(\beta)$ have complex entries. This is the reason we do not write $d^j_{mn}(\beta)^\dagger=d^j_{mn}(\beta)^T$.}  

A rotation around the $z$--axis, for instance, by an angle $\gamma$, is given by adding up many infinitesimal rotations, each given by the generator $J_z$. Formally,
\begin{equation}
\lim_{N \rightarrow \infty } ( 1 - i \gamma J_z/N )^N = e^{-i \gamma J_z}  
\end{equation}  
Thus, an arbitrary group element $h \in SO(3)$ may be written as 
\begin{equation}
h = h(\alpha,\beta,\gamma)=  e^{-i \alpha J_z}\,e^{-i \beta J_y}\,e^{-i \gamma J_z}  
\end{equation}
The matrix elements of Wigner's D-matrices (representation matrices for $SO(3)$ and $SU(2)$), in this parametrization, read
\begin{eqnarray} \label{eq:WignerDEulerParam}
D^j_{mn} ( h(\alpha,\beta,\gamma) & \equiv & \bra{jm} h(\alpha,\beta \gamma) \ket{jn} \nonumber \\ 
&=& \bra{jm} e^{-i \alpha J_z}\,e^{-i \beta J_y}\,e^{-i \gamma J_z} \ket{jn}   \nonumber \\
&=& e^{-i \alpha m} \bra{jm}\,e^{-i \beta J_y} \ket{jn} e^{-i \gamma n}
\end{eqnarray}
where we used that $J_z \ket{jm} = m \ket{jm}$. The matrix $\bra{jm}\,e^{-i \beta J_y} \ket{jn}$ is Wigner's small-d matrix, and we use the standard notation
\begin{equation}
d^j_{mn}(\beta) \equiv \bra{jm}\,e^{-i \beta J_y} \ket{jn}
\end{equation}
We will not need the explicit expression for $d^j_{mn}(\beta)$.\footnote{The explicit expression for $D^j_{mn}(h)$, analytically extended to $GL(2,\mathbb{C})$, is equation \eqref{eq:WignerDjjElement}} 

Most of the properties needed in practice can be derived from the geometrical interpretation of the Euler parametrization and the properties of the two groups. 

 For instance, $h(0,\beta,0)$ acting on a 3D vector, implements a rotation by $\beta$ in the plane perpendicular to the $y$-axis. Thus, 
\begin{equation}
 h(0,\beta,0) h(0,-\beta,0) = \ident_{SO(3)}
\end{equation}
from which it follows that
\begin{equation}
 \sum_{k=-j}^j d^j_{mk}(\beta) d^j_{kn}(-\beta) = \delta_{mn} = \ident_{j}
\end{equation}
where $\ident_{j}$ is the identity on $H^j$. Similarly, since $h(0,\beta,0) \in SU(2)$, we have that 
\begin{equation}
h(0,\beta,0) h(0,\beta,0)^\dagger =\ident_{SU(2)} 
\end{equation}
and thus
\begin{equation}
d^j_{km}(-\beta) =  d^j_{km}(\beta)^\dagger
\end{equation}

Integrating over $SO(3)$, amounts to integrating $h(\alpha,\beta,\gamma)$ over all values of the Euler angles $\alpha$,$\beta$,$\gamma$, that are in the ranges
\begin{equation}
\alpha \in [0,2 \pi)\, \ \ \beta \in [0, \pi]\, \ \ \gamma \in [0,2 \pi)
\end{equation}
The (left and right invariant) Haar measure $\mu(h)$ of $SO(3)$ in the Euler parametrization reads
\begin{equation} 
\mu_{SO(3)}(h) = \frac{1}{8 \pi^2}\dd \alpha \, \dd \beta \sin \beta \, \dd \gamma
\end{equation}
with the normalization $\frac{1}{8 \pi^2}$ so that the volume of $SO(3)$ is unit (the Haar measure is unique up to scaling). 

\subsubsection{$SU(2)$ and the three-sphere $S^3$}
The group $SU(2)$ \emph{is} the three-sphere $S^3$. The two spaces are diffeomorphic as smooth manifolds and their Lie algebras are isomorphic. One can anticipate this by writing an $SU(2)$ element $h$ as  
\begin{equation}
h = \begin{pmatrix}
a & -\bar{b}\\
b & \bar{a}
\end{pmatrix} 
\end{equation}
the condition that $\det(h)=1$ reads
\begin{equation}
1 = (\re a)^2 + (\im a)^2+(\re b)^2 + (\im b)^2
\end{equation}
which gives a map from $SU(2)$ to a point on the unit three-sphere, realised as the hypersurface
\begin{equation}
x^2+y^2+z^2+w^2=1
\end{equation}
in a 4D Euclidean space with a Cartesian coordinate system. 

In what follows, we will construct ``semiclassical 3D directions'', corresponding to points on the two-sphere $S^2$,  by acting with $SU(2)$ group elements on states in $H^j$. The procedure is explained below. Keep in mind that we have to somehow project $SU(2) \sim S^3$ to $S^2$.

\medskip

The group $SU(2)$ is parametrized similarly with Euler angles as in the $SO(3)$ case, with the difference being that the range of $\gamma$ is doubled  
\begin{equation}
\gamma \in [0,4 \pi)
\end{equation}
This is because $SU(2)$ covers $SO(3)$ twice. That is, as $SO(3)$ elements, 
\begin{equation}
h(\alpha,\beta,\gamma) = h(\alpha,\beta,\gamma+ 2 \pi)
\end{equation}
see \eqref{eq:WignerDEulerParam}. Thus, seen as $SU(2)$ group elements, in which case they are distinct, they are both mapped to the same $SO(3)$ element. That is, $SO(3) \sim SU(2)/\mathbb{Z}_2$. 

Since there are twice as many group elements, the volume has doubled, and the normalized Haar measure for $SU(2)$ is half that of $SO(3)$, 
\begin{equation} 
\mu_{SU(2)}(h) = \frac{1}{2} \mu_{SO(3)}(h) 
\end{equation}

\subsubsection{The two-sphere $S^2$ }
The two-sphere $S^2$ as a differentiable manifold is described by the line element 
\begin{equation}
\dd s^2 = \dd \Omega^2 = \dd \alpha ^2 + \sin^2 \beta  \dd \beta^2  
\end{equation}
where the polar angles $\alpha$ and $\beta$ are understood as the azimuth and zenith. The induced Haar measure on the $SU(2)$ proper subgroup spanned by $\alpha$ and $\beta$ in the Euler parametrization, which is isomorphic to $S^2$, coincides with the induced proper area of this line element
\begin{equation} 
\mu_{S^2}(h) = \frac{1}{4 \pi} \dd \alpha \, \dd \beta \sin \beta 
\end{equation}
where we normalized by the sphere area $4 \pi$. The angles $\alpha$ and $\beta$ take values as in $SO(3)$ and $SU(2)$.

\begin{figure}
\centering
\includegraphics[scale=0.8]{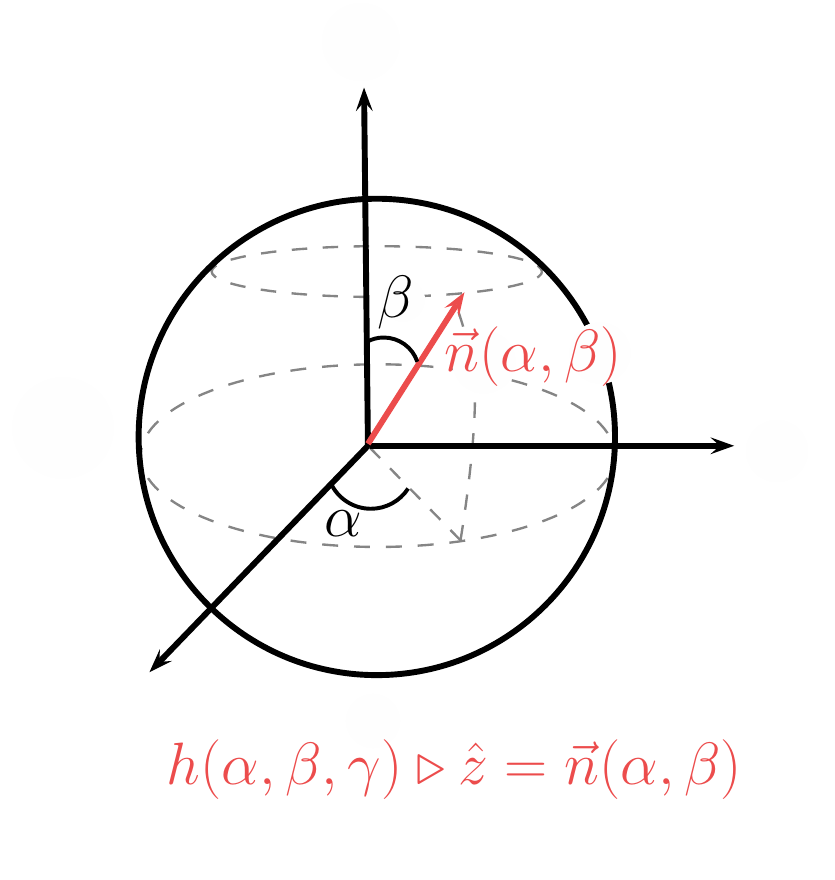}
\caption[Convention for elements of $S^2$]{Convention for elements of $S^2$ and relation to Wigner's matrices. $\vec{n}(\alpha,\beta)$ is defined by the action of an element of $SO(3)$ on the $z$-axis. The first rotation by $\gamma$ is around the $z$-axis, thus, for given $\alpha$ and $\beta$, any $\gamma$ maps to the same $\vec{n}(\alpha,\beta)$. For this reason, we may also understand $h(\alpha,\beta,\gamma)$ as an $SU(2)$ element. }
\label{fig:thetaPhiConvention}
\end{figure}

In order to fix conventions and to aid the geometrical understanding of the construction of $SU(2)$ coherent states, we take a moment to explicit our conventions, which are summarized in Figure \ref{fig:thetaPhiConvention}.

We embed $S^2$ as the unit sphere in a flat Euclidean 3D space with Cartesian $x,y,z$ coordinates. Given a point $\alpha,\beta$ on the sphere, we map it to a normalized 3D vector 
\begin{equation}
\vec{n}(\alpha,\beta) 
\end{equation}
such that $\beta$ is the angle of $\vec{n}(\alpha,\beta)$ with the positive $z$-axis and $\alpha$ is the angle of the projection of $\vec{n}(\alpha,\beta)$ in the $xy$-plane with the positive $x$-axis. That is, we define the vector $\vec{n}(\alpha,\beta)$, as the one resulting from an $SO(3)$ rotation of the vector $\hat{z}$, pointing in the positive $z$-axis direction. 

Notice that in the Euler angle parametrization, the first rotation is around the $z$-axis. Thus, $\vec{n}(\alpha,\beta)$ is defined by the action of an $SO(3)$ rotation $h(\alpha,\beta,\gamma)$, as 
\begin{equation} \label{eq:cohStateU1gamma}
h(\alpha,\beta,\gamma) \triangleright \hat{z} = \vec{n}(\alpha,\beta) 
\end{equation} 
\emph{for any} $\gamma$. Any fixed choice for $\gamma$ defines a projection from $S^3$ to $S^2$. We make a choice for $\gamma$ in the next section. We may then freely understand $h(\alpha,\beta,\gamma)$ as an $SU(2)$ element, with the same geometrical meaning.

\subsection{Semiclassical directions: $SU(2)$ coherent states }
\label{eq:su2CoherentStates}

With conventions as above, given a vector $\vec{n}(\alpha,\beta) \in S^2$, we define an embedding to $SU(2)$
\begin{equation}
S^2  \ni \; \vec{n}(\alpha,\beta) \rightarrow  n(\alpha,\beta,-\alpha) \; \in SU(2)
\end{equation}
Notice that in $n(\alpha,\beta,-\alpha)$ we have fixed the angle $\gamma$ to $-\alpha$. This choice is the standard convention introduced by Peremolov \cite{perelomov_coherent_1972}. Understanding $SU(2)$ as $S^3$, the Hopf fibration tells us that $S^3$ can in turn be understood as $S^2$ with a circle $S^1$ as a fiber on each point. We take each fiber to be parametrized by $\gamma$, and the choice $\gamma=-\alpha$ amounts to choosing a single point from each circle (in other words, a choice of section of the Hopf fibration). This projects $SU(2)\sim S^2 \times S^1 \rightarrow S^2$.  

$SU(2)$ coherent states (Peremolov coherent states \cite{perelomov_coherent_1972}) are  defined as 
\begin{eqnarray} \label{eq:defPeremolov}
\ket{j \, \vec{n}(\alpha,\beta) } &\equiv&
\vec{n}(\alpha,\beta) \triangleright \ket{jm} \nonumber \\ & \equiv &  n(\alpha,\beta,-\alpha)  \ket{jm} \nonumber \\ 
&\equiv & \sum_m D^j_{mj}(\alpha,\beta,-\alpha) \ket{j,m}
\end{eqnarray}
The bra is defined by\footnote{In calculations this is often a source of confusion and is good to keep in mind : in \eqref{eq:defPeremolov}, the $D^j_{mj}(\alpha,\beta,\gamma)$ are coefficients in a sum over states , not a matrix, and conjugation simply gives their complex conjugate in \eqref{eq:defPeremolovConj}. That is, we get $\overline{D^j_{mj}(\alpha,\beta,\gamma)}$ and not $D^{j\dagger}_{mj}(\alpha,\beta,\gamma)$.}
\begin{eqnarray} \label{eq:defPeremolovConj}
\bra{j \, \vec{n}(\alpha,\beta) } &\equiv & \sum_m \overline{D^j_{mj}(\alpha,\beta,-\alpha)} \; \bra{jm} \nonumber \\
&=& \sum_m  D^{j \, \dagger}_{jm}(\alpha,\beta,-\alpha) \bra{jm}
\end{eqnarray}
The states $\ket{j \, \vec{n}(\alpha, \beta)}$ are normalized so that
\begin{eqnarray}
\braket{j \, \vec{n}(\alpha, \beta) \vert j \, \vec{n}(\alpha, \beta)} &=& \sum_{m m' } \overline{D^{j}_{mj}(\alpha,\beta,-\alpha)} D^j_{m'j}(\alpha,\beta,-\alpha)  \braket{jm \vert jm'  } \nonumber \\
&=& h^\dagger(\alpha,\beta,-\alpha) \; h(\alpha,\beta,-\alpha)=1
\end{eqnarray}
where we used equation \eqref{eq:normalizHjBasis}, $\braket{jm \vert jm'  } =\delta_{m'm}$, and the fact that $h(\alpha,\beta,\gamma)$ is in $SU(2)$. 
As a display of the use of the notation in calculations, we may also show this as
\begin{equation}
1= \braket{j j \vert j j} = \braket{j j \vert n^\dagger\, n \vert j j} = \braket{j \vec{n} \vert j \vec{n}}
\end{equation}

\bigskip

\emph{Intuitively, $\ket{j,j}$ is a ``semiclassical normalized quantum vector'' pointing in the $z$-direction. }

\bigskip

Let us see this explicitly. The uncertainty of an operator on some state $\psi$ is defined by 
\begin{equation}
\Delta_\psi \mathcal{O} \equiv \sqrt{\braket{\mathcal{O}^2}_\psi -\braket{\mathcal{O}}_\psi^2 }
\end{equation}
A standard result from operator theory reads that for any two Hermitian operators $\mathcal{O}_1$ and $\mathcal{O}_2$, we have   
\begin{equation} \label{eq:uncertaintyRelation}
\Delta_\psi \mathcal{O}_1 \;  \Delta_\psi \mathcal{O}_2 \geq  \frac12 \braket{[\mathcal{O}_1,\mathcal{O}_2]}
\end{equation}
and which we understand in quantum mechanics as the uncertainty principle.

It is straightforward to check for instance that
\begin{eqnarray}
\Delta_{\ket{jm}} J_z &\equiv& \sqrt{\braket{J_z^2}_{\ket{jm}} -\braket{J_z}_{\ket{jm}}^2 } \nonumber \\
&=& \sqrt{\braket{jm\vert J_z^2 \vert jm} -\braket{jm \vert J_z \vert jm}^2 } \nonumber \\
&=& \sqrt{m^2 -m^2 } =0
\end{eqnarray}
that is, all states $\ket{jm}$ have zero uncertainty in the $z$-direction. 

\emph{This is the sense in which $\ket{jm}$ are peaked on a point in $S^2$}. In our conventions, $\ket{jm}$ is peaked on $\vec{n}(0,0) \equiv \hat{z}$.

The uncertainty principle for $J_x$ and $J_y$ gives
\begin{eqnarray} \label{eq:uncertaintyXY}
\Delta_{\ket{jm}} J_x \;  \Delta_{\ket{jm}} J_y & \geq &  \frac12 \braket{ [J_x, J_y]}_{\ket{jm}} \nonumber \\
&=& \frac12 \braket{ J_z}_{\ket{jm}}  = \frac{m}{2} 
\end{eqnarray}
By symmetry, since $\ket{jm}$ is an eigenvector of $J_z$, the uncertainties $\Delta_{\ket{jm}} J_x$ and $\Delta_{\ket{jm}} J_x$ cannot depend on $m$ and must be equal. Thus, since the left hand side does not depend on $m$, equality in \eqref{eq:uncertaintyXY} may only be achieved when $m$ takes its largest value, that is, when $m=j$. From this we guess that 
\begin{equation}
\Delta_{\ket{jm}} J_x = \Delta_{\ket{jm}} J_y =\sqrt{\frac{j}{2}}
\end{equation}
the veracity of which can be easily checked using equation \eqref{eq:J2Eigen}.

\emph{ This is the sense in which $\ket{jj}$ is the \emph{``most classical''} quantum state peaked on $\vec{n}(0,0)$; the uncertainty principle is saturated on $\ket{jj}$.}

Then, we see that
\begin{equation}
\frac{\Delta_{\ket{jm}} J_x}{\braket{J^2}_{\ket{jm} }} = \frac{\sqrt{j/2}}{\sqrt{j d_j /2}}= \frac{1}{d_j}
\end{equation}
where we used \eqref{eq:J2Eigen}. Thus, as we move to larger quantum numbers, the uncertainty vanishes and the state becomes completely classical i.e.\! it behaves as the classical vector $n(0,0)$. 

\emph{In conclusion, $\ket{jj}$ behaves as a geometrical wavepacket carrying the information of a ``semiclassical direction''}, corresponding to $n(0,0)$. 

\bigskip

These states are the building blocks from which we can construct wave--packets of geometry.
Specifically, we will construct semiclassical states describing a wavepacket of geometry, peaked on a discrete 3D Euclidean spacelike geometry, \emph{embedded} in a Lorentzian spacetime. 

The way the embedding will be achieved, although we have $SU(2)$ and not \slt states, is through the interpretation of the $SU(2)$ group elements as the holonomies of the Ashtekar-Barbero connection, which knows the extrinsic curvature $K$
\begin{equation}
h_{AB} \sim e^{\int_{\Upsilon} \Gamma + \gamma K}
\end{equation}
where $\Gamma$ is the spin connection, $\Upsilon$ a curve and $\gamma$ here is the Immirzi parameter, not to be confused with the Euler angle. See Section \ref{sec:holonAsParTransport} for the definition and intuition of the holonomy.

\begin{center}
\rule{0.5 \textwidth}{0.7pt}
\end{center}

The construction of a state peaked on an arbitrary element of $S^2$ follows easily.
The state, $\ket{j \, \vec{n}(\alpha, \beta)}$ is peaked in the direction $n(\alpha,\beta)$.  The angular momentum operator $J_{\vec{n}}$ in the $\vec{n}$ direction corresponds to a rotation of $J_z$ by $n \in SO(3)$. At the level of the Lie algebra, this operation corresponds to the conjugate action on $J_z$ by $n$
\begin{equation}
J_{\vec{n}} \ket{j \vec{n} }= n J_z n^\dagger \ket{j\vec{n}} = n J_z n^\dagger n \ket{jj} = n J_z \ket{jj} = n j \ket{jj} =j \ket{j\vec{n}}
\end{equation}
thus, $\ket{j \vec{n} }$ is an eigenstate of $J_{\vec{n}}$.

The semiclassical properties for $\ket{jj}$ carry through to $\ket{j \vec{n}}$, which we understand as a wavepacket corresponding to a classical direction $\vec{n}$ in 3D space.

 When doing calculations, properties of these states can be derived from their geometrical meaning. For instance, since $\ket{jj}$ corresponds to the z-axis, which is the vector $n(0,0)$, it must be that $n(0,\pi) \ket{jj}$ corresponds to the negative $z$-axis and thus to $-\ket{jj}$. In turn, we may deduce from this the property of the small-d Wigner matrix that $d^j_{mj}(\pi)=-\delta_{mj}$ et cetera.

Lastly, the coherent states $\ket{j \, \vec{n}(\alpha, \beta)}$, provide a resolution of the identity in $H_j$. This is easy to show by exploiting the relation of $S^2$ with $SU(2)$ and $SO(3)$. 

We notice that we may replace $-\alpha$ with an arbitrary $\gamma$ in
\begin{eqnarray}
\overline{D^j_{mj}(\alpha,\beta,-\alpha)} D^j_{m'j}(\alpha,\beta,-\alpha) &=& D^j_{jm}(\alpha,-\beta,-\alpha) D^j_{m'j}(\alpha,\beta,-\alpha) \nonumber \\
&=& D^j_{jm}(\alpha,-\beta,-\gamma) D^j_{m'j}(\gamma,\beta,-\alpha) \nonumber \\
\end{eqnarray}
see \eqref{eq:WignerDEulerParam}. We may promote the integration over $S^2$ to an integration over $SO(3)$ (or $SU(2)$) and use the orthogonality relation for Wigner's-D matrices
\begin{equation} 
\int_{SO(3)}\!\!\!\! \mu(h)\; D^{j}_{kl}(h^\dagger) D^{j'}_{mn}(h) = \frac{\delta_{kn} \, \delta_{lm} \,\delta_{j j'}}{d_{j}}
\end{equation}

Then, it is straightforward to see that 
\begin{eqnarray}
\int_{S_2} \mu(\vec{n}) \; \ket{j \, \vec{n}} \bra{j \, \vec{n}} &=& \frac{1}{2 \pi} \int^{2 \pi}_0 \dd \gamma \int_{S_2} \mu_{S^2}(\vec{n}) \; \ket{j \, \vec{n}} \bra{j \, \vec{n}} \nonumber \\
&=&  \frac{\ident_j}{2\pi \; d_j}
\end{eqnarray}

Thus, the states $\ket{j \, \vec{n}(\alpha,\beta)}$ resolve the identity, for an integration measure $2\pi \,d_j\, \mu(\vec{n}) $. That is, \emph{they provide an over--complete basis of semiclassical states for $H_j$}.

\subsubsection{$SU(2)$ coherent states with the $H$ map and relation to $\mathbb{CP}1$}
Here we prove a technical result, providing the Euler angle parametrization for the $SU(2)$ coherent states defined through spinors, via the $H$-map of Section \ref{sec:spinorForKrajHan}. This parametrization provides provides a strict equality between the two definitions and allows to combine results in the literature. These technicalities are important to nail down when constructing wave--packets of geometry starting from a triangulated hypersurface, for a concrete application of geometry transition.

We will see in Section \ref{sec:discretizationScheme} that in a physical application we are first given a triangulated surface. Then, we construct the normals to the faces explicitly as vectors $\vec{n}(\alpha,\beta)$, by fixing a gauge (triad) in a tetrahedron of the triangulation. From this information, we will want to construct a wavepacket peaked on the distributional geometry of the triangulation, using coherent states. 
 Thus, we need to know how to go unambiguously from a classical vector $\vec{n}(\alpha,\beta)$ to both definitions of the coherent states. 

The reader may skip this section since the intuition developed previously suffices for what follows.

\bigskip

The coherent states $\ket{j \, \vec{n}}$ may be defined also with respect to the $H$-map, equation \eqref{eq:HzMatrix} 
\begin{equation} \label{eq:cohStatesWithHmap}
\vec{n} \triangleright \ket{jj} = \ket{j \, \vec{n}} \equiv \sum_{m} D^j_{mj} \big(H(\tilde{n})\big) \ket{jj}
\end{equation}
where now $\tilde{n}$ in $H(\tilde{n})$ is a normalized \emph{spinor}. 

The coherent states $\ket{j \, \vec{n}}$ defined this way, satisfy all the properties of the previous section. The calculations are essentially identical, using the properties for the $H$--map of Section \ref{sec:spinorForKrajHan} to perform matrix multiplication. The coherent state $\ket{j \, \vec{n}}$ is peaked on the 3D Euclidean vector 
\begin{equation} \label{eq:vecnDef}
\vec{n} = \braket{\tilde{n} \vert \vec{\sigma} \vert \tilde{n}}
\end{equation}
where $\vec{\sigma}$ are the Pauli matrices.  This map from normalized spinors to the Riemann sphere is explained in Appendix \ref{app:spinorsAndRiemannSphere}, where conventions are also set. Notice that this map is not invertible, there is a $U(1)$ phase ambiguity in defining $n$ for a given $\vec{n}$

We would like to know how to \emph{unambiguously define} a normalized spinor
\begin{equation}
n(\alpha,\beta)
\end{equation}
\emph{starting} from an element of $S^2$, $\vec{n}(\alpha,\beta)$, so that the geometrical interpretation of the coherent state  $\ket{j \, \vec{n}(\alpha,\beta)}$ of equation  \eqref{eq:cohStatesWithHmap}, agrees with the conventions for the azimuth and zenith as in Figure \ref{fig:thetaPhiConvention}. Second, we would like to have a strict equality between the two definitions of the coherent states, which requires to fix the face ambiguity in equation \eqref{eq:vecnDef}.

Let us start by assuming we are given a vector 
\begin{equation}
\vec{n}(\alpha,\beta)
\end{equation}
We define a family of spinors $n(\alpha,\beta,\gamma)$ by 
\begin{equation} \label{eq:defFamOfSpinors}
\ket{n(\alpha,\beta,\gamma)} = e^{\frac{i}{2} \gamma} \; K(-1/\bar{\zeta}) \ket{+}
\end{equation}
where the $SU(2)$ matrix $K(\zeta)$ reads
\begin{equation}
K(\zeta)=\frac{1}{\sqrt{1+|\zeta|^2}} \begin{pmatrix}
1 & \zeta \\ -\bar{\zeta} & 1
\end{pmatrix}
\end{equation}
and
\begin{equation}
\zeta = \cot \frac{\beta}{2} e^{i \alpha}
\end{equation}
The phase $e^{\frac{i}{2} \gamma}$ in \ref{eq:defFamOfSpinors}, parametrizes the $U(1)$ ambiguity for some $\gamma \in (0,4 \pi)$ and $\ket{+} = \left(1 \ 0 \right)^T$.

We will see that defined this way $\gamma$ will be the equivalent of the corresponding Euler angle. We show in Appendix \ref{app:spinorsAndRiemannSphere}, that 
\begin{equation}
\braket{\overline{n(\alpha,\beta,\gamma)} \vert \vec{\sigma} \vert \overline{n(\alpha,\beta,\gamma)} } = \begin{pmatrix}
\cos\alpha \sin\beta \\ \sin\alpha \sin\beta \\ \cos\beta \end{pmatrix}  = \vec{n}(\alpha,\beta) 
\end{equation}
That is, the geometrical interpretation is exactly as in Figure \ref{fig:thetaPhiConvention}. The reason we need the complex conjugate is because the $y$-axis is inverted in the mapping to the Riemann sphere as defined by \eqref{eq:vecnDef}, see Appendix \ref{app:spinorsAndRiemannSphere}.

We now want to fix the face ambiguity and show that the two definitions are equivalent. We notice that 
\begin{equation}
H(z) \ket{+} = \ket{z} 
\end{equation}
for any spinor $z$ (the notation is confusing but common, both  $z$ and $\ket{z}$ denote a spinor). Thus,

\begin{eqnarray}
\ket{\overline{n(\alpha,\beta,\gamma)}}
 &=& H\bigg(\overline{n(\alpha,\beta,\gamma)}\bigg) \ket{+} \nonumber \\
 &=& H\bigg(e^{\frac{-i}{2} \gamma} \; \overline{n(\alpha,\beta,0)} \bigg) \ket{+}
\end{eqnarray}

 We notice that
\begin{eqnarray}
D^j_{jm}(h(\alpha,\beta,\gamma)) & \equiv & D^j_{jm}(\alpha,\beta,\gamma) \nonumber \\
&=& D^j_{jm}(\alpha,\beta,0) e^{i\gamma j}
\end{eqnarray}
Similarly, using the definition for the analytic continuation of $D^j_{jm}$ to $GL(2,\mathbb{C})$, equation \eqref{eq:WignerDjjElement}, it is easy to check that
\begin{equation}
D^j_{jm}\left( H(\overline{n(\alpha,\beta,0)}e^{\frac{i}{2}\gamma }\right) = D^j_{jm}\left( H(\overline{n(\alpha,\beta,0)})\right) e^{i \gamma j}
\end{equation}
Since $n(0,0,\gamma) \equiv \hat{z}$, we have that  $H(\overline{n(\alpha,\beta,0)}) \triangleright \hat{z}= \vec{n}(\alpha,\beta)$. The same applies for $h(\alpha,\beta,0)$. From the definition \eqref{eq:WignerDEulerParam} of the Euler parametrization, and the fact that $D^j_{jm}\left( H(\overline{n(\alpha,\beta,0)})\right)$ is defined through an analytic continuation, we conclude that as abstract elements of $SU(2)$, we have 
\begin{equation}
H(\overline{n(\alpha,\beta,0)}) =  h(\alpha,\beta,0) = e^{i \alpha J_z} e^{i \beta J_y} e^{i \gamma J_z}
\end{equation}

Thus,  
\begin{equation}
\boxed{
D^j_{jm}\big(\alpha,\beta,\gamma \big) = D^j_{jm}\bigg( H(\, \overline{n(\alpha,\beta,\gamma )} \, )  \bigg)
}
\end{equation}
This provides an Euler angle parametrization for the space of normalized spinors. For any choice of section (fixing $\gamma$) we get a correspondance between $S^2$ and the Riemann sphere $\mathbb{CP}1$, such that their $x-y-z$ components in the vector representation are strictly equal and given by the usual spherical coordinates $\alpha$ and $\beta$.

Fixing $\gamma=-\alpha$ gives us an equivalence with the coherent states as defined in the previous section. 

To summarize, as can be verified from the above, given a point $\vec{n}(\alpha,\beta)$ in $S^2$ the correspondance between the Euler angle parametrization of Wigner's D-matrices and the $H$ map, is given by the normalized spinor
\begin{equation} \label{eq:correspondanceSpinorsVectors}
n(\alpha,\beta,\gamma) = \begin{pmatrix}
\cos \frac{\beta}{2} & - \sin \frac{\beta}{2} e^{i \alpha} \\
\sin\frac{\beta}{2} e^{-i \alpha} &\cos \frac{\beta}{2} 
\end{pmatrix} e^{\frac{i}{2} \gamma}
\end{equation}
and 
\begin{equation}
\vec{n} \triangleright \ket{jj} = \ket{j \, \vec{n}} \equiv \sum_{m} D^j_{mj} \big(H(\,\overline{n(\alpha,\beta,-\alpha)}\,)\big) \ket{jj} = \sum_{m} D^j_{mj}(\alpha,\beta,\alpha)) \ket{jj}
\end{equation}

\subsection{Semiclassical tetrahedra}
In the previous section we have introduced coherent states describing a ``semiclassical direction'', i.e.\! the quantum equivalent of a normalized three--dimensional vector.

By itself, such a state has no geometrical meaning per se, because a single normalized vector $\vec{n}$ only holds information that is purely gauge. To give any meaning to $\vec{n}$, we need to fix a convention as in Figure \ref{fig:thetaPhiConvention}, which is the equivalent of fixing the tetrad.

However, we may combine such states, and construct geometrical shapes, polyhedra, see \cite{bianchi_polyhedra_2011} for a generic treatment.  

To construct the semiclassical equivalent of a tetrahedron $\nn^\star$, we associate a state $\ket{j_\ell \vec{n}_\ell}$ to each of the four triangles $\ell^\star$ of $\nn^\star$, such that the closure condition, equation \eqref{eq:closureCond},
is satisfied
\begin{equation} 
0= \sum_{\ell \in \nn} j_\ell \; \vec{n}_{\ell}
\end{equation}
That is, we associate to each tetrahedron $\nn^\star$ the tensor product 
\begin{equation}
 \bigotimes_{\ell \in \nn} \ket{j_\ell,k_{\ell}}  
\end{equation} 
which is a state in the tensor product of the corresponding $SU(2)$ spin representation spaces, $H^{j_1} \otimes H^{j_2} \otimes H^{j_3} \otimes H^{j_4}$. 

To take into account the orientation of the graph $\Gamma$, we introduce the notation $\ell \tar \nn$ and $\ell \sou \nn$, signifying that the half-link $\ell$ has the node $\nn$ as target and source respectively.

We define for each node the state
\begin{eqnarray}
v_\nn(j_\ell; \vec{k}_{\ell\nn})& \equiv & \bigotimes_{\ell \sou \nn} \bra{j_\ell,k_{\ell\nn}}   \ \bigotimes_{\ell \tar \nn} \ket{j_\ell,k_{\ell\nn}}  \\ \nonumber \\ \nonumber \\  
v_\nn(j_\ell; \vec{k}_{\ell\nn}) & \in & \bigotimes_{\ell \sou \nn} H^{j_\ell \, \star} \ \bigotimes_{\ell \tar \nn} H^{j_\ell} 
\end{eqnarray} 
Similarly to the definition of spin network states, we define functions $\psi (h_\ell,j_\ell; k_{\ell\nn})$, depending on group elements living in $L$ copies of $SU(2)$ ,
\begin{equation} \label{eq:gaugeVariantIntrinsicStates}
\psi(h_\ell,j_\ell; k_{\ell\nn}) = \bigotimes_{\ell} D^{j_\ell}(h_\ell) \rhd \bigotimes_{\nn } v_\nn(j_\ell; k_{\ell\nn})
\end{equation}
We can rearrange this expression as a product over links
\begin{eqnarray}
\psi_\Gamma(h_\ell,j_\ell; k_{\ell\nn}) &=& \bigotimes_{\ell} D^{j_\ell}(h_\ell) \rhd \bigotimes_{\nn } \left( \bigotimes_{\ell \sou \nn} \bra{j_\ell,\vec{k}_{\ell\nn}}    \bigotimes_{\ell \tar \nn} \ket{j_\ell,\vec{k}_{\ell\nn}}\right) \nonumber \\
&=& \bigotimes_{\ell } D^{j_\ell}(h_\ell) \rhd \ket{j_\ell,k_{\ell t(\ell)}} \otimes \bra{j_\ell,k_{\ell s(\ell)} }
\end{eqnarray}
Or more simply,
\begin{equation}\label{eq:intrinsicStatesLikeSpinNets}
\psi_\Gamma(h_\ell,j_\ell; k_{\ell\nn}) = \bigotimes_{\ell } \braket{j_\ell,k_{\ell s(\ell)} \vert D^{j_\ell}(h_\ell) \vert j_\ell,k_{\ell t(\ell)}} 
\end{equation}

These states are closely related to the Livine-Speziale states \cite{livine_new_2007} or intrinsic coherent states \cite{rovelli_covariant_2014}. 

The states \eqref{eq:intrinsicStatesLikeSpinNets} provide an overcomplete basis\footnote{It suffices to use the resolution of the identity for the Peremolov states to show this.} of the gauge--variant Hilbert space $\mathcal{L}(SU(2)^L)$ of square integrable functions on $L$ copies of $SU(2)$.

To get to the Livine-Speziale states we need to gauge average at each node by integrating over $SU(2)$. Since we will be considering them in contraction with a spinfoam amplitude it is redundant to gauge--average, due to the use of the Haar measure over $SL(2,\mathcal{C})$ in the definition of the vertex amplitude. The gauge invariance over its subgroup $SU(2)$ is imposed automatically. 

We may write an explicit expression for these states in terms of the definition of equation \eqref{eq:defPeremolov} or \eqref{eq:cohStatesWithHmap} for the Peremolov coherent states 
\begin{equation} 
\psi_\Gamma(h_\ell,j_\ell; k_{\ell\nn}) = \sum_{m_s m_t} D^{j_\ell}_{m_s j_\ell}(k_{\ell s(\ell)}^\dagger) \; D^{j_\ell}_{m_t j_\ell}(k_{\ell t(\ell)}) \;D^{j_\ell}_{m_s m_t}(h_\ell)
\end{equation}

\section{Wavepackets of an embedded 3-geometry}
\label{sec:boundaryState}
We now have at our disposal geometrical wave--packets peaked on 3D tetrahedral triangulations.

We want to think of them as plane--waves describing a quantum geometry with a sense of \emph{embedding} in a Lorentzian spacetime. Since they are peaked on the intrinsic geometry, we turn  them into plane waves, completely spread on the extrinsic geometry, by multiplying with
\begin{equation}
\left(\prod_\ell e^{i \zeta_\ell\!\, j_\ell}\right) \psi(h_\ell,j_\ell; k_{\ell\nn})
\end{equation}
where $\zeta_\ell$ is a boost angle, to be understood as the discrete equivalent of the extrinsic curvature. 

The defining expression we use for the boundary states is in the group representation and we employ the following notation
\begin{equation}
\Psi_\Gamma^t(h_\ell;\eta_\ell,\zeta_\ell,k_{\ell\nn}) \equiv \braket{h_\ell|\Psi_t(\eta_\ell,\zeta_\ell,k_{\ell\nn})}
\end{equation}
with one $SU(2)$ group element $h_\ell$ per link.   We have 
\begin{equation} \label{eq:BoundaryStateDef}
\Psi^{t}_\Gamma(h_\ell;\eta_\ell,\zeta_\ell,k_{\ell\nn}) = \sum_{\{j_\ell\}} \prod_\ell d_{j_\ell} e^{-(j_\ell-\omega_\ell\!(\eta_\ell,t)\,)^2 t + i \zeta_\ell\!\, j_\ell} \;  \psi (h_\ell,j_\ell; k_{\ell\nn}) 
\end{equation}
The Gaussian weight in the spins is peaked on
\begin{equation}
\omega_\ell(\eta_\ell,t)=\frac{\eta_\ell-t}{2t}
\end{equation}
This quantity will be identified with the areas resulting from the discretization scheme.

These states were first introduced heuristically in \cite{rovelli_graviton_2006} and have been used extensively in the literature to study the graviton propagator \cite{rovelli_graviton_2006,bianchi_lqg_2009,
bianchi_lorentzian_2012,shirazi_hessian_2016,
bianchi_graviton_2006,alesci_complete_2007,
alesci_complete_2008}. They have since been understood to be the semi--classical limit of Thiemann's heat--kernel states. 

\subsection{Coherent spin--network states}

 The boundary states we consider in this manuscript are Thiemann's heat kernel states \cite{thiemann_gauge_2001,thiemann_gauge_2001-1,
 thiemann_gauge_2001-2,thiemann_gauge_2001-3,
 thiemann_gauge_2001-4}, in the twisted-geometry parametrization introduced in \cite{freidel_twistors_2010,freidel_twisted_2010}  by Freidel and Speziale. These states, also known as extrinsic coherent states or coherent spin-networks, were systematically introduced as ``wavepackets of an embedded spacelike 3--geometry'' in \cite{bianchi_coherent_2010}. 
 
 They are labelled by points in the gauge invariant classical phase space $SU(2)^L / SU(2)^N$, corresponding to the  the Hilbert space $\mathcal{H}_\Gamma$ associated to the boundary graph $\Gamma$. We briefly review the main construction. 

The states are defined by 
\begin{equation}\label{eq:ThiemannState}
	\Psi_{\Gamma}^t(h_\ell, H_{\ell}) = \int_{SU(2)}\dd h_{\ell \nn}\, \prod_{\ell} K_{\ell}^{t} (h_\ell, h_{\ell s(\ell)}\, H_{\ell}\, h_{\ell t(\ell)}^{-1}),
\end{equation}
where $K^{t}(h, H)$ is the $SU(2)$ heat kernel with a complexified $SU(2)$ element as second argument. Complexifying $SU(2)$ yields \slt.

$SL(2,\mathbb{C})$ is isomorphic to $SU(2) \times su(2) \simeq T^*SU(2)$ which corresponds to the (linkwise, non gauge invariant) classical phase \footnote{We recall that the configuration variables for Loop Quantum gravity are the holonomies of the Ashtekar-Barbero connection, living in $SU(2)$. The fluxes are the conjugate variables, co-triads smeared along a two-surface, and which we can think of as smeared ``area--elements'' for $SU(2)$ spinors, naturally living in the co-tangent bundle.
} space associated to the Hilbert space on a graph. 

Since $SU(2)^{\mathbb{C}}\simeq SL(2, \mathbb{C})$, $H\in SL(2, \mathbb{C})$. The Wigner D-matrices of the $SU(2)$ heat kernel are the analytical extension to $GL(2,\mathbb{C})$ of equation \eqref{eq:WignerDjjElement}.

The boundary states $\Psi_{\Gamma}^t(h_\ell; H_{\ell})=\ket{\Psi_{\Gamma}^t(H_{\ell})}$ depend on group elements $h_\ell$ living on the links $\ell$ of the boundary graph, that serve as ``handles'' for a contraction with a spinfoam amplitude defined on a two-complex with boundary $W_\mathcal{C}(h_\ell)=\bra{W_\mathcal{C}}$. Contracting a spinfoam amplitude with a boundary states means integrating out the $h_\ell$, resulting in a \emph{transition} amplitude $\braket{W_\mathcal{C} \vert \Psi_{\Gamma}^t(H_{\ell})}$, see next section.

The boundary states $\ket{\Psi_{\Gamma}^t(H_{\ell})}$ also depend on \emph{data} $H_{\ell}$. As mentioned above, these label points in the classical phase space, and are the analogue of position and momentum data $x_0$ and $p_0$ of coherent states for a free particle, peaked on a position $x_0$ and a momentum $p_0$. The situation here is analogous, with the spins playing the role of the momentum $p$. 

Finally, $\ket{\Psi_{\Gamma}^t(H_{\ell})}$ are labelled by a real number $t$, the \emph{semiclassicality parameter}. We have one \emph{family} of coherent states for each value of $t$. The semiclassicality parameter controls the peakedness properties of the coherent states. It is a small number, and in physical applications, it is to be taken proportional to a positive power of $\hbar$, divided by a characteristic physical scale of the problem, such that $t$ is dimensionless. 

Keep in mind that no dimensionful quantities appear in the boundary state (nor the spinfoam amplitude).

Concretely, $K^{t}(h, H)$ is given in the spin-representation by
\begin{equation} \label{eq:tempHeatKern}
	K^{t}(h, H) = \sum_{j} d_j e^{-j(j+1) t} \: \tr D^j(h H^{-1}).
\end{equation}

The states $\Psi_{\Gamma}^t(H_\ell)$ provide a resolution of the identity on the kinematical Hilbert space $\mathcal{H}_{\Gamma}$. Suppressing the link subscript momentarily, we write the resolution of the identity as
\begin{equation}
	\delta(hh'^\dagger)_\Gamma = \int_{SL(2, \mathbb{C})}\dd H\, \alpha(H) K^{t}(h, H) \overline{K^{t}(h', H)},
\end{equation}
where $\delta(h, h')$ is the delta distribution on $SU(2)^L / SU(2)^N$, see for instance \cite{bianchi_coherent_2010}. 

The states we consider are the gauge--variant version
\begin{equation}\label{eq:ThiemannState}
	\Psi_{\Gamma}^t(h_\ell, H_{\ell}) = \prod_{\ell} K_{\ell}^{t} (h_\ell, {H_{\ell}}^{-1}),
\end{equation}

The measure function $\alpha$ for which the resolution of identity holds is explicitly known \cite{thiemann_gauge_2001-3,
thiemann_gauge_2001,thiemann_gauge_2001-2,
thiemann_gauge_2001-1,bianchi_coherent_2010}.\newline
As was first done in \cite{bianchi_coherent_2010}, the $SL(2, \mathbb{C})$ element $H_\ell$ can be written in the twisted geometry parametrization $(\eta_\ell, \zeta_\ell, k_{\ell \nn})$ of the classical phase space corresponding to $H_\Gamma$, introduced in \cite{freidel_twisted_2010,freidel_twistors_2010}. In this parametrization, $H_\ell$ takes the form
\begin{equation}\label{eq:HParam}
	H_\ell = k_{\ell s(\ell)}\, e^{i \gamma \xi_\ell \frac{\sigma_3}{2}} e^{\eta_\ell \frac{\sigma_3}{2}}\, k_{\ell t(\ell)}^{-1},
\end{equation}
where $\eta_\ell$ is a positive real number, $\xi_\ell\in [0, 4\pi)$ and $k_{\ell s(\ell)}$, $k_{\ell t(\ell)}$ are two unrelated $SU(2)$ elements corresponding to two unit vectors.

The twisted geometry parametrization is tailor--made to correspond to the geometrical meaning of the phase-space variables. The two unit vectors, correspond to normals to the triangles of the tetrahedra sharing the face dual to the link $\ell$. The area information is to be encoded in the variable $\eta_\ell$. The area and normals are interpreted as the smeared densitized triads. The angle $\xi_\ell$ corresponds to a discretization of the Ashtekar-Barbero holonomy, and is split in two parts $\alpha$ and $\zeta$. The former corresponds to the discrete equivalent of the spin-connection and the latter to the extrinsic curvature. That is, we have the correspondence
\begin{eqnarray} \label{eq:simonePhase}
\xi &=& \alpha + \gamma \zeta \nonumber \\
A_{AB} &=& \omega + \gamma K
\end{eqnarray}
The phase $\alpha$ is thus pure gauge and is absorbed in the boundary state definition, see \cite{christodoulou_realistic_2016} for an example. Henceforth, we neglect the phase $\alpha$.

 Replacing $H_\ell$ by its twisted geometry parametrization in \eqref{eq:ThiemannState} allows us to construct coherent states with a prescribed discretized 3D geometry. The boundary data $(\eta_\ell, \zeta_\ell, k_{\ell\nn})$ encode the discretized intrinsic and extrinsic 3D geometry.

The semiclassical limit of the states \eqref{eq:HParam} (for large $\eta_\ell$), reads\footnote{Use the  heighest weight approximation $D^j_{nm}(e^{\eta\sigma_3/2}) \approx \delta^j_n \delta ^j_m e^{\eta j} + \mathcal{O}(e^{-\eta})$ in equation \eqref{eq:tempHeatKern} to get this expression. } \begin{equation}\label{eq:ExtrinsicStates}
	\Psi_{\Gamma}^t( h_\ell;\eta_\ell, \zeta_\ell, k_{\ell \nn}) = \sum_{\{j_\ell \}} \prod_\ell d_{j_\ell} e^{-\left(j_\ell - \omega_\ell(\eta_\ell, t)\right)^2 t \,+\, i \zeta_\ell j_\ell}\, \psi_{\Gamma}(j_\ell, k_{\ell \nn}; h_\ell),
\end{equation}
This was first shown in \cite{bianchi_coherent_2010}. The Gaussian weight factor in the spins is peaked on the value
\begin{equation}
\omega_\ell\!(\eta_\ell, t) = \frac{\eta_\ell-t}{2t}
\end{equation}
and it will later be identified with the \emph{areas} resulting from the discretization of the boundary surface. 

The states $\psi_{\Gamma}(j_\ell, k_{\ell \nn}; h_\ell)$ are the intrinsic (non gauge-invariant) states of the previous section. 

Before closing this section we note that the resolution of identity does not hold anymore for the states \eqref{eq:ExtrinsicStates}, since they were obtained from \eqref{eq:ThiemannState} by taking an approximation. The next section is devoted to finding an approximation of the measure function $\alpha$ in the semiclassical limit.

\subsection{The Measure Function $\alpha$ and the Semi-Classicality Condition}
We are looking to find an approximate expression for the measure, expressed in terms of the classical data $(\eta_\ell, \zeta_\ell, k_{\ell\nn})$. That is, an approximation of  $\tilde{\alpha}(\eta_\ell, \zeta_\ell, k_{\ell\nn})$ in 
\begin{equation} \label{eq:resIdCalc}
\delta(h_\ell \, h'^{\dagger}_\ell) = \int \! \dd \eta_\ell \, \dd \zeta_\ell\, \dd k_{\ell \nn} \; \tilde{\alpha}(\eta_\ell, \zeta_\ell, k_{\ell\nn}) \,\Psi_{\Gamma}^t(\eta_\ell, \zeta_\ell, k_{\ell\nn}; h_\ell) \,\overline{\Psi_{\Gamma}^t(\eta_\ell, \zeta_\ell, k_{\ell\nn}; h'_\ell)}  
\end{equation}

Making the ansatz $\tilde{\alpha} = \tilde{\alpha}(\eta_\ell)$, to be justified à posteriori, we can readily integrate over $\zeta_\ell$
\begin{equation} \label{zetaIntegration}
\int^{4 \pi}_0 \! \dd \zeta_\ell \, e^{i \zeta_\ell(j_\ell-j'_\ell)} = 4\pi \, \delta_{j_\ell j'_\ell}.
\end{equation}

Using \eqref{zetaIntegration} we are left with a single sum over the spins and the resolution of the identity  reads
\begin{align} 
& \delta(h_\ell h'^{\dagger}_\ell ) = \int_{0}^{\infty} \! \dd \eta_\ell \,  \tilde{\alpha}(\eta_\ell) \, \sum_{\{j_\ell\}} \prod_\ell d^2_{j_\ell} e^{-2 \left(j_\ell-\omega_\ell\!(\eta_\ell, t)\right)^2 t} \; \times \nonumber \\ \times
& \int_{S^2} \! \dd k_{\ell \nn} \, \psi_{\Gamma}(j_\ell, k_{\ell \nn}; h_\ell) \overline{\psi_{\Gamma} (j_\ell, k_{\ell \nn}; h'_\ell)}
\end{align}

The integration over the gauge data $k_{\ell\nn}$ is performed using the resolution of the identity for the Peremolov coherent states,
\begin{equation} \label{gaugeIntegration}
\int_{S^2} \! \dd k_{\ell \nn} \, \psi_{\Gamma}(j_\ell, k_{\ell \nn}; h_\ell) \overline{\psi_{\Gamma} (j_\ell, k_{\ell \nn}; h'_\ell)} = \prod_\ell \frac{1}{d^2_{j_\ell}} \tr^j (h_\ell h'^{\dagger}_\ell).
\end{equation} 

Employing the Peter-Weyl theorem to expand the $SU(2)$ delta function $\delta(h_\ell \,h'^{\dagger}_\ell)$ we get
\begin{equation} 
\sum_{\{j_\ell\}} \prod_\ell d_{j_\ell}  \tr^j (h_\ell h'^{\dagger}_\ell) = \sum_{\{j_\ell\}} \prod_\ell \left(
\int_{0}^{\infty} \! \dd \eta_\ell \,  \tilde{\alpha}(\eta_\ell) e^{-2 (j_\ell-\omega_\ell\!(\eta_\ell, t)\,)^2 t} \right)  \tr^j (h_\ell h'^{\dagger}_\ell)
\end{equation}
which translates to a condition for $\tilde{\alpha}$ 
\begin{equation} \label{eq:IntCond}
\int_{0}^{\infty} \! \dd \eta_\ell \,  \tilde{\alpha}(\eta_\ell) e^{-2 (j_\ell-\omega_\ell\!(\eta_\ell, t)\,)^2 t} = d_{j_\ell}.
\end{equation}
This equation is solved for large spins by
\begin{equation} \label{eq:measureEta}
\tilde{\alpha}(\eta_\ell) = \frac{\eta_\ell}{\sqrt{2 \pi} \; t^{3/2}}.
\end{equation}

This is an approximate expression for the measure providing the resolution of the identity. An exact expression in the twisted geometry parametrization has recently become available and will appear in a common work by the author and Fabio d'Ambrosio \cite{mariosFabio}. 

There is however an interesting lesson to be drawn from this approximate form. Thinking of $\tilde{\alpha}(\eta_\ell)$ in equation \eqref{eq:measureEta} as providing an ``approximate'' resolution of the identity, we can read-off a semiclassicality condition for the semi-classicality parameter $t$ and the area data $\omega$. 

To see this, we perform the integration \eqref{eq:IntCond} with the $\tilde{\alpha}$ as in \eqref{eq:measureEta}. Dropping the $\ell$ subscript momentarily, we have 
\begin{equation}
\int_0^\infty \! \dd \eta \,  \tilde{\alpha}(\eta)  e^{-2 \left(j-\omega\!(\eta,t) \right)^2 t} = \frac{d_{j} }{2} \left(1 +  \,\erf\left[\sqrt{\frac{t}{2}} d_j\right] \right)+ \frac{e^{-\frac{d^2_j t}{2}}}{\sqrt{2 \pi t}}.
\end{equation}

In the limit where the product $d_j \sqrt{t}$ is large (an assumption that is justified due to the Gaussian weight factors) we can neglect the exponential term and the error function $\erf[\sqrt{\frac{t}{2}} d_j] \approx 1$, yielding the desired result. Assuming $d_j \sqrt{t}$ to be large, we get
\begin{equation}
j \sqrt{t} \gg 1
\end{equation}
Since the spins are peaked on the area data $\omega$, we have
\begin{equation} \label{eq:semiclassicalityCondition}
\boxed{
\omega \sqrt{t}  \gg 1
}
\end{equation}

This is the \emph{semiclassicality} condition. We will understand its meaning in the following section, when it will reappear in the context of the asymptotic analysis of the transition amplitudes. We will see that it imposes that the \emph{fluctuations} of the spins are such that when interpreted as areas, we always have a simplicial triangulation. That is, the quantum fluctuations are restricted such that the spins along with the normals $k_{\ell \nn}$ satisfy the closure condition, which guarantees the existence of a geometrical tetrahedron with the spins as areas and the face normals to be $k_{\ell \nn}$. 

This is consistent with the intuitive picture of a wavepacket of an embedded 3-geometry, constructed from a superposition of plane-waves describing the intrinsic spacelike 3-geometry: imposing the semiclassicality condition is equivalent to demanding that each intrinsic coherent state $\psi_\Gamma$ in the superposition is peaked on a classical triangulation, a three-dimensional simplicial manifold.

\section{Gravitational Tunneling from covariant Loop Quantum Gravity}
\label{sec:gravTunneling}

We emphasize that the \emph{spinfoam} amplitude, defined by the EPRL ansatz of equation \eqref{eq:theEprlAnsatz}, takes in as ``data'' only \emph{combinatorics} of the 2-complex $\mathcal{C}$. In other words, only the topological information. In the limit of large-spins, we recover a notion of simplicial geometry, which we understand as the result of having imposed at the quantum level the simplicity constraint, reducing a quantization of a topological BF-theory to a quantization of discrete General Relativity.

However, spins are \emph{dimensionless}. Their \emph{interpretation} as areas, comes from the fundamental result of Loop Quantum Gravity, shown by Rovelli and Smolin in \cite{rovelli_discreteness_1995}, that the area opearator has the discrete spectrum 

\begin{equation} \label{eq:areaSpectrum}
A_j= 8 \pi G \, \hbar \gamma \sqrt{j (j+1)}
\end{equation} 

This is a result regarding the \emph{kinematics}, that is, the boundary states. The results from the asymptotic analysis of the EPRL model presented in the previous chapter are at the level of \emph{fixed} spins and regard the so-called partial amplitude, which has no physical interpretation. 

To attribute an interpretation as areas to the spins, one has to \emph{perform the spin-sum}. Integrating out the spins, we will be left with the value on which they were peaked, the $\omega$ data. It is these that will then be identified with physical areas, through equation \eqref{eq:areaSpectrum}.

This is the procedure followed in the graviton propagator calculations. There is an important difference in our case: we are not interested in a perturbative calculation. Thus, we expect the physical transition amplitudes to vanish in the limit $\hbar$. In particular, we expect them to decay exponentially with the large parameter $\lambda$, the uniform large scaling of the spins.

We will show in this section that a class of physical transitions amplitudes (with no interior faces) decay exponentially in the semi-classical limit $\hbar \rightarrow 0$, in the mismatch between the extrinsic geometry of the boundary and that of the geometrical simplicial complex preferred by the dynamics.

\subsection{Contraction of boundary state with spinfoam amplitude} 
We consider a 2-complex $\mathcal{C}$ (dual to a topological simplicial manifold) \emph{with a boundary} $\partial \mathcal{C}$. The boundary $\partial \mathcal{C}$  of the 2-complex is a four-valent graph $\Gamma$. To get a \emph{transition amplitude} we contract a \emph{spinfoam amplitude} with a coherent spin network boundary state of equation \eqref{eq:ExtrinsicStates}, built on the graph $\Gamma \equiv \partial \mathcal{C}$

\begin{equation}
W(\omega_\ell,\zeta_\ell,\vec{k}_{\ell\nn}, t) \equiv \braket{W_\mathcal{C} \vert \Psi^t_\Gamma(\omega_\ell,\zeta_\ell,\vec{k}_{\ell\nn}, t) }
\end{equation}
\bigskip

The contraction is performed as follows. We mentioned at the end of Section \ref{sec:BFSpinfoamQuantization} that the $SU(2)$ elements living on the fiducial boundary around each vertex will be used to contract with a boundary state when a boundary is introduced. 

The definition of the EPRL amplitudes when a boundary is present, comes down to a supplementary definition of the face holonomy in terms of $SU(2)$ elements, equation \eqref{eq:faceHolonomySU2}, for a face that includes a link $\ell$ in its boundary. That is, for faces that are not \emph{interior}, bounded only by edges $e$. Recall equation \eqref{eq:faceHolonomySU2}
\begin{equation} 
h_f \equiv  \prod_{vf} h_{vf}
\end{equation}
When the face $f$ has a boundary link $\ell$ (there can only be at most one boundary link in faces of 2-complexes dual to simplicial triangulations), we define
\begin{equation} 
h_f \equiv  h_\ell \prod_{vf} h_{vf} 
\end{equation}
where no $SU(2)$ element is assigned at the nodes attached to $\ell$.

No integration is performed over the boundary $SU(2)$ elements $h_\ell$. With the above definition, the EPRL ansatz of equation \eqref{eq:theEprlAnsatz} remains almost the same, with the difference that the amplitude now depends on $h_\ell$

\begin{equation}  
W_\mathcal{C}(h_\ell) = \sum_{\{j_f\}} \mu(j_f) \; \int_{\sl2c}  \! \mu(g_{ve}) \; \prod_f A_f(g_{ve},j_f,h_\ell) 
\end{equation}
where a face amplitude $A_f$ introduces explicit dependence on an $h_\ell$ if the face $f$ is not interior.

Similarly, the boundary state $\Psi_{\Gamma}^t( h_\ell; \eta_\ell, \zeta_\ell, k_{\ell \nn})$ in equation \eqref{eq:ExtrinsicStates} also has $SU(2)$ elements $h_\ell$ living on each boundary link. The contraction consists of integrating out $h_\ell$

\begin{eqnarray}
W(\omega_\ell,\zeta_\ell,\vec{k}_{\ell\nn}, t) &\equiv & \braket{W_\mathcal{C} \vert \Psi^t_\Gamma(\omega_\ell,\zeta_\ell,\vec{k}_{\ell\nn}, t) } \nonumber \\
& \equiv & \! \int \left( \prod \dd h_\ell \right) \, W_\mathcal{C}(h_\ell)  \; \Psi^t_\Gamma(h_\ell;\omega_\ell,\zeta_\ell,\vec{k}_{\ell\nn}, t) 
\end{eqnarray}

The $SU(2)$ integrations are straightforward to carry out using the orthogonality relation for the matrix elements of Wigner's D-matrices 
\begin{equation} 
\int_{SU(2)}\!\!\!\!\! \dd h\;\; D^{j}_{kl}(h^\dagger) D^{j'}_{mn}(h) = \frac{\delta_{kn} \, \delta_{lm} \,\delta_{j j'}}{d_{j}}
\end{equation}
and the properties of the Haar measure.

In terms of the spinor representation, after the contraction, equation
 \eqref{eq:Ffunction} for  the $F_f(g,z)$ function when the face $f$ is not interior becomes
\begin{align} 
& F_f(\{g_f\},\{z_f\};\vec{k}_{\ell_t},\vec{k}_{\ell_s}) =  \log \frac{ \braket{k_{\ell_s}\vert Z_{v_s e_s f} }^2 \braket{Z_{v_t e_t f}\vert k_{\ell_t} }^2}{\braket{Z_{vef}\vert Z_{vef} }  \braket{Z_{ve'f}\vert Z_{ve'f} }} +i \gamma \log \frac{\braket{Z_{ve'f}\vert Z_{ve'f}}}{\braket{Z_{vef}\vert Z_{vef}} } \nonumber \\
&+ \sum_{v \in f} \log \frac{  \braket{Z_{vef} \vert Z_{ve'f}}^2}{\braket{Z_{vef}\vert Z_{vef} }  \braket{Z_{ve'f}\vert Z_{ve'f} }} +i \gamma \log \frac{\braket{Z_{ve'f}\vert Z_{ve'f}}}{\braket{Z_{vef}\vert Z_{vef}} } \nonumber \\
\end{align}
The second line is the same definition as in equation \eqref{eq:Ffunction}, where we changed the sumation from edges $e$ in the boundary of $f$, to a summation over vertices $v$ in the boundary of $f$. The first line takes into account that there is a boundary present. In particular, in each boundary face there is a vertex $v_s$ attached to an edge $e_s$, that is in turn attached to the node $n_s$ which is the source of the link $\ell$. We remind that $\ell_s$ is the half--link attached to $n_s$. The notation is similar for the target node $n_t$ of $\ell$. The correspondance beetween the normalized vector $\vec{k}$ and the normalized spinor $k$ is as in equation \eqref{eq:correspondanceSpinorsVectors}.

\bigskip

In what follows, we will only consider 2-complexes with no interior faces. In this case, each face $f$ has exactly one link $\ell$. The combinatorics of the 2-complex are coded in the boundary graph, and we may label faces instead with their corresponding link $\ell$. 

Putting the above together, the transition amplitude reads 

\begin{align} \label{eq:contractedAmplitude}
W(\omega_\ell,\zeta_\ell,\vec{k}_{\ell\nn}, t) &= \sum_{\{j_\ell\}}  \! \mu(j_\ell) \ \left(\prod_{\ell}  e^{- t_{\ell} (j_\ell-\omega_\ell)^2}\right)\left( \prod_\ell e^{i j_{\ell} \gamma \zeta_\ell} \right) \times \nonumber \\ &\times \int \! \mu( g) \; \mu( z) \; \prod_{\ell} e^{j_\ell\; F_\ell(g,z;\vec{k}_{\ell\nn})} \;
\end{align}

\subsection{Heuristic derivation}
The idea behind the derivation is simple. Subtleties have to be taken into account and are discussed in the next section where we give the detailed derivation. These results will appear in \cite{mariosFabio}.  

In this section, we give a heuristic derivation in a simplified setting. We leave the product over links and link subscripts implicit and use the simplified notation of Section \ref{sec:emergenceOfGR}.

The transition amplitude is of the form 
\begin{equation} 
W(\omega,\zeta,\vec{k},t) = \sum_{\{j\}} e^{- t (j-\omega)^2} e^{i j \zeta} \int\!\! \mu( g) \, \mu( z)\; e^{j\; F(g,z;\vec{k})} \;
\end{equation}
where $\omega,\zeta,\vec{k}$ are the classical data from the coherent state, $t$ the semiclassicality parameter, and $g$ and $z$ the \slt and $\mathbb{CP}1$ spinfoam variables. 

We saw in Section \ref{sec:summaryFixedSpins}, that at a critical point corresponding to a simplicial geometry,  $F_\ell(g,z,k_{\ell \nn})$ becomes the Palatini deficit angle. Ignoring the cosine feature, considered in the following section, we bring in the Regge deficit angle by  
\begin{equation}
F(g,z,\vec{k}) \xrightarrow{j\rightarrow \lambda j, \; \lambda \gg 1 } i \gamma \phi(j;\vec{k}) 
\end{equation}
see Section \ref{sec:summaryFixedSpins}. The amplitude now reads
\begin{equation} \label{eq:stepOneHeuristic}
W(\omega,\zeta,\vec{k},t) \sim \sum_{\{j\}} e^{- t (j-\omega)^2} e^{i j \gamma  \left( \zeta -\phi(j;\vec{k} ) \right)} 
\end{equation}

This step is not justified. We have performed asymptotic analysis on $g$ and $z$, at \emph{fixed spins}, and ignored the fact that there is now also the summation over the spins. Furthermore, we have assumed that for any spin configuration in the spin sum there corresponds a critical point that has an interpretation as a simplicial geometry, and that a deficit angle $\phi(j;\vec{k})$ can be defined. We will see in the following section that it is not difficult to handle this properly and indeed it works out. 

Remember that we emphasized in Section \ref{sec:HEtoPal} a similar thing happening for the Palatini action. The field equations of General Relativity can be derived from the Palatini action, by first taking the variation over the spin connection $\omega$ (the equivalent here being the group variables $g$) and recover the Hilbert-Einstein action by replacing the solution of this equation of motion, the Levi-Civita connection $\omega(e)$, in the Palatini action.

 The variation over the tetrad $e$, which is here the equivalent of the spins $j$, can be carried out \emph{after} plugging--in the solution of ``half'' the equations of motion, which is in general the wrong procedure, yielding the correct result. 

The expression \eqref{eq:stepOneHeuristic} is simple. We may be cavalier and neglect the precise dependence of $\phi(j;\vec{k} )$ on the spins, simply replacing them by their value $\omega$ at the peak of the (highly peaked) gaussian weight. We also treat the spin--sums as integrals:
\begin{equation} 
W(\omega,\zeta,\vec{k},t) \sim \int \! \! \mu(j) \; e^{- t (j-\omega)^2} e^{i j \gamma  \left( \zeta -\phi(\omega;\vec{k} ) \right)} 
\end{equation}
This is a just gaussian integral. Performing it yields (up to normalization)

\begin{equation} \label{eq:heuristicResult}
\boxed{
W(\omega,\zeta,\vec{k},t) \sim e^{- (\gamma  \zeta - \gamma \phi(\omega;\vec{k} )\,)^2 /{4t}} e^{ i \omega \gamma  ( \zeta -\phi(\omega;\vec{k})\,) }
}
\end{equation}

Now, remember that the semiclassicality parameter $t$ is proportional to a positive power of $\hbar$. Thus, we have an \emph{exponentially  decaying amplitude in the limit $\hbar \rightarrow 0$.}

This is the gist of the result we will be arriving at in the next section. We caution that there are several details to be added. Before proceeding to the calculation, let us understand the meaning of this expression. We are neglecting the Immirzi parameter $\gamma$ for the rest of this section.

What we have done here is simply to replace the partial amplitude $I(j_f)$ with the Regge action, with the spins replaced by the area data $\omega$. We are considering spinfoams with no interior faces, defined on 2-complexes dual to a simplicial triangulation. Then, each link $\ell$ corresponds to a face $f$, dual to a geometrical triangle at a geometrical critical point (we are assuming this to be the case here). 

That is, bringing back the product over links, the part $e^{ - i \omega\phi(\omega;\vec{k})\ }$, becomes
\begin{equation}
\prod_\ell e^{ - i \omega_\ell \phi_\ell(\omega_f;\vec{k}) } = \exp{ - i \sum_f \omega_f \phi_f(\omega_f;\vec{k}) }
\end{equation}
Identifying $\omega$ with a dimension--full area by 
\begin{equation}
\omega \sim A/\hbar
\end{equation}
gives the exponential of the Regge action
\begin{equation}
\exp{ - \frac{i}{\hbar} \sum_f A_f \phi_f(\omega_f)}
\end{equation}
which are the fixed--spin asymptotics of the partial amplitude, the summand for the spins in the partition function. 

The meaning of \eqref{eq:heuristicResult} is then transparent. Had we matched the extrinsic data $\zeta$ with $\phi$, the amplitude at the semiclassical limit would be of order unit in $\hbar$. This would be the case in the setting of the gravition propagator, where one would then put insertions of operators in the path integral, prior to performing the spin-sum, to calculate expectation values of components of the quantum gravitational field. For the Lorentzian graviton propagator of the EPRL model see \cite{bianchi_lorentzian_2012}.

However, in a non-perturbative process, there is no reason why $\zeta$ would match $\phi$. If it did, then there would be a notion of an interpolating discrete geometry, naively solving the equations of discrete General Relativity in the form of the Regge action. 

Let us look at the decaying factor
\begin{equation}
e^{- ( \zeta -\phi(\omega;\vec{k} ) )^2 /{4t}}
\end{equation}
The semiclassicality parameter $t$ is to be equal to a characteristic scale of the problem and be proportional to a positive power of $\hbar$. 

This scale would in general be the common large scale $\lambda$ with which we take the spins to scale. We recall from the previous chapter that asymptotic analysis at fixed-spins is done by splitting $\omega_f =\lambda \delta_f$ with $\lambda \gg1$. Let us write $\omega = \lambda \delta_f =  \frac{A}{\hbar} \delta_f $, where we have introduced a dimension--full area scale $A$.

In the Regge action, written in terms of areas, the only characteristic scale would then be the scale of the areas. In a tunneling phenomenon, we expect the probability to decay with a quantity that has units of action. The \emph{only} possibility, is to set 
\begin{equation}
t = \frac{\hbar}{A}= \frac{1}{\omega} 
\end{equation} 
and we have a probability amplitude decaying as 
\begin{equation}
e^{- \frac{A}{\hbar} ( \zeta -\phi(\delta)^2)}
\end{equation}
We see below that the value $t = \frac{\hbar}{A}$ is in very good agreement with the semiclassicality condition. The angle $\phi$ can only depend on $\delta$, because changing $\lambda$ corresponds to a dilatation of the simplicial geometry (it is a global conformal transformation).

\bigskip

\emph{
Such amplitudes describe gravitational tunneling:  $\frac{A}{\hbar}$ is a large but finite number. Quantum theory ascribes a non--vanishing probability for a classically forbidden geometry evolution to take place.}

\medskip

\begin{figure}
\centering
\includegraphics[scale=0.4]{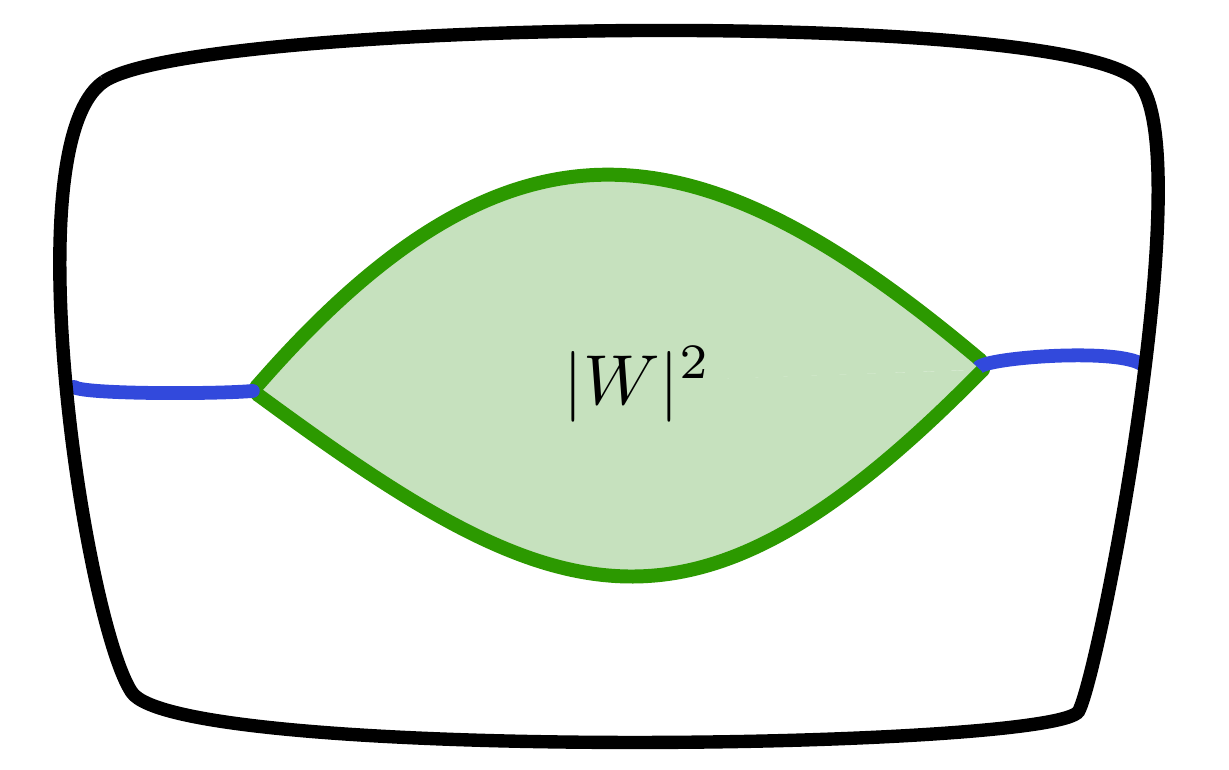}
\caption[The amplitudes of covariant LQG as probabilities for gravitational tunnelling.]{The amplitudes of covariant LQG as probabilities for gravitational tunnelling.  Solving an initial value problem for Einstein's equations with Cauchy data the intrinsic and extrinsic geometry of the hypersurface formed by the blue hypersurfaces and the upper boundary surface, and evolving towards the direction in which the foliation time increases, gives the upper--half of the spacetime. This is connected to the lower--half, which is the evolution in the past with Cauchy data the intrinsic and extrinsic geometry of the hypersurface formed by the blue hypersurfaces and the lower boundary surface.
}
\label{fig:conectingGRsolutionsCHAP4}
\end{figure}

See Figure \ref{fig:conectingGRsolutionsCHAP4}.

\subsection{Performing the spin--sum}
\label{sec:decayingAmplitudesCalc}

In this Section we show how the results from the fixed spins asymptotics can be used to estimate decaying amplitudes. Our starting point, is the transition amplitude \eqref{eq:contractedAmplitude}, repeated here for convenience
\begin{align} \label{eq:tempContAmp}
W(\omega_\ell,\zeta_\ell,k_{\ell\nn}, t) &= \sum_{\{j_\ell\} \in \Omega(\{j_\ell\},t,K) } \!\!\!\!\!\! \mu(j) \ \left(\prod_{\ell}  e^{- t_{\ell} (j_\ell-A_\ell)^2}\right)\left( \prod_\ell e^{i j_{\ell} \gamma \zeta_\ell} \right) \times \nonumber \\ &\times \int_{\Omega(g,z)}\!\!\!\! \mu( g) \; \mu( z) \prod_{\ell} e^{j_\ell\; F_\ell(g,z;k_{\ell\nn})} \;
\end{align}
where we have denoted $\Omega$ the ranges of integration and summation. 

 Notice that we have replaced $\zeta$ with $\gamma \zeta$. This is because $\zeta$ here is the discrete equivalent of the holonomy of the Ashtekar-Barbero connection which we wrote as $\xi$ in equation \eqref{eq:simonePhase}. It is written as $\xi = \alpha + \gamma \zeta$, with the phase $\alpha$ corresponding to the discrete version of the levi-civita spin connection. The phase $\alpha$ which we neglected is gauge and must be appropriately dealt with, by fixing a phase in the boundary states, also to ensure the proper gluing of the tetrahedra and avoid having twisted geometries \cite{freidel_twistors_2010,freidel_twisted_2010}. See \cite{christodoulou_realistic_2016} for an explicit example.  We are using the simplified notation of Section \ref{sec:emergenceOfGR}.

 We remind that we are considering spinfoam amplitudes defined on a fixed 2-complex $\mathcal{C}$ without interior faces and dual to a simplicial complex, with its boundary coinciding with the graph on which the boundary state is defined, i.e.\! $\partial \mathcal{C}=\Gamma$. 

\bigskip

The summation range for each spin is restricted to be at most $K$ times the standard deviation away from the peak of the gaussian. We leave $K$ unspecified for the moment, which is taken to be of order unity. The gaussian weights play the role of a regulator and $K$ can be understood as playing the role of a cut--off. The summation range is then the direct product space of the truncated ranges of individual spins
\begin{equation} \label{eq:spinsRange}
\Omega(\{j_\ell\},t,k) \equiv \bigotimes_\ell \; \{A_\ell - \frac{k}{\sqrt{2 t}}, A_\ell + \frac{k}{\sqrt{2 t}}\}
\end{equation}
This step is a good approximation: the spinfoam amplitudes are oscillating in the spins and are finite and extensive numerical tests show that the gaussian weights strongly dominate. Further justification is provided by the procedure performed in \cite{han_spinfoams_2016}. The calculation is along similar lines, in the context of studying the possibility of phase transitions in large spinfoams, where the authors argue that an exponential in the spins $\sim e^{-j}$ is sufficient as a regulator. Here, because we are using coherent spin-network states, they naturally provide a stronger regulator $\sim e^{-j^2}$.

The integration range over $g$ and $z$ is simply the direct product space of $N_{ve}-N_v$ copies of $SL(2,\mathbb{C})$ (the number $N_{ve}$ of half edges, where the $g_{ve}$ variables live, minus the redundant integrations that equal the number $N_v$ of vertices) and $N_{vf}$ copies of $\mathbb{C}\mathbb{P}1$ (the number $N_{vf}$ of pairs of vertices and faces, where the $z_{vf}$ variables live)
\begin{equation}
\Omega(g,z) \equiv \bigotimes_{N_{ve}-N_v} SL(2,\mathbb{C}) \; \bigotimes_{N_{vf}} \mathbb{C}\mathbb{P}1
\end{equation}

The summand in equation \eqref{eq:tempContAmp} will be forced to be zero for some set of spin configurations $\{j_\ell\}$, by the integration over $g$ and $z$. Specifically, the summand will be zero if the $SU(2)$ intertwiner space from the spins of the four faces attached to each edge is of zero dimension, see Appendix \ref{appsec:integratingDeltas}. The four faces correspond to four links on the boundary graph $\Gamma$.

We define the  set $\Omega_\Gamma$ of spin configurations for which this is not the case  by
\begin{equation}
\Omega_\Gamma \equiv \{ \{j_\ell\} : \forall e, \{\max(|j_i-j_k|),\min(j_i+j_k) \} \neq \emptyset, (i,j) \; \in e \}
\end{equation}
and introduce the characteristic function of $\Omega_\Gamma$
\begin{equation}
\chi_{\Omega_\Gamma} \equiv \chi(\{j_\ell\})=
 \begin{cases} 
      1 &, \{j_\ell\}\in \Omega_\Gamma \\[0.5cm]
      0 &,  \{j_\ell\}\notin \Omega_\Gamma \\
   \end{cases}
\end{equation}
We introduce the characteristic function $\chi(\{j_\ell\})$ explicitly in the amplitude
\begin{align} \label{ampBH2WHchar}
W((\omega_\ell,\zeta_\ell,k_{\ell\nn}, t) &= f(\omega_\ell,\zeta_\ell) \!\!\!\!\!\! \sum_{\{j_\ell\} \in \Omega(j_\ell,t,K) } \!\!\!\!\!\! \mu(j_\ell) \left(\prod_{\ell}  e^{- t (j_\ell-\omega_\ell)^2}\right)\left( \prod_\ell e^{i j_{\ell} \gamma \zeta_\ell} \right) \times \nonumber \\ & \times \;\int_{\Omega(g,z)} \!\!\!\! \mu( g) \,\mu(z)\; \prod_{\ell} e^{j_\ell\; F_\ell(g,z;\vec{k}_{\ell\nn})} \; \chi(\{j_\ell\})
\end{align}
where we defined
\begin{equation}
f(\omega_\ell,\zeta_\ell) \equiv e^{i\gamma \sum_\ell \omega_\ell \zeta_\ell}
\end{equation}
With $\chi(\{j_\ell\})$ explicit we can safely treat the sums as free and since they are finite sums they can be exchanged with integrals. 

Before doing that, we change variables, from the spin variables $j_\ell$ to the the spin \emph{fluctuations} $s_\ell$ about the peak $\omega_\ell$
\begin{eqnarray}
j_\ell=\omega_{\ell} +s_\ell \\
s_\ell \in  \{- \frac{K}{\sqrt{2 t}}, \frac{K}{\sqrt{2 t}}\}
\end{eqnarray} 
We define the summation range for the fluctuation variables $s_\ell$, which follows from
equation \eqref{eq:spinsRange}
\begin{equation} \label{eq:fluctrange}
\Omega(\{s_\ell\},t,K) \equiv \bigotimes_\ell \; \{- \frac{K}{\sqrt{2 t}}, + \frac{K}{\sqrt{2 t}}\}
\end{equation}
By exchanging summation and integration the amplitude takes the form
\begin{eqnarray} \label{eq:ampExchange}
W(\omega_\ell,\zeta_\ell,t) &=& f(\omega_\ell,\zeta_\ell) \! \int_{\Omega(g,z)} \!\!\!\! \mu( g) \; \mu( z) \prod_{\ell}e^{\omega_\ell\; F_\ell(g,z;\vec{k}_{\ell\nn})} \ \times \nonumber \\ &\times & \sum_{\{s_\ell\} \in \Omega(\{s_\ell\},t,K)} \left(\prod_{\ell}  e^{- s_\ell^2 t + i s_\ell \gamma \zeta_\ell + s_\ell F_\ell(g,z;\vec{k}_{\ell\nn})} \right)  \chi(\{\omega_\ell+s_\ell\})  \nonumber   \\
\end{eqnarray}
To bring this integral to a form suitable for stationary phase asymptotic analysis, and use the fixed--spins asymptotics, we split $\omega_\ell$ as
\begin{eqnarray}
\omega_{\ell}= \lambda \, \delta_\ell \\ \nonumber \\
\lambda >>1
\end{eqnarray}
where $\delta_\ell$ is of order unity in both the large and the small dimensionless parameter of the problem; the large spin scale $\lambda$ and the semi--classicality parameter $\sqrt{t}$. The condition $\lambda >>1$ implies that the quantum numbers are large with respect to the minimal non-zero value of the spins, one--half. This translates to the condition that the areas be macroscopic, much larger than $\hbar$. Notice that $\lambda$ is taken here to be a \emph{fixed}, large, dimensionless number, related to the area data and is not a scaling of the spins i.e.\! we are not taking $j\rightarrow \lambda j$. 

 The amplitude takes the form
\begin{eqnarray} \label{eq:ampStationaryChi}
W(\lambda,\delta_\ell,\zeta_\ell,t_{\ell}) &=&  f(\omega_\ell,\zeta_\ell) \! \int_{\Omega(g,z)} \!\!\!\! \mu( g) \; \mu( z) \; e^{\lambda \, \Sigma(g,z;\delta_\ell,\vec{k}_{\ell\nn}) }  \nonumber \\ 
&\times & \sum_{\{s_\ell\} \in \Omega(\{s_\ell\},t,K)}   \!\!\!\!\!\!\!\!\! \mu(s_\ell;\delta_\ell,\lambda) \; \mathcal{U}(s_\ell,g,z;t,\zeta_\ell )  \nonumber \\  &\times & \; \chi(\{\lambda \delta_\ell+s_\ell\}) \nonumber \\
\end{eqnarray}
where we have defined
\begin{equation} \label{eq:sigmaAction}
\Sigma(g,z;\delta_\ell,\vec{k}_{\ell\nn})= \sum_\ell \delta_\ell\; F_\ell(g,z;\vec{k}_{\ell\nn})
\end{equation}
and
\begin{equation}
\mathcal{U}(s_\ell,g,z;t,\zeta_\ell )=\prod_{\ell}  e^{- s_\ell^2 t + i s_\ell \gamma \zeta_\ell + s_\ell F_\ell(g,z,\vec{k}_{\ell\nn})}
\end{equation}
The semicolon separates the variables that are summed or integrated from the boundary data. Keep also in mind that we have not made the dependence on the Immirzi parameter $\gamma$ explicit. 

We have almost arrived to the point where we can apply the stationary phase theorem and use the fixed--spins asymptotics of the EPRL model. The point here is that the first line of equation \eqref{eq:ampStationaryChi}, is exactly the starting point for the fixed--spins asymptotics, see Section \ref{sec:critPointEqs}.

To apply the stationary phase theorem, we need to remove the dependence on $\lambda$ from the second and third lines of equation \eqref{eq:ampStationaryChi}. It appears in two places,  the characteristic function $\chi(\{\lambda \delta_\ell+s_\ell\})$ and the measure $\mu(s_\ell;\delta_\ell,\lambda)$.

Let us start from $\chi(\{\lambda \delta_\ell+s_\ell\})$. We want to derive a necessary and sufficient condition such that the fluctuations $s_\ell$ never take values such that the configuration $\{j_\ell\} =\{\lambda \delta_\ell + s_\ell\} \notin \Omega_\Gamma$. 

This is the case if the summation range $\Omega(\{s_\ell\},t,k)$ of the fluctuations is a subset of $\Omega_\Gamma$. The set $\Omega_\Gamma$ in terms of the fluctuations $s_\ell$ reads
\begin{eqnarray}
\Omega_\Gamma &\equiv & \{ \{s_\ell\} : \forall e, \{\max(|\lambda_0 (\delta_i-\delta_k) + (s_i-s_k) |), \nonumber \\ &{}& \min(\lambda_0 (\delta_i+\delta_k) + (s_i+s_k)) \} \neq \emptyset, (i,j) \; \in e \}
\end{eqnarray}
Let us fix a tetrahedron $e$. We call the maximum of the difference $\vert \delta_i-\delta_k \vert$, $\tilde{\delta}$, and the minimum of $\delta_i+\delta_k$, $\hat{\delta}$.

From equation \eqref{eq:fluctrange}, we see that the condition 
\begin{equation} \label{eq:fluctuationsCondition}
\Omega(\{s_\ell\},t,K) \subset \Omega_\Gamma
\end{equation}
is equivalent to
\begin{equation}
\lambda \tilde{\delta} + \frac{2K}{\sqrt{2t}} < \lambda \hat{\delta} - \frac{2K}{\sqrt{2t}} 
\end{equation}
which reads
\begin{equation}
\lambda \sqrt{t} > K\, \frac{2 \sqrt{2}}{\hat{\delta}-\tilde{\delta}}
\end{equation} 
This is the \emph{semiclassicality condition} of equation \eqref{eq:semiclassicalityCondition}. The difference $\hat{\delta}-\tilde{\delta}$ is of order unit and can be negative or positive. The cut-off $K$ is of order unit as well. We conclude that when 
\begin{equation}
\omega \sqrt{t} \gg 1
\end{equation}
the condition of equation \eqref{eq:fluctuationsCondition} is satisfied, and
\begin{equation} \label{eq:fluctuationsCondition}
\chi\big(\, \Omega(\{s_\ell\},t,K)\,\big)=1
\end{equation}
We now know the meaning of the semiclassicality condition: it imposes that all the intrinsic states $\psi_\Gamma(j_\ell,\vec{k}_{\ell\nn})$ superposed in the boundary have spins that are well within the triangle inequalities. We may understand this as a geometricity condition, imposing that the wavepacket is composed of superpositions of $\psi_\Gamma(j_\ell,\vec{k}_{\ell\nn})$ that describe triangulations of a spacelike hypersurface. This is a reasonable condition given that we are working in the semiclassical limit. 

Let us now turn to the measure. The measure $\mu(s_\ell;\delta_\ell,\lambda)$ is in general a product of $d_j$ factors, possibly chosen to be to some power \cite{mikovic_finiteness_2013}. In the standard definition of the model, we have
\begin{eqnarray}
\mu(s_\ell;\delta_\ell,\lambda) &=& \prod_\ell d_{j_\ell} \nonumber \\
&=& \prod_\ell (2j_\ell+1) \approx 2^L \prod_\ell j_\ell \nonumber \\
&=& 2^L  \prod_\ell (\lambda \delta_\ell + s_\ell) \nonumber \\
&=& 2^L  \lambda^L  \bigg( \prod_\ell  \delta_\ell \bigg)\ \ \left(1+ \mathcal{O}(s_\ell/\lambda) \right)
\end{eqnarray}

Thus, to justify dropping the $\mathcal{O}(s_\ell/\lambda)$, it must be the case that 
$\vert s_\ell \vert \ll \lambda$, or equivalently, $ K\sqrt{2} \ll \sqrt{t}\lambda$, which is again the semiclassicality condition.

Thus, the summation measure for $s_\ell$ is trivial and we omit the overall factor $2^L  \lambda^L  \bigg( \prod_\ell  \delta_\ell \bigg)$ in what follows. Imposing the semiclassicality condition, the amplitude takes the form

\begin{equation} 
\boxed{
W(\lambda,\delta_\ell,\zeta_\ell,t_{\ell}) = f(\omega_\ell,\zeta_\ell) \! \int_{\Omega(g,z)} \mu( g) \; \mu( z) \; e^{\lambda \, \Sigma(g,z;\,\delta_\ell,\vec{k}_{\ell\nn}) }  \ \mathcal{\tilde{U}}(g,z;t,\zeta_\ell )
}
\end{equation}
where we have defined
\begin{equation}
\mathcal{\tilde{U}}(g,z;t,\zeta_\ell ) \equiv \sum_{\{s_\ell\} \in \Omega(\{s_\ell\},t,K)} \; \mathcal{U}(s_\ell,g,z;t,\zeta_\ell )  
\end{equation}

Brought in this form, the function $\mathcal{\tilde{U}}(g,z;t,\zeta_\ell )$ is a continuous function in $g$ and $z$ only and the stationary phase theorem may be applied \cite{hormander_linear_1969,han_spinfoams_2016}. The critical point equations
\begin{equation}
0=\re \Sigma(g,z;\delta_\ell) = \delta_g \Sigma(g,z;\delta_\ell) = \delta_z \Sigma(g,z;\delta_\ell)
\end{equation}
are exactly those of the fixed--spins asymptotics of Section \ref{sec:critPointEqs}.

\begin{center}
\rule{0.5 \textwidth}{0.7pt}
\end{center}

Because we have contracted the amplitude with a semiclassical bounary state, we may neglect the possibility of a critical point that yields a vector geometry. The choices we have made for the $SU(2)$ coherent states in the previous section guarantee that there are only two possibilities for a given set of boundary data: either they will be Regge-like \cite{barrett_asymptotic_2009,han_asymptotics_2012,barrett_lorentzian_2010,barrett_quantum_2010}, in which case we will have a geometrical critical point corresponding to one of the three types of possible simplicial geometries, or there will be no critical point.

We henceforth assume that the data $\delta_\ell$ and $\vec{k}_{\ell\nn}$ are chosen so that they correspond to a set of $2^N$ geometrical critical points for the fixed--spin asymptotics, related by the orientation of the tetrad, see Section  \ref{sec:emergenceOfSimpGeoms}

\begin{equation}
(g_c,z_c) = \big(\, g_c(\delta_\ell,k_{\ell\nn}),z_c(\delta_\ell,k_{\ell\nn})\, \big) \ \ , \, c=1,\ldots,2^N
\end{equation}
We then have the following estimation for the amplitude
\begin{eqnarray} \label{eq:tempEstimCritPointW}
W(\lambda,\delta_\ell,\zeta_\ell,t_{\ell}) &=& f(\omega_\ell,\zeta_\ell) \, \sum_c  \lambda^{M^c_\mathcal{C}} \,\mathcal{F}_c(\delta_\ell, \vec{k}_{\ell\nn} ) \ e^{\lambda \, \Sigma(g_c,z_c;\,\delta_\ell,\vec{k}_{\ell\nn}) } \times  \nonumber \\ &\times & \; \mathcal{\tilde{U}}(g_c,z_c;t,\zeta_\ell ) \big(1 + \mathcal{O}(1/\lambda)\big)
\end{eqnarray}
where the power $M^c_\mathcal{C}$ depends on the combinatorics of the 2-complex $\mathcal{C}$ and $\mathcal{F}_c(\delta_\ell, \vec{k}_{\ell\nn} )$ includes the determinant of the Hessian. We do not need here the exact expressions, see \cite{mariosFabio,han_asymptotics_2013} for the details. 

The important point to keep for physical applications is that in the first order approximation, the scale $\lambda$ appears only as an overall scaling factor $\lambda^{M^c_\mathcal{C}}$ and as a linear term in the exponential. In particular, $\mathcal{F}_c(\delta_\ell, \vec{k}_{\ell\nn} )$ does not depend on $\lambda$.

We proceed to evaluate $\mathcal{\tilde{U}}$ at the critical point. Heuristically, it is evident that we need only perform a gaussian integral to get a simple expression for $\mathcal{\tilde{U}}$. We have yet however to remove the regulator $K$ and we take a moment to explain the approximation carefully. 

The function $\mathcal{\tilde{U}}$ of $g$ and $z$ , reads
\begin{equation}
\mathcal{\tilde{U}}(g,z;t,\zeta_\ell ) \equiv \sum_{\{s_\ell\} \in \Omega(\{s_\ell\},t,K)}  \prod_{\ell}  e^{- s_\ell^2 t + i s_\ell \gamma \zeta_\ell + s_\ell F_\ell(g,z,\vec{k}_{\ell\nn})}
\end{equation}
A a critical point $c$, we have

\begin{equation}
F_\ell(g_c,z_c,\vec{k}_{\ell\nn}) \rightarrow - i \, \phi_\ell(s_c(v) ; \delta_\ell, \vec{k}_{\ell\nn}) 
\end{equation}
where $\phi_\ell(s_c(v) ; \delta_\ell, \vec{k}_{\ell\nn})$ is the Palatini deficit angle of equation \eqref{palatiniAngle}. Thus, $\tilde{U}$ evaluated at $c$, reads 
\begin{equation} \label{eq:tempUtilde}
\mathcal{\tilde{U}}(g_c,z_c;t,\zeta_\ell ) = \sum_{\{s_\ell\} \in \Omega(\{s_\ell\},t,K)}  \prod_{\ell}  e^{- s_\ell^2 t_{\ell }}  e^{i s_\ell \left( \gamma \zeta_\ell -  \phi_\ell(s_c(v) ; \delta_\ell, \vec{k}_{\ell\nn}) \right)}
\end{equation}
Let us return to the semiclassicality condition and the semiclassicality parameter $t_\ell$. In a physical problem for geometry transition, treated through the techniques presented here, the only available physical scale will be the area scale that will be identified with $\lambda$. Let us call the dimensionfull area scale $A$,

\begin{equation}
A = \hbar \,\lambda
\end{equation}

Thus, the semiclassicality parameter will be set to 
\begin{equation}
t = \frac{\hbar^n}{A^n}= \frac{1}{\lambda^n}
\end{equation}
to some positive power $n$. The semiclassicality condition then imposes that 
\begin{equation}
\lambda^{1-n/2} \gg 1
\end{equation}
which implies
\begin{equation}
n < 2
\end{equation}
A further requirement on $n$ comes from demanding that the spin flucuations are larger than Planckian, so that we have a large number of intrinsic states $\psi_\Gamma$ in the superposition $\Psi_\Gamma$. 

That is, that $s_\ell$ are allowed to take values in a discrete set much larger than $\{-\frac{1}{2},0,\frac{1}{2}\}$. We may encode this condition by demanding that the standard deviation $\frac{1}{\sqrt{2 t}}$ is larger than order unit in $\lambda$, that is,  
\begin{equation}
n > 0
\end{equation}
Thus, we henceforth set 
\begin{equation}
t = \frac{1}{\lambda^n} \ \; , n \in (0,2) 
\end{equation}
but leave $t$ explicit in our expressions. From dimensional considerations (see previous section) , \emph{the natural value} for $t$ is for $n=1$, right in the middle of the allowed range above, that achieves good semiclassicality properties.

The above justify the following approximation for the sum over the spins:
\begin{equation}
\sum_{\{s_\ell\} \in \Omega(\{s_\ell\},t,K)} \approx \prod_\ell \int_{-\frac{K}{\sqrt{2t}}}^{\frac{K}{\sqrt{2t}}} \dd s_\ell \approx \prod_\ell \int_{-\infty}^{\infty} \dd s_\ell
\end{equation}
We have thus removed the regulator $K$. Returning to equation \eqref{eq:tempUtilde} and performing the gaussian integrals, we have

\begin{equation} 
\mathcal{\tilde{U}}(g_c,z_c;t,\zeta_\ell ) = \prod_{\ell} \exp\;{- \frac{1}{4t}\left( \gamma \zeta_\ell - \phi_\ell(s_c(v) ; \delta_\ell, \vec{k}_{\ell\nn}) \right)^2}
\end{equation}

At the critical point $c$, as we saw in the previous chapter,
\begin{equation} \label{eq:sigmaAction}
\Sigma(g_c,z_c;\delta_\ell,\vec{k}_{\ell\nn})= -i \sum_\ell \delta_\ell\; \phi_\ell(s_c(v) ; \delta_\ell, \vec{k}_{\ell\nn})
\end{equation}

\subsection{Summary}
\label{sec:summaryGravTunn}

Given the data $\zeta_\ell$, $\omega_\ell$, $k_{\ell\nn}$, and defining $\lambda$, we have the following estimate for the amplitude
\begin{eqnarray} \label{eq:ampEstimOne}
\vspace*{-0.4cm}	W \sim \left[ \sum_{s(v)} \; ( \lambda )^{N} \mu( \delta_\ell ) \prod_\ell \; \exp{\left( -\frac{1}{4t}\left(\gamma \zeta_\ell - \beta \phi_\ell(s(v) ; \delta_\ell, \vec{k}_{\ell\nn}) +  \Pi_\ell \right)^2\, + \, i \, \gamma \, \zeta_\ell \, \omega_\ell \right)} \right] \nonumber \\  
\vspace*{-0.9cm} \left( 1 +O(1/\lambda) \right) \nonumber \\
\end{eqnarray}
We take a moment to go through the various quantities appearing in this formula. We have introduced a few important subtleties regarding the different kinds of geometrical critical points that we neglected in the derivation above. The $\Pi_\ell$ contribution accounts for an extra phase in the Lorentzian intertwiners, see \cite{bianchi_lorentzian_2012,
barrett_lorentzian_2010}.

 The power $N$ is in general a half integer that depends on the rank of the hessian at the critical point and the combinatorics of the 2-complex $\mathcal{C}$. The function $\mu( \delta_\ell )$ includes the summation measure over the spins and the Hessian evaluated at the critical point. Neither the summation measure nor the Hessian scale with $\lambda$.

The estimation \label{eq:ampEstimOne} is valid for all three types of possible geometrical critical points. If $\omega_\ell$ and $k_{\ell\nn}$ specify a Lorentzian geometry, then 
\begin{equation}
	\beta=\gamma
\end{equation}
and
\begin{equation}
	\Pi_\ell = 
		\begin{cases} 
      		0 & \text{thick wedge}\\
      		\pi & \text{thin wedge}
   \end{cases}
\end{equation}
If $\omega_\ell$ and $k_{\ell\nn}$ specify a degenerate geometry, then the dihedral angles $\phi_\ell(\omega_\ell,k_{\ell\nn})$ either vanish or are equal to $\pi$, according to whether we are in a thick or thick wedge. By abuse of notation, we express this simply by setting $\beta=0$ in this case and keeping $\Pi_\ell$ defined as above.

If $\omega_\ell$ and $k_{\ell\nn}$ specify a Euclidean geometry, then we have
\begin{equation}
	\beta=1
\end{equation}
and
\begin{equation}
	\Pi_\ell = 0 
\end{equation}

The function $\phi_\ell(s_c(v) ; \delta_\ell, \vec{k}_{\ell\nn})$ denotes the Palatini deficit angle. 

We discuss the interplay between the different geometrical points and the sum over the multiple semiclassical limits in Chapter \ref{ch:calculationOfAnObservable} where we apply the above to a concrete geometry transition problem.

\chapter{Exterior spacetime and definition of a complete observable} \label{ch:exteriorSpacetime}
In this chapter we introduce the exterior spacetime, which we call the Haggard-Rovelli (HR) spacetime,  modelling the spacetime surrounding the geometry transition region.  We provide a hypersurface--independent formulation of the HR spacetime, introduced originally in \cite{haggard_black_2014}. Similar ideas have been independently and concurrently been put forward by Garay, Barcelo et al \cite{barcelo_exponential_2016,barcelo_black_2016,
barcelo_lifetime_2015,barcelo_mutiny_2014}.

We introduce the relevant classical and quantum observables, the mass $m$ and the bounce time $T$, and the lifetime $\tau(m)$ respectively. We propose an interpretation for the transition amplitudes and formulate the problem such that a path-integral over quantum geometries procedure can be naturally applied. 

\section{Black holes age} \label{sec:blackholesAge}

Black holes are not eternal, mathematical objects. They are astrophysical objects and as such they are expected to go through phases, evolve and age. 

The most widely accepted feature of black hole evolution is Hawking evaporation, with complete evaporation predicted by Hawking theory (assuming the semiclassical approximation holds for late times) to be of order $\sim m^3$. Indications that ``something weird'' happens in shorter time scales include the Page time \cite{page_average_1993,page_black_1993,
page_information_1993-1,page_time_2013}, and the debate on the ``firewall' paradox \cite{perez_no_2014}.

 For lack of space we do not go through the list of proposals for how black holes may end their life, and refer to the literature. Such proposals typically attempt to provide a resolution for the information paradox \cite{hawking_breakdown_1976}, see \cite{hossenfelder_conservative_2010} for a (non--exhaustive) review.
 
\bigskip

Another clear indication that black holes age comes from studying the properties of their interior volume. The maximal volume inside the horizon is given at the dominant order in the mass $m$ by 
\begin{equation}
V(v) \sim m^2 v
\end{equation}
where $v$ is the (advanced) time as measured by a far away observer from the moment of the formation. This result was shown in a paper by Carlo Rovelli and the author for the static spherically symmetric case, including the Reissner-Nordstrom charged black hole \cite{christodoulou_how_2015}. 

In a follow--up work by the author and Tommasso de Lorenzo, we extended this result to the Hawking evaporating black hole \cite{christodoulou_volume_2016}, surprisingly yielding an identical formula where now the mass is the initial mass $m_0$ \emph{at the moment of formation}
\begin{equation}
V(v) \sim m_0^2 v
\end{equation}
Subsequent studies by other authors generalised this result in Anti-de Sitter, Kerr and other more exotic kind of black holes \cite{chen_black_2015,bengtsson_black_2015,
ong_never_2015,ong_persistence_2015} suggesting that this volume monotonicity appears to be a general feature of black hole solutions.

The volume of black holes is a good geometrical measure of their age. Black holes then build up immense volumes in their lifetime. For instance Saggitarius $A^\star$ currently has interior spacelike hypersurfaces large enough to fit the solar system millions of times. 

What happens to all this volume it the black hole completely evaporates? If a solar mass black hole almost completely evaporated, say to a few grams, it would contain volumes equivalent to $10^5$ times our observable universe, behind a minuscule horizon of the order of $10^{-50}$ squared meters. 

The process described by Haggard and Rovelli appears natural in this geometrical aspect: the inverse measure of age applies for a white hole, and the volume will deflate.

In related work, we show that under minimal assumptions, the Von Neumann entropy for volumes from the Hilbert space spanned by 4-valent spin networks is well behaved and finite\cite{astuti_volume_2016}, based on results from the spectrum of the volume in \cite{brunneman_properties_2008,
brunnemann_properties_2008}. 

These ideas can be relevant for future studies of the phenomenon we discuss in the rest of this chapter.

\begin{center}
\rule{0.5 \textwidth}{0.7pt}
\end{center}

 In the renowned 1974 letter titled ``Black hole explosions?'' \cite{hawking_black_1974}, Hawking argues that quantum physics can be relevant in space-time regions where curvature is low. The author begins with the observation that when the large time scale involved in near horizon physics is taken into account, the lifetime of the universe, significant departures from classical physics may take place. It may take even longer: Hawking's theory implies that a solar mass black hole is expected to radiate a Planck mass worth of Hawking photons in approximately $10^{15}$ times the Hubble time. This time is of order $ \sim m^2$ in Planck units, where $m$ is the mass of the hole.

 Hawking closes with the comment that quantum fluctuations of the metric are neglected in calculations of near-horizon physics and taking them into account ``might alter the picture''. Haggard and Rovelli, combining these two ideas, argued in \cite{haggard_black_2014} for the possibility that when enough time, as measured far from the hole, is allowed to lapse, and quantum fluctuations of the metric near the horizon become significant, a geometry transition of the trapped region to an anti-trapped region takes place, from which the matter trapped inside the hole is released. 

In \cite{haggard_black_2014}, the authors give a simple model of the global spacetime picture away from the region where the transition takes place. The existence of the exterior metric, which we call in what follows the Haggard--Rovelli metric (HR-metric), lends plausibility to the process, since general Relativity need only be violated in a compact spacetime region. The stability of the exterior spacetime after the quantum transition has taken place has been studied in \cite{de_lorenzo_improved_2016}, where the known instabilities of white hole spacetimes were shown to limit the possible duration of the anti-trapped phase, but do not otherwise forbid this process taking place. 
\section{Haggard-Rovelli Spacetimes} \label{sec:RH}  
In this section we construct the Haggard-Rovelli (HR) spacetime \cite{haggard_black_2014,de_lorenzo_improved_2016}. The original construction is based on a partly one to many (one to two) mapping from the Kruskal manifold. The relation between the two constructions is explained in Section \ref{sec:crossedFingers}.  The construction presented here avoids using specific hypersurfaces and helps reveal the insights of the original paper. 

\subsection{Physical picture and the Penrose diagram}

The physical phenomenon modelled by an HR-spacetime is \emph{the transition, via quantum gravitational effects that are non-negligible only in a compact region of spacetime, of a trapped region formed by collapsing matter to an anti-trapped region, from which matter is released}. 
 
A compact region is excised from spacetime, by introducing a \textit{spacelike compact interior} boundary of the spacetime. Outside of this non-classical region the usual spacetime picture applies and the metric solves Einstein's field equations \textit{exactly everywhere}, including on the interior boundary. 

The HR-spacetime is constructed by taking the following simplifying assumptions:
\begin{itemize}
\item Collapse and expansion of matter are modelled by thin shells of null dust of mass $m$. 
\item Spacetime is spherically symmetric. 
\end{itemize}

These assumptions determine the local form of the metric by virtue of Birkhoff's theorem, which can be stated as follows \cite{hawking_large_2011} : Any solution to Einstein's equations in a region that is spherically symmetric and empty of matter is \emph{locally} isomorphic to the Kruskal metric in that region.

It therefore follows that the metric inside the null shells is flat (i.e. Schwartzschild with $m=0$), the metric outside the shells is locally Kruskal with $m$ being the mass of the shells and the spacetime is asymptotically flat. Moreover, it follows that the trapped and anti-trapped regions in the spacetime can only be portions of the black and white hole regions of the Kruskal manifold, respectively. In particular, the marginally trapped and anti-trapped surfaces bounding these regions can only be portions of the $r=2m$ hypersurfaces.

\begin{figure} 
\centering
\includegraphics[scale=1]{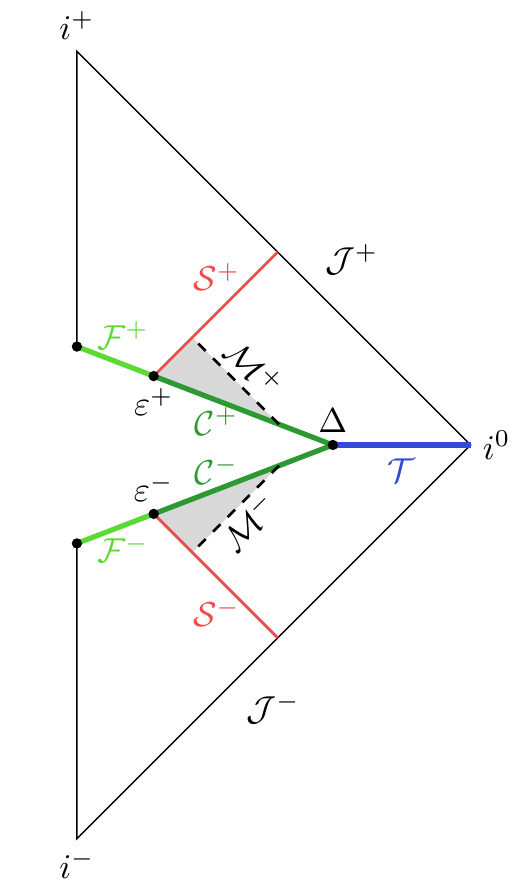} 
\caption{The Haggard-Rovelli spacetime.}
\label{fig:ansatz}
\end{figure}

The above are summarized in the Penrose diagram of an HR-spacetime given in Figure \ref{fig:ansatz}. Compare this diagram with figure \ref{fig:lens}. The metric, energy-momentum tensor and expansions of null geodesics are given explicitly in Eddington-Finkelstein coordinates in the next section. The surfaces and regions in Figure \ref{fig:ansatz} have the following properties :

\begin{itemize}
\item $\mathcal{S}^-$ and $\mathcal{S}^+$ are null hypersurfaces. The junction condition on the intrinsic metric holds. Their interpretation as thin shells of null dust of mass $m$ follows : the allowed discontinuity in their extrinsic curvature results in a distributional contribution in $T_{\mu\nu}$ on $\mathcal{S}^-$ and $\mathcal{S}^+$, see next Section. This is standard procedure in Vaidya null shell collapse models\cite{vaidya_gravitational_1951}, see for instance \cite{poisson_relativists_2004}. $T_{\mu\nu}$ vanishes everywhere else. 

\item The surfaces $\mathcal{F^+}$, $\mathcal{F^-}$, $\mathcal{C^+}$, $\mathcal{C^-}$ depicted in Figure \ref{fig:ansatz} are spacelike. Their union $\mathcal{F}^-\cup\mathcal{C}^-\cup\mathcal{C}^+\cup\mathcal{F}^+$ constitutes the interior boundary. The intrinsic metric is matched on the spheres $\Delta$ and $\varepsilon^{\pm}$. The extrinsic curvature is discontinuous on $\varepsilon^{\pm}$, see previous point, and is also discontinuous on $\Delta$ because of the requirement that $C^\pm$ are spacelike.

\item $\mathcal{T}$ is a spacelike surface. The junction conditions for both the intrinsic metric and extrinsic curvature, hold, including on the sphere $\Delta$. As we see below, $\mathcal{T}$ plays only an auxiliary role and need not be further specified. 
\item $\mathcal{M^-}$ and $\mathcal{M^+}$ are marginally trapped (anti-trapped) surfaces and the shaded regions are trapped (anti-trapped). That is, the expansion of outgoing (ingoing) null geodesics vanishes on $\mathcal{M^-}$ ($\mathcal{M^+}$), is negative inside the shaded regions and positive everywhere else. 
\end{itemize}

Before explicitly giving the metric, let us comment on the necessity of extending the interior boundary outside the (anti-)trapped regions. The excised region removes from the spacetime the region where strong quantum gravitational effects are expected to be non-negligible. It excludes the region where the singularity surface would form and, crucially, it extends outside the marginally trapped surfaces. The need to do the latter can also be understood from Birkhoff's theorem, we refer the reader again to Figure \ref{fig:ansatz}. The marginally trapped and anti-trapped surfaces $\mathcal{M^-}$ and $\mathcal{M^+}$ can only be realized as being portions of the $r=2m$ surfaces of the Kruskal manifold. If we restrict the interior boundary to be inside the marginally trapped surfaces, $\mathcal{M^-}$ and $\mathcal{M^+}$ will intersect. Then, the metric around their intersection point (sphere) \emph{must} be locally isomorphic to the Einstein-Rosen bridge and the Penrose diagram would be radically altered: $\mathcal{M^-}$ and $\mathcal{M^+}$ will extend to null infinity and become past and future event horizons.
 
\subsection{The HR-metric} \label{sec:EFRH}
In this section we explicitly construct the HR-metric in Eddington-Finkelstein (EF) coordinates, in which it takes a particularly simple form.
We need to introduce two charts. Referring to Figure \ref{fig:ansatz}, we define four regions. Regions I and IV are the past and future flat regions inside the shells respectively. Regions II and III are the Kruskal regions outside the shells separated by the fiducial surface $\mathcal{T}$. Region II includes the trapped region and region III the anti--trapped region.

The first chart provides ingoing EF coordinates $(v,r)$ for the union of regions I and II.  The second chart provides outgoing EF coordinates $(u,r)$ for the union of regions III and IV. There is only the junction condition on $\mathcal{T}$ to be considered, which we give below. The radial coordinate $r$ will be trivially identified in the two coordinate systems. For regions I and II the metric reads
\begin{equation} \label{eq:downMetricEF}
\dd s^2 = -\left(1-\frac{2m}{r} \Theta(v-v_0) \right)\dd v^2 + 2 \dd v \, \dd r+ r^2 \dd \Omega^2
\end{equation}

and for regions III and IV	
\begin{equation} \label{eq:upMetricEF}
\dd s^2 = -\left(1-\frac{2m}{r} \Theta(u-u_0)\right)\dd u^2 -2\dd u\, \dd r + r^2 \dd \Omega^2
\end{equation}
where $\Theta$ is the Heaviside step function. 

The ingoing and outgoing EF times $v_0$ and $u_0$ denote the position of the shells $\mathcal{S^-}$ and $\mathcal{S^+}$, in these coordinates.

The two junction conditions on $\mathcal{T}$ are satisfied by the identification of the radial coordinate along $\mathcal{T}$ and the condition
\begin{equation} \label{eq:junctT}
v - u \stackrel{\mathcal{T}}{=} 2r^*(r)
\end{equation}
where $r^*(r)=r + 2m \log{\left|r/2m-1 \right|}$. We emphasize that we need not and will not in what follows choose the hypersurface $\mathcal{T}$ explicitly. The HR-metric is independent of any such explicit choice. The reason it is necessary to consider it as an auxilliary structure is explained in \cite{haggard_black_2014} : the spacetime doubly covers a portion of one of the asymptotic regions of the Kruskal manifold and thus we need to use at least two separate charts describing a Kruskal line element. Where we take the separation of these charts to be, that is, the explicit choice of $\mathcal{T}$, is irrelevant.

To explicitly define the metric we need to give the range of coordinates. Let us assume a choice of boundary surfaces has been given. Having covered every region of spacetime by a coordinate chart, we can describe embedded surfaces. Since all surfaces $\Sigma$ appearing in Figure \ref{fig:ansatz} are spherically symmetric, it suffices to represent the surfaces as curves in the $r-v$ and $r-u$ planes. Using a slight abuse of notation we write $v=\Sigma(r)$ or, in parametric form, $(\Sigma(r), r)$. The range of coordinates is in the intersection of the conditions:
\begin{center}
\medskip
\begin{tabular}{ r l l l l l}
  I $\cup$ II : & $v \in (-\infty, +\infty)$ & $r \in (0, +\infty)$ &  $v \leq \mathcal{F^-}(r)$ & $v \leq \mathcal{C^-}(r)$ & $v \leq \mathcal{T}(r)$\\
  III $\cup$ IV  : & $ u \in (-\infty, +\infty) $ &  $r \in (0, +\infty)$ & $u \geq \mathcal{F^+}(r)$ & $u \geq \mathcal{C^+}(r)$ & $u \geq \mathcal{T}(r)$
\end{tabular}
\medskip
\end{center}

What remains is to ensure that we kept trapped and anti-trapped regions where we want them, as depicted in Figure \ref{fig:ansatz}. This is equivalent to the geometrical requirement that the spheres $\epsilon^\pm$ have proper area less than $4 \pi (2m)^2$ while the sphere $\Delta$ has proper area larger than $4 \pi (2m)^2$. We may write this in terms of the radial coordinate as
\begin{eqnarray}
r_{\epsilon^\pm} < 2m \nonumber\\
r_{\Delta} > 2m
\end{eqnarray} 
Since $\epsilon^\pm$ and $\Delta$ are specified once the boundary is explicitly chosen, this is a condition on the allowed boundary surfaces that can be used as an interior boundary of an HR-spacetime : $\mathcal{C^\pm}$ can be any spacelike surfaces that have their endpoints at a radius less and greater than $2m$, crossing in the latter endpoint. Since $\mathcal{C}^\pm$ are spacelike, it follows that we necessarily have a portion of the $r=2m$ surfaces in the spacetime along with trapped and anti-trapped regions in their interior. 

This completes the construction of the HR-metric. To complete the definition of the family HR-spacetimes, identify its coordinate-independent parameters and restrict to their physical range. The HR-metric is a two-parameter family of metrics. The geometry of the spacetime, up to the choice of interior boundary is fully determined once two dimension-full coordinate independent quantities are specified. Obviously, one parameter is the mass $m$ which we take to be positive. The second parameter is the \emph{bounce time} $T$, the meaning of which is discussed in the following section. We can express $T$ in terms of $u_0$ and $v_0$ simply by 

\begin{equation} \label{eq:BounceTimeEF}
T = u_0-v_0
\end{equation}
Similarly to $m$, we take also $T$ to be positive. Let us repeat for emphasis that the Haggard-Rovelli geometry has two characteristic physical scales : a length scale and a time scale. The subject of this article is to define and compute a probabilistic correlation between these two scales from quantum theory. The boundary conditions for the path-integral will be the geometry of the interior boundary. 

\medskip
The role of the bounce time $T$ as the second spacetime parameter is obscured in the line elements \eqref{eq:downMetricEF} and \eqref{eq:upMetricEF}. In equation \eqref{eq:BounceTimeEF}, we expressed the bounce time in terms of the null coordinates at the collapsing and anti-collapsing shells. The bounce time $T$, is then encoded implicitly in the line element through the ranges $v \geq v_0 $ and $u \leq u_0$, via the Heaviside functions that specify the part of the spacetime that is curved.

We may make $T$ appear explicitly as a dimension-full parameter in the metric components. This is achieved by time shifting both coordinates $u$ and $v$ by $\frac{v_0+u_0}{2}=T/2$ 
\begin{eqnarray} \label{eq:timeShift}
v \rightarrow v- \frac{v_0+u_0}{2} \nonumber \\ \nonumber \\
u \rightarrow u- \frac{v_0+u_0}{2}\end{eqnarray}
The above isometry ($(\partial_v)^\alpha$ and $(\partial_u)^\alpha$ are the timelike Killing fields in each patch) simply amounts to shifting simultaneously the origin of the two coordinates systems. The line elements \eqref{eq:upMetricEF} and \eqref{eq:BounceTimeEF} now read

\begin{equation} \label{eq:downMetricNew}
\dd s^2 =  - \left(1- \frac{2m\,\Theta(v+\frac{T}{2})}{r} \, \right)\dd v^2 + 2 \dd v \, \dd r+ r^2 \dd \Omega^2 
\end{equation}
and 
\begin{equation} \label{eq:upMetricNew}
\dd s^2 = -\left(1-\frac{2 m\,\Theta(u-\frac{T}{2})}{r} \right)\dd u^2 - 2 \dd v \, \dd r+ r^2 \dd \Omega^2 
\end{equation}
The role of $T$ as a spacetime parameter is manifest in this form of the metric. It is instructive to compare with the Vaidya metric for a null shell collapse-model, describing the formation of an eternal black hole by a null shell collapsing from past null infinity. Setting the null shell to be at $v=v_0$, the line element would be identical to \eqref{eq:downMetricEF}, with the difference that the range of the $v$ coordinate is in $(-\infty,\infty)$. The choice $v=v_0$ for the position of the null shell is immaterial and we can always remove $v_0$ from the line element by shifting the origin with $v \rightarrow v-v_0$. In the HR-metric, the two coordinate charts are related by the junction condition \eqref{eq:junctT}. It is impossible to make both $v_0$ and $u_0$ disappear from the line element by shifting the origins of the coordinate charts, the best we can do is remove one of the two or, as we did above, a combination of them. This observation emphasizes that the bounce time $T$ is a free parameter of the spacetime. Notice that the junction condition \eqref{eq:junctT} is unaffected by a simultaneous shifting of the form \eqref{eq:timeShift}.

\medskip

The field equations are solved for the energy momentum tensor \cite{poisson_relativists_2004}
\begin{center}
\medskip
\begin{tabular}{ r l }
  I $\cup$ II : &   $T_{\mu\nu} = + \frac{\delta(v+T/2)}{4 \pi r^2} \delta^v_\mu \delta^v_\nu $ \\ & \\
  III $\cup$ IV : & $T_{\mu\nu} = - \frac{\delta(u-T/2)}{4 \pi r^2} \delta^u_\mu \delta^u_\nu $
  \end{tabular}
\medskip
\end{center}

The expansion $\theta^-$ of outgoing null geodesics in the patch I $\cup$ II and the expansion $\theta^+$ of ingoing null geodesics reads in the patch III $\cup$ IV, read 
\begin{center}
\medskip
\begin{tabular}{ r l }
  I $\cup$ II : &   $\theta^- \equiv \nabla_\mu k^{-\mu} = \Gamma^- \left(1- \frac{2 m}{r} \Theta(v+T/2)\right) $
   \\ & \\
  III $\cup$ IV : &   $\theta^+ \equiv \nabla_\mu k^{+\mu} = \Gamma^+ \left(1- \frac{2 m}{r} \Theta(u-T/2)\right) $
    \end{tabular}
\medskip
\end{center}
where $k^{-\mu}$ and $k^{+\mu}$ are affinely parametrized tangent vectors of the null geodesics and $\Gamma^-$ and $\Gamma^+$ are positive scalar functions, see \cite{poisson_relativists_2004}. It follows directly from these expressions that the spacetime possesses a trapped and an anti-trapped surface, where the expansions $\theta^\pm$ vanish, and which we identify with $\mathcal{M^+}$ and $\mathcal{M^-}$ in Figure \ref{fig:ansatz}. Thus, in EF coordinates, $\mathcal{M^\pm}$ are defined by 
\begin{center}
\medskip
\begin{tabular}{ l l l}
  $\mathcal{M^-}$ : &   $r=2m$ & $v\in \left(-T/2, \,\mathcal{C^-}(2m)\right)$ 
   \\ & \\
  $\mathcal{M^+}$ : &   $r=2m$ & $u\in \left(\mathcal{C^+}(2m), \,T/2\right)$
    \end{tabular}
\medskip
\end{center}

 As explained above, it will always be the case that these surfaces are present in the spacetime, along with a trapped and anti-trapped regions where $\theta^\pm$ are negative, if $r_{\epsilon^\pm} < 2m$ and $r_{\Delta} > 2m$. We may describe explicitly the  trapped region as the intersection of $r<2m$, $v\in (-T/2,\mathcal{C^-}(2m))$ and $v \leq \mathcal{C^-}(r)$ and for the anti-trapped region $r<2m$, $u\in \left(\mathcal{C^+}(2m),T/2\right)$ and $u \geq \mathcal{C^+}(r)$.  $\theta^\pm$ are positive in the remaining spacetime.

\section{The Bounce Time $T$} \label{sec:BounceTime}
In this section we discuss the meaning of the bounce time $T$ as a time scale that characterises the geometry of the HR-spacetime. Intuitively, $T$ controls the time separation between the two shells. We will then see that the bounce time can be understood as an equivalent concept to the Hawking evaporation time for the HR-spacetime.

In equation \eqref{eq:BounceTimeEF}, we expressed the bounce time in terms of the null coordinates at the collapsing and anti-collapsing shells. As explained in \cite{haggard_black_2014}, the bounce time has a clear operational meaning in terms of the proper time $\tau_R$ along the wordline of a stationary observer, that is, an observer at a constant radius $R$, between the events at which he intersects with the collapsing and anti-collapsing shells $S^\pm$. A straightforward calculation gives $\tau_R$ as 
\begin{equation}
\tau_R= \sqrt{f(R)} ( u_0-v_0 + 2 r^*(R) )
\end{equation}
where $f(R)=1-2m/R$. To get this expression, we have to add the contributions from the two line elements \eqref{eq:downMetricNew} and \eqref{eq:upMetricNew} and use the junction condition \eqref{eq:junctT}. Inverting this equation and using \eqref{eq:BounceTimeEF}, we have for the bounce time
\begin{equation}
T = \frac{\tau_R}{\sqrt{f(R)}}-2 r^*(R) 
\end{equation}
Thus, the bounce time may be directly measured by an observer equipped with a clock, provided he has measured the mass $m$ of the black hole and his distance from the black hole. 

Let us rephrase the above equation to emphasize the role of $T$ as a spacetime \emph{parameter}, a coordinate and observer independent quantity, and its relation with the symmetries of the spacetime. 

The exterior spacetime described by the HR-metric has the three Killing fields of a static spherically symmetric spacetime, a timelike Killing field generating time translation and two spacelike Killing fields that together generate spheres. The orbits $\gamma$ of the timelike Killing field are labelled by an area $A_\gamma$ : the area of a sphere generated by the two spacelike Killing fields on any point on $\gamma$. This is of course the geometrical meaning of the coordinate $r$. 

We can thus avoid mention of any coordinates or observers and define $T$ through the following geometrical construction. Consider any orbit $\gamma$ that does not intersect with the interior boundary. The proper time $\tau_\gamma$ is an invariant integral evaluated on the portion of $\gamma$ that lies between its intersections with the null hypersurfaces $S^\pm$. For any such $\gamma$, we have

\begin{equation} \label{eq:defBounceTime}
T = \frac{\tau_\gamma}{\sqrt{f(A_\gamma)}}-r^*(A_\gamma) 
\end{equation}   
$T$ is independent of the chosen orbit $\gamma$ and is expressed only in terms of geometrical quantities, an area and a proper time. This expression can be taken to be the \emph{definition} of $T$. 

 \subsection{The Bounce time as the evaporation time} \label{sec:bounceTimeAsEvapTime}
The HR-spacetime neglects Hawking radiation. This assumption should be justified \`a posteriori, by comparing the lifetime, defined in the next section as the expected value of $T$ from quantum theory for a given mass, with the evaporation time scale $\sim m^3$. The degree to which it is justified to compare the bounce time with the Hawking evaporation time has been a subject of discussion. In these paragraphs, we clear up this question and show that the bounce time $T$ is an equivalent concept to the Hawking evaporation time for the HR-spacetime. Thus, it can be directly compared to the evaporation time scale $\sim m^3$. 

Often, an evaporation time is defined as the interval of the affine parameter $u$ at future null infinity $\cal{J}^+$ between the ``first'' and ``last'' Hawking photon, with an ambiguity typically of order $m$ in these times. The analogous to the ``last'' Hawking photon in our setting is the anti-collapsing null shell $S^+$. The first Hawking photon will be emitted when the collapsing shell $S^-$ reaches some radius which we call $r_{fhp}$ (first Hawking photon). Let us call $u_{fhp}$ the retarded time at $\cal{J}^+$ corresponding to the sphere with ingoing coordinates $(r_{fhp},v_0)$. When we take $r_{fhp} = 2m(1+W(1/e))\approx 2.56 m $, where $W$ is the Lambert function, then the evaporation time of the HR-spacetime equals the bounce time

\begin{equation}
T = u_0 - u_{fhp}
\end{equation}

\begin{equation}
r_{fhp} \approx 2.56 m
\end{equation}

The value $r_{fhp} \approx 2.56 m$ is a very reasonable value for the emission of the first Hawking photon; as mentioned above, typically an ambiguity of order $m$ is involved in such calculations. Comparing with literature, in \cite{bianchi_entanglement_2015} the authors studied the production of entanglement entropy from the emission of Hawking radiation and defined $u_{fhp}$ to be when the entanglement entropy starts to significantly depart from zero. They estimated the radius at which it is most likely for the first Hawking photon to be emitted to be roughly when the collapsing shell reaches a radius $\sim 3m$.

We advertise that we will be seeing again the value $r_{fhp} = 2m(1+W(1/e)m$ in Section \ref{sec:crossedFingers}, which has a special meaning, when we discuss the ``crossed fingers'' representation of the HR-spacetime on the Kruskal manifold.

\section{Lifetime $\tau(m)$} \label{sec:Lifetime}
In this manuscript we consider the quantum transition of a trapped region to an anti-trapped region as described in the framework of covariant LQG. In this section we propose a probabilistic interpretation to the transition amplitudes ascribed by the EPRL model to this process and define an observable, the value of which a quantum theory of gravity should be in a position to predict.

Credit to the ideas presented in this section goes mainly to Robert Oeckl, who explained in an email conversation how the positive formalism \cite{Oeckl:2016tlj,Oeckl:2012ni,Oeckl:2005bv} may be implemented in a quantum bounce of a black hole to a white hole. We do not go here into the details of this framework, the precise application of which to this phenomenon is currently a work in progress, see also \cite{OecklPres} for a recent discussion. 

The observable defined below first appeared in \cite{christodoulou_realistic_2016}. We will revisit its physical interpretation in Section \ref{sec:LifetimeIsCrossing}

\subsection{Transition Amplitudes as Conditional Probability Distributions} \label{sec:ProbInterpretation}

The precise definition of the boundary states we will use is given in Chapter \ref{ch:gravTunneling}. In this section, we neglect unnecessary details in order to keep the notation light. 

Consider a family of coherent states $\Psi(\omega,\zeta)$, providing an over-complete basis of the Hilbert space of interest, with the resolution of the identity given by a measure $\alpha(\omega,\zeta)$. The \emph{transition} amplitude $W(\omega,\zeta) = \braket{W|\Psi(\omega,\zeta)}$ is then a function of these data. We make a distinction between the physical transition amplitude and the \emph{spinfoam} amplitude $\bra{W}$, that will only know of the combinatorial structure of the 2-complex on which it is defined.

The labels $(\omega, \zeta)$ have the geometrical meaning of areas and 4D hyperobolic angles respectively and will become functions $\omega(m, T), \zeta(m, T)$ of the spacetime parameters when identified with the discrete distributional geometry that approximates the geometry of the interior boundary, see \cite{christodoulou_realistic_2016} and Chapter \ref{ch:calculationOfAnObservable}. We may then exchange the labels $\omega$ and $\zeta$ with $m$ and $T$ and write $\Psi(m, T)$ for the state and $W(m,T) \equiv \braket{W|\Psi(m,T)}$ for the transition amplitude.

We \emph{define} a probability distribution by 
\begin{equation} \label{eq:defCondProbDistr}
P(T|m) = \frac{A(m, T)}{A_{T}(m)} 
\end{equation}
where
\begin{equation} \label{eq:defMarg}
A_{T}(m) = \int \dd T \;\alpha(m,T) \,|W(m,T)|^2
\end{equation}
and 
\begin{equation}\label{eq:probdens}
A(m, T) = \alpha(m, T)\, \vert W(m, T) \vert^2
\end{equation}

The function $\alpha$ used in the definition of the probability density \eqref{eq:probdens}, must be derived from the measure providing the resolution of the identity of the class for boundary states considered here and satisfies
\begin{equation}
\int \dd m\, \dd T \; \alpha(m, T) \ket{\Psi(m, T)} \bra{\Psi(m, T)} = \ident.
\end{equation}
 There are subtleties in defining $\alpha$, because in general the states $\Psi(\omega,\zeta)$ are labelled by more than two parameters $\omega,\zeta$. We then have to find the induced measure on a (proper) two-dimensional subspace of the larger Hilbert space. We ignore this difficulty here and assume that such a measure can be found. Equivalently, the discussion here may be understood as being at the level of a state defined on a single link, when $\alpha(m, T)$ is simply $\alpha(\omega, \zeta)$ times the Jacobian of the parameter change  $\omega, \zeta \rightarrow m,T$. 

The conditional probability $P(T|m)$ is automatically normalized by these definitions. We attach to it the following interpretation : it is \emph{the conditional probability distribution for measuring $T$ for a given fixed mass $m$}.

Note the similarity of equations \eqref{eq:defCondProbDistr}, \eqref{eq:defMarg} and \eqref{eq:probdens} with the usual definition of conditional probability distributions, in which case $A(m,T)$ and $A_{T}(m)$ are respectively the joint (in $T$ and $m$) and marginal (in $m$) distributions. The difference here is that we do not require $A(m,p)$ and $A_{m}(p)$ to be normalized and so they are not themselves distributions. 

\medskip

$P(T|m)$ \emph{is a prediction of the quantum theory.}

\medskip

This definition of $P(T|m)$ overcomes an important technical difficulty in \cite{christodoulou_realistic_2016}, were we  directly interpreted the amplitude $\vert W(m, T) \vert^2$ as proportional to the physical joint probability distribution in $m$ and $T$. This would imply that the spinfoam amplitude must be normalized\footnote{To see this, formally write $|W(m, T)|^2 = \braket{W|\Psi(m, T)}\braket{\Psi(m, T)|W}$ and keep in mind that $\ket{W}$ knows nothing about $m$ and $T$, only about the combinatorial structure of the 2-complex on which it is defined.}
, i.e.\! that
$\braket{W|W}=1$.

Spinfoam amplitudes are defined up to an arbitrary normalization. The probability distribution $P(T|m)$ as defined in \eqref{eq:defCondProbDistr} does not depend on the normalization of the spinfoam amplitude. Furthermore, a conditional probability seems to be in a relativistic or relational setting the natural object to work with. We comment further on this point at the end of the next section.

\subsection{Lifetime $\tau(m)$ of the HR-Spacetime}  \label{sec:lifetime}

 The probability distribution \eqref{eq:defCondProbDistr} is a physical prediction of the theory. We define the \emph{lifetime} of the HR-spacetime to be  the expectation value of $T$ on $P(T|m)$
 
\begin{equation} \label{eq:lifetime}
\tau(m) \equiv\braket{T}_{P(T|m)} = \int \dd T\; T P(T|m).
\end{equation}

The point of view taken in  \cite{christodoulou_realistic_2016}, is that the mass $m$ and the bounce time $T$ serve the role of partial observables in the sense of \cite{rovelli_partial_2002}. Quoting from this work,  

\begin{quote}
\say{A partial observable is a physical quantity to which we can associate a (measuring) procedure leading to a number.} 
\end{quote}

 The conditional probability distribution $P(T\vert m)$, or, more precisely, the value of $T$ for a given $m$, is then a complete observable of the HR-spacetimes. Quoting again from \cite{rovelli_partial_2002}, 

\begin{quote} 
\say{A complete observable is a quantity whose value can be predicted by the theory (in classical theory); or whose probability distribution can be predicted by the theory (quantum theory).}
\end{quote}

\bigskip

\emph{In summary, the HR-spacetime provides a prototypical setup for geometry transition in the covariant framework. The geometry of the spacetime depends on two classical physical scales, which become encoded in the geometry of the interior boundary -- the boundary condition for the path-integral -- and in turn quantum theory correlates the two scales in a probabilistic manner.}

\bigskip

\begin{center}
\rule{0.5 \textwidth}{0.05cm} 
\end{center}

Before closing this section, we discuss how in this scheme and for this physical application, the set of issues known as the ``problem of observables'' and the ``problem of time'', are sidestepped.

 The problem of observables in gravity, quantum or classical, is a set of central conceptual and technical issues, stemming from  general relativity. They have been much discussed the past century and are still a subject of active debate  
 \cite{bergmann_observables_1961,dittrich_perturbative_2007,dittrich_chaos_2015,
 perez_observables_2001,rovelli_partial_2002,dittrich_partial_2006-2}
. The essential issue is that by observables, in the classical sense and strictly speaking, we mean functionals of the metric that are constant on the equivalence classes that define the same solution of the field equations. That is, on the metrics that are related by diffeomorhisms. In the phase space sense, this problem comes down to the fact that it is not known how to explicitly describe the physical phase space, that is, the space of gauge orbits, which are now assumed to be projected to points. The mechanism for carrying out the projection for a general solution has not been explicated. 
 
 In the case where the spacetime is asymptotically flat, we do have at our disposal a few of these ``true'' observables, the ADM charges. In the simple spherically symmetric case, the only such charge is the mass $m$. In coordinate language, the components of $g_{\mu\nu}$ or the coordinates themselves, will depend on this same parameter $m$ in any coordinate system, which serves as a label of the equivalence class. 

This provides one natural candidate for a good ``partial'' observable. The quotes are because a partial observable need not be a diffeomorphism invariant quantity such as $m$. The mass $m$ has this extra good property.

The second partial observable, is provided by construction, the bounce time $T$ which arises naturally in the HR-spacetime. $T$ is also a diffeomorphism invariant quantity.

Both $T$ and $m$ serve as both ``partial'' and as ``true'' observables in our case. This need not be the case since the fact that they are diffeomorphism invariant quantities is not required. It is however the reason that the complete observable is a manifestly cordinate/observer independent quantity.

 A generalization of the interpretation given here, applicable to more general spacetimes, could perhaps be achieved by working with partial observables that have an operational meaning, but refer to specific observers. This is a project for the future, when, for instance, attempting to take into account realistic matter dissipation.

Finally, the set of issues known as ``problem of time'' arise from the fact that in a Hamiltonian formulation of General Relativity, the Hamiltonian vanishes and one has to carefully define the meaning of ``dynamical evolution''. Taking a relational point of view, that true dynamical evolution is evolution of partial observables with respect to one another, provides a way to at least sidestep these issues, in concrete applications.

\section{Crossed fingers: mapping the HR-metric on the Kruskal manifold} \label{sec:crossedFingers}
In this section we briefly explain how the HR-metric can be represented on the Kruskal manifold and the relation of the original ``crossed-fingers'' construction.

In Section \ref{sec:RH}, we gave the HR-metric using two different patches from the Kruskal manifold. This is necessary, there is no one-to-one mapping from the HR-spacetime to (a region of) the Kruskal manifold. The easiest way to see this is through a figure. 

\begin{figure} 
\centering
\includegraphics[scale=0.7]{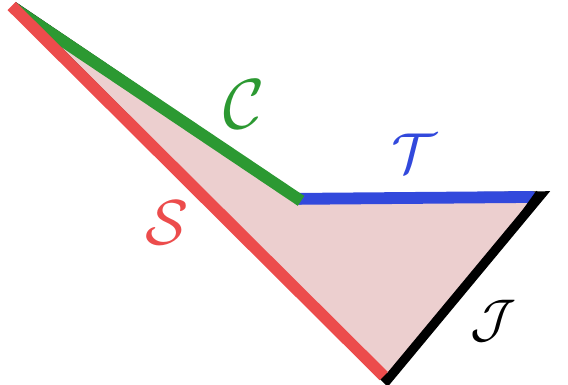}
\caption[One Kruskal patch of the Haggard-Rovelli spacetime]{One Kruskal patch of the Haggard-Rovelli spacetime. $\mathcal{S}$ is the outgoing ($+$) or ingoing ($-$) null shell $\mathcal{S^\pm}$. Respectively, $\mathcal{C}$ is the boundary surface $\mathcal{C^\pm}$ in the curved part of the HR-spacetime, and {$\cal{J}$} is the future and past asymptotic null infinity {$\cal{J^\pm}$}. $\mathcal{T}$ is the common spacelike hypersurface on which the two patches are identified, yielding the two junction conditions : the identification of the radial coordinate $r$ and the junction condition \eqref{eq:junctT}. See also Figure \ref{fig:crossedFingers}.   }

\label{fig:fireworksPatch}
\end{figure}

One Kruskal patch of the HR-spacetime is depicted in  Figure \ref{fig:fireworksPatch}. In Figure \ref{fig:ansatz}, a patch corresponds to the part of the HR-spacetime above or below the fiducial surface $\mathcal{T}$, described by the line element in equation \eqref{eq:downMetricNew} or \eqref{eq:upMetricNew}.

\begin{figure} 
\centering
\includegraphics[scale=0.6]{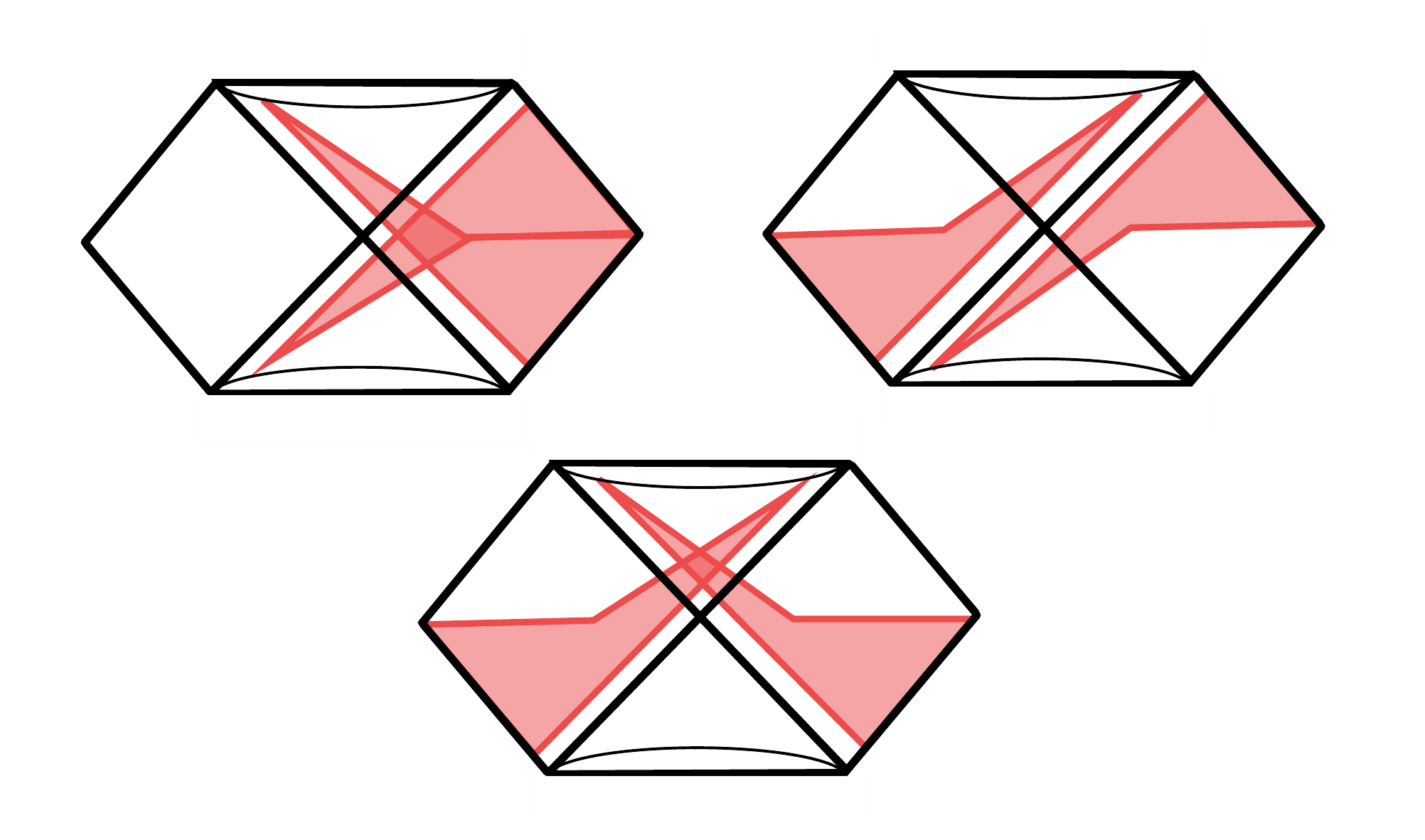}
\caption[Mappings of the HR-spacetime to the Kruskal manifold.]{Some of the possible mappings of the two Kruskal patches of the HR-spacetime to the Kruskal manifold. See Figure \ref{fig:fireworksPatch} for a breakdown of a single patch. It is impossible to map the HR-spacetime to the Kruskal manifold using a single patch : the patches either overlap or are disjoint. Thus, we need to use two distinct patches. The upper-left case is the ``crossed-fingers'' construction that was used in the original presentation (cite Fireworks). }
\label{fig:crossedFingers}
\end{figure}

Some possible ways to map the two patches of the HR-spacetime in the Kruskal manifold are depicted in Figure \ref{fig:crossedFingers}. It takes a moment of reflection to be convinced that any such mapping will either have overlapping regions, or be disjoint. Thus, at least two patches of the Kruskal manifold are necessary to describe the HR-spacetime. 

The original construction in \cite{haggard_black_2014} corresponds to the top left ``crossed fingers'' mapping. Let us now take the point of view that this mapping is a representation of the HR-spacetime on the Kruskal manifold. 

The Kruskal manifold is a one--parameter family of spacetimes, the parameter being the mass $m$. The HR-spacetime is a two-parameter family of spacetimes, in the sense in which we defined it here, that is, with an unspecified interior boundary satisfying the constraints of Section \ref{sec:RH}.

Where is the bounce time $T$ encoded in the ``crossed--fingers'' representation? It is encoded in the radius where the two null shells meet. We call this radius $r_\delta$ and the point (sphere) of their intersection $\delta$.

The easiest way to see this is to use the definition of the bounce time when the HR-metric is written as in the line elements in equations \eqref{eq:downMetricEF} and \eqref{eq:upMetricEF}.

The null shells are at $u_0$ and $v_0$. The bounce time is given by equation \eqref{eq:BounceTimeEF}, repeated here for convenience
\begin{equation}
T=u_0-v_0
\end{equation}
The point where the two shells intersect in the crossed--fingers representation has some Schwarzschild coordinates $(t_\delta,r_\delta)$. In ingoing Eddington--Finkelstein coordinates (in the lower patch, line element in equation \eqref{eq:downMetricEF}) its coordinates are $(t_\delta,v_0)$. In the upper patch, in outgoing EF--coordinates, its coordinates are $(t_\delta,u_0)$. The EF--coordinates are related to the Schwartzschild coordinates by
\begin{eqnarray}
v_0=t_\delta + r^\star(r_\delta) \nonumber \\
u_0=t_\delta - r^\star(r_\delta)
\end{eqnarray}
Thus,
\begin{equation}
T=-2 r^\star(r_\delta)
\end{equation}
We conclude that in the crossed--fingers representation, it is equivalent to consider the radius $r_\delta$ as the second spacetime parameter for the HR-metric.

By abuse of notation, as we did with the parameter $\Delta$ and the sphere $\Delta$, we introduce the parameter $\delta > 0$ for the sphere $\delta$ at radius $r_\delta$ by 
\begin{equation}
r_\delta = 2m (1+\delta)
\end{equation}  
The bounce time $T$ and $\delta$ are related by  
\begin{equation}
e^{-\frac{T}{4m}} = \delta e^{\delta+1} 
\end{equation}
where we used $r^\star(r)= r+2m \log \vert \frac{2m}{r}-1 \vert$.
In terms of the Lambert function, the relation is inverted to
\begin{equation}
\delta =W\left( \frac{e^{-\frac{T}{4m}}}{e}  \right) \
\end{equation}
The bounce time $T$, as discussed in Section \ref{sec:RH}, is constrained to be positive, as is the mass $m$. This translates to the condition 
\begin{equation}
\delta > W\!\left( \frac{1}{e}  \right) \approx 0.28
\end{equation}
The radius $r_\delta$ at this value, corresponding to $T=0$, is the same value we got in Section \ref{sec:bounceTimeAsEvapTime} for the first hawking photon radius along the null shell, by interpreting $T$ as an evaporation time. 

Lastly, an infinite bounce time corresponds to a vanishing $\delta$. Thus, in the crossed--fingers representation of the HR-spacetime, we may use as parameters the mass $m$, constrained to be positive, and the parameter $\delta$, constrained to be in the interval

\begin{equation} \label{eq:deltaIntervalAllowedValues}
\delta \in (0, W\!\left( \frac{1}{e}  \right))
\end{equation}

\section{A few more properties}
In this chapter we have presented the HR-spacetime as a well-defined setup for studying geometry transition. Before closing this chapter, we discuss a few further properties that are of interest for the following chapter. 

We discussed the definition of Lorentzian (boost) angles in Section \ref{sec:lorReggeAction}. They are defined as inverse hyberbolic cosines of inner products between normalized timelike vectors at a point. A straightforward calculation that involves a bit of algebra shows that all boost angles in the HR-spacetime scale monotonically with the bounce time $T$ and the combination $T/m$.

They \emph{increase} in the exterior of the trapped and anti-trapped regions and they \emph{decrease} in their interior, as we increase $T$ or $T/m$. This intriguing property can be useful for future attempts at encoding the presence of the trapped surface in the boundary state.

Another relevant feature is that for a large class of interior boundary choices, the bounce time is \emph{only encoded in the area of the sphere $\Delta$}. Such boundary choices arise whenever the two hypersurfaces are drawn from foliations in the ``crossed finger'' representation that are generated by time translation. For instance, in Eddington Finkelstein coordinates, whenever we use as interior boundary two surfaces from the families
\begin{eqnarray}
v-C^-(r)=c^- \nonumber \\
u-C^+(r)=c^+
\end{eqnarray}
with $C^\pm$ having no dependence on the spacetime parameters. 

It is not hard to see that when attempting to do such a choice, $c^\pm$ cannot be constants, they have to depend on $T$ because their intersection defines $\delta$. 

In turn, because the radius of the spheres $\epsilon^\pm$ is taken fixed, the bounce time is only encoded in the intrinsic and extrinsic metric via $T/m$ through the parameter $\Delta$ (which is dimensionless). This effect was noted accidentally in \cite{christodoulou_realistic_2016} where such a boundary choice was made. 

Finally, we note that at no point did we mention the symmetry or asymmetry of the space--time. The notions of time--symmetry and time--asymmetry in \cite{haggard_black_2014,
de_lorenzo_improved_2016-1}, refer to properties of the boundary. That is, the hypersurface $C^-$ may be chosen such that the light-tracing of its endpoints $\epsilon^-$ and $\Delta$ to $\cal{J^-}$ corresponds to an affine interval $\delta v$ and similarly for $C^+$ for an interval $\delta u$. In the terminology of \cite{de_lorenzo_improved_2016-1}, choosing a boundary such that $\delta u \neq \delta v$ corresponds to an asymmetric spacetime. This gives a notion of the duration of the white and black hole phase at asymptotic infinity.

\chapter{Calculation of an observable} 
\label{ch:calculationOfAnObservable}
In this chapter, we estimate the scaling of the lifetime $\tau(m)$ with the mass $m$. 

We start by commenting on the discretization procedure. We prove a simple relation between the extrinsic curvature of a surface and the boost relating the normal to its parallel transport along a curve in that surface. 

We proceed to make an explicit boundary hypersurface choice, carry out a discretization and derive an explicit expression for a transition amplitude describing this process. The resulting formula is cumbersome and its numerical analysis a daunting task. Currently, much effort is put in the Centre de Physique Th\'eorique in Marseille, by a group led by S.Speziale for the development of numerical methods suitable for such amplitudes. 

We then proceed to use the semiclassical approximation developed in Chapter \ref{ch:gravTunneling}, and estimate the scaling of the lifetime with the mass of the collapsed object forming the trapped region. We do this first for the explicit setup of the previous section and then for a class of amplitudes and for an arbitrary choice of boundary conditions. 

We close by discussing the physical interpretation of our results and the difficulties that arise.  

\section{Discretization}
\label{sec:Discretization}
In this section we discuss the discretization of a continuous surface to a piecewise flat simplicial manifold. We see how the resulting distributional geometry is interpreted in terms of holonomy-flux data. We start by proving a simple relation between the extrinsic curvature and the boost relating the normal of a hypersurface to its parallel transport along a curve. This result serves as a justification for the discretization procedure and could be interesting in its own right.

\subsection{What is the relation between boosts, the extrinsic curvature and parallel transport? } 
\label{sec:zetaIsK}

We will prove the following statement, true for both Euclidean and Lorentzian metrics. Consider a curve $\Upsilon$ in a spacelike hypersuface $\Sigma$ and the normal field $n$ to $\Sigma$: 

{\quote \emph{Given two points $s$ and $t$ on $\Upsilon$, the angle between the normal vector $n_t$ at $t$ and its parallel transport $Pn_s$ to $t$, is the invariant line integral of the extrinsic curvature $K$ along $\Upsilon$}.} \\

By angle for the Lorentzian case, we mean the rapidity of the boost taking $Pn_s$ to $n_t$, see below and Section \ref{sec:lorReggeAction}. The proof works for any dimension and in Lorentzian signature. 

For definiteness, we restrict to the case relevant to us, four dimensions and a spacelike three dimensional hypersurface. We keep both signatures for easier comparison of Euclidean and Lorentzian angles. This comes down to keeping the normalizeation $\epsilon \equiv n^\alpha n_\alpha$ of the normal, which is $\epsilon=-1$ for the Lorentzian case and $\epsilon=1$ for the Euclidean case.

Consider a manifold $M$, with a coordinate choice $x^\alpha$ and a hypersurface $\Sigma$ with coordinates $y^a$ (lower-case latin letters $a$ take three values). 

A basis of tangent vectors on $\Sigma$ is provided by\footnote{The basis of tangent vectors $\chi^\alpha_a$ on $\Sigma$ is often denoted $e^\alpha_a$ because it can be also understood as a choice of triad. This is a bad notation which we avoid and stress that $\chi^\alpha_a$ is a coordinate based choice of basis of $T\Sigma$. Furthermore, note that $\chi^\alpha_a$ are rectangular, non-invertible matrices. }

\begin{equation}
\chi^\alpha_a \equiv \frac{\partial{x}^\alpha}{\partial{y}^a}
\end{equation}

The curve $\Upsilon$ is given parametrically by a map $\lambda \rightarrow x^\alpha(\lambda)$, with $x^\alpha(\lambda) \in \Sigma$. 

From the line element
\begin{equation}
\dd s^2 \equiv g_{\alpha\beta} \dd x^\alpha \dd x^\beta = g_{\alpha\beta}\chi^\alpha_a \chi^\beta_b dy^a dy^b 
\end{equation}
we read the induced metric 
\begin{equation}
h_{ab} = \chi^\alpha_a \chi^\beta_b g_{\alpha \beta}
\end{equation}
with the completeness relation for the inverse metric 
\begin{equation}
g^{\alpha \beta} = \epsilon \, n^\alpha n^\beta+ h^{ab} \, \chi^\alpha_a \, \chi^\beta_b 
\end{equation}
easily checked and memorised by the equality of dimensions and co-dimensions: $g^{\alpha \beta}g_{\alpha \beta}=4$, $h^{ab}h_{ab}=3$, $n^\alpha n_\alpha=\epsilon^2=1$.

The extrinsic curvature of $\Sigma$ is given by 
\begin{equation}
K_{ab}= \chi^\alpha_a \chi^\beta_b \nabla_\alpha n_\beta
\end{equation}

We may choose the coordinate system $y^a$ on $\Sigma$ to be such that, in a neighbourhood of $\Upsilon$, $\lambda$ along  $\Upsilon$ is one of the coordinates. The basis vector $\chi^\alpha_\lambda$ then has an intrinsic meaning on $\Upsilon$, it is its tangent vector $V^\alpha = \chi^\alpha_\lambda \equiv \ \frac{\dd x^\alpha(\lambda)}{\dd \lambda}$. We define the \emph{extrinsic curvature $K(\lambda)$ along $\Upsilon$} by
\begin{equation} \label{eq:meanKonUpsilon}
K(\lambda) = \frac{ V^\alpha V^\beta \nabla_\alpha n_\beta  }{ V^\rho V_\rho  }
\end{equation}
This is a spacetime scalar. For convenience, we henceforth take $\lambda$ to be an affine parameter. Consider two points $s$ and $t$ on $\Upsilon$. We saw in Section \ref{sec:holonAsParTransport} that the parallel transport equation along $\Upsilon$ for any vector at $s$ is solved by the image of the holonomy $P$. Thus, the parallel transport $(P n_s) ^\alpha (\lambda)$ of the normal at $s$ up to the point labelled by $\lambda$, satisfies 
\begin{equation} \label{eq:parTranspV}
V^\beta \nabla_\beta (P n_s) ^\alpha = 0
\end{equation}

We are interested in the change of the inner product between the parallel transported $(P n_s)$ and the normal $n$ at some point $\lambda$, with respect to $\lambda$. We have
\begin{equation} \label{eq:tempParTran}
\frac{\dd }{\dd \lambda} (P n_s)^\alpha n_\alpha = V^\beta (P n_s)^\alpha \nabla_\beta  n_\alpha 
\end{equation}
where we used that $(P n_s)^\alpha n_\alpha$ is a scalar  and the parallel transport equation \eqref{eq:parTranspV}.

Now, the vector $(P n_s)^\alpha$ written as a linear combination of $n^\alpha$ and $V^\alpha$ reads
\begin{equation}  \label{eq:temp2ParTran}
(P n_s)^\alpha = \epsilon (P n_s)^\rho n_\rho \; n^\alpha + \frac{(P n_s)^\mu V_\mu}{V^\nu V_\nu } \; V^\alpha  
\end{equation}
This can be checked by contracting with $n_\alpha$ and $V_\alpha$. Replacing $(P n_s)^\alpha$ in equation \eqref{eq:tempParTran}, the first term vanishes because $n^\alpha$ is normalized, i.e.\! $n^\alpha \nabla_\beta n_\alpha = \frac12 \nabla_\beta n^\alpha n_\alpha = 0$. In the second term, we recognise $K(\lambda)$ from equation \eqref{eq:meanKonUpsilon}. Thus, we have  
\begin{equation}
\frac{\dd }{\dd \lambda} (P n_s)^\alpha n_\alpha = (P n_s)^\mu V_\mu \; K(\lambda)
\end{equation}
We are almost done. The inner product with the metric is invariant under parallel transport with the Levi-Civita connection, and thus $(P n_s)^\alpha (P n_s)_\alpha=\epsilon$. Contracting equation \eqref{eq:temp2ParTran} with $(P n_s)_\alpha$ we have 
\begin{equation}
\epsilon = \epsilon \bigg( (P n_s)^\alpha n_\alpha \bigg)^2 + \bigg( \frac{ (P n_s)^\mu V_\mu}{V^\rho V_\rho} \bigg)^2  
\end{equation}
To ease notation, let us call 

\begin{equation}
\epsilon \, I(\lambda) \equiv (P n_s)^\alpha n_\alpha
\end{equation}
 The minus sign for the Lorentzian case is because $(P n_s)^\alpha$ and $n^\alpha$ are in the same light-cone, thus their contraction with the metric is negative and then $I(\lambda)$ is positive in both cases.\footnote{The parallel transport cannot take $(P n_s)^\alpha$ outside the light cone. This follows from the preservation of the inner product under parallel transport: by continuity, $(P n_s)^\alpha$, would have to turn null and then spacelike, but, $(P n_s)^\alpha(P n_s)_\alpha=-1$.}
 We want to then solve the differential equation 
\begin{equation} \label{eq:tempDiff}
\frac{\dd  }{\dd \lambda} \epsilon \,I(\lambda) = \bigg(\epsilon-\epsilon \,I(\lambda)^2 ) \bigg)^{1/2} \sqrt{V_A V^A} \; K(\lambda)
\end{equation}
The angles are defined as in Section \ref{sec:lorReggeAction}. The Lorentzian angle $\zeta(\lambda)$ is the rapidity of the $SO^+_\uparrow (1,1)$ boost (proper hyperbolic rotation) taking $(P n_s)^\alpha$ to $n^\alpha$, in the plane defined by $(P n_s)^\alpha$ to $n^\alpha$. The Euclidean angle $\phi(\lambda)$ is similarly the angle of the $SO(2)$ rotation taking $(P n_s)^\alpha$ to $n^\alpha$. Remembering that $(P n_s)^\alpha$ and $(P n_s)^\alpha$ are normalized, these are given by
\begin{eqnarray}
I(\lambda) &=& \cosh \zeta(\lambda) \ \,\ , \; \ \epsilon=-1 \nonumber \\
I(\lambda) &=& \cos \phi(\lambda) \ \ ,\; \ \epsilon=1  
\end{eqnarray}
Let us do the Lorentzian case, the Euclidean will then be evident. Replacing in  \eqref{eq:tempDiff}, we have
\begin{equation} 
- \frac{\dd  }{\dd \lambda} \cosh \zeta = \bigg(\cosh \zeta^2 - 1 \bigg)^{1/2} \sqrt{V_\alpha V^\alpha} \; K(\lambda)
\end{equation}
Note that $\cosh^2 \zeta$ is always larger than unit. Thus,
\begin{equation} 
- \frac{\dd  }{\dd \lambda} \zeta(\lambda) = \sqrt{V_\alpha V^\alpha} \; K(\lambda)
\end{equation}
where we have used that $\cosh^2 \zeta - \sinh^2 \zeta=1$ and $\frac{\partial \cosh \zeta}{\partial \zeta}= \sinh \zeta$. We can now integrate for $\zeta(\lambda)$. The parallel transport of the normal at the starting point $s$ is by definition itself, we have $I(\lambda)=\cosh \zeta = 1$, thus, $\zeta(\lambda_s)=0$. Then, the Lorenzian angle $\zeta(\lambda)$ at a point $\lambda$ on $\Upsilon$, between $n_s$ and its parallel transport $Pn_s(\lambda)$ reads
\begin{equation}
\zeta(\lambda_t) = -\int_{\lambda_s}^{\lambda} \dd \lambda \sqrt{g_{\alpha\beta} \dot{x}^\alpha \dot{x}^\beta} \; K(\lambda)
\end{equation}
where we reintroduced the notation $V^\alpha = \dot{x}^\alpha$, to recognise the invariant line integral. The minus sign is a matter of convention, and we may redefine $K_{ab} \rightarrow -K_{ab}$ to write, more compactly

\begin{equation}
\boxed{
\zeta = \int_{\Upsilon} \; K
}
\end{equation}

\bigskip

\subsection{Discretization scheme, discretization ambiguities and boundary data}
\label{sec:discretizationScheme} 

\subsubsection{Coarse triangulations can only capture a few qualitative degrees of freedom}
In a physical problem treated as a geometry transition, the path integral is conditioned to a quantum boundary geometry. The path to the quantum boundary geometry is to start from a continuous surface, discretize it on a triangulation, and encode the discrete distributional geometry in a coherent spin-network state of Section \ref{sec:boundaryState}.

In the following section, we exploit the spherical symmetry of the spacetime and the specific choice of hypersurfaces, that will be intrinsically flat, to carry out the discretization in a simple manner. This will result in the path--integral being peaked on a degenarate configuration (4D simplicial geometry of zero four--volume). This will allow us to investigate the interplay between the possible geometrical configurations: by slightly modifying the boundary data, the path-integral will be dominated by a 4D Lorentzian or Euclidean simplicial geometry instead.

In turn, we will proceed to discuss a general expression independent of the choice of hypersurface.

\medskip

 In this section we briefly comment on the arbitrariness of carrying out such discretizations. We have in mind ``rough'' discretizations, suitable for spinfoam amplitudes built on coarse 2-complexes, where only a few degrees of freedom of the gravitational field may be captured. The discussion here concerns the passage from a continuous surface to a triangulation and is strictly classical.

In precise discrete evolution schemes, such as the Sorkin evolution scheme for Regge calculus \cite{sorkin_time-evolution_1975}, we have to discretize in a consistent way, such that there is a sense in which the discrete Einstein's equations are approximately solved to a better approximation as the triangulation is refined. 

There is little utility in using such schemes for spinfoam amplitudes defined on coarse 2-complexes. Unless a refinement procedure is to be performed subsequently, the precise way in which the discretization is carried out will be largely amibiguous in any scheme.

However, spinfoam amplitudes defined on coarse 2-complexes \emph{are} expected to be physically relevant, in particular for treating problems where few-degrees of freedom are sufficient to capture some of the relevant physics.

This is assumed to be the case in our setting, where we are only looking to get a scaling, relating two characteristic physical scales, and there is a high degree of (spherical)symmetry. This is also typically assumed to be the case for treating the cosmological singularity, in spinfoams, and in general.

We note that the above observations hold regardless of whether the amplitudes defined on the 2-complex are to be understood as describing fundamental degrees of freedom or coarse-grained degrees of freedom.

At the level of coarse triangulations of a continuous surface, \emph{any} discretization scheme would fail to capture details of the geometry. Thus, \emph{the result of the calculation should not depend on the details of the discretization scheme}. We present evidence for this in the interplay between the different kinds of geometric critical points when we estimate the lifetime in Section \ref{sec:lifetimeEstim}.

To summarize, for say an area data $A$, it is not the precise \emph{numerical value}, to be four or eight meters squared, that is relevant. It will be the \emph{dependence} of the data as functions of the relevant physical scales that is of importance, in our case, on the mass $m$ and the bounce time $T$. Numerical coefficients reflecting the arbitrariness of the discretization scheme \emph{should not} play a role, if the calculation is to be consistent.

\subsection*{A simple discretization scheme}

We are initially given a continuous spacelike boundary $B$, possibly disconnected, composed of a number of spacelike hypersurfaces. The exterior spacetime metric $g_{\mu\nu}$ induces an intrinsic metric $h_{ab}$ and an extrinsic curvature tensor $K_{ab}$ on $B$. 

 The geometry of the boundary $B$ will be approximated by a discretization. The discretized geometry will be distributional and defined on a piece-wise flat 3D simplicial manifold $\Gamma^*$. That is, $\Gamma^*$ will be a spacelike tetrahedral triangulation of $B$.

Information on the extrinsic and intrinsic curvature will be concentrated on the interfaces between the flat pieces, the triangles $\ell^\star$ shared by two tetrahedra $\nn^\star$ and $\nn'^\star$, the source and target of the link. To avoid confusion, we stress that the 3D simplicial complex $\Gamma^*$ will \emph{not} in turn be embedded in a simplicial triangulation of spacetime. The definition of the embedding only requires a local extension, see below.

The dual graph $\Gamma$ (which will be four-valent) has links $\ell$ and nodes $\nn$ dual to triangles and tetrahedra in $\Gamma^*$. The boundary state is built on this graph. The discrete geometry of $\Gamma^*$ will be fully specified by giving a set of data $A_\ell =\hbar \omega_\ell$, $\zeta_\ell$, and $\vec{k}_{\nn\ell}$, corresponding (up to the gauge phase which we ignore here) to the twisted-geometry parametrization of the coherent spin-network states, see Section \ref{sec:boundaryState}. 

\bigskip

  Below we sketch the steps for carrying out explicitly such a discretization. The first step is to choose a \emph{topological} triangulation $\Gamma^\star$. At this point, $\Gamma^\star$ is a topological 3D simplicial-complex and we have yet to ascribe any geometry to it.
 
 We then \emph{embed} $\Gamma^\star$ inside $B$. We now have triangular 2-surfaces $\ell^\star$ and tetrahedral 3-regions $\nn^\star$ in the boundary $B$. Notice that we may not understand the state as being invariant under spatial diffeomorphisms once this step is taken.  Diffeomorphism invariance, as well as gauge invariance, is understood to be imposed by the spinfoam.  
 
 The precise embedding is a matter of choice. In practice a reasonable choice will be dictated by the symmetries of the problem. In a space-time without symmetries, one should try different embeddings and see whether the functional dependence of the data on the physical scales of the problem changes.
 
The area data are identified with the proper areas of the triangular 2-surfaces $\ell^\star$ of $\Gamma^\star$, embedded in $B$ and dual to the link $\ell$, and are interpreted as the norms of the fluxes. The spins of the boundary state will be peaked on these areas.
  
 The area data, along with the boost angles $\zeta_\ell$ encoding the extrinsic geometry, are the two quantities with which we will be primarily concerned. They are manifestly gauge--invariant (and covariant) data. They are geometrical quantities, proper areas and boost angles, that do not depend on a choice of a triad frame in $\Gamma^\star$. 

The boost angle $\zeta_\ell$ can be calculated in various manners. The reason we present below different ways of calculating the boost data is to emphasize that the result should not depend on the choice of discretization scheme, thus, use of any of the following procedures should yield the same result for the physical observable.

 One way is to use the result of the previous section. We choose ``representative'' points in the two tetrahedra $n^\star$ and $n'^\star$ that share the triangle $f^\star$ and a curve $\Upsilon$ that joins them. We then define 
\begin{equation}
\zeta_\ell \sim \int_\Upsilon K_\Upsilon
\end{equation}
with $K_\Upsilon$ the extrinsic curvature along $\Upsilon$ as defined in the previous section and the subscript $\Upsilon$ on the integral means that integration is performed with the invariant line element (infinitesimal proper length) along $\Upsilon$. This boost angle will only see one component of the extrinsic curvature, the one in the plane defined by the normal to the hypersurface and the tangent to $\Upsilon$.

 Alternatively, we may instead define the boost angle by smearing along $\Upsilon$ the \emph{mean} extrinsic curvature, 
\begin{equation}
{\bf K}=K_{ab}q^{ab}=\nabla_\mu n^\mu
\end{equation}
which smears the extrinsic curvature in all infinitesimal directions along $\Upsilon$. 

A third way to do things is the following: choose a region $\mathcal{R}_{\ell^\star}$ of $B$ that includes the triangular surface $\ell^\star$. We want to capture the average extrinsic curvature in $\mathcal{R}_{\ell^\star}$ and distribute it only on $\ell^\star$. This implies the definition

\begin{equation}
\frac{K_{\mathcal{R}_{\ell^\star}}}{A_{\ell^\star}} \equiv \frac{1}{A_{\ell^\star}} \int_{\mathcal{R_{\ell^\star}}} \dd {\bf vol}_B \; {\bf K}
\end{equation}
which again defines a covariant dimensionless quantity that is readily interpreted as an ``averaged'' boost angle. 

In practice, one should try different curves, regions and the definitions above, to verify that the functional dependence of $\zeta_\ell$ on the physical scales does not depend on these choices. Note that both ${\bf K}$ and $K_\Upsilon$ are straighforwardly interpreted as measuring the infinitesimal change in the normal, and their units are inverse length. The integration with the infinitesimal proper length yields a dimensionless quantity. We showed in the previous section that $K_\Upsilon$ precisely corresponds to the boost relating the parallel transport between the two regions. 
\medskip

We now impose that $\Gamma^\star$ be piecewise-flat. We are missing two reals for each tetrahedron $v^\star$: we have specified four variables per tetrahedron and need to fix two more to get classical geometry. The remaining missing information, in order to fully specify the geometry of $\Gamma^\star$, is purely directional \cite{rovelli_geometry_2010}: the relative local orientation of the face dual to $\ell$ with respect to the adjacent faces. This information will be in the normal data $k_{\nn\ell}$. 

 Given the areas, the closure condition in each flat tetrahedron must be satisfied  
\begin{equation}
0 = \sum_{\ell^\star \in v^\star} A_{\ell^\star} \; k_{\ell^\star \, \nn^\star}
\end{equation}

A simple counting argument can convince the reader that this completes the intrinsic geometrical information for $\Gamma^\star$: for each tetrahedron, there are four $k_{\nn\ell}$, thus eight real numbers. Understood as normalized 3D vectors, they must satisfy closure (minus three). Arbitrary rotations of the triad (gauge) remove another three degrees of freedom. Thus we are left with two reals which correspond to having specified two 3D dihedral angles of the tetrahedron, plus the four areas; in all six numbers for each tetrahedron, that fully specify the intrinsic geometry of $\Gamma^*$. The fact that they do so uniquely up to inversion is known as Minkowski's theorem.

The distributional extrinsic geometry is encoded on the interfaces between tetrahedra, the triangles $\ell^\star$. This is equivalent to demanding that the tetrahedra be \emph{flatly} embedded \emph{locally}, so that the normal is parallel transported to itself. We stress again that this does not imply a triangulation of spacetime, we need only assume a local extension of $\Gamma^*$.\footnote{This is also the sense in which isolated continuous hypersurfaces can have an extrinsic curvature, that can serve for instance as Cauchy data in an initial value problem.}

Thus, we need only a single number, the boost angle, and the directions that specify which is the component of the extrinsic curvature that is non-zero.

\medskip

There remains to specify the normal data $\vec{k}_{\nn\ell}$.  They are understood as points in $S^2$ or in the Riemann sphere $\mathbb{CP}1$ and are represented as components in an orthonormal frame (triad) in a tetrahedron $v^\star$ , which corresponds to fixing the $SU(2)$ local gauge, see Figure \ref{fig:gaugeChoice}. We will use the Euler angle parametrization and the boundary states of Section \ref{sec:boundaryState}.

Ideally, what one would wish for is a sense in which the smearing of the intrinsic curvature is captured by the 3D deficit angles of the hypersurface. Then, calculate the 4D deficit angles and then compare with the gauss-codazzi equations that relate the three curvatures.

For the explicit choice of hypersurfaces in the following section, we do not perform such a step. We exploit a convenient choice of hypersurface and the spherical symmetry of the spacetime, to deduce a sense in which the intrinsic geometry is captured by the 3D deficit angles. This will result in the dominant contribution to the path integral coming from a degenerate 3D geometry.

\section{A well-defined, explicit expression for the amplitude} \label{sec:explicitAmp}

In this section we outline the choice of hypersurface and discretization and give the explicit form of the transition amplitude defining the lifetime for an explicit choice of surface. The derivation is given in Appendix \ref{app:derivationTreeAmplitudes}.

\subsection{Choice of hypersurface and triangulation}
In this section we briefly review the setting of \cite{christodoulou_realistic_2016}. We refer the reader to this article for more details. We discuss here the main assumptions and some important features.

 We choose the boundary $B$ to be composed by two Lemaître $t_L=const.$ surfaces \cite{lemaitre_expansion_1931}. These are the $\mathcal{C}^\pm$ surfaces of Figure \ref{fig:ansatz}. The pieces in the flat part of spacetime are ignored and we take $\mathcal{C}^\pm$ to extend from $\Delta$ up to $r=0$.

This approximation can be understood as the assumption that the timescale of the physical process is dominated by the tunneling probability of the degrees of freedom that are in the low curvature region. That is, by the intuitive understanding that it is ``easier'' for the geometry to tunnel in the strong curvature regime.

Another justification comes from the fact that we \emph{can} choose boundary surfaces, such that the spacetime parameters $m$ and $T$ are encoded only at the tip $\Delta$ of the quantum region, see the final section of the previous chapter. By the assumption that the amplitude does not depend on the hypersurface choice, further justified in the following section by a calculation for an arbitrary surface, refining the triangulation in the strong curvature region would not have added any information to the amplitude. In conjunction with the coarse triangulation we will use, the strong curvature inside the hole is essentially ignored in this setup.

\medskip

The constant Lemaître time surfaces are \emph{intrinsically flat}. As spacelike hypersurfaces in a three--dimensional spherically symmetric spacetime, with the radial coordinate taking values from $r=r_\Delta$ to $r=0$, they are topologically 3-balls. That is, they are foliated by 2-spheres with proper area $ 4 \pi r^2$ for each value of $r\in(0,r_\Delta)$. Geometrically, since they are intrinsically flat, they are the equivalent of a cone, in three dimensions: they are intrinsically flat but there is a curvature singularity at $r=0$ (a point in $\mathcal{C}^\pm$). 

This property and the spherical symmetry suggests a straightforward discretization. We first triangulate the boundary of $\mathcal{C}^\pm$, where the two hypersurfaces are identified. This is the 2-sphere $\Delta$. We triangulate the sphere at $r_\Delta$ with a \emph{regular} tetrahedron. That is, the areas of the four triangles are all equal. Let us call this area $a$. These four triangles are referred to as \emph{angular}, since they triangulate a sphere which is at constant radius.

We triangulate the interior of $\mathcal{C}^\pm$ with four tetrahedra. Their base triangle, is one of the triangles triangulating $\Delta$ and is of area $a$. By spherical symmetry we take each of these tetrahedra to be iscosceles and call the area of each of the remaining three triangles $b$. We refer to these triangles as \emph{radial} because they extend from $r=r_\Delta$ to $r=0$. We stress however the discussion in the previous section, one should not imagine this triangulation as embedded back in the spacetime of Figure \ref{fig:ansatz}. This is the 3D simplicial complex $\Gamma^\star$.

To visualize the analogous discretization in 3D, imagine a cone, discretized to a hollow triangular pyramid: the circle that is the boundary of the cone becomes a regular triangle. This is the equivalent of triangulating the sphere $\Delta$ with a tetrahedron. The surface of the cone is triangulated by three iscosceles triangles, which is the equivalent of the four tetrahedra triangulating the interior of $\mathcal{C}^\pm$.

Up to the extrinsic data, we have almost completely specified the discretization. We embed $\Gamma^\star$ inside $\mathcal{C^\pm}$. We need not calculate the precise values for the areas here, see \cite{christodoulou_realistic_2016} for a calculation. 

It is obvious that any reasonable embedding will result in both $a$ and $b$ being proportional to $\sim m^2$. A further dependence on $T/m$ may be present, which will be weak. The explicit form resulting from the calculation is given in the following subsection. 

Say we have calculated $a$, which will be approximately equal to $\sim \pi m^2$, one fourth of the area of the sphere at $\Delta$.  

How can we impose that the intrinsic curvature be zero? By demanding that the triangulation of $\Gamma^\star$ is a triangulation in a flat 3D Euclidean space. This is equivalent to demanding that the volumes of the four isosceles tetrahedra be equal to the volume of a regular tetrahedron with triangle areas $a$ and is a condition on $a$ and $b$. Since $a$ is specified, we have fixed $b$, which will equal $a$ up to a numerical factor of order one. 

As a consequence of this choice of discretization (and hypersurface), the volume of the geometrical 4-simplices dual to the 2-complex on which the spinfoam amplitude is peaked vanishes and corresponds to a 3D geometry (4D of zero four--volume).\footnote{This is checked by calculating a Cayley--Menger determinant.} We will see in Section \ref{sec:explicitSetup} that this is not a bad feature of the discretization. 

\begin{figure}
\centering
\includegraphics[scale=0.3]{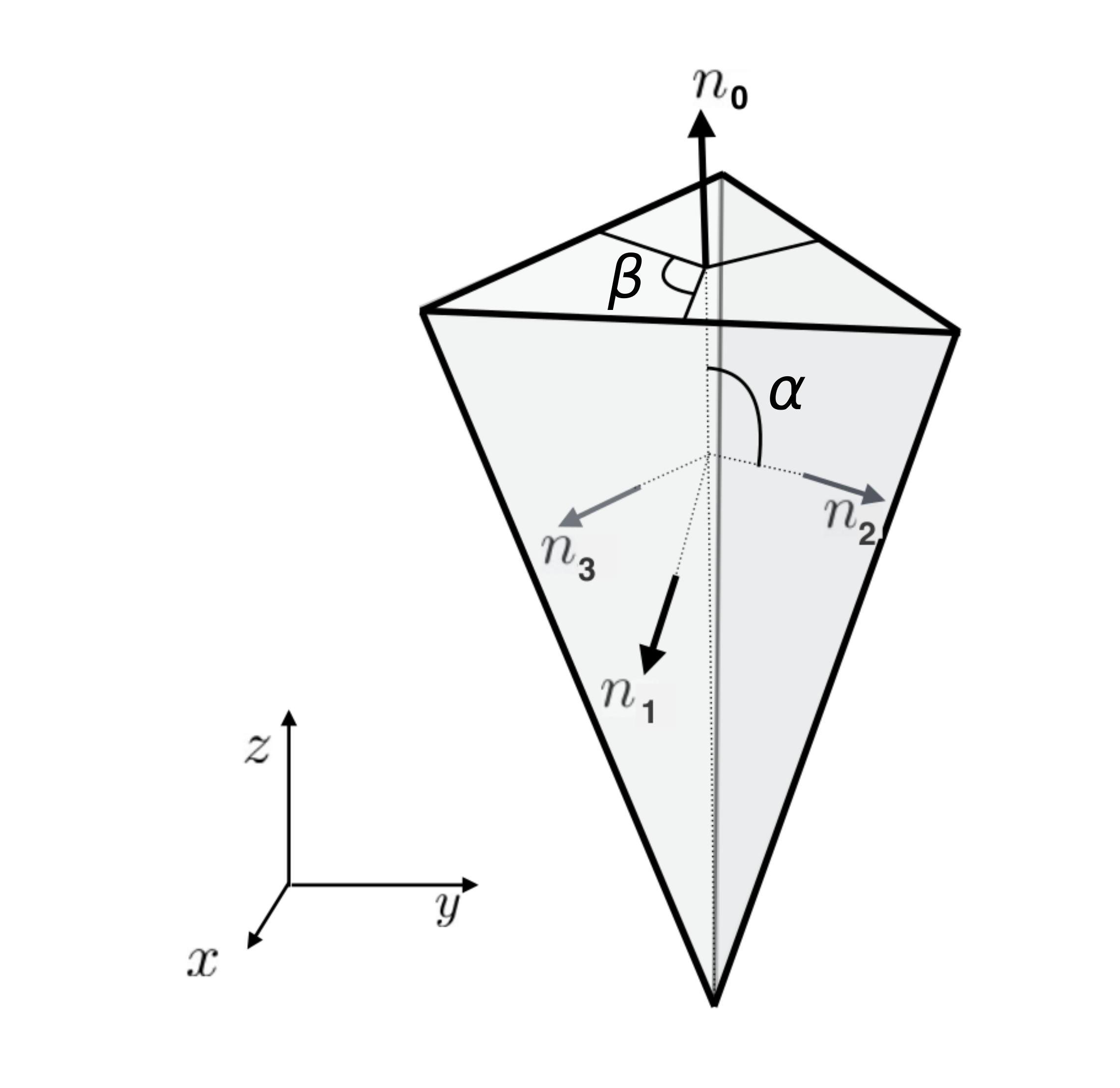}
\caption[A choice of orthonormal frame]{The choice of orthonormal frame in each of the four iscosceles tetrahedra composing the triangulation $\Gamma^\star$. Four such tetrahedra triangulate each surface $\mathcal{C^\pm}$ of Figure \ref{fig:ansatz}. The angles $\alpha$ and $\beta$ correspond to the Euler angle parametrization of the $SU(2)$ coherent states of Section \ref{sec:intrinsicCoherentStates}. The $x--y--z$ axis correspond to the choice of orthonormal frame. They imply that the components of the normal vectors may be read as in Cartesian coordinates but we caution that they are to be understood as the triad vectors $e_a=(e_1,e_2,e_3)$.}
\label{fig:gaugeChoice}
\end{figure}

Having calculated all the area data $A_\ell$, for a given choice of frame in each tetrahedron, and having imposed that all of them are iscosceles, we have also fixed all the normals $k_{\ell\nn}$. The gauge choice used in \cite{christodoulou_realistic_2016} to express $k_{\ell\nn}$ in components on a triad basis is depicted in Figure \ref{fig:gaugeChoice}.

There remains to calculate the extrinsic data. There are two kinds: one is the boost angle at the tip $\Delta$, ascribed to the links dual to the four angular triangles of the regular tetrahedron. These data will be living on a thin--wedge of the simplicial geometry. The calculation of a boost angle at a thin wedge is explained in Section \ref{sec:lorReggeAction}.

The second kind are the boost angles associated to the radial triangles of the iscosceles tetrahdra. For this we may use any of the procedures of the previous section.

In summary, and recalling the closing comments of the last chapter, the bounce time is only encoded in the combination $T/m$, through the area of the sphere $\Delta$. To avoid confusion, we recall the abuse of notation in Chapter \ref{ch:exteriorSpacetime}: $\Delta$ is denotes both the sphere at the tip of the quantum region and the dimensionless quantity related to its radius by $r_\Delta=2m(1+\Delta)$. 

The areas are given as 
\begin{eqnarray}
a \propto 2m(1+\Delta(T/m)) \nonumber \\
b \propto 2m(1+\Delta(T/m)) 
\end{eqnarray}
The radial boost angles by 
\begin{eqnarray}
\zeta \propto 1
\end{eqnarray} 
and the angular boost angle by
\begin{eqnarray}
\zeta \propto \arccosh \sqrt{ 1 + \frac{1}{\Delta(T/m)} }
\end{eqnarray} 

\begin{figure}
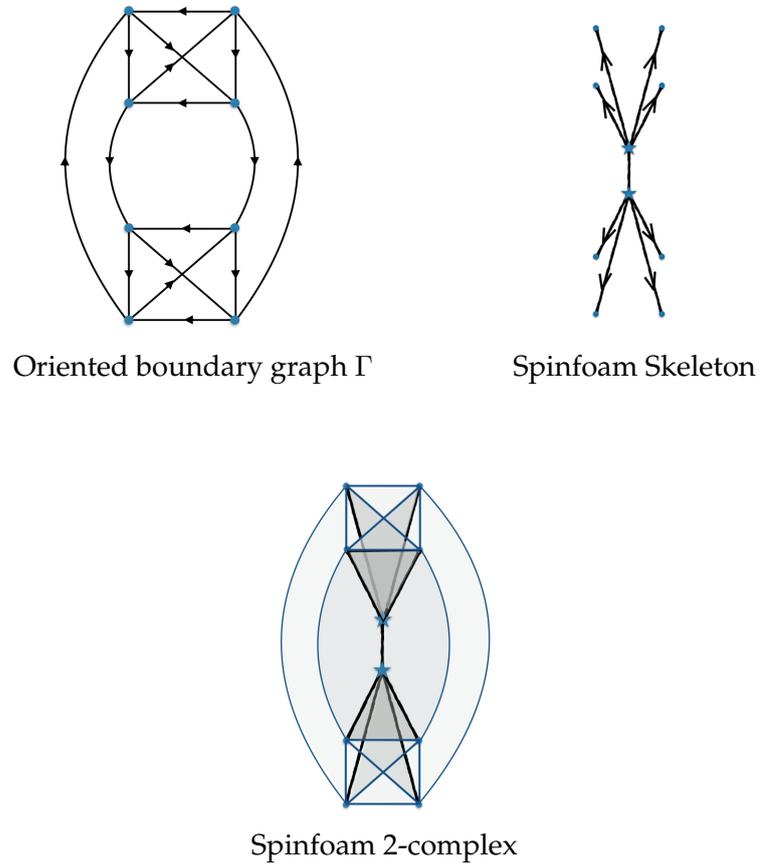

\centering
\begin{tabular}{rcl}
  \includegraphics[scale=0.2]{BH2WHorientedGamma} &   {\hspace*{1cm}} & \includegraphics[scale=0.15]{BH2WHspinfoamOrientedSkeleton} \\
Oriented boundary graph $\Gamma$ & {\hspace*{1cm}} & Spinfoam Skeleton \\[30pt]
\multicolumn{3}{c}{\includegraphics[scale=0.25]{BH2WHspinfoam} }\\
\multicolumn{3}{c}{Spinfoam 2-complex} \\
\multicolumn{3}{c}{}
\end{tabular}
\caption[Spinfoam with its oriented boundary and skeleton.]{The 2-complex with its oriented boundary and skeleton, on which the transition amplitude $W_\mathcal{C}(m,T)$ is defined. Notice that $\Gamma=\partial\mathcal{C}$. The spinfoam is composed of sixteen faces. The four ``middle'' include both vertices and are bounded by three edges and one boundary link. These have one interior edge, the \slt group elements $g_{ve}$ living on this edge are fixed to $\ident_\sl2c$, exploiting the fact that one integration per vertex is redundant. The twelve other faces (six on the top vertex and six on the bottom vertex) are of the wedge form (that is, their boundary is composed by two edges and one link) and are bounded by two edges and one boundary link. }

\label{fig:spinfoam}
\end{figure}

\subsection{Summary of the amplitude}

This is a summary in self-contained form of all the formulas defining the transition amplitude $W(m,T)$ for the choice of hypersurface and discretization discussed in the previous section and the 2-complex of Figure \ref{fig:spinfoam}. All labels refer to the oriented boundary that determines the pattern of contraction. We refer to \cite{christodoulou_realistic_2016} for the details and notation. See Appendix \ref{app:derivationTreeAmplitudes} for the derivation and Appendix \ref{app:recoupTheory} for a presentation of the basic recoupling techniques used here. 

For lack of space, we do not explain the details of the objects appearing in this expression. The amplitude as defined below is finite and expressed in terms of explicitly known analytic functions. 

Much effort has gone into the numerical investigation of the amplitude. Asymptotics of the various bits and pieces are available, but there is no guarantee that putting these together yields a meaningful approximation. Interpolations between the asymptotic expressions and other approximations and techniques to deal with such amplitudes are currently under development at the Center for Theoretical Physics in Marseille. 

The numerical study of such objects is resource intensive and unstable. The biggest difficulty is to perform the spin--sum. Currently, performing the spin--sum even at the level of spins as low as less than ten is infeasible. 

We derived in Section \ref{sec:decayingAmplitudesCalc} a simple estimate for such amplitudes.
In the following section we will estimate the lifetime using these results.

\bigskip

The amplitude is given by 

\medskip

\begin{eqnarray}
W(m,T)&=& \sum_{\{ j_a,j^\pm_{ab} \}} w(z_0,z_\pm,j_a,j^\pm_{ab})\ 
 \times
\\ & \times &  \sum_{\{J_a^\pm,{K}_a^\pm ,l_a ,l_{ab}^\pm\}} \left( \bigotimes _{a,\pm}N^{J_a^\pm}_{\{j_a^\pm\}}(\nu_{\ell \in a^\pm}) \; f^{J_a^\pm,K_a^\pm}_{\{j_a^\pm\}\{l_a^\pm\}}\ \ 
\right)\ \ 
 \times
\\ & \times &  \left(  \bigotimes_{a,\pm} i^{K_a^\pm,\{l_a^\pm\}}\right)_\Gamma\ \ . 
\end{eqnarray}
\bigskip
where the weight function is
\begin{eqnarray}
w(z_0,z_\pm,j_a,j^\pm_{ab}) &=& c(\eta,\eta_0) 
\left( \prod_a d_{j_a} e^{-\frac{1}{2 \eta}(j_a - \frac{(2 \eta^2 -1)}{2} )^2} e^{i \gamma \zeta j_a }\right) \ \times \nonumber \\ & \times & 
\left( \prod_{ab,\pm} d_{j_{ab}^\pm} e^{-\frac{1}{2 \eta_0}(j_{ab}^\pm - \frac{(2 \eta_0^2 -1)}{2} )^2} e^{i \gamma \zeta_0 j_{ab}^\pm}\right)
\end{eqnarray}
with
\begin{equation}
c(\eta,\eta_0) =
\left(e^{\frac{1}{2 \eta_0}\left(\frac{(2 \eta_0^2 -1)}{2}\right) ^2}\right)^4 
\left(e^{\frac{1}{2 \eta}\left(\frac{(2 \eta^2 -1)}{2}\right) ^2}\right)^{12}
\end{equation}
The normals are in
\begin{equation}
N^{J_a^\pm}_{\{j_a^\pm\}} = \left( \overleftarrow{\bigotimes_{\ell \in a^\pm}} D^{j_\ell}_{m_\ell j_\ell} (\nu_\ell) \right) \; i^{\,J_a,\, \{j_a^\pm\}}_{\ \ \ \{\overrightarrow{m}_a^\pm\}}
\end{equation}
The arrowed product indicates that the magnetic indices of the representation matrices on the half links outgoing from the node come with a minus sign. The boost part is
\begin{equation}
f^{K_a^\pm,J_a^\pm}_{\{j_a^\pm\}\{l_a^\pm\}} \equiv \; d_{J_a^\pm} \; i^{\,J_a,\, \{j_a^\pm\}}_{\ \ \ \{\overrightarrow{p}_a^\pm \}}\;\;\left( \int dr_a^\pm \, \frac{\sinh ^2 r_a^\pm}{4 \pi} \, \overrightarrow{\bigotimes_{\ell \in a^\pm}} d_{j_\ell l_\ell p_\ell}\!(r_a^\pm) \right)\;\;i^{\,K_a,\, \{l_a^\pm \}}_{\ \ \ \{\overleftarrow{p}_a^\pm \}}\;d_{K_a^\pm}
\end{equation}
The arrow in the tensor product of the $d_{jlp}(r)$ indicates that those living on links ingoing to the node appear as $d_{ljp}(-r)$. The ranges on the $l$ and $p$ indices are $l\leq j$ and $p$ is summed over the range $|p|\leq j$. The functions $d_{jlp}(r)$ are given by the integral 
\begin{eqnarray}
d_{jlp}(r) 
&=& \sqrt{d_j} \sqrt{d_l}\int_0^1 dt \ 
d^l_{jp}\!\left(\frac{te^{-r}-(1-t)e^r}{te^{-r}+(1-t)e^r}\right)  \\ \nonumber && \ \ \ \times \ \ \  d^j_{jp}(2t-1) \ (te^{-r}+(1-t)e^r)^{i\gamma j-1},
\end{eqnarray}
where $d^j_{mn}(\cos \beta)$ are Wigner's $SU(2)$ matrices.  
The 24--j symbol is given by
\begin{equation}
\left(\bigotimes_{a,\pm} i^{K_a^\pm,\{l_a^\pm\}}\right)_\Gamma \! = \! \sum_{\{h_a,h_{ab}^\pm \}} \! \! (-1)^{\sum_{\ell \in \Gamma} h_\ell} \prod_{a,\pm} \! \,i^{K_a^\pm,\, \{l_a^\pm\}}_{\ \ \ \{\overleftarrow{h}_a^\pm\}}
\end{equation}
The four-valent intertwiners are defined as 
\begin{align}
& i^{J,\; j_1,j_2,j_3,j_4}_{\ \ \ m_1,m_2,m_3,m_4}=  \\  \nonumber
& = (-1)^{j_1-j_2+\mu}
\left(\begin{array}{lcl}j_1&j_2&J \\ m_1&m_2&\mu\end{array}\right)\left(\begin{array}{lcl}j_3&j_4&J \\ m_3&m_4&-\mu\end{array}\right)
\end{align}
with $\mu=-m_1-m_2=m_3+m_4$ and $\left(\begin{array}{lcl}j_1&j_2&j_3 \\ m_1&m_2&m_3\end{array}\right)$ is the Wigner 3--j symbol. Finally
\begin{eqnarray}
z_o =\eta_0 + i \gamma \zeta_0 &=&  \frac{2m(1+e^{-T/2m})}{\sqrt{2 \gamma \hbar G}}  +i \frac{T}{2m}.
\\
z_\pm = \eta_0 \pm i \gamma \zeta_0 &=&  \frac{2m(1+e^{-T/2m})}{\sqrt{\sqrt{6}2 \gamma \hbar G}}  \mp i \arctan4\sqrt6.
\end{eqnarray}

\section{Estimation of lifetime} 
\label{sec:lifetimeEstim}

\subsection{Estimation for the explicit setup} \label{sec:explicitSetup}
\begin{figure}
\centering
\includegraphics[scale=0.3]{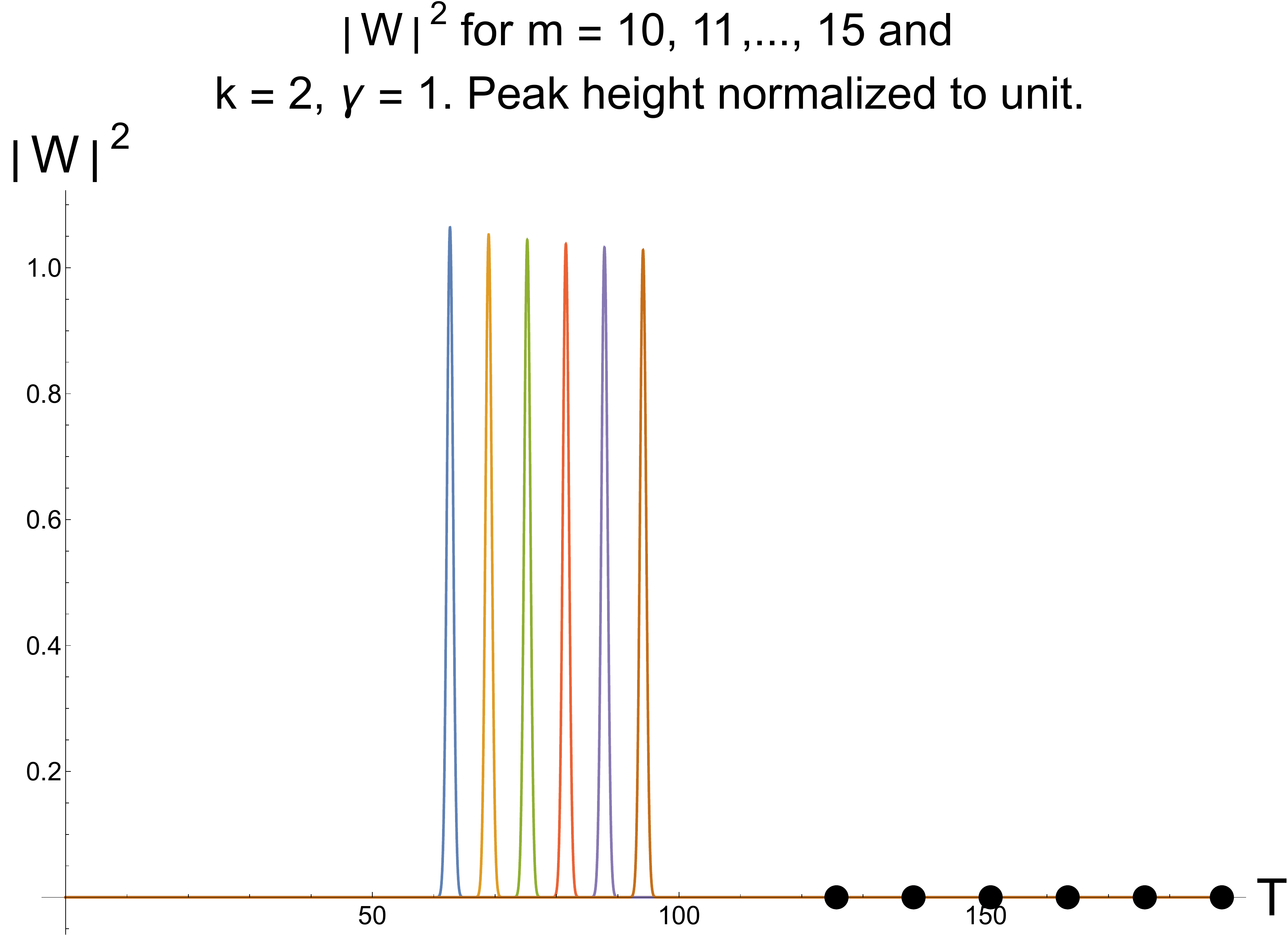}
\caption[Explicit estimation of a characteristic time--scale for a black to white transition.]{The vertical axis is the squared modulus of the transition amplitude and the horizontal axis is the bounce time $T$. The different plots correspond to different values for the mass. The amplitude is periodic in the extrinsic data $\zeta$. The period for each value of the mass is marked with a black dot. }
\label{fig:numerics}
\end{figure}

To estimate the lifetime from the setup of the previous section, we investigated numerically the estimation of Section \ref{sec:summaryGravTunn}, by plugging in the boundary data. A representative result is plotted in Figure \ref{fig:numerics}

After extensive tests, we conclude that quantity we have called the lifetime scales as 
\begin{equation} \label{eq:scalingOfTau}
\tau(m) \sim \frac{m}{\gamma}
\end{equation}
 The physical interpretation of our results is evolving at the time of writing. In the next section we explain that in the author's opinion, this result implies that the physical meaning of $\tau(m)$ is different than the one implied in Chapter \ref{ch:exteriorSpacetime}, which reflects the ideas in the work \cite{christodoulou_realistic_2016}.

For lack of space, and because we will understand this result analytically in a more general setting in the following section, we do not present more details on the numerical investigations here and refer to \cite{mariosFabioCarlo}. We give a qualitative understanding of why this turns out to be the case. 

\bigskip

The result \ref{eq:scalingOfTau} is easily understood as follows. To invert the relation $\Delta(T/m)$ we used the approximation $\Delta(T/m) \ll 1$, that is, we took the sphere $\Delta$ to be close to the horizon, see \cite{christodoulou_realistic_2016} for a discussion.\footnote{This approximation has been subsequently lifted and numerical analysis yields identical results.}

The area data in this approximation depend weakly on the bounce time, as $\alpha \propto m^2 (1+e^{-T/2m})^2$  while the radial boost angles do not depend on $T$. The only dependence on $T$ at play is then from the angular boost angles, that sit on a thin wedge, at the sphere $\Delta$, and read $\zeta \propto T/m$.

The simplicial geometry dominating the spinfoam amplitude according to the estimation of Section \ref{sec:summaryGravTunn}, is the degenarate geometry determined by the area data. Thus, all the dihedral angles vanish on a thick wedge, and the angles at a thin wedge all equal $\pi$. We coded this in Section \ref{sec:summaryGravTunn} by setting $\beta=0$ and $\Pi_\ell=\pi$ for the degenarate configurations. 

The amplitude is then dominated by a product of four exponentials of the form

\begin{equation} \label{eq:lifetimeestim}
P(T \vert m) \propto \vert W \vert ^2 \sim e^{-(\gamma \frac{T}{m} - \pi)^2/t}
\end{equation} 

The expectation value of $T$ in the distribution $P(T \vert m)$ is then highly peaked on $m \pi /\gamma$. 

\subsubsection*{The role of Euclidean, degenerate and Lorentzian simplicial configurations}

We discuss this result further in what follows. Before continuing, we want to emphasize the interplay between the different geometrical critical points and give evidence for the independence of the result on the discretization procedure.

In this specific example, our choices resulted to in a degenerate configuration, where both 4-simplices dual to the 2-complex of Figure \ref{fig:spinfoam} have zero four-volume. 

We may parametrize the ambiguity in the area data by allowing them to vary such that the geometry is no longer degenerate. When we allow for small variations, of the order of unit, we scan a range of Lorentzian and Euclidean simplicial geometries. \footnote{It is possible to fall on a non--geometric configuration but we do not consider this case here.} That is, $\phi(\omega)$ will not vanish and will be interpreted as either a Lorentzian or Euclidean dihedral angle. 

To see this, imagine a tetrahedron with three of its triangles squashed on its fourth triangle. The tetrahedron is thus of zero three-volume.  Lifting the apex at which the three triangles meet, the segment lengths increase and we have a Euclidean tetrahedron. In the Euclidean case, for small variations,  the dihedral angle itself is approximately equal to $\pi$. This is because in Euclidean geometry, two normals pointing in the opposite direction form an angle $\pi$ amongst them.

 Similarly, lifting the apex in a Minkowski time direction (thus the segment lengths decrease) results in a Lorentzian tetrahedron. The dihedral boost angle for the Lorentzian case will be small, see section \ref{sec:lorReggeAction}. However, because of the presence of $\Pi_\ell=\pi$ in the Lorentzian case, the estimate of equation \eqref{eq:lifetimeestim} given above is not altered.

\subsection{Estimate for arbitrary choice of hypersurface and 2-complex} \label{sec:EstimArbitr}
In this section, we analyse the lifetime for an arbitrary choice of boundary surface. 

We will give an estimate for the lifetime based on general grounds and within the assumption that the area and normal boundary data $\omega$ and $k$ are Regge-like. 

We will not make an explicit choice of graph $\Gamma$ or 2-complex $\mathcal{C}$ on which the boundary state and the spin-foam amplitude is defined. The analysis presented below applies to the transition amplitudes studied in Chapter \ref{ch:gravTunneling}. In summary, we are considering spinfoam amplitudes, defined on a fixed 2-complex $\mathcal{C}$ without interior faces and dual to a simplicial complex, with its boundary coinciding with the graph on which the boundary state is defined, i.e.\! $\partial \mathcal{C}=\Gamma$. 

The lifetime, defined in equation \eqref{eq:lifetime}, is given in terms of the measure $\alpha(m,T)$ and the transition amplitude $W(m,T)$ through equations \eqref{eq:defCondProbDistr}, \eqref{eq:probdens} and \eqref{eq:defMarg}.

At the time of writing this thesis, we do not have an explicit expression for the measure $\alpha(m,T)$. We will show that in the above setting, the lifetime will scale linearly with the mass, unless the measure brings in a non-trivial dependence on the areas.

The amplitude $W(m,T)$ will be approximated by a stationary phase evaluation, using the results of Chapter \ref{ch:gravTunneling}. We consider that an explicit choice of interior boundary $B$ has been made. The continuous intrinsic and extrinsic geometry of $B$ is then approximated by a piece-wise flat distributional geometry, as explained in Section \ref{sec:Discretization}. The spacelike surface $B$ is triangulated (in tetrahedra) and the metric information is encoded on the faces of the triangulation. To each face corresponds an area $A$, a boost angle $\zeta$ and two normalized 3D normals $k$. Once the discretization has been carried out, all of these boundary data will in general depend on $m$ and $T$. Only $A$ and $\zeta$ appear explicitly in the approximation for $W(m,T)$.

Areas of the discretization will in general scale  as $m^2$. A dimensionless function of the two spacetime  parameters can only be a function of $T/m$.\footnote{Dimensionless quantities including $\hbar$ (for example $T/\hbar$) are excluded because the dependence of the boundary data on $T$ and $m$ is calculated from a classical discretization procedure and there is no isolated $\hbar$ that could come into play from classical general relativity.}

The areas will then in general have the following functional form in terms of the two spacetime parameters
\begin{equation} \label{eq:areaForm}
A(m,T)=m^2 \tilde{A}(T/m)
\end{equation}
and the boost angles will be of the form
\begin{equation} \label{eq:boostAngleForm}
\tilde{\zeta}(T/m)
\end{equation}
where tilded quantities are dimensionless.

The identification with the labels of the coherent states is immediate for the boost angles and we simply set $\zeta(T/m) \equiv \tilde{\zeta}(T/m)$. The areas are identified with $\omega$ on which the Gaussian weights of the boundary state are peaked. The area boundary data $\omega$ are given by $\omega = \frac{\eta}{2t}-\frac{1}{2}$. The semiclassicality condition $\sqrt{t}\omega \gg 1$ implies that $\eta /t \gg 1/\sqrt{t}$ and since the semiclassicality parameter $\sqrt{t}$ is small, then $\eta /t \gg 1$. Thus, it is a good approximation to take $\omega \approx \frac{\eta}{2t}$.

For the labels $\eta$, we then have the identification
\begin{equation}
\omega = \frac{\eta}{2t} \equiv \frac{m^2}{\hbar} \tilde{A}(T/m)
\end{equation}
The factor $\frac{m^2}{\hbar}$ serves as the large parameter $\lambda$ with respect to which we take the stationary phase approximation. The functions $\tilde{A}(T/m)$ along with the 3D normal data $k$ determine the dominant contribution to the spinfoam amplitude.

We recall that we are working in geometrical units ($G=c=1$), where lengths, times and mass all have the dimensions $\sqrt{\hbar}$. The spins $j$ are dimensionless and so are $\omega$, $\eta$ and $t$. 

\medskip

We now turn our attention to the amplitude. Given only the data $\omega$ and $k$, there are two distinct possibilities: they are either Regge-like or not. We only consider here the case where $\omega$ and $k$ are Regge-like, which means that they specify either a 4D Lorentzian, 4D Euclidean or 3D degenerate (Euclidean) geometry for the simplicial complex dual to the 2-complex $\mathcal{C}$. The possibility for a vector geometry, that does not admit an interpretation as a simplicial complex, is excluded by the choice of semi--classical boundary state. The picture is then the following. The spinfoam amplitude is dominated by the configuration which corresponds to the geometrical simplicial complex that is determined only by $\omega$ and $k$. 

On this configuration, once the spin-sum is performed, the data $\omega$ and $k$ determine the dihedral angles $\Theta(\omega,k)$ between tetrahedra that share a face. In turn, these determine the (Palatini) deficit angles $\phi(\omega,k)$. These are all boost angles among tetrahedra on the boundary because the spinfoams we consider do not have interior faces (on the boundary, the deficit angles are exterior dihedral angles, i.e.\! boost angles among tetrahedra). The $\phi(\omega,k)$ are the exact analog of $\zeta$, and thus determine an embedding $\phi(\omega,k)$ preferred by the dynamics.

  The angles $\phi(\omega,k)$ would have to match the data $\zeta$ in order for the amplitude to not be suppressed in the semiclassical limit. Choosing the boost data $\zeta$ to match $\phi(\omega,k)$ is equivalent to the procedure followed in a perturbative setting, i.e.\! in the graviton propagator calculations, see for instance Section V.B of \cite{bianchi_lorentzian_2012}.
  
  In general, when an interpolating solution to Einstein's field equations is not to be expected, this will not be the case and the amplitude will be suppressed exponentially in the mismatch between $\zeta$ and $\phi$. 
  
Given the data $\zeta$, $\omega$, $k$, and defining $\lambda \equiv m^2/\hbar$, we have the following estimation for the amplitude, repeated here from Section \ref{sec:summaryGravTunn} for convenience,

\begin{eqnarray} \label{eq:ampEstimOne}
\vspace*{-0.4cm}	W &\sim & \bigg[ \sum_{s(v)} \; ( \lambda )^{N} \mu( \omega/\lambda ) \times \nonumber \\ &\times & \prod_\ell \; \exp{\left( -\frac{1}{4t}\left(\gamma \zeta_\ell - \beta \phi_\ell(\omega/\lambda,k) +  \Pi_\ell \right)^2\, + \, i \, \gamma \, \zeta_\ell \, \omega_\ell \right)} \bigg] \nonumber \\  
\vspace*{-0.9cm} \left( 1 +O(1/\lambda) \right) \nonumber \\ 
\end{eqnarray}
where again that $\beta= \gamma,0,1$ when a Lorentzian, degenerate or Euclidean simplicial geometry dominates the path--integral respectively. The function $s(v)$ encodes the co-frame orientation and takes values $+1$ or $-1$ on each vertex. The presence of multiple semiclassical critical points, that is, the sum over $s(v)$ configurations can be neglected, since for a given choice of $\zeta$, one configuration $s(v)$ will be dominant. Calling $P_{\text{full}}(T|m)$ the probability distribution resulting from \eqref{eq:ampEstimOne} and $P(T|m)$ the probability distribution resulting from keeping in \eqref{eq:ampEstimOne} only the contribution from the dominant critical point, their ratio is of the form

\begin{equation}
\frac{P_{\text{full}}(T|m)}{P(T|m)} = 1+ e^{-h(T/m) \frac{m^2}{\hbar} }
\end{equation}
where $h(T/m)$ is a positive function. Thus, we may neglect the sum over the $2^N$ critical points related by the different co-frame orientations.  \emph{However, notice that any one of the $2^N$ critical points may be the dominant one, and not necessarily the one corresponding to the two Regge configurations ($s(v)=\pm1$ everywhere)}. This implies that the recently proposed EPRL model and the proper vertex model \cite{engle_lorentzian_2016,engle_lorentzian_2016-1,engle_spin-foam_2013,shirazi_hessian_2016}, which modifies the EPRL model so as to remove the multiple semiclassical limits, will differ in their predictions regarding this effect in the subdominant contribution to the lifetime.

\medskip

The exponential $\prod_\ell e^{i \gamma \zeta_\ell \omega_\ell}$, is then a pure phase and the estimation of the amplitude is of the form 
\begin{equation} \label{eq:ampEstimArbitrary}
\vert W(m, T/m, t) \vert^2 \sim m^N
\mu(T/m) \prod_\ell e^{-( \gamma \zeta_\ell(T/m) - \beta \phi_\ell(T/m) + \Pi)^2/4t}
\end{equation}
Defining 
\begin{equation}
x \equiv  \frac{T}{m}
\end{equation} 
we are then left with the following expression for the lifetime
\begin{equation} \label{eq:tauX}
\boxed{
\tau(m) = m \frac{ 
\int \dd x \; x \,  F(x) \prod_\ell e^{-(\gamma \zeta_\ell(x) - \beta \phi_\ell(x) + \Pi_\ell)^2/4 t} 
}
{
\int \dd x \; F(x) \prod_\ell e^{-(\gamma \zeta_\ell(x) - \beta \phi_\ell(x)^2 + \Pi_\ell)^2/4 t} 
}
}
\end{equation}
where we defined $F(x) \equiv \alpha(x) \mu(x)$ and the power $m^N$ cancels out between nominator and denominator. From this expression, it is manifest that $\tau(m)$ is $m$ times a function that only depends on $\gamma$ and $t$, which we denote as $f(\gamma,t)$.

Thus, we conclude that 
\begin{equation} \label{eq:mainResult}
\tau(m) = m \, f(\gamma,t)
\end{equation}

\medskip

Let us discuss this result. The first thing is that other than the explicit linear dependence of $\tau(m)$ on $m$, the mass may enter the scaling only through the semiclassicality parameter $t$. 

It is not clear whether a dependence of $\tau(m)$ on $t$ is reasonable. As we saw above, taking $t=\hbar/m^n$, the allowed range for $n$, for the approximations taken here to be valid, is $0<n<4$. This would imply that the scaling of $\tau(m)$ on $m$ in general depends on the chosen $n$. 

However, formula \eqref{eq:tauX} suggests that, $t$ will not be present in the leading order in $m$ in a generic situation. To see this, let us omit the product over links and consider that there is only one gaussian in \eqref{eq:tauX}. The behaviour of the integrand will plausibly be dominated by the gaussian weight over the function $F(x)$, unless something exceptional happens with the Hessian. We assume here that $F(x)$ can be neglected. The standard deviation of the gaussian is proportional to the square root of $t$, which is a very small number for a macroscopic black hole: taking for instance $k=2$, $ \sqrt{t} = \sqrt{\hbar}/m $ (for example, for a small, lunar mass black hole, this number is $\sim 10^{-30}$). The integrand, since it is also normalized in the denominator, will behave like a Dirac delta, evaluating the integrand on the solution of the equation
\begin{equation}
\gamma \zeta(x) - \gamma \phi(x) + \Pi =0
\end{equation}
which we call $x_0(\gamma)$. There are two possibilities. The peak is in the range of integration, that is $x_0(\gamma) \leq 0$, or not. In the former case, we expect 
\begin{eqnarray} 
\tau(m) &\sim & m \int \dd x \; x \frac{ 
 \,  e^{-(\gamma \zeta_\ell(x) - \gamma \phi_\ell(x) + \Pi_\ell)^2/4 t} 
}
{
\int \dd x \; e^{-(\gamma \zeta_\ell(x) - \gamma \phi_\ell(x)^2 + \Pi_\ell)^2/4 t} 
} \nonumber  \\
&\sim & m \int \dd x \; x \, \delta(x-x_0(\gamma)) \\
&\sim & m \; x_0(\gamma)
\end{eqnarray} 
More precisely, we have 
\begin{equation}
\tau(m) \propto m \; \left( \; x_0(\gamma) + O(t) \right)
\end{equation}

In the case when $x_0(\gamma) < 0$, we will have\footnote{There is a third possibility that is also suppressed: if it happens to be the case that there is a global extremum and thus no solution $x_0(\gamma)$.}
\begin{equation}
\tau(m) \approx m \; O( e^{- x_0(\gamma)^2/t} )
\end{equation}
In this case, the scaling of $\tau(m)$ on $m$ depends on the choice of $t$ and further suppresses the scaling with $m$. When we consider the product over links, $\tau(m)$ will be generically a sum over such terms. Thus, we conclude that  
\begin{equation} \label{eq:crossingTime}
\boxed{
\tau(m) \approx m f(\gamma) 
}
\end{equation}
with the leading order coefficient $f(\gamma)$ only depending on the Immirzi parameter. 

The estimate given here neglects the truncation issue, which is discussed in Section \ref{sec:issues}. At the time of writing, an exact estimate of the transition amplitudes that takes this issue into account has become available and results will be reported in \cite{mariosFabio}.

\section{Discussion of result and intepretation of lifetime}
In the previous section we saw that employing the estimation of Section \ref{sec:summaryGravTunn} for the transition amplitudes, we arrive at a simple expression, equation \eqref{eq:ampEstimArbitrary}, which provides an estimate for modulus squared of the transition amplitude. 

From this expression, we deduced that the lifetime, as defined in Section \ref{sec:Lifetime}, scales linearly with the mass as in equation \eqref{eq:crossingTime}. The estimate for the transition amplitudes of Section \ref{sec:summaryGravTunn} concerns 2-complexes without interior faces and assumes that the boundary data are Regge-like. Both restrictions are taken because currently we do not have an estimation for the transition amplitudes when either of these two conditions is not satisfied. Whether the estimate changes when they are lifted is a subject for future investigations.

Within these limitations, we have
an estimate for an otherwise arbitrary 2-complex, an arbitrary choice of boundary surface and of discretization scheme. 

 Choosing a one-vertex spinfoam instead of a two-vertex spinfoam like in Figure \ref{fig:spinfoam} would not yield a different estimation for $\tau(m)$. We note that we may create 2-complexes with an arbitrary number of vertices while not having interior faces. 
 
 The precise choice of surface does not play a role in the estimate of $\tau(m)$. Furthermore, the discussion at the end of Section \ref{sec:explicitSetup} applies to the estimate presented here: the linear scaling in the mass $m$ does not depend on whether the boundary data are such that the path--integral is dominated by a Lorentzian, Euclidean or 3D degenerate simplicial configuration.

\subsubsection{Physical interpretation} 

We estimated above that the lifetime scales linearly with the mass. This result contradicts appears too short to be physically correct.

The lifetime as defined in Section \ref{sec:Lifetime}, is the characteristic duration of the phenomenon as deduced by a far-away observer. This was understood physically in the sense: ``how long do we have to wait until a black hole transitions to a white hole''.

In this sense, the expectation would be a lifetime that is far larger than $m$ and hopefully shorter than $\tau \sim m^3$. The latter would be a consistency check for the exterior spacetime, in order for the approximation of neglecting Hawking evaporation to be valid. Furthermore, an intermediate scaling of the order $m^2$ would imply that the phenomenon is possibly astrophysically relevant, on the assumption that small (lunar mass) promordial black hole were formed in the early universe, see \cite{barrau_fast_2014-1,barrau_planck_2014,vidotto_quantum-gravity_2016}. 

Reasonably, it should take a long time for such a geometry transition process to take place. An estimate implying too long a time for the phenomenon to be of astrophysical relevance would be a dissapointing, but reasonable result. 

Instead, we have an estimation that is too short. For a solar mass black hole, a time of order $m$ is in the microseconds.  

The lifetime estimate also suffers from the fact that there is no $\hbar$ appearing. The time-scale we have estimated appears to be a classical quantity, with corrections including $\hbar$ that vanish as $\hbar \rightarrow 0$. 

We would expect the process to be switched off when $\hbar \rightarrow 0$ (the transition probability to go to zero) and the lifetime to go to infinity. What has gone wrong?

\subsection{The lifetime is the crossing--time}
\label{sec:LifetimeIsCrossing}
We propose below that the issue here is that the time-scale we defined as $\tau(m)$ and called the lifetime, corresponds to a different time--scale relevant to the phenomenon, which is expected to be of order $m$, and is best called the \emph{crossing--time}.

The linear scaling with the mass estimated above corroborates two other studies, outside the context of LQG, which are relevant and we briefly mention here before proposing an alternative interpretation of our results. 

The first concerns the study initiated by Kiefer and Hajicek in \cite{hajicek_singularity_2001} (2001), where the authors argued that collapse is succeeded by anti--collapse in the context of null--shell quantization. In a subsequent study \cite{ambrus_quantum_2005-1} (2005) by Hajicek and Ambrus, the authors calculated a characteristic time scale of the phenomenon, which they called scattering time, and concluded that the scaling is linear in the mass. The authors found this result troubling. It was too short for it to be meaningful, on similar grounds as those discussed above. 

\medskip

The second set of results concerns the recent and concurrent studies by Garay, Barcel\'o, Carballo et al \cite{barcelo_lifetime_2015,barcelo_mutiny_2014,barcelo_black_2016,barcelo_exponential_2016} , where the authors independently used an exterior spacetime closely resembling the HR-spacetime presented in Chapter \ref{ch:exteriorSpacetime}. The path--integral is performed using a wick--rotation (thus the integrand is non--oscillating) and performing a functional integration over interpolating Euclidean geometries. This is a feature that is not present in our setting (we saw in Chapter \ref{ch:AsympAnalysis} that at the level of fixed--spins, the asymptotic behaviour of the EPRL amplitudes is oscillatory for both Euclidean and Lorentzian critical points).

The above calculation results in a scaling linear in the mass $m$, for the quantity we have called here the lifetime $\tau(m)$.

In other words, the three calculations that we are aware of, that attempted to calculate a characteristic time--scale for this phenomenon,  reach the same conclusion, although the techniques used and physical pictures in mind are different. 

It is of interest to briefly mention the physical picture that Garay, Barcel\'o, Carballo et al  have in mind, because it is radically different than the physical picture presented here, but, at the technical level, the quantity of which a calculation is attempted in both approaches is the same:  the value of the bounce time at which the modulus squared of the transition amplitude is peaked.\footnote{The author had the opportunity to exchange views recently on these matters with Garay and Carballo, in insightful discussions during a visit in Complutense University of Madrid.} General relativity is deduced from this result to be strongly violated in short time--scales, essentially resulting in the non--existence of anything resembling (geometrically) an apparent horizon or trapped--surface at the gravitational radius. Instead, as soon as the collapse has taken place, an oscillatory phase begins alternating between collapse and anti--collapse, with energy gradually dissipated until a notion of equilibrium is reached. The end result are conjectured to be long--lived extremely compact objects called ``black stars'', with their surface protruding a Planckian distance outside the Schwartzschild radius of the remaining mass.

The point of view taken here, is that the event horizon present in exact solutions of Einstein's field equations will be approximately manifested in nature by a long--lived apparent horizon. In other words, we do not expect General Relativity to be violated near the Schwartzschild radius of the collapsed object, unless a large time scale with respect to a time--scale of the phenomenon (the lifetime of the universe with respect to $m$ ) allows for the fluctuations of the metric near the Schwartzschild radius to become significant. 

\bigskip

In our opinion, although the physical picture in mind may be different, the results from these three studies are not contradictory nor do they imply a violation of General Relativity in short time--scales. While the definition in Section \ref{sec:Lifetime} does define in principle an observable, we propose here that its physical interpretation is different than the one initially envisioned. 

The lifetime calculated here should be understood as the crossing--time: the characteristic time scale of the \emph{duration} of the transition \emph{when it does take place}.

The naming ``lifetime'' alludes to an atomic decay process and is the characteristic time in which a decay becomes likely. This implies that the \emph{probability} of decay per time interval has been estimated (a small number) and of which roughly the inverse gives the lifetime (a large number). 

There is however a second characteristic time--scale for such phenomena: the time it takes for the decayed particle to cross the nucleus. This time is proportional to the radius of the nucleus times a characteristic classical speed deduced from the energy gap. It is given in the dominant order by a ``classical'' contribution, in the sense that if the barrier was not present it is roughly the time it would take for the classical particle to travel from the center of the nucleus to the apparatus where it was detected.

To emphasize this point, we consider a rectangular potential tunneling experiment. Imagine an electron gun shooting electrons from the left hand side of a potential barrier at regular time intervals, at some energy. The transition probability gives an estimate for the time we have to wait on the other side of the barrier until one of the electrons tunnels.

When it does and is detected, we may also measure its crossing time, the time it took from its emission from the gun until its detection on the other side of the barrier. This will be roughly equal to the classical distance between the two apparatuses times the speed; with some quantum corrections. 

In the propagator between two coherent states built with gaussians, peaked on the position of emission  $x_L$ left of the barrier and the position of detection $x_R$ right of the barrier (and on the same momentum $p_0$), we expect to see a gaussian peaked on the crossing--time

\begin{equation}
\braket{x_L,p_0 \vert e^{i H t /\hbar} \vert x_R,p_0 } \sim e^{-(t - (x_R-x_L) p_0/m )^2/\hbar}
\end{equation}

\bigskip

In the Penrose diagram of Figure \ref{fig:ansatz} we essentially \emph{assume} the process to take place. To deduce the \emph{probability} of the process to take place, we need to compare with the probability of ``anything else'' taking place. A first step in this direction is to assume that either the black to white process takes place, or, that the black hole remains a black hole. Current investigations are focused on providing an estimation for the lifetime (not the crossing--time) along these lines.

In summary, we appear to have calculated the crossing--time. We conclude that a prediction of the EPRL model is that \emph{when the transition takes place}, the crossing--time is of order $m$, with sub--dominant corrections involving $\hbar$. Indeed, a crossing--time of order $m$ is to be expected as it is the time--scale in which timelike geodesics crossing the event horizon in a Schwartzschild geometry reach the singularity.

The lifetime should be deduced from the \emph{probability} for the process to take place, which should not be confused with he value of the bounce time on which the amplitude is peaked and which gives the crossing--time. The lifetime, is of course the desired quantity in order to conclude whether this phenomenon is predicted to be of astrophysical relevance or not.

\subsection{Current developments and open issues}
\label{sec:issues}
\subsubsection{The truncation issue}

To get the estimate \ref{eq:crossingTime}, we normalised the probability distribution neglecting the fact that the truncation allows  for meaningful boost angles values up to $4 \pi m / \gamma$. This difficulty arises from the intepretation of the holonomies as discretizations of the Ashtekar--Barbero connection, which is an $SU(2)$ connection. Neglecting the phase $\alpha$ in the discretization of the 3D spin connection (it is pure gauge) \cite{freidel_twistors_2010,freidel_twisted_2010}, the discrete extrinsic curvature is coded in the twisted geometry parametrization as
\begin{equation}
h_{AB}(\zeta) \sim e^{i \gamma \zeta \frac{\sigma_3}{2}}
\end{equation}
Thus, $h_{AB}(\zeta)$ is periodic in $\gamma \zeta$ with a period $4 \pi$. This is the maximal boost angle that can be meaningfully ascribed at each link. There is thus a \emph{maximum boost} that can be described between two adjacent space-time quanta by kinematical states at the level of a fixed graph. 

The amplitude is a function of these boundary data and thus it is also periodic. This effect and a possible relation between the maximum boost and a bound in the spectrum of the extrinsic curvature is presented in Appendix \ref{app:maximumBoost}, were we reproduce an unpublished note by C.Rovelli, S.Speziale and the author.

The periodicity of the holonomy and the degree to which we can describe a discrete geometry at the level of a fixed graph is discussed in detail by E.Livine in \cite{charles_ashtekar-barbero_2015}, where the Immirzi parameter is claimed to act as an effective cut-off scale for the excitations of the extrinsic curvature \cite{charles_ashtekar-barbero_2015}, for the reasons also explained above and in Appendix \ref{app:maximumBoost} (taking $\gamma \rightarrow 0$ allows for arbitrary large boosts $\zeta$).

\subsubsection{The semiclassical limit}

The role of the Immirzi parameter in physical predictions from the EPRL model is not well understood. 
The results of this manuscript have been extensively based on previous results obtained from the semiclassical limit of the EPRL amplitudes at fixed--spins. The limit $\gamma \rightarrow 0$ has been claimed to be necessary in order to recover the emergent behaviour of a path--integral over geometries, on which our analysis is based \cite{perini_holonomy-flux_2012,magliaro_emergence_2011,
han_semiclassical_2013}. 

Furthermore, the two-point correlation function for the Lorentzian EPRL spinfoam model for a single vertex, another central success of the model, is known to match with the one in Regge calculus in the double limit $j\rightarrow \infty \;,\; \gamma \rightarrow 0$ \cite{bianchi_lorentzian_2012,rovelli_euclidean_2011} while keeping the product $j \gamma$ fixed. 

Another indication comes from LQC, where the critical density $\rho_{max}$ at which the bounce occurs is given by $\rho_{max} = \rho_{max} (18 \pi / l_0^3) \rho_{Pl} $ where $\rho_{Pl}$ is the Planck density and $l_0 = 4 \sqrt{3} \pi \gamma$ is the \emph{area gap}, the smallest non--zero eigenvalue for the area operator. The semiclassical limit in LQC is recovered when the area gap is taken to zero, which is mathematically the same as taking gamma to zero because it depends linearly on gamma. 

Conceptually, gamma is treated as a mathematical parameter of the theory, that may be of order unit, but to get the semiclassical limit, we need to take the relevant physical scale, the area gap which happens to be proportional to $\gamma$, to zero \cite{ashtekar_overview_2017}. 

This is not unlike saying that $\hbar$ has the value it has and which is fixed by experiment, but to get classical physics back we need to take a physical scale, such as the quantum of the oscillator which is proportional to $\hbar$, to zero.

Taking the limit $\gamma \rightarrow 0$ in our context allows for arbitrary boosts and overcomes the truncation issue. This avenue is currently under investigation.

\subsubsection{The probability for small $T$}
Qualitative information on the lifetime can be extracted by interpreting the amplitude for small values of the bounce time $T$ with respect to the mass $m$ ( $\gamma \zeta \sim T/m << 4 \pi$) as a transition probability per unit of time (in Planck units) when the spacetime is still well-approximated by the classical metric.

The transition probability then goes as 
\begin{equation}
P \sim e^{-1/t}
\end{equation}
with the semi--classicality parameter taking values in 
\begin{equation}
1 \ll t \ll \frac{\hbar^2}{m^4}
\end{equation}
This implies a lifetime of the order $1/P$, that is
\begin{equation}
1 \ll \tau \ll e^{m^4}
\end{equation}
These bounds include and are compatible with observed physics, albeit largely inconclusive. Also, on dimensional grounds and within the semiclassical approximation for gravitational phenomena, we would expect the probability for such a transition to decay with
\begin{equation}
P \sim e^{-m^2/\hbar}
\end{equation}
so as to have a dimension--full quantity that has units of inverse action. The behaviour of the transition probability estimate appears reasonable on these grounds. An important possibility not excluded in this work is whether a more detailed analysis will significantly increase the transition probability.

\subsubsection{The measure}

We showed in Section \ref{sec:EstimArbitr} that the parameter $\lambda$ may only enter the lifetime through the measure $\alpha(m,T)$.

At the time of writing, we do not have an explicit expression for the measure $\alpha(m,T)$. We have thus not excluded in this manuscript the possibility that 
$f(\gamma,t,\lambda)$ depends on $\lambda$ or $t$ in a way that changes the scaling of the lifetime.

The measure at a level of a state built on a single link can be easily calculated and does not bring a dependence on $\lambda$ in $\tau$. Preliminary results indicate that the dependence on $\lambda$ in $\tau$ from the complete measure will be subdominant and will not alter the results here. An explicit expression for the measure is expected to be available soon. The current expectation is that the results will not be altered at the dominant order in $m$.

\subsubsection{The Hessian}
We have not calculated the Hessian on the critical points. The Hessian for the critical points of the EPRL amplitude is unfortunately given by long formulas that are difficult to manipulate. Work is currently underway in Marseille by S.Speziale and collaborators as well as by B.Bahr and collaborators \cite{
bahr_investigation_2016-1,
bahr_numerical_2016-1}, to simplify the expressions for the Hessian. We do not expect the Hessian to play an important role. It does not depend on $\lambda$ nor on $t$.

\subsubsection{The classical setup}
The Rovelli-Haggard spacetime is a minimalistic model of the phenomenon we wish to describe. Important simplifications are 
\begin{itemize}
\item Collapsing and anti-collapsing matter modelled by null-shells. 
\item Hawking radiation neglected
\item Spherical symmetry, no rotation and no charge
\end{itemize}
To the degree that the lifetime turns out to be an intermediate scale between $\sim m$ and $\sim m^3$, the first two simplifications are  reasonable and appear sufficient to capture the necessary features of the spacetime. 

 Spherical symmetry and no rotation is not a reasonable assumption for an astrophysical black hole. I find unlikely that our estimations will be changed if slow rotation is taken into account, although a dependence on the angular momentum is to be expected that could significantly alter the picture close to extremality. 

\medskip
 
 Research in this direction is also paramount for the plausibility of the process. That is, if an HR type spacetime can be shown to be impossible to construct, at least for a slowly rotating black hole, then the process may be excluded from taking place \emph{in principle}. Significant progress in this direction has been made recently by H. Haggard and collaborators and an HR--spacetime for a Kerr black hole will possibly be available soon. 

\chapter{Conclusions}
\label{ch:Conclusions}
\begin{figure}
\centering
\includegraphics[scale=0.4]{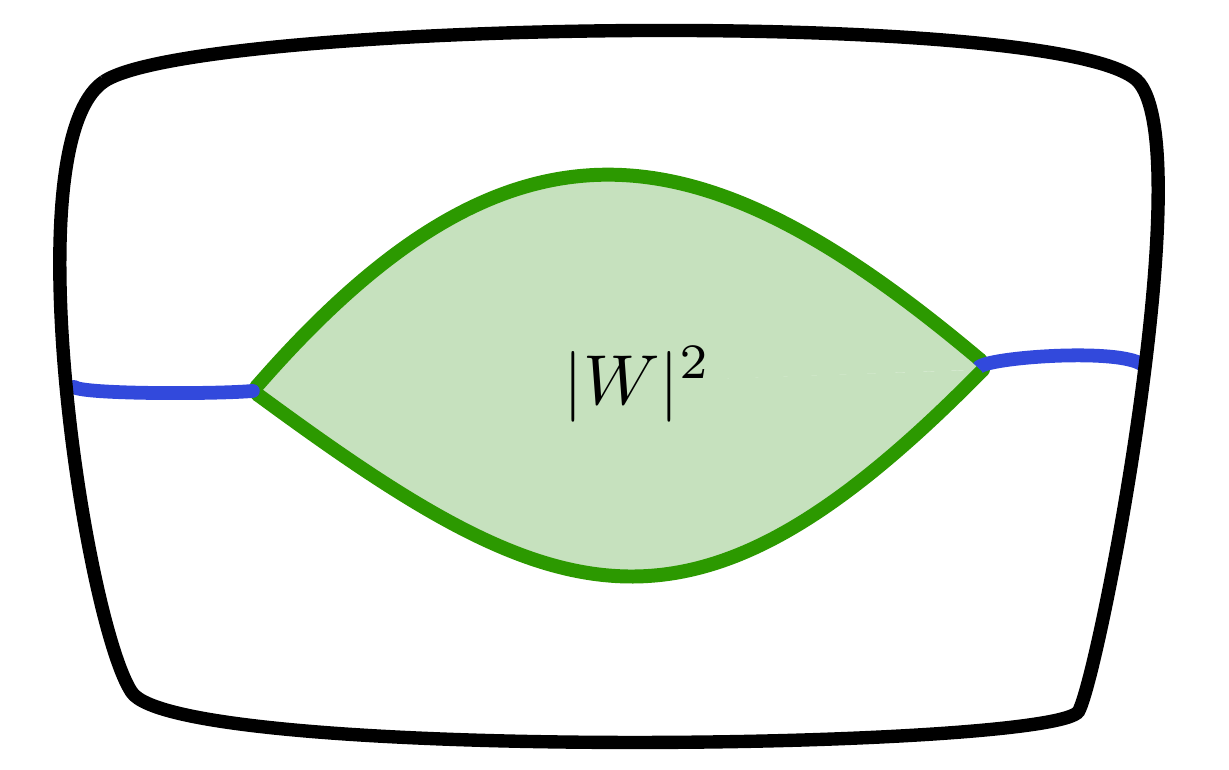}
\caption[]{The amplitudes of covariant LQG as probabilities for gravitational tunnelling.  Solving an initial value problem for Einstein's equations with Cauchy data the intrinsic and extrinsic geometry of the hypersurface formed by the blue hypersurfaces and the upper boundary surface, and evolving towards the direction in which the foliation time increases,  gives the upper--half of the spacetime. This is connected to the lower--half, which is the evolution in the past with Cauchy data the intrinsic and extrinsic geometry of the hypersurface formed by the blue hypersurfaces and the lower boundary surface.
}
\label{fig:conectingGRsolutions}
\end{figure}

In this manuscript, tools and ideas developed over the past decade were combined and supplemented in order to complete a first attempt at using covariant LQG to treat non-perturbatve singularity resolving phenomena as quantum transitions between geometries.

We have argued that transition amplitudes of covariant LQG describing geometry transition for non-perturbative phenomena are expected to behave as a product of terms
\begin{equation}
\vert W \vert^2 \sim e^{-(\gamma \zeta - \phi(\omega) )^2 / t}
\end{equation}
where $\zeta$ are the boost data and  $\omega$ the area data of the boundary state. The dihedral angles $\phi(\omega)$ code the extrinsic geometry of the geometrical configuration dominating the path integral.

The dimensionless semiclassicality parameter $t$ is proportional to a positive power of $\hbar$ and divided by the corresponding dimension--full area scale, so as to keep $t$ dimensionless. 

Thus, the transition amplitudes decay exponentially in the limit $\hbar \rightarrow 0$, unless the exponents vanish. This will be the case in a perturbative setting. For a non-perturbative phenomenon, the transition amplitude will decay exponentially in the semiclassical limit.

\medskip

In a realistic application, $t$ is a small,\emph{ finite} number. 

\medskip

We then have a picture of gravitational tunneling. The amplitudes do not vanish identically but give a small propability for the process to happen. Thus, solutions of general relativity generated by evolving in the future and in the past a cauchy surface that includes the upper and lower spacelike surface forming the boundary respectively, are connected by quantum theory. This is summarized in Figure \ref{fig:conectingGRsolutions}, repeated here for convenience from Chapter \ref{ch:gravTunneling}.

\begin{center}
\rule{0.5 \textwidth}{0.7pt}
\end{center}

The motivations leading to the current definition for the amplitudes of covariant LQG and a formal derivation of the EPRL ansatz where presented and discussed in Chapter \ref{ch:topBottom}.

 We proceeded to a review of results from the fixed--spins asymptotic analysis of the EPRL model and gave an independent derivation of the Han--Krajewski representation in Chapter \ref{ch:AsympAnalysis}.

We entered the main part of this thesis in Chapter \ref{ch:gravTunneling}. We reviewed and extended techniques that allow to construct a wave-packet of an embedded 3-geometry, peaked on both the intrinsic and extrinsic geometry of a given discrete geometry.

The above were put together in the second half of Chapter \ref{ch:gravTunneling}, where we showed how the fixed--spin asymptotics of spinfoam amplitudes can be used to perform the spin--sum, and estimate amplitudes decaying exponentially in the limit of large quantum numbers.

We saw that gravitational tunnelling emerges naturally from covariant LQG.

In Chapter \ref{ch:exteriorSpacetime} we constructed the Haggard-Rovelli spacetime. We gave an alternative formulation with a simplified form of the metric that reveals the role of the bounce time $T$ as a spacetime parameter. We discussed the relation with the original construction and pointed out some interesting properties of the spacetime.

We adopted and developed the point of view that the HR-spacetime presents an à priori well-defined setup for the calculation of a good observable from non-perturbative Quantum Gravity, the lifetime of the HR-spacetime. The lifetime is the characteristic time scale in which the spacetime metric fluctuates sufficiently for the transition of a trapped to an anti--trapped region to become likely. The question is well--posed and quantum theory should be in a position to provide an answer.

In Chapter \ref{ch:calculationOfAnObservable} we completed a first attempt at calculating the lifetime. We started by discussing the discretization procedure and the construction of a transition amplitude for a given choice of boundary hypersurfaces. 

Subsequently, we made an explicit choice of boundary hypersurfaces and 2-complex, to set-up a transition amplitude. The amplitude is finite, expressed and defined explicitly in terms of analytic functions. This effort has been part of a longer-term project to analyse numerically such amplitudes, led by S. Speziale in the Centre de Physique Th\'eorique in Marseille.

The results of Chapter \ref{ch:gravTunneling} were then used to estimate the lifetime. We gave an explicit estimate, demonstrating that such a calculation can be successfully completed with the currently available technology. 

We arrived at the conclusion that the lifetime scales linearly with the mass. We proposed that the quantity we have called the lifetime corresponds instead to a different physical timescale, the crossing--time, which is expected to scale linearly with the mass.

  A set of technical difficulties in interpretation were identified, due to an interplay between the truncation, the normalization, the semiclassical limit of LQG and the physical role of the Immirzi parameter. We discussed that our result may have to be understood in the limit where the Immirzi parameter is taken to be small. This is a subject currently under investigation.

The physical interpretation of our results is thus inconclusive and is discussed at the end of Chapter \ref{ch:calculationOfAnObservable}. The truncation issue, the existence of a maximum boost at the level of a link in the kinematical state space of LQG on a fixed graph, is further discussed in Appendix \ref{app:maximumBoost}. 

 We demonstrated how a calculation can be carried out while keeping the choice of boundary surface arbitrary, under the assumption that the boundary data are such that they determine a geometrical critical point at the level of the partial amplitude. While much remains to be understood, this provides evidence that the choice of boundary does not alter the physical conclusions drawn from a transition amplitude. 
 
   The choice of surface and discretization may result in the path--integral being dominated by configurations that admit an interpretation as either 4D Euclidean, 4D Lorentzian or 3D degenerate geometries. We saw that this surprising feature of the EPRL amplitudes results in a consistent estimate as one traverses all three kinds of geometries. 
   
Finally, we gave partial evidence for our results being independent of the choice of 2-complex, within the class of spinfoams that do not have interior faces.

\subsubsection{Future directions}
Many refinements and extensions of the calculation presented here are possible. Also, many questions have been left open and they deserve to be better understood. 

In closing, we present below a selected list of the directions that could be achievable in the short-term, in order to have a convincing prediction from covariant LQG for the scaling of the characteristic time in which quantum fluctuations of the metric make a transition of a black to a white hole geometry likely. 

\begin{figure}
\centering
\includegraphics[scale=0.6]{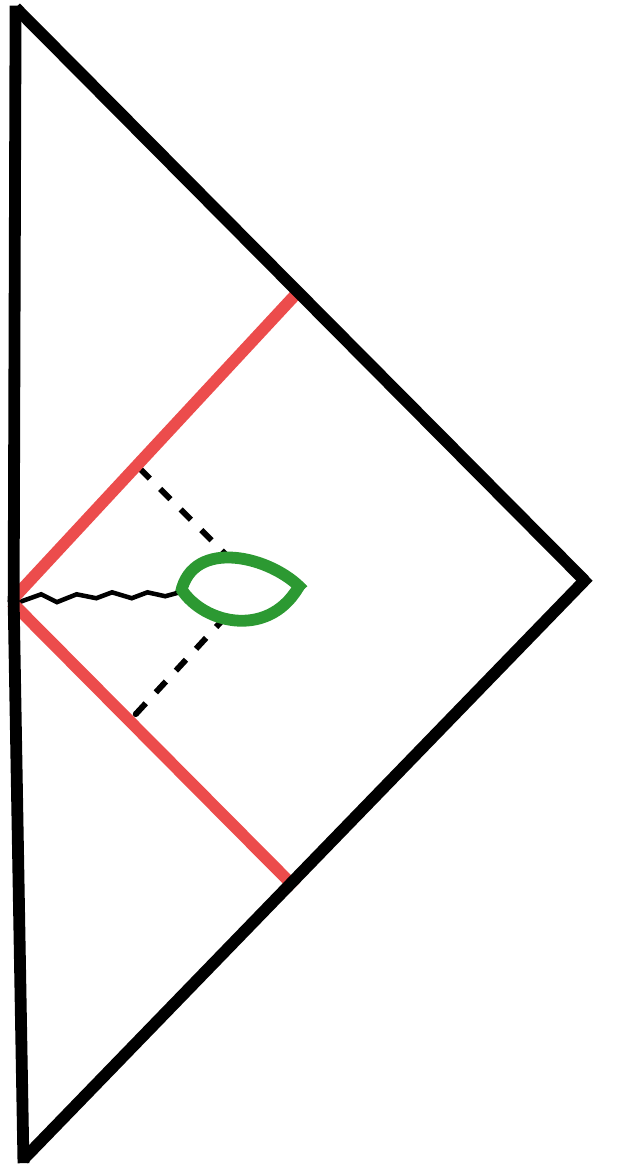}
\caption[A project for the future]{ The singularity surface can be allowed to be present in the HR-spacetime. Geodesics can be unambiguously connected across the singularity. The non-perturbative phenomenon appears to concern a region close to the horizon.
}
\label{fig:singeFireworks}
\end{figure}

\begin{itemize}
\item Provide a rigorous interpretation of the transition amplitude as a conditional probability with a measure $\alpha(m,T)$ providing the resolution of the identity on the subspace of physical states under consideration.

\item Investigate the role of the Immirzi parameter in macroscopic physics and in conjuction with the double scaling limit $\gamma \rightarrow 0$, $j \rightarrow \infty$.

\item Extend the techniques presented here to spinfoams with interior faces. Doing so will require better control of the so-called flatness issue. Interior faces will also allow the investigation of whether the existence of multiple semiclassical critical points, the cosine feature, alters the picture. The latter is expected to significantly affect the dynamics \cite{christodoulou_divergences_2013-1}. All but one semiclassical critical points are suppressed by the boundary state at the level of amplitudes without interior faces, but this is not expected to be the case for larger spinfoams.

\item Use a simplified symmetry reduced model. 
The use of cuboidal triangulations can be convenient. Currenlty, the group led by B. Bahr has studied extensively such configurations in the context of homegeneity and isotropy \cite{bahr_towards_2017,bahr_investigation_2016}. The asymptotics are essentially identical and the same qualitative behaviour is expected. Symmetry reduced models would simplify numerics and perhaps make them feasible, considerably simplify the Hessian and allow the study of the behaviour under 2-complex refinements \cite{bahr_numerical_2016}.

\item The bounce time is encoded only at the radius of the sphere at the tip of the quantum region for a family of natural boundary choices. In conjuction with the assumption and expectation that physical predictions from transition amplitudes do not depend on the choice of boundary hypersurface, this is an indication that for a more precise treatment, focus should be given in a region close to the horizon.

 A singularity hypersurface can be allowed to be explicitly present in the HR-spacetime without presenting interpretational difficulties. Synge pointed out as early as the 1950's that the Kruskal manifold is \emph{not} the maximal extension for the spherically symmetric solution in the sense of geodesic completeness\cite{synge_gravitational_1950}. Geodesics can be unambiguously continued across the singularity.
 
 This understanding was clarified in \cite{peeters_extended_1995} where the authors explain that it suffices to allow for a slight generalization of the notion of a manifold, that can accomodate geometries exhibiting singularities of the type present at the tip of a geometrical cone. The Schwartzschild singularity is of this type. The idea for this setup is sketched in Figure \ref{fig:singeFireworks}.
 
\item Encode the presence of a trapped and anti-trapped surface in the boundary state. The alternate monotonic behaviour of the boost angles with respect to the bounce time $T$ inside and outside the trapped and anti-trapped regions may provide a convenient way to do so. This step can be taken independently or in conjuction with the previous point.
\end{itemize}

\appendix 


\chapter{Elements of recoupling theory }
\label{app:recoupTheory}

\section{Integrating products of deltas and a flavour of recoupling theory} \label{appsec:integratingDeltas}
In this sub--appendix we demonstrate with a simple example how to integrate products of deltas over a compact group, using recoupling invariants. The procedure for \slt and the EPRL model are explained in the following two sub--appendices. These tools are used in Appendix \ref{app:derivationTreeAmplitudes} to derive the explicit form of the transition amplitude in Section \ref{sec:explicitAmp}.
 
In an expression of the form
\begin{equation} 
\int \left( \prod_e \dd g_e \right) \;\; \prod_f \tilde{\delta} \left( \prod_{e \in f} g_e \right)
\end{equation}
we will have integrations of the form 
\begin{equation}  \label{eq:fourDeltas}
\int \dd g_e \; \delta \left( g_e \, A \right) \delta \left( g_e \, B\right) \delta \left( g_e \,C\right) \delta \left( g_e \, D\right) 
\end{equation}
where $e$ is now fixed and $A,B,C,D$ will be products of the remaining group variables $g_{e'}$ that bound each of the four faces. An analogy on the real line would be 
\begin{equation}
\int \dd x \; \delta \left( x - A \right) \delta \left( x - B\right) \delta \left( x - C\right) \delta \left( x - D\right) 
\end{equation}
Such expressions are generally ill-defined over the reals. 

To demonstrate the general idea, and to see how \eqref{eq:fourDeltas} is well-defined and the difference between the two cases, we take $g_e$ to be an $SU(2)$ element and consider the product of only two deltas. 

We introduce a test function $f(x)$, to attempt to give meaning to the ``bad expression''
\begin{equation} 
\int \dd x \; \delta \left( x - A \right) \delta \left( x - B\right) f(x)
\end{equation}
Naively, we can guess the result by setting as a test function $\delta \left( x - B\right) f(x)$, giving
\begin{equation} 
f(A)\delta(A-B)
\end{equation}
which is clearly an ill-defined expression \footnote{In our analogy, there would be other integrations over $A$ and $B$ and one might attempt to remedy this problem by taking into account those integrations as well. However, there would also be other delta functions with $A$ and $B$ as arguments and we will run into the same problem.}. To see what goes wrong, use the fourier transform. We go through the steps since we will do the analogous procedure in the group variables below. For simplicity, we are neglecting constant factors in the definition of the fourier transform. We have

\begin{eqnarray} 
\int \! \dd x \; \delta \left( x - A \right) \delta \left( x - B\right) f(x) &=& \int \! \dd x \int \! \dd p_1 \, e^{i p_1(x-A)}\int \, \dd p_2 \, e^{i p_2(x-B)} f(x)   \nonumber \\
&=& \int \! \dd p_1 \dd p_2\, e^{-i p_2 B}\ e^{-i p_1 A}  \int \, \dd x \, e^{i (p_1 +p_2)x} f(x) \nonumber \\ 
&=& \int \! \dd p_2 \, e^{-i p_2 B} \int \, \dd p_1 \, e^{-i p_1 A} \hat{f}(p_1+p_2)
\nonumber  \\
&=& \int \! \dd p_2 \, e^{-i p_2 (B-A)} \int \, \dd p_1 \, e^{-i p_1 A} \hat{f}(p_1)
 \nonumber \\
&=& f(A) \int \! \dd p_2 \, e^{i p_2(A- B)} = f(A) \, \delta(A-B) \nonumber 
\end{eqnarray}  
where in the third equality we introduced the Fourier transform $\hat{f}$, in the fourth equality we shifted $p_1 \rightarrow p_1+p_2$, and in the fifth we took the inverse Fourier transform. 

The step that cannot be done when working with the group element as a configuration variable is the second to third equality, when we eliminated $x$. We do not get back the inverse fourier transform: we get \emph{intertwiners}, recoupling the different representations.

Let us do the analogous manipulation for the case where we are integrating over $SU(2)$, starting from 
\begin{equation} \label{eq:twoDeltasSU2}
\int \dd g \; \delta \left( g - A \right) \delta \left( g - B\right) f(g)
\end{equation}
As before, we employ harmonic analysis, which in this case is to use the Peter-Weyl theorem, which for our purposes can be stated as follows : square integrable functions of the group decompose as a direct sum of all unitary irreducible representations, the latter serving as an orthogonal basis for this space of functions. Then, $f$ can be expanded in Wigner's D-matrices (that provide the unitary irreducible representations for $SU(2)$) as 

\begin{equation} \label{eq:Peter-Weyl}
f(g) = \sum_{j} \sum_{mn} f^{j}_{mn} D^j_{mn}(g)
\end{equation}
The Peter-Weyl expansion of the delta functions over $SU(2)$ is given by the coefficients $f^{j}_{mn}$ set to the dimension $d_j=2j+1$ of the representation space $\mathcal{H}^j$ : 
\begin{equation} \label{eq:deltaPeter-Weyl}
f(g) = \sum_{j} \sum_{mn} d_j D^j_{mn}(g)
\end{equation}
These expansions are the analogue of Fourier expanding in the momentum representation, where the momentum variables are the spins $j$, which are now discrete. 

We may now expand the two delta functions and the function $f(h)$ and rewrite \eqref{eq:twoDeltasSU2} as

\begin{eqnarray} 
&\int & \!\! \dd g \; \delta \left( g - A \right) \delta \left( g - B\right) f(g) = \nonumber \\ &=& \!\!\! \int \!\! \dd g  \sum_{\{j\}}  \sum_{\{m n\}} \! d_{j_1} D^{j_1}_{m_1 n_1}\!(g) D^{j_1}_{m_1 n_1}\!(A) d_{j_2} D^{j_2}_{m_2 n_2}\!(g) D^{j_2}_{m_2 n_2}\!(B) f^{j_3}_{m_3 n_3} D^{j_3}_{m_3 n_3}\!(g) \nonumber \\
&=& \!\!\! \sum_{\{j\}}\! d_{j_1}\! d_{j_2}\! \sum_{\{m n\}}\! f^{j_3}_{m_3 n_3} \! \left( \int \!\! \dd g D^{j_1}_{m_1 n_1}\!(g)  D^{j_2}_{m_2 n_2}\!(g) D^{j_3}_{m_3 n_3}\!(g) \right)\! D^{j_1}_{m_1 n_1}\!(A)  \!D^{j_2}_{m_2 n_2}(B)  \nonumber \\
\end{eqnarray} 
We have introduced the notation $\sum_{k}$, which means summation over all configurations of the indices $k_i$ appearing in the expression, that is  $\sum_{k}= \sum_{k_1}\sum_{k_2}\sum_{k_3} \ldots$. T

This is how we end up with a spin state-sum model. The sum over configurations ${j}$ is the analogue of the sum $sum_{l}$ over length configurations in the naive Regge calculus path integral.

This is a simple instance of the appearance of a recoupling theory invariant. Because we are using an invariant measure, the Haar measure, the term in parenthesis is invariant under the action of $SU(2)$, meaning
\begin{equation}
\int  \dd g \,D^{j_1}(g g')\,  D^{j_2}(g g')\, D^{j_3} (g g') =
\int  \dd g \,D^{j_1}\!(g) \,  D^{j_2}\!(g ) \, D^{j_3} (g )
\end{equation}
where we skipped the magnetic indices for conciseness.

This follows from the properties of the Haar measure (it is the unique measure that has these properties and is determined by them for compact groups)
\begin{equation}
\int \dd g f(g) = \int \dd g f(g g') =\int \dd g f(g' g) = \int \dd g f(g^-1)
\end{equation}
where $g'$ is any other group element. The Haar measure is repeatedly used in the calculations of Section \ref{sec:explicitAmp} and is central to recoupling theory and the study of invariants. These properties can be expressed by slight abuse of notation as
\begin{equation}
\dd g   = \dd (g g')  = \dd (g' g) = \dd {g^{-1}}
\end{equation}
from where the name invariant measure is perhaps more apparent : the integration measure is invariant under the actions of the group; inversion, and left and right multiplication.

The last equality, although useful to keep in mind in practice, is an instance of either of the first two, by setting $g'= g^-1 g^-1$.
Compare these properties with the usual Lebesgue integration measure over the reals, which is also 
a Haar measure when addition is taken as the group composition 
\begin{equation}
\dd x  = \dd (x \pm y)  = \dd (y \pm x) 
\end{equation}

The integral 
\begin{equation} \label{eq:threeDmatrices}
\int \!\! \dd g D^{j_1}_{m_1 n_1}\!(g)  D^{j_2}_{m_2 n_2}\!(g) D^{j_3}_{m_3 n_3}\!(g)
\end{equation}
should be an object with nine indices, $\{j\}$,$\{m\}$ and $\{n\}$, that is invariant by the multiplication with three Wigner-D matrices $D^{j_i}_{m_i n_i}(g')$ and contraction in either set of magnetic indices. That is to say, by group multiplication by $g'$ on either side. The unique object with this property is called the 3-j symbol  
\begin{equation}
\sum_{\{m\}} D^{j_1}_{k_1 m_1}\!(g)  D^{j_2}_{k_2 m_2}\!(g) D^{j_3}_{k_3 m_3}\!(g) \begin{pmatrix}
j_1  & \; j_2 & \; j_3 \\
m_1  & \; m_2 & \; m_3 
\end{pmatrix} = \begin{pmatrix}
j_1  & \; j_2 & \; j_3 \\
k_1  & \; k_2 & \; k_3 
\end{pmatrix}
\end{equation}
and we may take this expression to be its definition.

 The integral in equation \eqref{eq:threeDmatrices}, since it has two sets of magnetic indices that can be simultaneously acted upon, is then just a product of two 3-j symbols
\begin{equation} 
\int \!\! \dd g D^{j_1}_{m_1 n_1}\!(g)  D^{j_2}_{m_2 n_2}\!(g) D^{j_3}_{m_3 n_3}\!(g) =  
\begin{pmatrix}
j_1  & \; j_2 & \; j_3 \\
m_1  & \; m_2 & \; m_3 
\end{pmatrix}
\begin{pmatrix}
j_1  & \; j_2 & \; j_3 \\
n_1  & \; n_2 & \; n_3 
\end{pmatrix}
\end{equation}
Anticipating the notation used in Appendix \ref{app:derivationTreeAmplitudes}, we also denote the $3-j$ symbol as 
\begin{equation}
i^{\{j\}}_{\{m\}} =
\begin{pmatrix}
j_1  & \; j_2 & \; j_3 \\
m_1  & \; m_2 & \; m_3 
\end{pmatrix}
\end{equation}

The fact that the $3-j$ symbol is unique implies that the dimension of the space of invariants intertwining between three representations is one-dimensional. This is no longer the case when tensoring four representations. 

Recall Figure \ref{fig:2Dtriang} and the discussion in Section secref: in four dimensions there are always exactly four faces attached to an edge of a 2-complex that is dual to a (topological) simplicial comlex. Thus, each \slt elements $g_e$ will appear exactly four times in the amplitude, becaus it is peaked up by four different delta functions that correspond to the four holonomies around the four faces. Similarly for  $g_{ve}$. As we see in the following two appendices, $g_{ve}$ is split into two $SU(2)$ elements and a boost. 

The integrations over the two pairs of four $SU(2)$ elements corresponding to a face, will give rise to two four-valent intertwiners. That is, objects in the space

\begin{equation}
\mathcal{V}^4_{j_i} = \rm{Inv}\left(\otimes_{i=1}^4 H^{j_i} \right) \ , \; i=1,2,3,4
\end{equation}

The basis for $\mathcal{V}^4_{j_i}$, an intertwiner basis, is labelled by another spin variable $K$,
taking values in 
\begin{equation}
\max[\vert j_1-j_2 \vert, \vert j_3-j_4 \vert] \leq K \leq \min[j_1+j_2,j_3+j_4]
\end{equation}
that is, the intertwiner space vanishes if the triangle inequalities between $j_1,j_2,K$ and $j_3,j_4,K$ are not satisfied.

Notice that for this we had to arrange the four spins in two pairs: this is a choice of recoupling scheme, in the case of particles it would be equivalent to saying that two particles with spin $j_1$ and $j_2$ interact to give two particles that transform under the $j_3$ and $j_4$ representations. Such choice is arbitrary but must be fixed in order to write the amplitudes explicitly in terms of intertwiners.

A basis for $\mathcal{V}^4_{j_i}$ is given by 

\begin{align}
& i^{J,\; j_1,j_2,j_3,j_4}_{\ \ \ m_1,m_2,m_3,m_4}=  \\  \nonumber
& = (-1)^{j_1-j_2+\mu}
\left(\begin{array}{lcl}j_1&j_2&J \\ m_1&m_2&\mu\end{array}\right)\left(\begin{array}{lcl}j_3&j_4&J \\ m_3&m_4&-\mu\end{array}\right)
\end{align}
with $\mu=-m_1-m_2=m_3+m_4$ and $\left(\begin{array}{lcl}j_1&j_2&j_3 \\ m_1&m_2&m_3\end{array}\right)$ are the Wigner 3j symbols.

Finally, the dimension of $\mathcal{V}_{j_i}$ is the range of values for $K$, that is
\begin{equation}
\dim\mathcal{V}^4_{j_i}= \leq \min[j_1+j_2,j_3+j_4]-\max[\vert j_1-j_2 \vert, \vert j_3-j_4 \vert]+1
\end{equation}

\section{Haar measure on \slt} \label{appsec:sl2cHaar}
 The integration measure $\mu(g_{ve})$ is the (left and right invariant) Haar measure over \slt. Because \slt is not compact, this measure is unique up to scaling (multiplication by a real number). In compact groups, we may include in the definition of the Haar measure the condition that the volume of the group is unit, or choose some other normalization. This would fix the scaling. In non-compact groups, this is impossible since the volume of the group is infinite. This ambiguity in the measure is absorbed in the arbitrary normalization $\mathcal{N}_\mathcal{C}$ of the EPRL amplitudes.
 
 The Haar measure $\mu(g)$ for $g\in\sl2c$ can be written explicitely as  (cite mux kraj)
 \begin{equation}
 \mu(g) = \frac{1}{\vert \delta \vert^2}\dd\beta \wedge \dd\bar{\beta} \wedge \dd\gamma \wedge  \dd\bar{\gamma} \wedge \dd\delta \wedge  \dd\bar{\delta} 
 \end{equation}
where $g = \begin{pmatrix}
  \alpha &\beta  \\
  \gamma & \delta  
 \end{pmatrix}$. The wedge product here should be understood in the usual sense of integrating over the parameter space in a specific parametrization of \slt. That is, we have eight real parameters for the four complex entries in $g$, and two conditions from $detg=1$, thus we are down to six parameters (\slt is six dimensional, an arbitrary Lorentz transformation is made of three rotations and three boosts). More explicitly, if $\alpha = a_R + i a_I $ with $a_R$ and  $a_I$ its real and imaginary parts, then 
\begin{equation}
  \dd a \wedge \dd \bar{a} = \dd (a_R + i a_I) \wedge \dd (a_R - i a_I) = -2 i \dd a_R \, \dd a_I
\end{equation}
In this parametrization, the invariance of $\mu(g)$ under $g \rightarrow g^\dagger$ is manifest.

Another form of the Haar measure, which is the one we use to get the explicit form of the amplitude in Section \ref{sec:explicitAmp}, uses the Cartan decomposition for \slt: for every $g \in \sl2c$, there exist $h$ and $h'$ in $SU(2)$, such that 
\begin{equation}
g= h A h'
\end{equation}  
where $A$ is a diagonal matrix with positive entries. Let us take a moment to show this and point out a subtlety.

Take any $g \in \sl2c$. Then $g g^\dagger$ is hermitian, and thus is unitarily equivalent to a diagonal matrix $A$ with real eigenvalues \footnote{This is the spectral theorem for hermitian operators, think observables in quantum theory.}. That is, there exists a $h \in SU(2)$ such that $g g^\dagger = h\, A\, h^\dagger$. Since $g$ and $h$ are of unit determinant, so is $A$. Thus, $A$ is of the form $A= \diag(a,1/a)$ with $a>0$. We thus may set $A(r)= \diag(e^r,e^{-r})$ with $r \in \mathbb{R}$. It is then straightforward to show that the group element $h'\equiv (h A(r/2))^{-1} g $ is in $SU(2)$, using that $A(r)A(r')=A(r+r')$ and $A(r)^{-1}=A(-r)$. Thus, we have shown that any $g \in a\sl2c$ can be written as 
\begin{equation} \label{eq:cartanDecompSL2C}
g= h A(r/2) h'
\end{equation}   
with $h,h' \in SU(2)$.

Then, the Haar measure $\mu(g)$ is two copies of the Haar measure $\mu(k)$ over $SU(2)$ and an integration measure over $r$. That is,

\begin{equation}
\mu(g)= \mu(h) \mu(h') \dd r f(r)
\end{equation}

where $f(r)$ is some real function of $r$. It turns out that $f(r)$ is given by 
\begin{equation}
f(r)=\sinh^2 r
\end{equation}
For the proof, see for instance \footnote{ http://www-users.math.umn.edu/~Garrett/m/v/SL2C.pdf}. Since this is an even function, we may integrate only over positive values of $r$ at the cost of a factor of two which we need not make explicit since $\mu(g)$ is only defined up to scaling. Thus, we may perform integrations over functions of \slt as

\begin{equation}
\int_\sl2c \!\!\! \mu(g) F(g) = \int_{SU(2)} \! \! \! \! \mu(h) \int_{SU(2)} \! \! \! \! \mu(h') \int_{\mathbb{R}^+} \! \! \dd r \; F(h \, A(r/2) \, h')
\end{equation}
where $A(r/2)=\diag(e^{r/2},e^{-r/2}) = e^{i r \frac{\sigma_3}{2}}$.

 Notice that there is something wrong with the dimensions in this formula. $SU(2)$ is three dimensional. We have integrations over two copies of $SU(2)$ and one integration over the reals, thus the parameter space is seven-dimensional. This is easily traced to the fact that that that the decomposition \eqref{eq:cartanDecompSL2C} is not uniquely defined. Simultaneously sending $h \rightarrow h e^{i \alpha \frac{\sigma_3}{2}}$ and $h' \rightarrow h e^{-i \alpha \frac{\sigma_3}{2}}$, defines the same $g$. 

This implies that one of the integrations is redundant. We should integrate only over $h$ and $h'$ that uniquely determine $g$.

We may see this explicitly by expanding $F(g)$ in unitary representations of the principal series. Then,  $F(g)$ will be a sum over representation matrices $D^{(p,k)}_{jmjm} (g)$. The Cartan decomposition of $D^{(p,k)}_{jmj'm'} (g)$ can be written explicitly in terms of Wigner-D matrices for $SU(2)$ as
\begin{equation}
D^{(p,k)}_{jmj'm'} (g) = \sum_n D^j_{mn}(h) d_{jj'n}^{(p,k)}(r) D^{j'}_{nm'}(h')
\end{equation}
and where the functions $d_{jj'm}^{(\rho,k)}$ are defined in Section \ref{sec:explicitAmp}, see (cite simone) for further details. 

Thus, the $SU(2)$ integrations may be isolated and we will have terms of the form
\begin{equation}
 \int_{SU(2)} \! \! \! \! \mu(h) \int_{SU(2)} \! \! \! \! \mu(h') \, D^j_{mn}(h) D^{j}_{nm'}(h') 
\end{equation}
Writing explicitly the Haar measure in the Euler angle parametrization, each $D^j_{mn}(h)$ can be written as $e^{-i m \alpha} d^j_{m n}(\beta) e^{-i n \gamma}$, with no summation implied and where $d$ are Wigner's small-d matrices. See Section \ref{sec:intrinsicCoherentStates} for a detailed explanation of this parametrization. Thus, the redundant integration can be remedied by inserting a delta function $\delta(\gamma-\gamma'+\tilde{\gamma})$ in the above equation when written in the Euler parametrization.

\section{\slt Clebsch-Gordan Coefficients for the EPRL model}
\label{appsec:CGsl2c}

In the EPRL model, \slt group elements appear only inside inner products of the form 
\begin{equation}
{}_\gamma\braket{jm \vert g_{ve'} g_{ve} \vert jm'}_\gamma
\end{equation}
see Section \ref{sec:ampsInAsAnform}.

In this appendix we see how the integrations over \slt can be explicitly performed and explain the relation of the objects appearing in Secion \ref{sec:explicitAmp} with the Clebsch-Gordan coefficients of \slt. 

 Unitary irreducible representations $H_{(k,\rho)}$ of \slt are labeled by a positive real number $\rho$ and a half-integer $k$, see also Section secref. 
The unitary representations of \slt and $SU(2)$ are related through the decomposition:

\[ H_{(k,\rho)} = \bigoplus_{j=k}^{\infty} H_j \]

The two Casimirs are $C_1 = L^2 - K^2$ and $C_2 = L \cdot K$ where $L^i$ and $K^i$   are rotation and boost generators respectively. We denote a basis of $H_{(k,\rho)}$ as $\ket{k\rho jm}$. On this basis, the Casimirs act as

\begin{align*}
C_1 \ket{k\rho jm} &= (k^2-\rho^2-1) \ket{k\rho jm} \\
C_2 \ket{k\rho jm} &= 2k\rho \ket{k\rho jm}
\end{align*}

The  simplicity constraints are applied, after the discretization, by the linear condition 

\[ D^i_f = K^i_f - \gamma L^i_f \]

on each face of the simplicial 2-complex.
The $SU(2)$ subgroup, analogue of a gauge in classical theory, is chosen arbitrarily at each edge of the 2-complex. These are called the linear simplicity constraints.

To impose these constraints we will (following Thiemann) use the master constraint

\[ M_f = D_f \cdot D_f = K^2 - 2\gamma K \cdot L + \gamma^2 L = (1+\gamma^2) L^2 -2\gamma C_2 - C_1 \]

We look for the minimum eigenvalue of the master constraint operator $M_f$, specifically:

\[ M_f \ket{k\rho jm} = \left((1+\gamma^2)j(j+1) -2\gamma k\rho + \rho^2+1-k^2\right) \ket{k\rho jm} \]

The minimum eigenvalue for fixed $j$ is attained when $\rho = \gamma j$ and $k=j$. Therefore, in what follows we will restrict the \slt representations to ones satisfying

\[ \rho = \gamma j \quad\quad k=j \]

for $j$ half-integer.

Now, consider the product space $H = H_{(k_1,\rho_1)} \otimes H_{(k_2,\rho_2)}$. In this product space there are two sets of orthogonal basis vectors. The first one is made up of the outer product of the basis vectors of the original spaces:

\[ \ket{k_1\rho_1 j_1 m_1, k_2 \rho_2 j_2 m_2} = \ket{k_1 \rho_1 j_1 m_1} \ket{k_2\rho_2 j_2 m_2} \]

The second is made up of the basis vectors $\ket{k\rho jm}$ that span $H_{(k,\rho)}$ contained in the tensor product space $H$.

The two bases are related by

\[ \ket{k\rho jm} = \sum_{j_1=k_1}^{\infty} \sum_{j_2=k_2}^{\infty} \sum_{m_1, m_2} \braket{k_1 \rho_1 j_1 m_1, k_2 \rho_2 j_2 m_2|k \rho j m} \ket{k_1 \rho_1 j_1 m_1} \ket{k_2 \rho_2 j_2 m_2} \]

The coefficients appearing in this formula are called the Clebsch-Gordan coefficients (CG coefficients). They can be defined in terms of an integral as follows:

\begin{align}
 &\braket{k_1 \rho_1 j_1 m_1, k_2 \rho_2 j_2 m_2|k \rho j m}  \braket{k_1 \rho_1 j'_1 m'_1, k_2 \rho_2 j'_2 m'_2|k \rho j' m'}^{*} = \nonumber \\
 &\int \dd g \overline{D^{k\rho}_{jmj'm'}(g)} D^{k_1 \rho_1}_{j_1 m_1 j'_1 m'_1}(g) D^{k_2 \rho_2}_{j_2 m_2 j'_2 m'_2}(g) 
\end{align}

To simplify the group integral, we use the decomposition $g = u e^{\frac{r}{2}\sigma_3} v$ where $u, v$ are $SU(2)$ elements and $r\geq0$. This decomposition written in terms of the Wigner matrices reads:

\[ D^{k\rho}_{jmj'm'} = \sum_p D^j_{mp}(u) d^{k\rho}_{jj'p}(r) D^{j'}_{pm'}(v) \]

Substituting and performing the $SU(2)$ integration we have

\begin{align} \label{eq:cgint}
&\braket{k_1 \rho_1 j_1 m_1, k_2 \rho_2 j_2 m_2|k \rho j m}  \braket{k_1 \rho_1 j'_1 m'_1, k_2 \rho_2 j'_2 m'_2|k \rho j' m'}^{*}  =  \nonumber \\ &\frac{1}{(2j+1)(2j'+1)} \threej{j_1}{m_1}{j_2}{m_2}{j}{m}  \threej{j_1}{m_1}{j_2}{m_2}{j'}{m'} \nonumber \\
& \sum_{n_1,n_2} \threej{j_1}{p_1}{j_2}{p_2}{j}{-p_1-p_2}  \threej{j_1}{p_1}{j_2}{p_2}{j'}{-p_1-p_2} \nonumber \\& \int \frac{\dd r \sinh^2 r}{4\pi} \overline{d^{k\rho}_{jj'p}(r)}d^{k_1 \rho_1}_{j_1j'_1p_1}(r)d^{k_2 \rho_2}_{j_2j'_2p_2}(r) \nonumber
\end{align}   


\chapter{Derivation of explicit form of amplitudes} 
\label{app:derivationTreeAmplitudes}
In this Appendix we show how to derive the final expression for the amplitude as in Section \ref{sec:emergenceOfGR}. The formula we arrive at applies to the following class of spinfoam 

\begin{itemize}
\item \emph{The 2-complex is dual to a topological simplicial complex.}
\item \emph{There are no integrations over $SL(2,\mathbb{C})$ elements living on internal edges} (internal edges are those that do not end on a node of the boundary but have two vertices of the bulk as endpoints). In the EPRL model we perform 
four integrations over $SL(2,\mathbb{C})$ at each five-valent vertex, with one of the group integrations being redundant and the group element set to $\ident$. We may choose not to perform the integration on an internal edge of the spinfoam. This is the case for the spinfoam in FIGREF.
\item  \emph{The boundary graph is four valent and can be oriented such that each node is the source of two links and the target of the two other links}. We assume such an orientation in what follows, see FIGREF for an example. This is not a crucial restriction for what follows and the notation is deliberately such that one can easily modify the formulas for boundary graphs that do not admit this orientation or have nodes of arbitrary valency on the boundary. All edges are oriented towards the boundary i.e. the vertex is their source, this is always possible for such spinfoams. These choices fix the orientation of the spinfoam.
\end{itemize}

 These spinfoams are such that all the combinatorics are delegated at the level of the boundary graph. This is because each $SL(2,\mathbb{C})$ element $g_e$ living on an edge $e$ naturally corresponds a node $\nn $, the target of the edge, and each face $f$ to the boundary link $\ell \in f$. We thus label the faces with their corresponding link $\ell$ (for a 2-complex dual to simplicial manifold there can be only one link per face).

\section*{Highest weight approximation}

We take as starting point the amplitude after the $SU(2)$ integrations have been performed and the arbitrary $SU(2)$ elements living on the boundary have been replaced with the coherent state data $H_\ell$: 

\begin{eqnarray}
W(H_\ell) &=& \int \! \left( \prod_\nn  \dd g_\nn  \right)\prod_\ell \sum_{j_\ell} w(j_\ell,\sigma_\ell) \tr^{j_\ell} \left( g^{-1}_{t(\ell)} \;H_\ell \; g_{s(\ell)} \right) 
\end{eqnarray}
where $w(j_\ell,\sigma_\ell)=d_{j_\ell} e^{-j_\ell(j_\ell+1)/2 \sigma_\ell}$. The amplitude is a function of the boundary data of the coherent states $H_\ell =  N_{t(\ell)}\; e^{z_\ell \sigma_3/2}\; N^{-1}_{s(\ell)}$ with $z_\ell = \eta_\ell + i \xi_\ell$. Note that as advertised all labels are boundary labels: the $g$'s and $N$'s are labelled by nodes and everything else by links. 

 Our goal is to rewrite the product over the group elements as a product over nodes. All the group elements we have are already labelled by nodes except the boost part $e^{z_\ell \sigma_3/2}$. We use the highest weight approximation (macroscopic areas)
\begin{equation} 
D^{j_\ell}_{a_\ell b_\ell}(e^{z_\ell \sigma_3/2}) \approx \delta^{j_\ell}_{a_\ell} \delta^{j_\ell}_{b_\ell} e^{z_\ell \,j_\ell\,/2}
\end{equation}
Then, $e^{z_\ell \sigma_3/2}$ simply contributes a multiplicative factor that can be pulled out  (it is easy to generalize these formulas for non-macroscopic areas by skipping this step) :
\begin{eqnarray}
\tr^{j_\ell} \left( g^{-1}_{t(\ell)} \;H_\ell \; g_{s(\ell)} \right) &=&  D^{j_\ell}_{h_\ell l_\ell a_\ell} \left(g^{-1}_{t(\ell)} \; N_{t(\ell)}\right)\;D^{j_\ell}_{a_\ell b_\ell}\left(e^{z_\ell  \sigma_3/2}\right) \;D^{j_\ell}_{b_\ell l_\ell h_\ell}\left( N^{-1}_{s(\ell)} \; g_{s(\ell)}\right) \\ 
&=& e^{z_\ell j_\ell /2}\;  D^{j_\ell}_{h_\ell l_\ell j_\ell} \left(g^{-1}_{t(\ell)} \; N_{t(\ell)}\right) \;D^{j_\ell}_{j_\ell l_\ell h_\ell}\left( N^{-1}_{s(\ell)} \; g_{s(\ell)}\right)
\end{eqnarray}
where the triple-index notation for the representation matrices is explained below. By pulling all the summations over the $j_\ell$ we rewrite the amplitude as:
\begin{align}
& W(H_\ell) =\sum_{\{j_\ell\}}\left( \prod_\ell  w(j_\ell,\sigma_\ell) e^{z_\ell j_\ell /2}\; \right) \; \nonumber \\ & \int \! \left( \prod_\nn  \dd g_\nn  \right)   \prod_\ell D^{j_\ell}_{l_\ell h_\ell j_\ell} \left(g^{-1}_{t(\ell)} \; N_{t(\ell)}\right) \;D^{j_\ell}_{j_\ell l_\ell h_\ell}\left( N^{-1}_{s(\ell)} \; g_{s(\ell)}\right) 
\end{align} 
We now have in the integrand a product of objects labelled by nodes. Since the nodes are four-valent and the links oriented as explained above, there are two $D^{j}_{l h j} \left(g^{-1} \; N \right)$  and two $D^{j}_{j l h } \left( N^{-1}\; g \right)$ terms corresponding to each node. Thus we have succeeded in rewriting the part of the amplitude involving group elements as a product of (functionally) identical integrals: 
\begin{equation} \label{eq:nodeProd}
\boxed{W(H_\ell) =\sum_{\{j_\ell\}}  w(\{j_\ell ,\sigma_\ell,z_\ell\}) \;\prod_\nn  \; \int \!  \dd g_\nn  \; \overrightarrow{\prod_{\ell \in \nn  }}\; D ^{(\ell)}\!\left(g_\nn  \; N_\nn \right)\;}
\end{equation}
where 
\begin{equation}
w(\{j_\ell ,\sigma_\ell,z_\ell\})=\prod_\ell  w(j_\ell,\sigma_\ell) e^{z_\ell j_\ell /2} = \prod_\ell d_{j_\ell} e^{z_\ell j_\ell /2-j_\ell(j_\ell+1)/2 \sigma_\ell}
\end{equation}
and the ``oriented product'' notation is a shortcut for
\begin{equation}
\overrightarrow{\prod_{\ell \in \nn  }}\; D^{(\ell)} \!\left(g_\nn  \; N_\nn \right) = \prod_{k=1}^2 
 D^{j_{\nn s_k}}_{ l_{\nn s_k} h_{\nn s_k} j_{\nn s_k}} \left(g^{-1}_\nn  \; N_\nn \right)
 \;D^{j_{\nn t_k}}_{j_{\nn t_k} l_{\nn t_k} h_{\nn t_k}}\left( N^{-1}_\nn  \; g_\nn \right) 
\end{equation}
where we are labelling each link with its source and target nodes, $\ell \equiv s(\ell) t(\ell) \equiv t(\ell) s(\ell) $ and the node $\nn $ is attached to four nodes $s_1 ,s_2 ,t_1 , t_2$ via links that have these nodes either as sources or as targets. 

Here is a good place to pause and explain how the multiplication of $SL(2,\mathbb{C})$ elements is performed. In each trace, we have initially two $SL(2,\mathbb{C})$ elements. When we split them we have 
\begin{equation}
D^{\gamma j,j}_{jmjn}\left(g g' \right) = D^{\gamma j,j}_{jmlh}\left(g \right) D^{\gamma j,j}_{lhjn}\left(g' \right) \equiv D^j_{mlh}\left(g \right) D^j_{lhn}\left(g'\right)
\end{equation}
with $l$ and $h$ summed (the ranges for the indices are $l\geq j$,\; $h \leq l$,\; $m,n \leq j$ with integer steps).
The triple-index notation for the multiplication of $SU(2)$ and $SL(2,\mathbb{C})$ elements means for instance 
\begin{equation}
D^{j}_{s l h}\left( N \; g\right) \equiv D^j_{s m}\left( N \right) D^{\gamma j,j}_{jmlh}\left( g \right) \equiv D^j_{s m}\left( N \right) D^j_{mlh}\left( g \right)
\end{equation}
 with summation over $m$.  
\section*{Decomposition of the $SL(2,\mathbb{C})$ integration}
The point of \eqref{eq:nodeProd} is that we can now focus on one of these integrals \footnote{However, do not forget that each representation matrix in \eqref{eq:nodeProd} has open indices that are contracted with those of another matrix that is in the integral over the corresponding node. For example, number the nodes by $\nn =1,2,\cdots$. If we are on the node ${\tt n =1}$ there is an index $\ell= ns_1$. Say $s_1=2$, then $\ell= 12=21$. Then there is another matrix in the integral over the node $n'=s_1=2$ with the same index where we would have for example $t'_1=\nn $ and $\ell'=\nn ' t'_1=21=12=s_1 \nn=\ell$.}:  
\begin{equation}
\int \!  \dd g_\nn  \; \overrightarrow{\prod_{\ell \in \nn  }}\; D ^{(\ell)}\!\left(g_\nn  \; N_\nn \right)
\end{equation}
What we will do next is to split the $SL(2,\mathbb{C}$ integration into two integrations over $SU(2)$ and one integration over a real parameter $r$. The $SU(2)$ integrations yield $SU(2)$ intertwiners, defined in terms of 3j-symbols, and we will be left with the integration over $r$.   

Each $SL(2,\mathbb{C})$ element can be written as $g_\nn  = u_\nn  e^{r_\nn  \sigma_3 /2} v^{-1}_\nn  $ with $u_\nn ,v_\nn  \in SU(2)$ and $r_\nn  \in(0,\infty)$ and the integration measure is given by $\dd g_\nn  = \dd u_\nn  \; \dd v_\nn  \; \dd r_\nn  \frac{\sinh^2 r}{4 \pi}  $ with $\dd u_\nn ,\, \dd v_\nn $ being Haar measures over $SU(2)$. We have in general:

\begin{equation}
D^{(\rho,k)}_{lnl'n'} = D^l_{np}(u) d_{ll'p}^{(\rho,k)} D^{l'}_{pn'}(v^{-1})
\end{equation}
where in the EPRL, $(\rho,k)=(\gamma j,j)$ (see Section secref and cite simone, crsv for the definition of $d_{jlp}^{(\rho,k)}$). 
We have for the ingoing links ( labelled as $\nn s_k$, thus $\nn$ is the target): 
\begin{eqnarray} \label{eq:inLinks}
D^{j}_{ l h j} \left(  g^{-1} N \right) &=& D^j_{l h m}\left( g^{-1} \right) D^j_{mj}\left( N \right) \nonumber \\
&=& D^l_{h p}\left(v \right) d_{ljp}(-r) D^j_{p m}\left(u^{-1} \right) D^j_{mj}\left( N \right) \nonumber \\
&=& (-1)^{m-p} D^l_{h p}\left(v \right) d_{ljp}(-r) D^j_{-m-p}\left(u \right) D^j_{mj}\left( N \right)
\end{eqnarray}
with $p$ summed and similarly for the outgoing links ( labelled as $\nn t_k$, thus $\nn$ is the source)
\begin{eqnarray} \label{eq:outLinks}
D^{j}_{ j l h} \left( N^{-1} g \right) &=&D^j_{jm}\left( N^{-1} \right) D^j_{m l h}\left( g \right) \nonumber \\
&=& D^j_{jm}\left( N^{-1} \right)\;D^j_{m p}\left(u \right) d_{jlp}(r) D^l_{p h}\left(v^{-1} \right)  \nonumber \\
&=& (-1)^{m-j} (-1)^{h-p} D^j_{-m-j}\left( N \right)\;D^j_{m p}\left(u \right) d_{jlp}(r) D^l_{-h-p }\left(v\right)  
\end{eqnarray}
We first do the integrations over $u$ and $v$:

\begin{align}
&\int_{{}_{SU(2)}} du_\nn   \prod_{k=1}^2 D^{j_{\nn s_k} }_{-m_{\nn s_k} -p_{\nn s_k}}(u) D^{j_{\nn t_k} }_{m_{\nn t_k} p_{\nn t_k}}(u)= \nonumber \\
 &=\sum_{J_\nn} (2J_\nn +1) \left( i^{J_\nn,\ \ j_{\nn s_1},\ \ \ j_{\nn s_2},\ \ j_{\nn t_1},\ \ j_{\nn t_2}}_{ \ \ -m_{\nn s_1},\ -m_{\nn s_2},\ m_{\nn t_1},\ m_{\nn t_2}} \right) \; \;\left(
 i^{J_\nn,\ \ j_{\nn s_1},\ \ \ j_{\nn s_2},\ \ j_{\nn t_1},\ \ j_{\nn t_2}}_{ \ \ -p_{\nn s_1},\ -p_{\nn s_2},\ p_{\nn t_1},\ p_{\nn t_2}} \right)\; \;   \nonumber \\
 & \equiv \sum_{J_\nn} d_{J_\nn}\; i^{J_\nn,\, j_\nn}_{\ \ \ \overrightarrow{m}_\nn} \; i^{J_\nn,\, j_\nn}_{\ \ \ \overrightarrow{p}_\nn}
\end{align}
where we have introduced the following notation: \emph{Lower case indices labelled by a node} as above are multi-indices of multiplicity equal to the node valency. Note that lower case (single) indices are labelled by a link (by two nodes) while upper case indices are labelled by a single node.  The arrow indicates that there is a sign rule pertaining to whether the node is the source or target to the link. A right arrow means that indices on links ingoing to $\nn$ (of the form $\nn s_k$) get a sign while a left arrow means that indices on links outgoing from $\nn$  (of the form $\nn t_k$) get a minus sign. Explicitly, $j_\nn=\{ j_{\nn s_1}, j_{\nn s_2}, j_{\nn t_1}, j_{\nn t_2} \}$, $\overrightarrow{p}_\nn=\{ -p_{\nn s_1}, -p_{\nn s_2}, p_{\nn t_1}, p_{\nn t_2} \}$ and $\overrightarrow{m}_\nn=\{ -m_{\nn s_1}, -m_{\nn s_2}, m_{\nn t_1}, m_{\nn t_2} \}$. 

\bigskip
For the $v$ integration we have:
\begin{align}
&\int_{{}_{SU(2)}} dv_\nn   \prod_{k=1}^2 D^{l_{\nn s_k} }_{h_{\nn s_k} p_{\nn s_k}}(v) D^{l_{\nn s_k} }_{-h_{\nn t_k} -p_{\nn t_k}}(v)= \\
 & = \sum_{K_\nn} d_{K_\nn}\; i^{K_\nn,\, l_\nn}_{\ \ \ \overleftarrow{h}_\nn}\; i^{K_\nn,\, l_\nn}_{\ \ \ \overleftarrow{p}_\nn}
\end{align}
with $\overleftarrow{p}_\nn=\{ p_{\nn s_1}, p_{\nn s_2}, -p_{\nn t_1}, -p_{\nn t_2} \}$ and $\overleftarrow{h}_\nn=\{ h_{\nn s_1}, h_{\nn s_2}, -h_{\nn t_1}, -h_{\nn t_2} \}$.

\bigskip
Grouping the $d(r)$ functions we have\footnote{We may exchange $d_{l_{\nn s_k} j_{\nn s_k} p_{\nn s_k}}\!(-r_\nn)$ with $d_{j_{\nn t_k} l_{\nn t_k} p_{\nn t_k}}\!(r_\nn)$ at the cost of some $\Gamma$ functions appearing cite ruhl. }

\begin{equation}
\int  \dd r_\nn \, \frac{\sinh^2r_\nn}{4 \pi}\, \prod_{k=1}^2 d_{j_{\nn t_k} l_{\nn t_k} p_{\nn t_k}}\!(r_\nn)\ \ d_{l_{\nn s_k} j_{\nn s_k} p_{\nn s_k}}\!(-r_\nn)
\end{equation}

\bigskip

\section*{Rearrangement}

\emph{Sign factors}: In \eqref{eq:inLinks} and \eqref{eq:outLinks} there is a $\; (-1)^{m_{\nn s_1}+m_{\nn s_2}-p_{\nn s_1}-p_{\nn s_2}}$ factor and a $\;(-1)^{m_{\nn t_1}+m_{\nn t_2}-p_{\nn t_1}+p_{\nn t_2}}$. Because $m$ and $p$ are magnetic indices of an SU(2) Wigner matrix, $m-p$ is an integer and thus $(-1)^{m-p}=(-1)^{-m+p}$. We can thus write them as $(-1)^M (-1)^{-P}$ with $M=-m_{\nn s_1}-m_{\nn s_2}+m_{\nn t_1}-m_{\nn t_2}$ and $P=-p_{\nn s_1}-p_{\nn s_2}+p_{\nn t_1}+p_{\nn t_2}$. The intertwiners impose that $P=0$ and $M=0$, so these factors are unit: we have that $i^{K_\nn,\, l_\nn}_{\ \ \ \overleftarrow{p}_\nn} = i^{K_\nn,\, l_\nn}_{\ \ \ \overleftarrow{p}_\nn} \delta(-P)$ and $i^{J_\nn,\, j_\nn}_{\ \ \ \overrightarrow{m}_\nn} = i^{J_\nn,\, j_\nn}_{\ \ \ \overrightarrow{m}_\nn} \delta(M)$.

\bigskip

Each node gives four intertwiners (they do not correspond to the four links) two from the $u$ integration and two from the $v$ integration. One from each has the $p$ indices that are contracted with the $d(r)$ functions. The other one from $v$ is contracted with the normals and the one left from $u$ has the $h$ indices that are contracted with intertwiners from \emph{other} nodes. This gives a $3(E-1)j$ symbol. We group all the objects, via some definitions, so that all the lower case (link) indices are hidden (except the $j$'s):   

\begin{equation}
N^{J_\nn \, j_\nn} \equiv \prod_{k=1}^2 \; D^{j_{\nn s_k}}_{m_{\nn s_k} j_{\nn s_k}}\!\left( N_{\nn s_k} \right)\; D^{j_{\nn s_k}}_{-m_{\nn t_k}-j_{\nn t_k}}\!\left( N_{\nn t_k} \right) \;\;i^{J_\nn,\, j_\nn}_{\ \ \ \overrightarrow{m}_\nn}
\end{equation}

\begin{align}
& f^{J_\nn \, K_\nn }_{l_\nn \, j_n} \equiv \; d_{J_\nn} \; i^{J_\nn,\, j_\nn}_{\ \ \ \overrightarrow{p}_\nn}\;\;\left( \int \dd r_\nn \, \frac{\sinh^2r_\nn}{4 \pi} \, \prod_{k=1}^2 d_{j_{\nn t_k} l_{\nn t_k} p_{\nn t_k}}\!(r_\nn)\ \ d_{l_{\nn s_k} j_{\nn s_k} p_{\nn s_k}}\!(-r_\nn) \right) \nonumber \\ & \;\;i^{K_\nn,\, l_\nn}_{\ \ \ \overleftarrow{p}_\nn}\;d_{K_\nn}
\end{align}

\bigskip

We can now rewrite $\eqref{eq:nodeProd}$ as 
\begin{equation} 
W(H_\ell) =\sum_{\{j_\ell\}}  w(\{j_\ell ,\sigma_\ell,z_\ell\}) \;\prod_\nn  (-1)^{J+H} N^{J_\nn \, j_\nn} \; f^{J_\nn \, K_\nn }_{l_\nn \,j_\nn}\; i^{J_\nn,\, l_\nn}_{\ \ \  \overleftarrow{h}_\nn}
\end{equation}
with summations implied over all indices at the level of the node, $J=j_{\nn t_1}+j_{\nn t_2}$ and $H=h_{\nn t_1}+h_{\nn t_2}$. Thus we finally have with all summations explicit

\begin{align} \label{eq:final}
& W(H_\ell) =\sum_{\{j_\ell\}}  w(\{j_\ell ,\sigma_\ell,z_\ell\}) (-1)^{\sum_\ell j_\ell} \; \sum_{{\{l_\ell\},\{J_\nn \},\{K_\nn \}}} \; \left(\prod_\nn  N^{J_\nn \, j_\nn} \; f^{J_\nn \, K_\nn}_{j_\nn \, l_\nn}\;\right) \ \times\nonumber \\  & \times \left(3(E-1)j\right)^{\{K_\nn \},\, \{l_\ell\}}
\end{align}  
where we defined 
\begin{equation}
\left(3(E-1)j\right)^{\{J_\nn \},\, \{l_\ell\}} =\sum_{\{h_\ell \}} (-1)^{\sum_\ell h_\ell} \prod_\nn \;\, \,i^{J_\nn,\, l_\nn}_{\ \ \ \overleftarrow{h}_\nn}
\end{equation}

The amplitude as in equation \eqref{eq:final} consists of summations and real integrations over functions which have known closed analytic forms. There is one infinite summation involved, the one over the $l$ indices. 

The integrations over the $SU(2)$ elements are performed as
\begin{align}
\int_{{}_{SU(2)}} dU  \otimes_a & D^{j_a}_{m_an_a}(U)= \\ &=\sum_J (2J +1) i^{J,j_1,j_{2},j_{3},j_{4}}_{\ \ \ m_1,m_2,m_3,m_4} i^{J,j_1,j_{2},j_{3},j_{4}}_{\ \ \ n_1,n_2,n_3,n_4}   \nonumber
\end{align}
where the basis of four-valent intertwiners is defined as 
\begin{align}
& i^{J,j_1,j_{2},j_{3},j_{4}}_{\ \ \ m_1,m_2,m_3,m_4}=  \\  \nonumber
& =\sum_{\mu} (-1)^{j_1-j_2+\mu}
\left(\begin{array}{lcl}j_1&j_2&J \\ m_1&m_2&\mu\end{array}\right)\left(\begin{array}{lcl}j_3&j_4&J \\ m_3&m_4&-\mu\end{array}\right)
\end{align}
and $\left(\begin{array}{lcl}j_1&j_2&j_3 \\ m_1&m_2&m_3\end{array}\right)$ are the Wigner 3j symbols.


\chapter{Coherent state representation} 
 \label{app:cohStateRepEPRL}
The coherent-state path-integral representation for the EPRL amplitudes is the usual form in which asymptotic analysis is undertaken. 

In the spinor representation, \eqref{eq:faceAmpSpinors}, the amplitude is composed of a product of terms of the form 
\begin{equation}
\braket{z\vert g \vert jm}_\gamma \equiv f^j_m(g^T z)
\end{equation} 
which we may write simply as 
\begin{equation}
\braket{Z \vert jm}_\gamma \equiv f^j_m(Z)
\end{equation} 
where $Z=g^T z$. Making the $Y_\gamma$ map explicit, we have 
\begin{equation}
\braket{Z \vert Y_\gamma \vert jm}
\end{equation}

We need only calculate
\begin{equation}
\braket{Z \vert j \vec{k}}_\gamma
\end{equation}
where $\vec{k}\in \mathbb{CP}^1$. We use the definition of coherent states $\vec{k} \triangleright \ket{ j j} \equiv \ket{\vec{k} j} $ of Section secref.

\begin{eqnarray}
\braket{Z \vert j \vec{k}}_\gamma &=& \braket{Z \vert Y_\gamma \bigg(k \triangleright \vert j j } \bigg) \nonumber \\
&=& \braket{Z \vert Y_\gamma \bigg(\sum_m D^j_{mj}(H(k)) \; \vert jm } \bigg) \nonumber \\
&=& \sum_m D^j_{mj}(H(k)) \braket{Z \vert jm }_\gamma \nonumber \\
&=& \frac{\sqrt{d_j}}{\pi} \braket{Z\vert Z}^{j(i\gamma-1)-1} \sum_m D^j_{mj}(H(k))  D^j_{mj}(H(Z)) \nonumber \\
&=& \frac{\sqrt{d_j}}{\pi} \braket{Z\vert Z}^{j(i\gamma-1)-1} \sum_m D^j_{jm}(H^\dagger(\bar{k}))  D^j_{mj}(H(Z)) \nonumber \\
&=& \frac{\sqrt{d_j}}{\pi} \braket{Z\vert Z}^{j(i\gamma-1)-1} D^j_{jj}(H^\dagger(\bar{k}) H(Z))\nonumber \\
&=& \frac{\sqrt{d_j}}{\pi} \braket{Z\vert Z}^{j(i\gamma-1)-1} \braket{\bar{k}\vert Z}^{2j} \end{eqnarray}

From this, it is straightforward to see that inserting the resolution of the identity for $\ket{\vec{k} j}$  in equation \eqref{eq:linkToCohStateRep} we get the amplitude in the coherent state representation with
\begin{equation}
F_f(\{g_f\},\{z_f\},\{k_f\}) = \sum_{v \in f} \log \frac{ \braket{k_{ef}\vert Z_{vef} }^2 \braket{Z_{ve'f}\vert k_{e'f}  }^2}{\braket{Z_{vef}\vert Z_{vef} }  \braket{Z_{ve'f}\vert Z_{ve'f} }} +i \gamma \log \frac{\braket{Z_{ve'f}\vert Z_{ve'f}}}{\braket{Z_{vef}\vert Z_{vef}} } 
\end{equation}
and the face amplitude will now include $E \times F$ integrations over $S^2$ (over the $k_{ef}$), on top of the $V \times E$ integrations over \slt  and the $V \times F$  integrations over $\mathbb{CP}1$.

\chapter{Spinors and the Riemann sphere} 
\label{app:spinorsAndRiemannSphere}

\newcommand{\col}[2]{
\ensuremath{
\begin{pmatrix}
{#1} \\ {#2}
\end{pmatrix}
}
}

\def\braj#1{\mathinner{[{#1}|}}
\def\ketj#1{\mathinner{|{#1}]}}
\def\brajket#1{\mathinner{[{#1}\rangle}}
\def\braketj#1{\mathinner{\langle{#1}]}}
\def\brajketj#1{\mathinner{[{#1}]}}

A spinor $z$ is an element of $\mathbb{C}^2$, which we think of as a column vector. We denote them in four different ways:
\begin{equation}
z = \ket{z} = \col{z^0}{z^1} = z^A
\end{equation}
where $z^0,z^1 \in \mathbb{C}$.

The space of spinors carries the fundamental representation of $SU(2)$. There are two (and only two) independent quadratic forms that are invariant under the action of $SU(2)$ on both arguments. The sesquilinear scalar product
\begin{equation}
\braket{z|w} = \bar{z}^0 w^0 + \bar{z}^1 w^1 = \bar{z}^A w^B \delta_{AB} 
\end{equation}
and an antisymmetric bilinear form
\begin{equation}
\brajket{z|w} = z^0 w^1 - z^1 w^0 = z^A w^B \epsilon_{AB}.
\end{equation}

The antisymmetric product $\brajket{z|w}$ can be written by introducing the parity map $J$ 
\begin{equation}
J \col{z^0}{z^1} = \col{-\bar{z}^1}{\bar{z}^0} = J \ket{z}= \ket{Jz} \equiv \ketj{z}  
\end{equation}

Then,
\begin{equation} \label{eq:parity}
\braket{z|\vec{\sigma}|z}=-\brajketj{z|\vec{\sigma}|z}
\end{equation} 
which earns $J$ its name, parity. As we see below, this relation implies that the spinors $\ket{z}$ and $\ket{J z}$ are mapped to a Euclidean 3D vector and its parity inverse respectively.

Evidently, $\braketj{z|w} = -\braketj{w|z}$. Further note that $\braketj{z|w} = -\overline{ \brajket{z|w} }$, which implies the relations $$\brajket{z|w} \braketj{z|w} = \braketj{z|w} \brajket{z|w} = -|\brajket{z|w}|^2 =-|\braketj{z|w}|^2 $$ 
Also, $\brajketj{w|z} =\braket{w|z}$.

A spinor $z$ can be traded for a 3D vector in Euclidean space by the definition
\begin{equation} \label{eq:Xzvector}
\vec{X}(z) = \frac{1}{2} \braket{z|\vec{\sigma}|z} = \frac{1}{2} \bar{z}_A \; \sigma^A_i{}_B \; z^B= \begin{pmatrix}
\re(z^0 \bar{z}^1)\\ -\im(z^0 \bar{z}^1) \\ \frac{|z^0|^2 - |z^1|^2}{2} 
\end{pmatrix}
\end{equation} 
The four independent real parameters in $z$ have been recast in the three components of $\vec{X(z)}$ and there is a phase ambiguity in defining $\ket{z}$ from $\vec{X(z)}$. The relation is not invertible unless we make a choice of section in the Hopf fibration, see Section secref. 

We will see how to to understand the last expression as the components of a vector $\vec{X(z)}$ in a Cartesian coordinate system.

\bigskip

Define the variable $\zeta \equiv \frac{z^0}{z^1}$ and the $SU(2)$ matrix
\begin{equation}
n(\zeta)=\frac{1}{\sqrt{1+|\zeta|^2}} \begin{pmatrix}
1 & \zeta \\ -\bar{\zeta} & 1
\end{pmatrix}
\end{equation}
In terms of the canonical basis of $\mathbb{C}^2$, $\ket{+}=\col{1}{0} \;,\; \ket{-}=\col{0}{1}$, we have
\begin{eqnarray} \label{eq:ketandketj}
\ket{z}&=& \sqrt{\braket{z|z}}\;e^{i \arg{z_1}}\;n(\zeta)\;\ket{-}   \\
\ket{z}&=& \sqrt{\braket{z|z}}\;e^{i \arg{z_0}}\;n(-1/\bar{\zeta})\;\ket{+}  
\end{eqnarray}
and
\begin{eqnarray} \label{eq:ketandketj}
\ketj{z}&=& \sqrt{\braket{z|z}}\;e^{-i \arg{z_0}}\;n(-1/\bar{\zeta})\;\ket{-}\\ 
\ketj{z}&=& - \sqrt{\braket{z|z}}\;e^{-i \arg{z_1}}\;n(\zeta)\;\ket{+}  
\end{eqnarray}

From these relations, it follows that $\braket{z|\vec{\sigma}|z}$ and its parity tranform $\brajketj{z|\vec{\sigma}|z}$ are related by $\zeta \rightarrow -1/\bar{\zeta}$ as:

\begin{eqnarray}
\frac{\braket{z|\vec{\sigma}|z}}{\braket{z|z}} &=&\braket{-|n^\dagger(\zeta) \; \vec{\sigma} \; n(\zeta)|-} = \braket{+|n^\dagger(-1/\bar{\zeta}) \; \vec{\sigma} \; n(-1/\bar{\zeta})|+}\nonumber \\
\\
-\frac{\braket{z|\vec{\sigma}|z}}{\braket{z|z}}=\frac{\brajketj{z|\vec{\sigma}|z}}{{\braket{z|z}}}&=&\braket{-|n^\dagger(-1/\bar{\zeta}) \; \vec{\sigma} \; n(-1/\bar{\zeta})|-}= \braket{+|n^\dagger(\zeta) \; \vec{\sigma} \; n(\zeta)|+}  \nonumber \\
\end{eqnarray}

To make contact with the usual Euclidean space in Cartesian coordinates, we define a normalized vector $\vec{N}(\zeta)$. To do the calculation, note the relation $ |z^1|^2=\frac{\braket{z|z}}{1+|\zeta|^2}$.

\begin{equation} \label{eq:normal}
\vec{N}(\zeta) = \frac{\braket{z|\vec{\sigma}|z}}{\braket{z|z}}=\braket{-|n^\dagger(\zeta) \; \vec{\sigma} \; n(\zeta)|-} = \frac{1}{1+|\zeta|^2} \begin{pmatrix}
2 \re \zeta \\ - 2 \im \zeta \\ |\zeta|^2 -1
\end{pmatrix}
\end{equation}
We see that we are in effect describing the Riemann sphere, with an annoying subtlety: the $y$-axis is inverted. 

\bigskip

Here is the usual Riemann sphere construction: Take a 3D Euclidean space in $x,y,z$ coordinates and identify $z=0$ with the complex plane. A complex number $\zeta'$ lives on $z=0$ with its real and imaginary part being on the positive $x$- and $y$-axis respectively. We place a unit sphere on the origin with its north pole $NP$ on the positive z-axis. 

Then, one geometrically defines the projection from the sphere to the complex plane via the straight line that passes from $NP$ and the point $\vec{x}=(x \; y \; z)^T$ on the unit sphere ( $\vec{x} \cdot \vec{x} = 1$ ). The intersection of this line with the plane defines $\zeta'$. The point on the sphere denoted as $(\beta,\alpha)$ in spherical coordinates. $\beta$ is the zenith, the angle with the positive $z$-axis, and $\alpha$ the azimith, the angle with the projection of the vector to the $x-y$ plane with the $x$-axis. See Figure \ref{fig:thetaPhiConvention} for this convention.

 Explicitly, the projection $\pi:S^2-\{NP\}\rightarrow\mathbb{C} $ is given by

\begin{equation}
\pi(\vec{x})=\pi(\beta,\alpha)= \frac{x+iy}{1-z} = \frac{\sin \beta\, e^{i\alpha}}{1-\cos\beta} = \cot\frac{\beta}{2}  e^{i\alpha}=\zeta'
\end{equation}

\begin{equation}
\pi^{-1}(\zeta')=\frac{1}{1+|\zeta'|^2} \begin{pmatrix}
2 \re \zeta' \\ 2 \im \zeta' \\ |\zeta'|^2 -1\end{pmatrix} = \begin{pmatrix}
\cos\alpha \sin\beta \\ \sin\alpha \sin\beta \\ \cos\beta \end{pmatrix}
\end{equation}
The first formula is immediate to see geometrically by using similar triangles. It is then easily inverted by noticing that $1+|\zeta|^2 = \frac{2}{1-z}$

By comparison with \eqref{eq:normal}, we now know how to make the correspondance with Euclidean vectors (directions): given the $\alpha,\beta$ as above, the same vector, with its components understood as $(x,y,z)$ components, is given by $N(\zeta)$  with 

\begin{equation}
\zeta = \bar{\zeta'} = \cot \frac{\beta}{2}  e^{-i\alpha}
\end{equation}
In other words, with $\zeta$ as above we have that 

\begin{equation}
\vec{N}(\zeta) = \begin{pmatrix}
\cos\alpha \sin\beta \\ \sin\alpha \sin\beta \\ \cos\beta \end{pmatrix} = \begin{pmatrix}
 n^x(\alpha,\beta)\\ -n^y(\alpha,\beta) \\ n^z(\alpha,\beta)  \end{pmatrix}
\end{equation}
Notice the minus in the $y$ component of $\vec{n}^y(\alpha,\beta)$.

\bigskip

\subsubsection{Conventions}
$$\sigma^A_B \sigma^C_D = \delta^A_B \delta^C_D - 2 \epsilon^{AC} \epsilon_{BD}$$ 
$$\epsilon_{AB}=\begin{pmatrix}
0 & 1\\-1 & 0\\ \end{pmatrix}=\epsilon^{AB}$$
$$\delta_{AB}\epsilon^{BC}\delta_{CD}=\epsilon_{AD}$$
$$\sigma_1^A{}_B=\begin{pmatrix}
0 & 1\\1 & 0\\ \end{pmatrix} \,,\; \sigma_2^A{}_B=\begin{pmatrix}
0 & -i\\i & 0\\ \end{pmatrix} \,,\; \sigma_3^A{}_B=\begin{pmatrix}
1 & 0\\0 & -1\\ \end{pmatrix}  $$

The ${}_A / {}^A$ rule:
$$ z_A = \epsilon_{AB}z^B$$
$$ z^A = z_B\epsilon^{BA}$$




\chapter{Maximum boost in LQG and a hint for maximum extrinsic curvature} 
 \label{app:maximumBoost}
In this appendix we reproduce an unpublished note by C. Rovelli, S. Speziale and the author.

\begin{center}
\rule{0.5 \textwidth}{0.4pt}
\end{center}
 
Loop quantum gravity predicts the existence of a maximum relative velocity between two adjacent space quanta. This can be taken as a hint that loop quantum gravity predicts the existence of a maximum value for the \emph{extrinsic} curvature.

\noindent The Planck scale provides a lower bound for the scale of physical geometry. This is clearly realised in loop quantum gravity, where, for instance, physical areas smaller than the Planck scale are forbidden, in the same sense in which angular momenta smaller than $\hbar/2$ are forbidden by conventional quantum theory. It is plausible that quantum gravity could equally provide an upper bound for physical curvature.  In loop quantum gravity, the  cosmological bounce \cite{ashtekar_loop_2009} and the possibility of a black-hole bounce appear to support the hypothesis of an upper bound to curvature in Nature.  

Here we analyse the \emph{extrinsic} curvature in the standard kinematics of loop quantum gravity, and find evidence that it is predicted to be bound. The mechanism yielding the bound is the characteristic interplay between the compactness of $SU(2)$ and the non compactness of $SL(2,\mathbb{C})$ on which the loop theory relies. 

More surprisingly, we also found an upper limit to  the relative velocity (the boost) between two adjacent space quanta.  This comes as a surprise because (in $c\!=\!1$ units) relative velocity is dimensionless and therefore independent of the Planck length, so that one might worry that this upper limit persists in the classical limit. This is not the case, however, because in the classical limit there can be an arbitrary number of space quanta between any two given points, and the upper limit on the boost between any two of them becomes irrelevant. 

\smallskip

We give evidence for these effects studying a simple configuration with extrinsic curvature. (In the Appendix we generalise the construction to a generic situation.) 
Consider a region $V$ of an intrinsically flat 3d physical space, in the shape of a square parallelepiped with area $\cal A$ and height $L$. Because of flatness, we can use Euclidean coordinates; adapt these so that the base is in the $(x^1,x^2)$ plane and the hight in the $x^3$ direction. Again for simplicity, say the extrinsic curvature in this region has only the  $k^3{}_3$ component, which we denote $k$, and this is constant. That is: the $V$ embedding of $V$ in 4d spacetime is uniformly curved.  Let us search for a state that approximates this geometry, in the conventional kinematics of loop quantum gravity. 

To this aim, choose a cellular decomposition of $V$. For simplicity, choose a cubic cellular decomposition, as in the left panel of Figure 1. (A triangulation gives the same results, in a more cumbersome manner.)  Decompose  $V$ into $M^2\times N$ small cubes, labelled by an index $n$, and split the extrinsic curvature $k$ uniformly on the faces $\Sigma_n$ of the cubes lying in the $(x^1,x^2)$ planes. That is, set
 
\begin{equation}
           k_{discrete}(x)=\sum_n \ \theta \int_{\Sigma} d^2\sigma\  \delta^3(x,x(\sigma)).   \label{1}
\end{equation}
where $x(\sigma)$ is a parametric description of the face $\Sigma$. 
The value of $\theta$ is fixed by requiring 
\begin{equation}
           \int_V  d^3x\   k(x) =      \int_V  d^3x\   k_{discrete}(x), \label{2}
\end{equation}
which gives immediately $k L {\cal A}  =    N M^2 ({\cal A}/M^2) \theta$, namely 
\begin{equation}
\theta = \frac{k\, L}{N}.    \label{3}
\end{equation}
Which simply means that the angle $\theta$ is the line integral of the extrinsic curvature in the vertical direction, divided by the number of discrete steps. Recall that the extrinsic curvature is in fact nothing else than the derivative of the normal to the surface; the angle $\theta$ is precisely the boost angle between two adjacent discrete flat cubes. See the right panel of Figure 1. 

\begin{figure}[t]
\centering
\includegraphics[height=2.7cm]{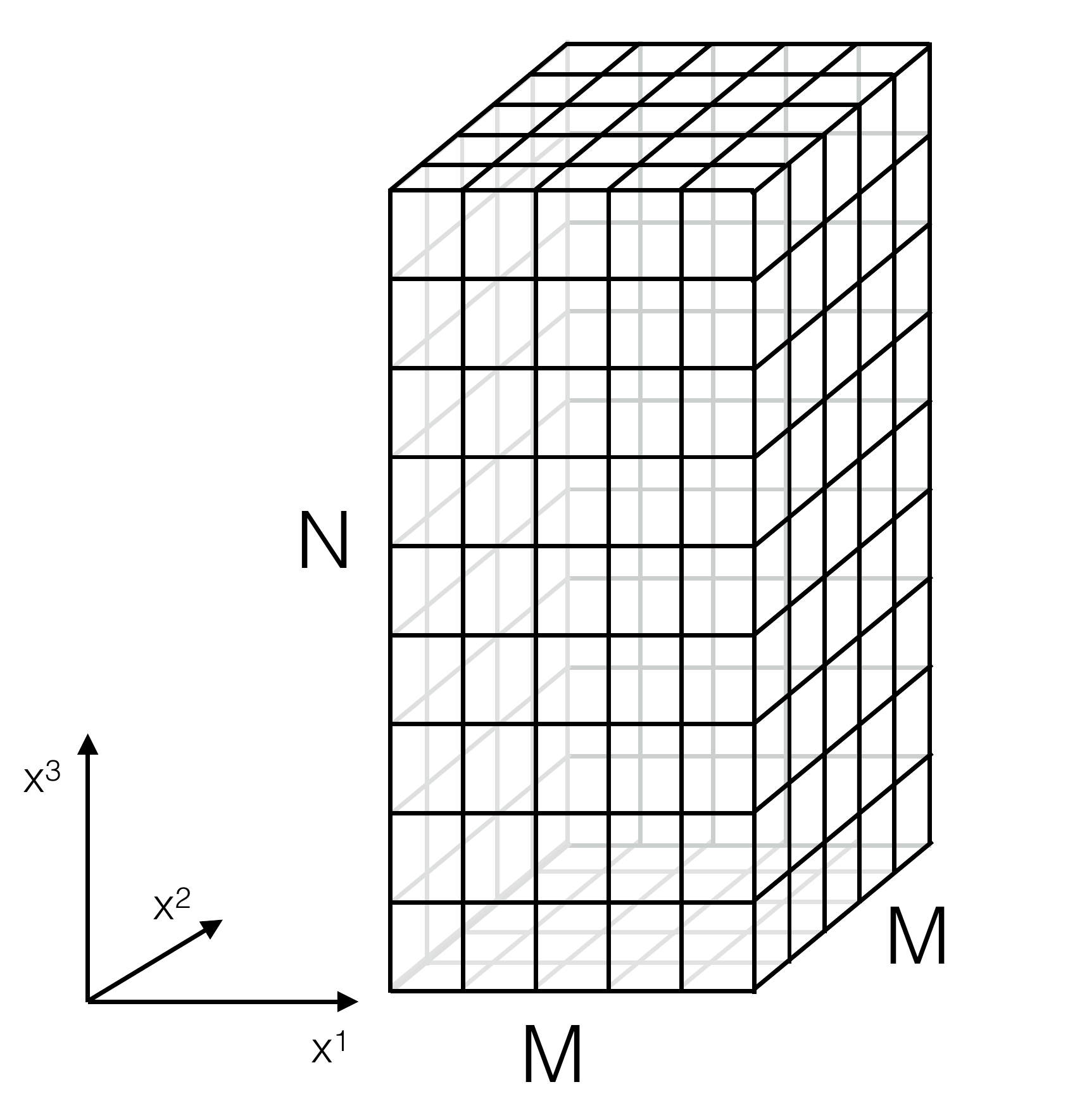}\ \ \  
\includegraphics[height=2.1cm]{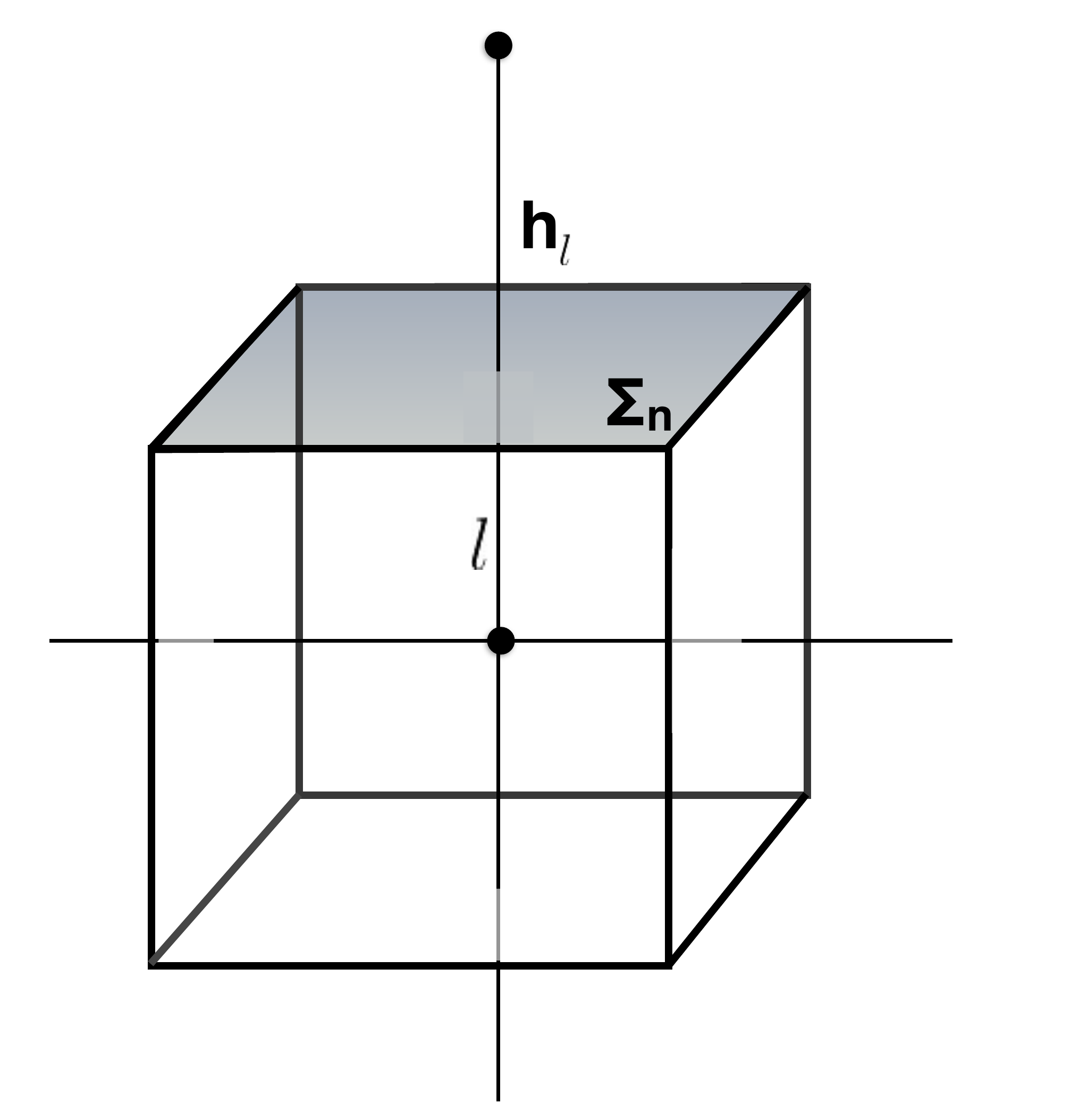}  
\includegraphics[height=2.4cm]{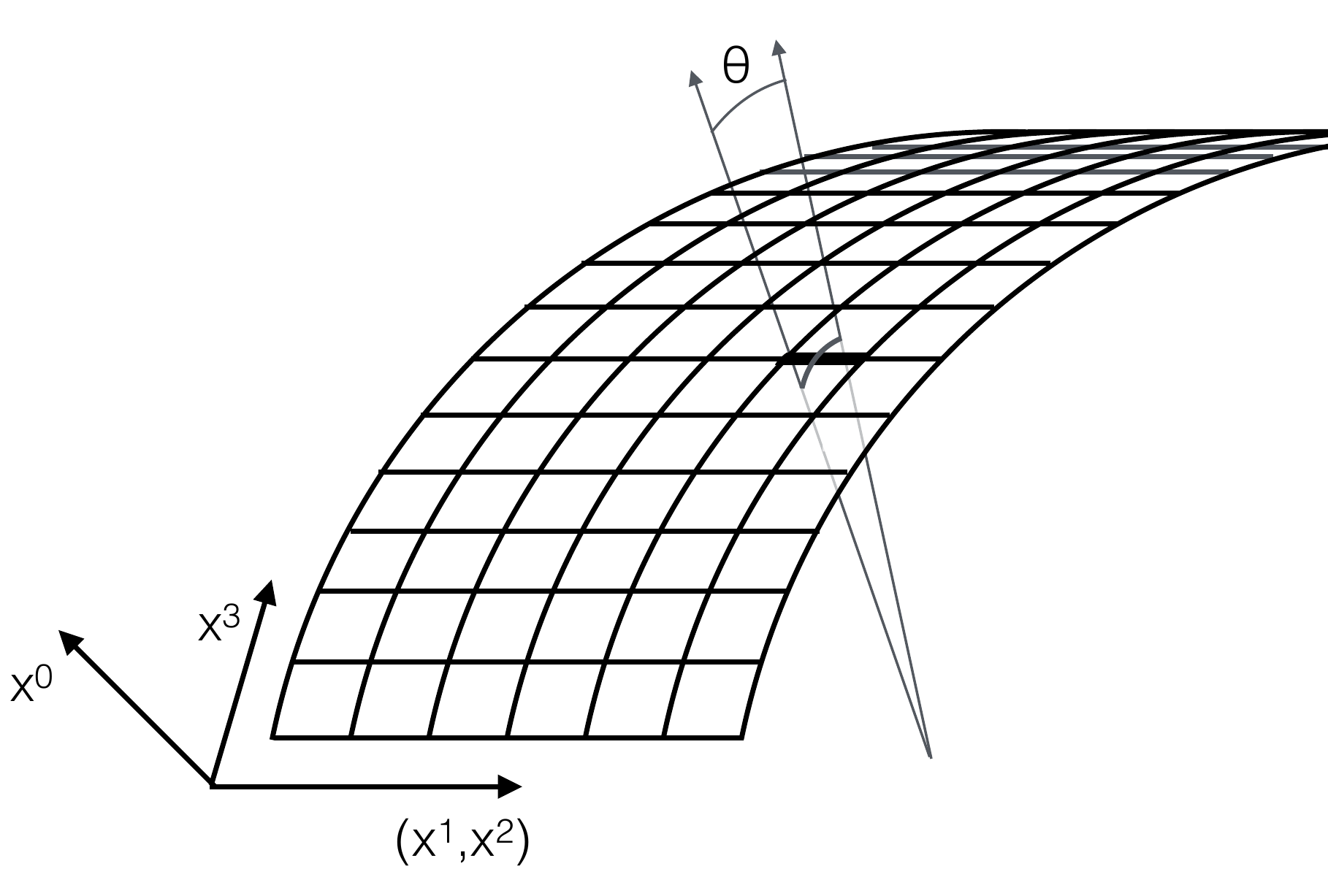}
\vspace{-1em}
\caption[]{Left: The cellular decomposition of $V$. Center: A single cell, with the links of the dual cellular decomposition; the extrinsic curvature is on the grey face and is read-out by the holonomy along the link. Right: The region $V$ immersed in 4d space (one dimension suppressed) and the angle $\theta$ between normals across a face $\Sigma$.}
\label{uno}
\end{figure}

Let us now build a quantum state that approximates the geometry described above. Consider a state defined on the graph dual to the cellular decomposition\footnote{More precisely, 
the graph is (topologically) isomorphic to the 2-skeleton of the 2-complex dual to the cellular decomposition, as is usual in LQG.}. In loop gravity, states can be written in the holonomy representation, as functions of one $SU(2)$ group element $h_\ell$ for each link $\ell$ of the graph. In our discretization, links are dual to the square faces of the small cubes. In particular, for each link in the $x^3$ direction (each square in a $(x^1,x^2)$ plane) we have a group element $h$. See the central panel of Figure1.  The physical interpretation of the group element is the holonomy 
\begin{equation}
    h_\ell=Pe^{\int_\ell A}. 
\end{equation} 
of the Ashtekar-Barbero connection $A=\Gamma+\gamma K$, which in turn  is the sum of two components: the 3d spin connection $\Gamma$ and the extrinsic curvature $K$, scaled by $\gamma$, the Barbero-Immirzi parameter. In the simple case considered here, the triad can be gauge-fixed to coincide with the coordinate directions, i.e. the triad is a Kronecker-delta, $\Gamma$ vanishes and $K= k^a{}_b e^i_a$ is the same matrix as $k^a{}_b$. Then, 
\begin{equation}
    h_\ell=Pe^{i\gamma \int_\ell k}. \label{5}
\end{equation} 
Inserting the value \eqref{1} for the discretized curvature gives, for all the vertical links, 
\begin{equation}
    h_\ell=h=e^{i\gamma \theta     \frac{\sigma_3}2 \int_l dl \int_{\Sigma_n} d^2\sigma\  \delta^3(x,x(\sigma))}= e^{i\gamma \theta     \frac{\sigma_3}2}
\end{equation} 
where the 3d delta function is properly cancelled by the three integrations. Using \eqref{3}, we have 
\begin{equation}
    h=e^{i\frac{ \gamma k\, L}{N} \frac{\sigma_3}2}. \label{7}
\end{equation} 

Now comes the key observation. A coherent state is going to be centered on some value $h$.  Let us choose the value of $h$ that is as far as possible from four dimensional flatness, namely from $h=1\!\!1$. The key point is that since $SU(2)$ is compact, this value \emph{exists}.  It is 
\begin{equation}
h_{max}=-1\!\!1=e^{i 2\pi \frac{\sigma_3}2}. \label{8}
\end{equation}
Setting (using \eqref{7})
\begin{equation}
h_{max}=e^{i\frac{ \gamma k_{max}\, L}{N} \frac{\sigma_3}2}
\end{equation}
we have then immediately from \eqref{8}
\begin{equation}
k_{max}=\frac{2\pi N}{\gamma L}.
\end{equation}
But $L/N$ is the actual \emph{size} of the triangulation cubes. Because of the discreteness of the geometry, this is bound from below by the minimal non-vanishing value of the area which in loop quantum gravity is $a_o=\sqrt3/2\gamma\hbar G$, which gives $(L/N)^2>{\sqrt3/2\gamma\hbar G}$.  With the last equation, we have 
\begin{equation}
k_{max}^2\sim\frac{8\pi^2 }{\sqrt3\gamma^3  \hbar G}.
\end{equation}
This is the main result. There is a maximal value for the extrinsic curvature. 

Let us now come to the maximum boost.  For this, consider an explicit geometrical interpretation to the discretised spacetime (recall that any such interpretation has a large degree of arbitrariness \cite{rovelli_geometry_2010}). The angle $\theta$ is the boost parameter between the Lorentz frames defined by two neighbouring flat discrete cells. See the right panel of Figure 1.  This follows from the fact that the extrinsic curvature can be written in the form $k_{ab}=D_{(a}n_{b)}$ and in locally flat 4d coordinates this can be identified with the derivative of the 4d normal $n$. Its integral gives the angle between the normals, namely the boost parameter. Integrating   \eqref{1} accross the boundary between one couple of cells gives  $\theta$, so accross each link only a \emph{finite} boost, with boost parameter $\theta_{max}=2\pi$ is possible.   Since the relation between the boost parameter and the velocity is 
\begin{equation}
\cosh[\theta]=\frac{1}{\sqrt{1-\frac{v^2}{c^2}}}
\end{equation}
the relative velocity between two adjacent cells cannot exceed the maximal velocity defined by $\theta_{max}$, namely
\begin{equation}
v_{max}=\sqrt{1-\frac{1}{(\cosh{2\pi})^2}}\ c \ \sim 
0.999993 \ c. 
\end{equation}
This is the maximal relative velocity between two space-time quanta. 

This does not imply that arbitrary boosts between two arbitrary cells are impossible, but only that larger boosts imply more quanta, or finer triangulations. 

The maximal boost derived here should not be confused with maximal acceleration. Maximal acceleration has been discussed by many authors, including in the context of loop quantum gravity (see \cite{rovelli_evidence_2013}  and references therein). Boost is dimensionless, while acceleration is dimension-full. The second follows from numerous hand waiving quantum gravity arguments, while the first appear to be characteristic of loop quantum gravity.

\subsection*{Schwarzschild geometry}

The result can be used in the context of Schwarzschild geometry a follows.  Chose equal-time surfaces in Lema\^itre coordinates \cite{lemaitre_expansion_1931}. These are 3d flat and both the metric and the extrinsic curvature are diagonal. In flat polar coordinates the extrinsic curvature is 
\begin{equation}
k_{ab}dx^adx^b= \sqrt{\frac{2m}{r^3}}  dr^2 - \sqrt{8m r} \, d\Omega^2.
\end{equation}
Fix a cellular decomposition formed by Planck size cells bounded by equal coordinate surfaces. Consider a radius $L_P\ll r \ll 2m$, where $L_P\sim\sqrt{\hbar G}$ is the Planck scale. 
At this scale the curvature is constant at the scale of the single cells and therefore we can repeat the derivation of the paper, obtaining a bound on the radial component of the extrinsic curvature:
\begin{equation}
k^{max}_{rr}=\sqrt{\frac{2m}{r_{min}^3}}\sim  \frac1{L_P},
\end{equation} 
which indicates that we should expect quantum effects modifying the classical geometry at the radius
\begin{equation}
r_{min}\sim \sqrt[3]{2m\, L^2_P}.
\end{equation}
This is of course the same radius at which the 4d Riemanninan curvature becomes Planckian. (Because of the Gauss-Codazzi equations, the 4d Riemannian curvature is determined by the 3d curvature and the extrinsic curvature).  Notice that quantum effects are expected at a radius much larger than $r\sim L_P$.





\bibliographystyle{JHEPs}
\bibliography{bib,extra}

\end{document}